\documentclass[rmp,aps,nofootinbib,floats,twocolumn]{revtex4}
\usepackage{amssymb}
\usepackage{amsmath}
\usepackage{graphics}
\usepackage{epsfig}
\usepackage{bm}
\usepackage[dvipdfm]{hyperref}

\begin{document}

\title{The Ginzburg-Landau Theory of Type II superconductors in magnetic
field}
\author{Baruch Rosenstein}
\email{vortexbar@yahoo.com} \affiliation{National Center for
Theoretical Sciences  and  Electrophysics Department, \\
National Chiao Tung University, Hsinchu, Taiwan, R.O.C.}
\altaffiliation[Permanent address ]{}
 \affiliation{Applied Physics
Department, University Center of Samaria, Ariel, Israel.}
\author{Dingping Li}
\email{lidp@phy.pku.edu.cn(correspondence author)}
\affiliation{Department of Physics, Peking University, 100871,
Beijing, China}

\begin{abstract}
Thermodynamics of type II superconductors in electromagnetic field based on
the Ginzburg - Landau theory is presented. The Abrikosov flux lattice
solution is derived using an expansion in a parameter characterizing the
"distance" to the superconductor - normal phase transition line. The
expansion allows a systematic improvement of the solution. The phase diagram
of the vortex matter in magnetic field is determined in detail. In the
presence of significant thermal fluctuations on the mesoscopic scale (for
example in high $T_{c}$ materials) the vortex crystal melts into a vortex
liquid. A quantitative theory of thermal fluctuations using the lowest
Landau level approximation is given. It allows to determine the melting line
and discontinuities at melt, as well as important characteristics of the
vortex liquid state. In the presence of quenched disorder (pinning) the
vortex matter acquires certain "glassy" properties. The irreversibility line
and static properties of the vortex glass state are studied using the
"replica" method. Most of the analytical methods are introduced and
presented in some detail. Various quantitative and qualitative features are
compared to experiments in type II superconductors, although the use of a
rather universal Ginzburg - Landau theory is not restricted to
superconductivity and can be applied with certain adjustments to other
physical systems, for example rotating Bose - Einstein condensate.
\end{abstract}

\date{March, 2009}
\maketitle
\tableofcontents

\section{Introduction}

Phenomenon of superconductivity was initially defined by two basic
properties of classic superconductors (which belong to type I, see below):
zero resistivity and perfect diamagnetism (or Meissner effect). The
phenomenon was explained by the Bose - Einstein condensation (BEC) of pairs
of electrons (Cooper pairs carrying a charge $-e^{\ast }=-2e,$constant $%
e^{\ast }$ considered positive throughout) below a critical temperature $%
T_{c}$. The transition to the superconducting state is described
phenomenologically by a complex order parameter field $\Psi \left( \mathbf{r}%
\right) =\left\vert \Psi \left( \mathbf{r}\right) \right\vert
e^{i\chi \left( \mathbf{r}\right) }$ with $\left\vert \Psi
\right\vert ^{2}$ proportional to the density of Cooper pairs and
its phase $\chi $ describing the BEC coherence. Magnetic and
transport properties of another group of materials, the type II
superconductors, are more complex. An external magnetic field $H$
and even, under certain circumstances, electric field do penetrate
into a type II superconductor. The study of \ type II superconductor
group is importance both for fundamental science and applications.

\subsection{Type II superconductors in magnetic field}

\subsubsection{Abrikosov vortices and some other basic concepts}

Below a certain field, the first critical field $H_{c1}$, the type II
superconductor is still a perfect diamagnet, but in fields just above $%
H_{c1} $ magnetic flux does penetrate the material. It is concentrated in
well separated "vortices" of size $\lambda $, the magnetic penetration
depth, carrying one unit of flux
\begin{equation}
\Phi _{0}\equiv \frac{hc}{e^{\ast }}.  \label{Fi0_I}
\end{equation}%
The superconductivity is destroyed in the core of a smaller width $\xi $
called the coherence length. The type II superconductivity refers to
materials in which the ratio $\kappa =$ $\lambda /\xi $ is larger than$\
\kappa _{c}=1/\sqrt{2}$ \cite{Abrikosov57}. The vortices strongly interact
with each other, forming highly correlated stable configurations like the
vortex lattice, they can vibrate and move. The vortex systems in such
materials became an object of experimental and theoretical study early on.

Discovery of high $T_{c}$ materials focused attention to certain particular
situations and novel phenomena within the vortex matter physics. They are
"strongly" type II superconductors $\kappa \sim 100>>\kappa _{c}$ and are
"strongly fluctuating" due to high $T_{c}$ and large anisotropy in a sense
that thermal fluctuations of the vortex degrees of freedom are not
negligibly, as was the case in "old" superconductors. In strongly type II
superconductors the lower critical field $H_{c1}$ and the higher critical
field $H_{c2}$ at which the material becomes "normal" are well separated $%
H_{c2}/H_{c1}\sim \kappa ^{2}$ leading to a typical situation $%
H_{c1}<<H<H_{c2}$ in which magnetic fields associated with vortices overlap,
the superposition becoming nearly homogeneous, while the order parameter
characterizing superconductivity is still highly inhomogeneous. The vortex
degrees of freedom dominate in many cases the thermodynamic and transport
properties of the superconductors.

Thermal fluctuations significantly modify the properties of the vortex
lattices and might even lead to its melting. A new state, the vortex liquid
is formed. It has distinct physical properties from both the lattice and the
\textquotedblright normal\textquotedblright\ metal. In addition to
interactions and thermal fluctuations, disorder (pinning) is always present,
which may also distort the solid into a viscous, glassy state, so the
physical situation becomes quite complicated leading to rich phase diagram
and dynamics in multiple time scales. A theoretical description of such
systems is a subject of the present review. Two ranges of fields, $H<<H_{c2}$
and $H>>H_{c1},$allow different simplifications and consequently different
theoretical approaches to describe them. For large $\kappa $ there is a
large overlap of their applicability regions.

\subsubsection{Two major approximations: the London and the homogeneous
field Ginzburg - Landau models}

In the fields range $H<<H_{c2}$ vortex cores are well separated and one can
employ a picture of line-like vortices interacting magnetically. In this
approach one ignores the detailed core structure. The value of the order
parameter is assumed to be a constant $\Psi _{0}$ with an exception of thin
lines with phase winding around the lines. Magnetic field is inhomogeneous
and obeys a linearized London equation. This model was developed for low $%
T_{c}$ superconductors and subsequently elaborated to describe the high $%
T_{c}$ materials as well. It was comprehensively described in numerous
reviews and books \cite{Brandt,Blatter,Kopnin,Tinkham}
and will not be covered here.

\begin{figure}[t]
\centering\rotatebox{270}{\includegraphics[width=0.3%
\textwidth,height=8cm]{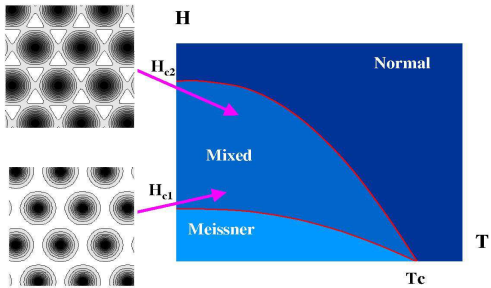}}
\caption{Schematic magnetic phase diagram of a type II superconductor.}
\label{figI3}
\end{figure}

The approach however becomes invalid as fields of order of $H_{c2}$ are
approached, since then the cores cannot be considered as linelike and
profile of the depressed order parameter becomes important. The temperature
dependence of the critical lines is sketched in Fig. \ref{figI3}. The region
in which the London model is inapplicable includes typical situations in
high $T_{c}$ materials as well as in novel "conventional" superconductors.
However precisely under these circumstances different simplifications are
possible. This is a subject of the present review.
When distance between vortices is smaller than $\lambda $ (at fields of
several $H_{c2}$) the magnetic field becomes homogeneous due to overlaps
between vortices. This means that magnetic field can be described by a
number rather then by a field. This is the most important assumption of the
Landau level theory of the vortex matter. One therefore can focus solely on
the order parameter field $\Psi \left( \mathbf{r}\right) $. In addition, in
various physical situation the order parameter $\Psi $ is greatly depressed
compared to its maximal value $\Psi _{0},$ due to various "pair breaking"
effects like temperature, magnetic and electric fields, disorder etc. For
example, in an extreme case of $H\sim H_{c2}$ only small \textquotedblleft
islands\textquotedblright\ between core centers remain superconducting, yet
superconductivity dominates electromagnetic properties of the material. One
therefore can rely on expansion of energy in powers of the order parameter,
a method known as the Ginzburg - Landau (GL) approach, which is briefly
introduced next.

To conclude, while in the London approximation one assumes constant order
parameter and operates with degrees of freedom describing the vortex lines,
in the GL approach the magnetic field is constant and one operates with key
notions like Landau wave functions describing the order parameter.

\subsection{Ginzburg - Landau model and its generalizations}

An important feature of the present treatise is that we discuss a great
variety of complex phenomena using a single well defined model. The
mathematical methods used are also quite similar in various parts of the
review and almost invariably range from perturbation theory to the so called
variational gaussian approximation and its improvements. This consistency
often allows to consider a smooth limit of a more general theory to a
particular case. For example a static phenomenon is obtained as a small
velocity limit of the dynamical one, the clean case is a limit of zero
disorder and the mean field is a limit of small mesoscopic thermal
fluctuations. The model is motivated and defined below, while methods of
solution will be subject of the following sections. The complexity increases
gradually.

\subsubsection{Landau theory near $T_{c}$ for a system undergoing a second
order phase transition}

Near a transition in which the $U\left( 1\right) $ phase symmetry, $\Psi
\rightarrow e^{i\chi }\Psi $, is spontaneously broken a system is
effectively described by the following Ginzburg-Landau free energy \cite%
{Mazenko}:
\begin{eqnarray}
F\left[ \Psi \right] &=&\int d\mathbf{r}dz{\Large \{}\frac{{\hbar }^{2}}{%
2m^{\ast }}\left\vert \mathbf{\nabla }\Psi \right\vert ^{2}+\frac{{\hbar }%
^{2}}{2m_{c}^{\ast }}\left\vert \mathbf{\nabla }\Psi \right\vert ^{2}
\label{GL_F_I} \\
&&+a^{\prime }|\Psi |^{2}+\frac{b^{\prime }}{2}|\Psi |^{4}{\LARGE \}}+F_{n}.
\notag
\end{eqnarray}%
Here $\mathbf{r}=\left( x,y\right) $ and we assumed equal effective masses
in the $x-y$ plane $m_{a}^{\ast }=m_{b}^{\ast }\equiv m^{\ast }$, both
possibly different from the one in the $z$ direction $m_{c}^{\ast }/m^{\ast
}=\gamma _{a}^{2}$. This anticipates application to layered superconductors
for which the anisotropy parameter $\gamma _{a}$ can be very large. The last
term, $F_{n}$, the "normal" free energy, is independent on order parameter,
but might depend on temperature. The GL approach is generally an effective
mesoscopic approach, in which one assumes that microscopic degrees of
freedom are \textquotedblright integrates out\textquotedblright . It is
effective when higher powers of order parameter and gradients, neglected in
eq.(\ref{GL_F_I}) are indeed negligible. Typically, but not always, it
happens near a second order phase transition.

All the terms in eq.(\ref{GL_F_I}) are of order $\left( 1-t\right) ^{2}$,
where $t\equiv T/T_{c},$ while one neglects (as \textquotedblright
irrelevant\textquotedblright ) terms of order $\left( 1-t\right) ^{3}$like $%
|\Psi |^{6}$ and quadratic terms containing higher derivatives. Generally
parameters of the GL model eq.(\ref{GL_F_I}) are functions of temperature,
which can be determined by a microscopic theory or considered
phenomenologically. They take into account thermal fluctuations of the
microscopic degrees of freedom (\textquotedblright integrated
out\textquotedblright\ in the mesoscopic description). Consistently one
expands the coefficients "near", with coefficient $a^{\prime }$ vanishing at
$T_{c}$ as $\left( 1-t\right) $:
\begin{eqnarray}
a^{\prime }(T) &=&T_{c}\left[ \alpha \left( 1-t\right) +\alpha ^{\prime
}\left( 1-t\right) ^{2}+...\right] ,\text{ }  \label{coefficients_I} \\
\text{ \ }b^{\prime }\left( T\right) &=&b^{\prime }+b^{\prime \prime }\left(
1-t\right) +...,  \notag \\
m^{\ast }\left( T\right) &=&m^{\ast }+m^{\ast \prime }\left( 1-t\right) +...
\notag
\end{eqnarray}%
The second and higher terms in each expansion are omitted, since their
contributions are also of order $\left( 1-t\right) ^{3}$ or higher.
Therefore, when temperature deviates significantly from $T_{c}$, one cannot
expect the model to provide a good precision. Minimization of the free
energy, eq.(\ref{GL_F_I}), with respect to $\Psi $ below the transition
temperature determines the value of the order parameter in a homogeneous
superconducting state:

\begin{equation}
|\Psi |^{2}=|\Psi _{0}|^{2}\left( 1-t\right) ,\text{ \ \ \ \ \ \ }|\Psi
_{0}|^{2}=\frac{\alpha T_{c}}{b^{\prime }}.  \label{psi0_I}
\end{equation}%
Substituting this into the last two terms in the square bracket in eq.(\ref%
{GL_F_I}), one estimates them to be of order $\left( 1-t\right) ^{2}$, while
one of the terms dropped, $|\Psi |^{6}$, is indeed of higher order. The
energy of this state is lower than the energy of normal state with $\Psi =0,$
namely, $F_{n}$ by

\begin{equation}
\frac{F_{0}}{vol}=-F_{GL}\left( 1-t\right) ,F_{GL}\ =\frac{b^{\prime }}{2}%
|\Psi _{0}|^{4}  \label{F0_I}
\end{equation}%
is the condensation energy density of the superconductor at zero temperature.

The gradient term determines the scale over which fluctuations are typically
extended in space. Such a length $\xi $, called in the present context the
coherence length, is determined by comparing the first two terms in the free
energy:
\begin{equation}
\mathbf{\nabla }^{2}\Psi \sim \xi ^{-2}\Psi \propto \left( 1-t\right) \Psi
,\xi =\frac{{\hbar }}{\sqrt{2m^{\ast }\alpha T_{c}}}.  \label{ksi_der_I}
\end{equation}%
So typically gradients are of order $\left( 1-t\right) ^{1/2}$, and the
first term in the free energy, eq.(\ref{GL_F_I}) is therefore also of the
order $\left( 1-t\right) ^{2}$. Since the order parameter field describing
the Bose - Einstein condensate of Cooper pair is charged, minimal coupling
principle generally provides an unambiguous procedure to include effects of
electromagnetic fields.

\subsubsection{Minimal coupling to magnetic field.}

Generalization to the case of magnetic field is a straightforward use of the
local gauge invariance principle (or the minimal substitution) of
electromagnetism. The free energy becomes%
\begin{eqnarray}
F\left[ \Psi ,\mathbf{A}\right] &=&\int d\mathbf{r}dz{\LARGE [}\frac{{\hbar }%
^{2}}{2m^{\ast }}\left\vert \mathbf{D}\Psi \right\vert ^{2}+\frac{{\hbar }%
^{2}}{2m_{c}^{\ast }}\left\vert D_{z}\Psi \right\vert ^{2}  \label{F_magn_I}
\\
&&+a^{\prime }|\Psi |^{2}+\frac{b^{\prime }}{2}|\Psi |^{4}{\LARGE ]}+G_{n}%
\left[ \mathbf{A}\right] ,  \notag
\end{eqnarray}%
while the Gibbs energy is

\begin{equation}
G\left[ \Psi ,\mathbf{A}\right] =F\left[ \Psi \right] +\int \frac{\left(
\mathbf{B}-\mathbf{H}\right) ^{2}}{8\pi }.  \label{G_I}
\end{equation}%
Here $\mathbf{B=\nabla \times A}$ and we will assume that "external"
magnetic field (considered homogeneous, see above) is oriented along the
positive $z$ axis, $\mathbf{H=}(0,0,H).$ The covariant derivatives are
defined by
\begin{equation}
\mathbf{D}\equiv \mathbf{\nabla }+i\frac{2\pi }{\Phi _{0}}\mathbf{A}.
\label{covariant_derivative_I}
\end{equation}%
\ The "normal electrons" contribution $G_{n}\left[ \mathbf{A}\right] $ is a
part of free energy independent of the order parameter, but can in principle
depend on external parameters like temperature and fields. Minimization with
respect to $\Psi $ and $\mathbf{A}$\ leads to a set of static GL equations,
the nonlinear Schr\"{o}dinger equation,%
\begin{equation}
\frac{\delta }{\delta \psi ^{\ast }}G=-\frac{{\hbar }^{2}}{2m^{\ast }}%
\mathbf{D}^{2}\Psi -\frac{{\hbar }^{2}}{2m_{c}^{\ast }}D_{z}^{2}\Psi
+a^{\prime }\Psi +b^{\prime }|\Psi |^{2}\Psi =0,  \label{GLeq1_I}
\end{equation}%
and the supercurrent equation,.%
\begin{equation}
c\frac{\delta }{\delta \mathbf{A}}G=\frac{c}{4\pi }\mathbf{\nabla \times B-J}%
_{s}-\mathbf{J}_{n}=0,  \label{GLeq2a_I}
\end{equation}%
where the supercurrent and the "normal" current%
\begin{eqnarray}
\mathbf{J}_{s} &=&\frac{ie^{\ast }{\hbar }}{2m^{\ast }}\left( \Psi ^{\ast }%
\mathbf{D}\Psi -\Psi \mathbf{D}\Psi ^{\ast }\right)  \label{Js_I} \\
&=&\frac{ie^{\ast }{\hbar }}{2m^{\ast }}\left( \Psi ^{\ast }\mathbf{\nabla }%
\Psi -\Psi \mathbf{\nabla }\Psi ^{\ast }\right) -\frac{e^{\ast 2}}{cm^{\ast }%
}\mathbf{A}\left\vert \Psi \right\vert ^{2}  \notag \\
\mathbf{J}_{n} &=&-\frac{\delta }{\delta \mathbf{A}}G_{n}\left[ \mathbf{A}%
\right] .  \notag
\end{eqnarray}%
$\mathbf{J}_{n}$ can be typically represented by the Ohmic conductivity $%
\mathbf{J}_{n}=\sigma _{n}\mathbf{E,}$ and vanishes if the electric field is
absent.

Comparing the second derivative with respect to $\mathbf{A}$ term in eq.(\ref%
{GLeq2a_I}) with the last term in the supercurrent equation eq.(\ref{Js_I}),
one determines the scale of typical variations of the magnetic field inside
superconductor, the magnetic penetration depth:
\begin{equation}
\mathbf{\nabla }^{2}\mathbf{A\mathbf{\sim }}\lambda ^{-2}\left( 1-t\right)
\mathbf{\mathbf{\mathbf{A}}\sim }\frac{4\pi e^{\ast 2}}{c^{2}m^{\ast }}%
\mathbf{A}\left( 1-t\right) \left\vert \Psi _{0}\right\vert ^{2}\text{%
\textbf{.}}  \label{lambda_der_I}
\end{equation}%
This leads to
\begin{equation}
\lambda =\frac{c}{2e^{\ast }}\sqrt{\frac{mb^{\prime }}{\pi \alpha T_{c}}}%
\text{.}  \label{lambda_I}
\end{equation}%
The two scales' ratio defines the GL parameter $\kappa \equiv \lambda /\xi $%
. The second equation shows that supercurrent in turn is small since it is
proportional to $\left\vert \Psi \right\vert ^{2}<\Psi _{0}^{2}$. Therefore
magnetization is much smaller than the field, since it is proportional both
to the supercurrent creating it and to $1/\kappa ^{2}$. Since magnetization
is so small, especially in strongly type II superconductors, inside
superconductor $B\approx H$ and consistently disregard the "supercurrent"
equation eq.(\ref{GLeq2a_I}). Therefore the following vector potential
\begin{equation}
\mathbf{A}=(-By,0,0)\simeq (-Hy,0,0)  \label{vector_potential_I}
\end{equation}%
(Landau gauge) will be use throughout. The validity of this significant
simplification can be then checked \textit{aposteriori. }

The upper critical field will be related in section II to the coherence
length eq.(\ref{ksi_der_I}) by
\begin{equation}
H_{c2}=\frac{\Phi _{0}}{2\pi \xi ^{2}}\text{.}  \label{Hc2_I}
\end{equation}%
The energy density difference between the superconductor and the normal
states $F_{GL}$ in eq.(\ref{GL_F_I}) can therefore be reexpressed as%
\begin{equation}
F_{GL}=\frac{H_{c2}^{2}}{16\pi \kappa ^{2}}.  \label{delta_final_I}
\end{equation}

\subsubsection{Thermal fluctuations}

Thermal fluctuations on the microscopic scale have already been taken into
account by the temperature dependence of the coefficients of the GL free
energy. However in high $T_{c}$ superconductors temperature can be high
enough, so that one cannot neglect additional thermal fluctuations which
occur on the mesoscopic scale. These fluctuations can be described by a
statistical sum:
\begin{equation}
Z=\int \mathcal{D}\Psi \left( \mathbf{r}\right) \mathcal{D}\Psi ^{\ast
}\left( \mathbf{r}\right) \exp \left\{ -\frac{F\left[ \Psi ^{\ast },\Psi %
\right] }{T}\right\} \text{,}  \label{Z_I}
\end{equation}%
where a functional integral is taken over all the configurations of order
parameter. In principle thermal fluctuations of magnetic field should be
also considered, but it turns out that they are unimportant even in high $%
T_{c}$ materials \cite{Halperin74,Dasgupta81,Herbut96,Herbut07,Lobb87} .

Ginzburg parameter, the square of the ratio of $T_{c}$ to the superconductor
energy density times correlation volume,
\begin{equation}
Gi=2\left( \frac{T_{c}}{16\pi F_{GL}\xi ^{2}\xi _{c}}\right) ^{2}=2\left(
\frac{4\pi ^{2}T_{c}\kappa ^{2}\xi \gamma _{a}}{\Phi _{0}^{2}}\right) ^{2}%
\text{,}  \label{Gi_I}
\end{equation}%
generally characterizes the strength of the thermal fluctuations on the
mesoscopic scale \cite{Levanyuk,Ginzburg60,Larkin-Varlamov}
and where $\Phi _{0}\equiv \frac{hc}{e^{\ast }}$. The definition of $Gi$ is
the standard one as in \cite{Blatter} contrary to the previous definition
used early in our papers, for example in \cite{Li02,Li03}. Here $\xi
_{c}=\gamma _{a}^{-1}\xi $ is the coherence length in the field direction.
The Ginzburg parameter is significantly larger in high $T_{c}$
superconductors compared to the low temperature one. While for metals this
dimensionless number is very small (of order $10^{-6}$ or smaller), it
becomes significant for relatively isotropic high $T_{c}$ cuprates like $%
YBCO $ ($10^{-4}$) and even large for very anisotropic cuprate $BSCCO$ (up
to $Gi=0.1-0.5$). Physical reasons behind the enhancement are the small
coherence length, high $T_{c}$ and, in the case of $BSCCO,$ large anisotropy
$\gamma _{a}\sim 150$. Therefore the thermal fluctuations play a much larger
role in these new materials. In the presence of magnetic field the
importance of fluctuations is further enhanced. Strong magnetic field
effectively suppresses long wavelength fluctuations in direction
perpendicular to the field reducing dimensionality of the fluctuations by
two. Under these circumstances fluctuations influence various physical
properties and even lead to new observable qualitative phenomena like the
vortex lattice melting into a vortex liquid far below the mean field phase
transition line.

Several remarkable experiments determined that the vortex lattice melting in
high $T_{c}$ superconductors is first order with magnetization jumps \cite%
{Zeldov95,Nishizaki00,Willemin98,Zeldov05,Zeldov07},
and spikes in specific heat (it was found that in addition to the spike
there is also a jump in specific heat which was measured as well) \cite%
{Schilling96,Schilling97,Nishizaki00,Bouquet01,Lortz06,Lortz07}.
These and other measurements like the resistivity and shear modulus point
towards a need to develop a quantitative theoretical description of thermal
fluctuations in vortex matter \cite{Pastoriza,Liang96,Matl02}
To tackle the difficult problem of melting, the description of both the
solid and the liquid phase should reach the precision level below 1\% since
the internal energy difference between the phases near the transition
temperature is quite small.

\subsubsection{Quenched Disorder.}

In any superconductor there are impurities either present naturally or
systematically produced using the proton or electron irradiation. The
inhomogeneities both on the microscopic and the mesoscopic scale greatly
affect thermodynamic and especially dynamic properties of type II
superconductors in magnetic field. Abrikosov vortices are pinned by
disorder. As a result of pinning the flux flow may be stopped and the
material restores the property of zero resistivity (at least at zero
temperature, otherwise thermal fluctuations might depin the vortices) and
make various quantities like magnetization becomes irreversible.\ \ Disorder
on the mesoscopic scale can be modeled in the framework of the Ginzburg -
Landau approach adding a random component to its coefficients \cite{Larkin70}%
. The random component of the coefficient of the quadratic term $W(r)$ is
called $\delta T$ disorder, since it can be interpreted as a local deviation
of the critical temperature from $T_{c}$. The simplest such a model is the
"white noise" with local variance:
\begin{equation}
a^{\prime }\rightarrow a^{\prime }\left[ 1+W\left( r\right) \right] ;\text{
\ }\overline{W\left( r\right) W\left( r^{\prime }\right) }=n\xi ^{2}\xi
_{c}\delta \left( r-r^{\prime }\right) .  \label{disorder_I}
\end{equation}%
A dimensionless disorder strength $n$, normalized to the coherence volume,
is proportional to the density of the short range point - like pinning
centers and average "strength" of the center. The disorder average of a
static physical quantity $A$, denoted by $"\overline{A}"$\ in this case, is
a gaussian measure $p\left[ W\right] $
\begin{eqnarray}
\overline{A} &=&\int \mathcal{D}W\left( r\right) A\left[ W\right] p\left[ W%
\right] ,  \label{average_I} \\
p\left[ W\right] &=&Ne^{\frac{\int_{r}W\left( r\right) ^{2}}{2n\xi ^{2}\xi
_{c}}},N^{-1}\equiv \int \mathcal{D}W\left( r\right) e^{\frac{%
\int_{r}W\left( r\right) ^{2}}{2n\xi ^{2}\xi _{c}}}.  \notag
\end{eqnarray}

The averaging process and its limitations is the subject of section
IV, where the replica formalism is introduced and used to describe
the transition to the glassy (pinned) states of the vortex matter.
They are characterized by irreversibility of various processes. The
quenched disorder greatly affects dynamics. Disordered vortex matter
is depinned at certain "critical current" $J_{c}$ and the flux flow
ensues. Close to $J_{c}$ the flow proceeds slowly via propagation of
defects (elastic flow) before becoming a fast plastic flow at larger
currents. The I-V curves of the disordered vortex matter therefore
are nonlinear. Disorder creates a variety of "glassy" properties
involving slow relaxation, memory effects etc. Thermal fluctuations
in turn also greatly influence phenomena caused by disorder both in
statics and dynamics. The basic effect is the thermal depinning of
single vortices or domains of the vortex matter. The interrelations
between the interactions, disorder and thermal fluctuations are
however very complex. The same thermal fluctuations can soften the
vortex lattice and actually can also cause better pinning near peak
effect region . Critical current might have a "peak" near the vortex
lattice melting.

\subsection{Complexity of the vortex matter physics}

\begin{figure}[t]
\centering \rotatebox{270}{\includegraphics[width=0.3%
\textwidth,height=8cm]{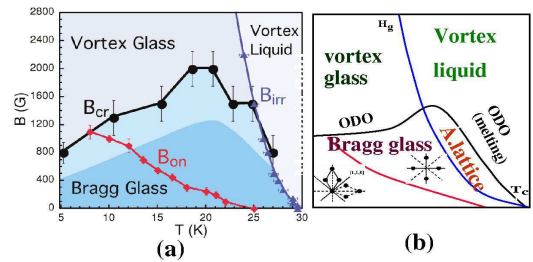}} \caption{Magnetic phase
diagram of high $T_{c}$. (a) Experimentally determined phase diagram
of $LaSCO$\cite{Divakar04}. (b) Theoretical phase diagram advocated
in this article.} \label{figI7}
\end{figure}

In the previous subsection we have already encountered several major
complications pertinent to the vortex physics: interactions, dynamics,
thermal fluctuations and disorder. This leads to a multitude of various
"phases" or states of the vortex matter. It resembles the complexity of
(atomic) condensed matter, but, as we will learn along the way, there are
some profound differences. For example there is no transition between liquid
and gas and therefore no critical point. A typical magnetic ($T-B$) phase
diagram advocated here\cite{Vinokur06} is shown on Fig. \ref{figI7}b. It
resembles for example, 
an experimental phase diagram of high $T_{c}$ superconductor \cite%
{Divakar04,Sasagawa00} $LaSCO$ Fig. \ref{figI7}a.
Here we just mention various phases and transitions between them and direct
the reader to the relevant section in which the theory can be found. Let us
start the tour from the low $T$ and $B$ corner of the phase diagram in
which, as discussed above, vortices form a stable Abrikosov lattice. Vortex
solid might have several crystalline structures very much like an ordinary
atomic solid. In the particular case shown at lower fields the lattice is
rhombic, while at elevated fields in undergoes a structural transformation
into a square lattice (red line on Fig. \ref{figI7}). These transitions are
briefly discussed in section II. Thermal fluctuations can melt the lattice
into a liquid (the "melting" segment of the black line), section III, while
disorder can turn both a crystal and a homogeneous liquid into a "glassy"
state, Bragg glass or vortex glass respectively (section IV). The
corresponding continuous transition line (blue line on Fig.\ref{figI7}) is
often called an irreversibility line since glassiness strongly affects
transport properties leading to irreversibility and memory effects.

To summarize we have several transition lines

1) The first order \cite{Zeldov95,Schilling96,Schilling97,Bouquet01} melting
line due to thermal fluctuations was shown to merge with the "second
magnetization peak" line due to pinning forming the universal order -
disorder phase transition line \cite{Fuchs98,Radzyner02}. At the low
temperatures the location of this line strongly depends on disorder and
generally exhibits a positive slope (termed also the "inverse" melting \cite%
{Paltiel00,Paltiel00b}), while in the "melting" section it is dominated by
thermal fluctuations and has a large negative slope. The resulting maximum
at which the magnetization and the entropy jump vanish is a Kauzmann point
\cite{Li03}. This universal "order - disorder" transition line (ODT), which
appeared first in the strongly layered superconductors ($BSCCO$ \cite%
{Fuchs98}) was extended to the moderately anisotropic superconductors ($%
LaSCCO$ \cite{Radzyner02}) and to the more isotropic ones like $YBCO$ \cite%
{Li03,Pal01,Pal02}. The symmetry characterization of the transition is
clear: spontaneous breaking of both the continuous translation and the
rotation symmetries down to a discrete symmetry group of the lattice.

2) The "irreversibility line" or the "glass" transition (GT) line, which is
a continuous transition \cite{Deligiannis00,Taylor03,Taylor07,Senatore08}.%
The almost vertical in the $T-B$ plane glass line clearly represents effects
of disorder although the thermal fluctuations affect the location of the
transition due to thermal depinning. Experiments in $BSCCO$ \cite%
{Fuchs98,Zeldov05,Zeldov07} 
indicate that the line crosses the ODT line right at its maximum, continues
deep into the ordered (Bragg) phase. This proximity of the glass line to the
Kauzmann point is reasonable since both signal the region of close
competition of the disorder and the thermal fluctuations effects. In more
isotropic materials the data are more confusing. In $LaSCCO$ \cite%
{Divakar04,Sasagawa00} the GT line is closer to the "melting" section of the
ODT line still crossing it. It is more difficult to characterize the nature
of the GT transition as a "symmetry breaking". The common wisdom is that
"replica" symmetry is broken in the glass (either via "steps" or via
"hierarchical" continuous process) as in the most of the spin glasses
theories \cite{Fischer,Dotsenko}. The dynamics in this phase exhibits zero
resistivity (neglecting exponentially small creep) and various irreversible
features due to multitude of metastable states. Critical current at which
the vortex matter starts moving is nonzero. It is different in the
crystalline and homogeneous pinned phases.

3) Sometimes there are one or more structural transitions in the lattice
phase \cite%
{Keimer94,Gilardi02,Johnson99,McKPaul98,Divakar04,Eskildsen01,Sasagawa00,Jaiswal-Nagar06,Li06a}%
. 
They might be either first or second order and also lead to a peak in the
critical current \cite{Chang98,Chang98b,Klironomos03,Park98,Rosenstein99b}.

\subsection{Guide for a reader.}

\subsubsection{Notations and units}

Throughout the article we use two different systems of units. In
sections not dealing with thermal fluctuations, namely in section II
and section IVA we use units which do not depend on "external"
parameters $T$ and $H$, just on material parameters and universal
constants (for example the unit of length is the coherence length
$\xi $). More complicated parts of the review involving thermal
fluctuations utilize units dependent on $T\,\ $and $H$. For example
the unit of length in directions perpendicular to the field
direction becomes magnetic length $l=\xi \sqrt{\frac{H_{c2}}{B}}$.
However throughout the review basic equations and important results,
which might be used for comparison with experiments and other
theories, are also stated in regular physical units.

\paragraph{The mean field units and definitions of dimensionless parameters}

Ginzburg - Landau free energy, eq.(\ref{GL_F_I}), contains three material
parameters $m^{\ast },m_{c}^{\ast }$ (in the $a-b$ directions perpendicular
to the field and in the field direction respectively), $\alpha
T_{c},b^{\prime }$. If in addition the $\delta T_{c}$ disorder, introduced
in eq.(\ref{disorder_I}), is present, it is described by the disorder
strength $n$. These material parameters are usually expressed via physically
more accessible lengths and time units $\xi ,\xi _{c},\lambda $.

\begin{equation}
\ \ \xi _{c}=\frac{{\hbar }}{\sqrt{2m_{c}^{\ast }\alpha T_{c}}}.
\label{t_GL_AppA}
\end{equation}%
Despite the fact that one often uses temperature dependent coherence length
and penetration depth, which as seen in equation eqs.(\ref{ksi_der_I}) and (%
\ref{lambda_der_I}) might be considered as divergent near $T_{c}$, we prefer
to write factors of $\left( 1-t\right) $ explicitly.

From the above scales can form the following dimensionless material
parameters $Gi$ and

\begin{equation}
\kappa =\lambda /\xi ,\text{ }\gamma _{a}^{2}=m_{c}^{\ast }/m^{\ast }.
\label{parameters_AppA}
\end{equation}

From the scales one can form units of magnetic and electric fields, current
density and conductivity:

\begin{equation}
H_{c2}=\frac{\Phi _{0}}{2\pi \xi ^{2}},  \label{J_GL_AppA}
\end{equation}%
as well as energy density $F_{GL}$. These can be used to define
dimensionless parameters, temperature $T,$ magnetic and electric fields $H$,
$E$%
\begin{equation}
t=\frac{T}{T_{c}}.\text{ \ \ }b=\frac{B}{H_{c2}},\text{ \ \ }h=\frac{H}{%
H_{c2}},
\end{equation}%
from which other convenient dimensionless quantity describing the proximity
to the mean field transition line are formed%
\begin{equation}
\text{ }a_{H}=\frac{1-t-b}{2}.  \label{v_AppA}
\end{equation}%
The unit of the order parameter field (or square root of the Cooper pairs
density) is determined by the mean field value\ $|\Psi _{0}|^{2}=\frac{%
\alpha T_{c}}{b^{\prime }}$:%
\begin{equation}
\overline{\Psi }=\frac{\Psi }{\sqrt{2}|\Psi _{0}|}=\left( \frac{b^{\prime }}{%
2\alpha T_{c}}\right) ^{1/2}\Psi .  \label{psi_MFunits_I}
\end{equation}

and the Boltzmann factor and the disorder correlation in the physics units
(length is in unit of $\xi $ in $x-y$ plane and in unit of $\xi _{c}$ along $%
c$ axis, order parameter in unit as defined by the equation above) is
\begin{gather*}
\frac{F\left[ \Psi ^{\ast },\Psi \right] }{T}=\frac{1}{\omega _{t}}\int
d^{3}x\{\frac{1}{2}|D\psi |^{2}+ \\
\frac{1}{2}|\partial _{z}\psi |^{2}-\frac{1-t}{2}\left( 1+W\left( r\right)
\right) |\psi |^{2}+\frac{1}{2}|\psi |^{4}\}. \\
\frac{G\left[ \Psi ,\mathbf{A}\right] }{T}=\frac{F\left[ \Psi ^{\ast },\Psi %
\right] }{T}+\frac{1}{\omega _{t}}\int d^{3}x\frac{(\mathbf{b-h)}^{2}}{4} \\
\overline{W\left( r\right) W\left( r^{\prime }\right) }=n\delta \left(
r-r^{\prime }\right) ,\omega _{t}=\sqrt{2Gi}\pi t.
\end{gather*}

\paragraph{The LLL scaled units}

When dealing with thermal fluctuations, the following units depend on
parameters $T,$ $H$ and $E$. Unit of length in directions perpendicular to
the field can be conveniently chosen to be the magnetic length,
\begin{equation}
l=\xi \sqrt{\frac{H_{c2}}{B}},  \label{magn_length_I}
\end{equation}%
in the field direction, while in the field direction it is different:
\begin{equation}
\xi _{c}\left( \frac{\sqrt{Gi}tb}{4}\right) ^{-1/3}.
\label{z_direction_unit_I}
\end{equation}%
Motivation for these fractional powers of both temperature and magnetic
field will become clear in section III. we rescale the order parameter to $%
\psi $ by an additional factor:
\begin{equation}
\Psi =\Psi _{0}\left( \frac{\sqrt{Gi}tb}{4}\right) ^{1/3}\psi .
\label{psi_LLLunits_I}
\end{equation}%
Instead of $a_{H}$ or $a_{H,E}$ it will be useful to use "Thouless LLL
scaled temperature":\cite{Thouless75,Ruggeri76,Ruggeri78}

\begin{equation}
a_{T}=-\frac{a_{H}}{\left( \frac{\sqrt{Gi}tb}{4}\right) ^{\frac{2}{3}}}=-%
\frac{1-t-b}{2\left( \frac{\sqrt{Gi}tb}{4}\right) ^{2/3}}.  \label{aT_I}
\end{equation}

The scaled energy is defined by
\begin{equation}
\mathcal{F}=\frac{H_{c2}^{2}}{2\pi \kappa ^{2}}\left( \frac{\sqrt{Gi}tb}{4}%
\right) ^{4/3}\emph{f}\left( a_{T}\right) ,  \label{F_scaled_I}
\end{equation}%
and magnetization by%
\begin{eqnarray}
\frac{M}{H_{c2}} &=&\frac{1}{4\pi \kappa ^{2}}\left( \frac{\sqrt{Gi}tb}{4}%
\right) ^{2/3}m\left( a_{T}\right) ,  \label{magnetization_scaled_I} \\
m\left( a_{T}\right) &=&-\frac{d}{da_{T}}f\left( a_{T}\right) .  \notag
\end{eqnarray}%
The disorder is characterized by the ration of the strength of pinning to
that of thermal fluctuations%
\begin{equation}
r=\frac{\left( 1-t\right) ^{2}}{\pi Gi^{1/2}t}n\text{.}
\label{r_parameter_I}
\end{equation}

\bigskip

\subsubsection{Analytical methods described in this article}

Discussion of properties of the GL model in magnetic fields utilizes a
number of general and special theoretical techniques. We chose to describe
some of them in detail, while others are just mentioned in the last section.
We do not describe numerous results obtained using the elasticity theory or
numerical methods like Monte Carlo and molecular dynamics simulations,
although comparison with both is made, when possible.

The techniques and\ special topics include:

1) Translation symmetries in gauge theories (electro - magnetic
translations) in IIA. Their representations, the quasi - momentum basis
(IIIB) is used throughout to discuss excitations of vortex matter either
thermal or elastic.

2) Perturbation theory around a bifurcation point of a nonlinear PDE
(differential equations containing partial derivatives). This is very
different from the perturbation theory used in linear systems, for example
in quantum mechanics

3) Variational gaussian approximation to field theory \cite{Kleinert} is
widely used in III to IV. It is defined in IIIC in the path integral form
and subsequently shown to be the leading order of a convergent series of
approximants, the so called optimized perturbation series (OPE). The next to
leading order, the post gaussian approximation, is related to the Cornwall
-Jackiw -Tomboulis method is sometimes used, while higher approximants are
difficult to calculate and are obtained to date for the vortex liquid only.

4) Ordinary perturbation theory in field theory is developed in the
beginning of every section with enough details to follow. Spatial attention
is paid to infrared (IR) and sometimes ultraviolet (UV) divergencies. We
generally do not use the renormalization group (RG) resummation, except in
subsection IIID, where it is presented in a form of Borel - Pade
approximants.

5) Replica method to treat quenched disorder is introduced in IVB and used
to describe the static and the thermodynamic properties of pinned vortex
matter. Most of the presentation is devoted to the replica symmetric case,
while more general hierarchial matrices are introduced in IVD following
Parisi's approach \cite{Parisi80,Mezard91}.

Some technical details are contained in Appendix. We compare with available
experiments on type II superconductors in magnetic field, while application
or adaptation of the results to other fields in which the model can be
useful (mentioned in summary) are not attempted..

\subsubsection{Results}

All the important results (in both regular physical units and the special
units described above) are provided in a form of Mathematica file, which can
be found on our web site.

\section{Mean field theory of the Abrikosov lattice}

In this section we construct, following Abrikosov original ideas \cite%
{Abrikosov57}, a vortex lattice solution of the static GL equations eq.(\ref%
{GLeq1_I}) "near" the $H_{c2}\left( T\right) $ line. In a region of the
magnetic phase diagram in which the order parameter is significantly reduced
from its maximal value $\Psi _{0}$, eq.(\ref{psi0_I}), one does not really
see well separated "vortices" since, as explained in the previous section,
their magnetic fields strongly overlap. Very close to $H_{c2}\left( T\right)
$ even cores approach each other and consequently the order parameter is
greatly reduced. Only small \textquotedblleft islands\textquotedblright\
between the core centers remain superconducting. Despite this
superconductivity dominates electromagnetic, transport and sometimes
thermodynamic properties of the material. One still has a well defined
"centers" of cores: zeroes of the order parameter. They still repel each
other and thereby organize themselves into an ordered periodic lattice.

To see this we first employ a heuristic Abrikosov's argument, based on
linearization of the GL equations and then develop a systematic perturbative
scheme with a small parameter - the "distance" from the $H_{c2}\left(
T\right) $ line on the $T-H$ plane. The heuristic argument naturally leads
to the lowest Landau level (LLL) approximation, widely used later to
describe various properties of the vortex matter. The systematic expansion
allows to ascertain how close one should stay from the $H_{c2}$ line in
order to use the LLL approximation. Having established the lattice solution,
spectrum of excitations around it (the flux waves or phonon) are obtained in
the next subsection. This in turn determines elastic, thermal and transport
properties of vortex matter.

\subsection{Solution of the static GL equations. Heuristic solution near $%
H_{c2} $}

\subsubsection{Symmetries, units and expansion in $\protect\kappa ^{-2}$}

\emph{\textbf{Broken and unbroken symmetries}}

Generally, before developing (sometimes quite elaborate) mathematical tools
to analyze a complicated model described by free energy eq.(\ref{GL_F_I})
and its generalizations, it is important to make full use of various
symmetries of the problem. The free energy (including the external magnetic
field) is invariant under both the three dimensional translations and
rotations in the $x-y$ ($a-b$) plane. However some of the symmetries in the $%
x-y$ plane are broken spontaneously below the $H_{c2}(T)$ line. The symmetry
which remains unbroken is the continuous translation along the magnetic
field direction $z$. As a result the configuration of the order parameter is
homogeneous in the $z$ direction $\Psi \left( \mathbf{r},z\right) =\Psi
\left( \mathbf{r}\right) $, $\mathbf{r\equiv }\left( x,y\right) $. Hence the
gradient term can be disregarded and the problem becomes two dimensional
(here we consider the mean field equations only, when thermal fluctuations
or point - like disorder are present the simplification is no longer valid.).

\emph{\textbf{Units, free energy and GL equations}}

To describe the physics near $H_{c2}(T),$ it is reasonable to use the
coherence length $\xi ={\hbar }/\sqrt{2m^{\ast }\alpha T_{c}}$ as a unit of
length (assuming for simplicity $m_{a}^{\ast }=m_{b}^{\ast }\equiv m^{\ast }$%
) and the value of the field $\Psi _{0}$ at which the "potential" part is
minimal, eq.(\ref{psi0_I}), (times $\sqrt{2}$) will be used as a scale of
the order parameter field
\begin{equation}
\overline{x}=x/\xi ,\ \overline{y}=y/\xi ;\ \overline{\Psi }=\left( \frac{%
b^{\prime }}{2\alpha T_{c}}\right) ^{1/2}\Psi ,  \label{psi_units_II}
\end{equation}%
while the (zero temperature energy) density difference between the normal
and the superconductor states $F_{GL}$ of eq.(\ref{delta_final_I})
determines a unit of energy density. Therefore dimensionless 2D energy $%
\overline{F}\equiv \frac{F}{8L_{z}\xi ^{2}F_{GL}}$, where $L_{z}$ is the
sample's size in the field direction, and eq.(\ref{G_I}) takes a form:

\begin{equation}
\overline{F}=\int d\overline{x}d\overline{y}\left[ \overline{\Psi }^{\ast }%
\widehat{H}\overline{\Psi }-a_{H}|\overline{\Psi }|^{2}+\frac{1}{2}|%
\overline{\Psi }|^{4}+\frac{\kappa ^{2}\left( \mathbf{b}-\mathbf{h}\right)
^{2}}{4}\right] \text{.}  \label{F_II}
\end{equation}%
Dimensionless temperature and magnetic fields are $t\equiv T/T_{c},$ $%
b\equiv B/H_{c2},\,h\equiv H/H_{c2}$ and $\kappa \equiv \lambda /\xi $. The
units of temperature and magnetic field are therefore $T_{c}$ and $%
H_{c2}\equiv \frac{\Phi _{0}}{2\pi \xi ^{2}}$.

The linear operator$\ \widehat{H}$ is defined as

\begin{equation}
\widehat{H}=-\frac{1}{2}\left( D_{x}^{2}+\partial _{y}^{2}+b\right) \text{.}
\label{H_II}
\end{equation}%
It coincides with the quantum mechanical operator of a charged particle in a
constant magnetic field. The covariant derivative (with all the bars omitted
from now on) is $D_{x}=\partial _{x}-iby$ and the constant is defined as%
\begin{equation}
a_{H}=\frac{1-t-b}{2}\text{.}  \label{ah_II}
\end{equation}%
It is positive in the superconducting phase and vanishes on the $%
H_{c2}\left( T\right) $ line, as will be shown in the next subsection. The
reason why $\widehat{H}$ is "shifted" by a constant $-b/2$ compared to a
standard Hamiltonian of a particle in magnetic field will become clear
there. In rescaled units the GL equation takes a form:

\begin{equation}
\widehat{H}\overline{\Psi }-a_{H}\overline{\Psi }+|\overline{\Psi }|^{2}%
\overline{\Psi }=0\text{.}  \label{GLeq_II}
\end{equation}%
The equation for magnetic field takes a form%
\begin{equation}
\kappa ^{2}\varepsilon _{ij}\partial _{j}b=\frac{i}{2}\overline{\Psi }^{\ast
}D_{i}\overline{\Psi }+c.c.  \label{GLeq2_II}
\end{equation}%
with boundary condition involving the external field $h$.

\emph{\textbf{Expansion in powers of $\kappa ^{-2}$}}

In physically important cases one is encounters strongly type II
superconductors for which $\kappa >>1$. For example all the high $T_{c}$
cuprates have $\kappa $ of order $100$, and even low $T_{c}$ superconductors
which are useful in applications have $\kappa $ of order $10$. In such cases
it is reasonable to expand the second equation in powers of $\kappa ^{-2}$:

\begin{eqnarray}
b &=&h+\kappa ^{-2}b^{\left( 1\right) }+...;  \label{kappa_II} \\
\overline{\Psi } &=&\overline{\Psi }^{\left( 0\right) }+\kappa ^{-2}%
\overline{\Psi }^{\left( 1\right) }+...\text{ .}  \notag
\end{eqnarray}%
It can be seen from eqs.(\ref{GLeq_II}) and (\ref{GLeq2_II}) that to leading
order in $\kappa ^{-2}$ magnetic field $b$ is equal to the external field $h$
considered constant. Therefore one can ignore eq.(\ref{GLeq2_II}) and use
external field in the first equation. Corrections will be calculated
consistently. For example magnetization will appear in the next to leading
order.

From now on we drop bars over $\Psi $ and consider the leading order in $%
\kappa ^{-2}$. Even this nonlinear differential equation is still quite
complicated. It has an obvious normal metal solution $\Psi =0$, but might
have also a nontrivial one. A simplistic way to find the nontrivial one is
to linearize the equation. Indeed naively the nonlinear term contains the
"small" fields $\Psi $ compared to one in the linear term. This assumption
is problematic since, for example the coefficient of the $\Psi $ term is
also small, but will follow this reasoning nevertheless leaving a rigorous
justification to subsection B.

\subsubsection{Linearization of the GL equations near $H_{c2}$\textit{.}}

Naively dropping the nonlinear term in eq.(\ref{GLeq_II}), one is
left with the usual linear Schr\"{o}edinger eigenvalue equation of
quantum mechanics for a
charged particle in the homogeneous magnetic field%
\begin{equation}
\widehat{H}\Psi =a_{H}\Psi .  \label{linearizedGL_II}
\end{equation}%
The Landau gauge that we use, defined in eq.(\ref{vector_potential_I}),
still maintains a manifest translation symmetry along the $x$ direction,
while the $y$ translation invariance is \textquotedblleft
masked\textquotedblright\ by this choice of gauge. Therefore one can
disentangle the variables:
\begin{equation}
\Psi (x,y)=e^{ik_{x}x}f\left( y\right) ,  \label{wavef_decomposition_II}
\end{equation}%
resulting in the shifted harmonic oscillator equation:%
\begin{equation}
\left[ -\frac{1}{2}\partial _{y}^{2}+\frac{b^{2}}{2}\left( y-Y\right) ^{2}%
\right] f=\frac{1-t}{2}f,  \label{wave_eq_II}
\end{equation}%
where $Y\equiv k_{x}/b$ is the $y$ coordinate of the center of the classical
Larmor orbital. For a finite sample $k_{x}$ is discretized in units of $%
\frac{2\pi \xi }{L_{x}}$, while the Larmor orbital center is confined inside
the sample $-L_{y}/2<Y\xi <L_{y}/2$ leading to $\frac{BL_{x}L_{y}}{\Phi _{0}}%
\equiv N_{L}$ values of $k_{x}$.\

Nontrivial $f\left( y\right) \neq 0$ solutions of the \textit{linearized}
equation exist only for special values of magnetic field, since the operator
$\widehat{H}$ has a discrete spectrum
\begin{equation}
E_{N}=Nb  \label{EN_II}
\end{equation}%
for any $Y$ (the Landau levels are therefore $N_{L}$ times degenerate).
These fields $b_{N}$ satisfy%
\begin{equation}
\frac{1-t}{2}=\left( N+\frac{1}{2}\right) b_{N}.  \label{bN_II}
\end{equation}%
and the eigenfunctions are:

\begin{eqnarray}
\phi _{Nk_{x}}(\mathbf{r}) &=&\pi ^{-1/4}\sqrt{\frac{b}{2^{N}N!}}H_{N}\left[
b^{1/2}\left( y-k_{x}/b\right) \right]  \label{HLL_functions_II} \\
&&\times e^{ik_{x}x-\frac{b}{2}\left( y-k_{x}/b\right) ^{2}}\text{,}  \notag
\end{eqnarray}%
where $H_{N}\left( x\right) $ are Hermit polynomials. As we will see
shortly, the \textit{nonlinear} GL equation eq.(\ref{GLeq_II}) acquires a
nontrivial solution also at fields different from $b_{N}$. The solution with
$N=0$ (the lowest Landau level or LLL, corresponding to the highest $b_{N}=1$%
) appears at the bifurcation point
\begin{equation}
1-t-b_{0}\left( t\right) =0  \label{bifurcation_II}
\end{equation}%
or $a_{H}=0$. It defines the $H_{c2}\left( T\right) =H_{c2}\left(
1-T/T_{c}\right) $ line.

For yet higher fields the only solution of nonlinear GL equations is the
trivial one: $\Psi =0$. This is seen as follows. The operator $\widehat{H}$
is positive definite, as its spectrum eq.(\ref{EN_II}) demonstrates.
Therefore for $a_{H}<0$ all three terms in the free energy eq.$\,$(\ref{F_II}%
) are non - negative and in this case the minimum is indeed achieved by $%
\Psi =0$. For $a_{H}>0$ the minimum of the nonlinear equations should not be
very different from a solution of the linearized equation at $B=H_{c2}\left(
T\right) $.

Since the LLL, $B=H_{c2}\left( T\right) $, solutions

\begin{equation}
\phi _{k_{x}}(\mathbf{r})=\frac{b^{1/2}}{\pi ^{1/4}}e^{ik_{x}x-\frac{b}{2}%
\left( y-k_{x}/b\right) ^{2}}\text{,}  \label{LLLfunctions_II}
\end{equation}%
are degenerate, it is reasonable to try the most general LLL function%
\begin{equation}
\Psi \left( \mathbf{r}\right) =\sum_{k_{x}}C_{k_{x}}\phi _{k_{x}}(\mathbf{r})
\label{varlattice_II}
\end{equation}%
as an approximation for a solution of the \textit{nonlinear} GL equation
just below $H_{c2}\left( T\right) $. However how should one chose the
correct linear combination? Perhaps the one with the lowest nonlinear
energy: the quartic term in energy eq.(\ref{F_II}) will lift the degeneracy.
Unfortunately the number of the variational parameters in eq.(\ref%
{varlattice_II}) is clearly unmanageable. To narrow possible choices of the
coefficients, one has to utilize all the symmetries of the lattice solution.
Therefore we digress to discuss symmetries in the presence of magnetic
field, the magnetic translations, returning later to the Abrikosov solution
equipped with minimal group theoretical tools.

\subsubsection{Digression: translation symmetries in gauge theories}

\emph{\textbf{Translation symmetries in gauge theories}}

\begin{figure}[t]
\centering \rotatebox{270}{\includegraphics[width=0.15%
\textwidth,height=4cm]{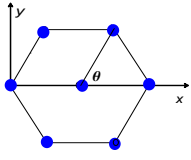}}
\caption{Symmetry of the vortex lattice. Unit cell. }
\label{figII2}
\end{figure}

Let us consider a solution of the GL equations invariant under two arbitrary
translations vectors. Without loss of generality one of them $\mathbf{d}_{1}$%
can be aligned with the $x$ axis. Its length will be denoted by $d$. The
second is determined by two parameters:
\begin{equation}
\mathbf{d}_{1}=d\left( 1,0\right) ;\text{ \ }\mathbf{d}_{2}=d\left( \rho
,\rho ^{\prime }\right) \text{.}  \label{translation_def_II}
\end{equation}%
We consider only rhombic lattices (sufficient for most applications), which
are obtained for $\rho =1/2$. The angle $\theta $ between $\mathbf{d}_{1}$%
and $\mathbf{d}_{2}$ is shown on Fig. \ref{figII2}. Flux quantization
(assuming one unit of flux per unit cell) will be
\begin{equation}
d^{2}\rho ^{\prime }b=2\pi ;\rho ^{\prime }=\frac{1}{2}\tan \theta .
\label{flux_quantization_II}
\end{equation}
Generally an arbitrary translation in the $x$ direction in the particular
gauge that we have chosen, eq.(\ref{vector_potential_I}), is very simple
\begin{equation}
T_{\mathbf{d}_{1}}\Psi \left( x,y\right) =\Psi \left( x+d,y\right) =e^{i%
\widehat{p}_{x}d}\Psi \left( x,y\right) ,
\end{equation}%
where $\widehat{\mathbf{p}}\equiv -i\mathbf{\nabla }$ is the "momentum"
operator. Periodicity of the order parameter in the $x$ direction with
lattice constant $d$ (in units of $\xi $, as usual) means that the wave
vector $k_{x}$ in eq.(\ref{LLLfunctions_II}) is quantized in units of $\frac{%
2\pi }{d}$: $k_{x}=\frac{2\pi }{d}n,$ $\ n=0,\pm 1,\pm 2,...$and the
variational problem of eq.(\ref{varlattice_II}) simplifies considerably:%
\begin{eqnarray}
\Psi \left( \mathbf{r}\right) &=&\sum_{n}C_{n}\phi _{n}(\mathbf{r})
\label{varlat_discrete_II} \\
\phi _{n}(\mathbf{r}) &=&\frac{b^{1/2}}{\pi ^{1/4}}e^{i\frac{2\pi }{d}nx-%
\frac{b}{2}(y-\frac{2\pi }{d}\frac{1}{b}n)^{2}}.  \notag
\end{eqnarray}%
Periodicity with lattice vector $\mathbf{d}_{2}$ is only possible only when
absolute values of coefficients $\left\vert C_{n}\right\vert $ are the same
and, in addition, their phases are periodic in $n$.

\emph{\textbf{Hexagonal lattice}}

In this case the basic lattice vectors are $\mathbf{d}_{1}=d_{\triangle
}\left( 1,0\right) $, $\mathbf{d}_{2}=d_{\triangle }\left( 1/2,\sqrt{3}%
/2\right) ,$ see Fig. \ref{figII2}, $\theta =60^{o}$. As a next simplest
guess to construct a lattice configuration out of Landau harmonics one can
try a two parameter Ansatz $C_{n+2}=C_{n}$:
\begin{eqnarray}
\Psi \left( x,y\right) &=&C_{0}\sum\limits_{n\text{ even}}\pi
^{-1/4}b^{-1/2}e^{i\frac{2\pi }{d_{\triangle }}nx-\frac{b}{2}(y-\frac{2\pi }{%
d_{\triangle }}\frac{1}{b}n)^{2}}  \label{two_parameter_II} \\
&&+C_{1}\sum\limits_{n\text{ odd}}\pi ^{-1/4}b^{-1/2}e^{i\frac{2\pi }{%
d_{\triangle }}nx-\frac{b}{2}(y-\frac{2\pi }{d_{\triangle }}\frac{1}{b}%
n)^{2}}.  \notag
\end{eqnarray}%
For the hexagonal (also called sometimes triangular) FLL $C_{1}=iC_{0}=iC$.
Geometry and the flux quantization gives us now $\xi ^{2}d_{\triangle }^{2}=%
\frac{2\Phi _{0}}{\sqrt{3}B},$ which becomes (in rescaled units of $\xi $)%
\begin{equation}
d_{\triangle }^{2}=\frac{2\pi }{b}\frac{2}{\sqrt{3}}.
\label{hexagonal_spacing_II}
\end{equation}%
We are therefore left again with just one variational parameter%
\begin{eqnarray}
\varphi _{\triangle }\left( x,y\right) &=&\frac{C}{b^{\frac{1}{2}}\pi ^{%
\frac{1}{4}}}{\LARGE \{}\sum\limits_{n\text{ odd}}e^{i\frac{2\pi }{%
d_{\triangle }}nx-\frac{b}{2}(y-\frac{\sqrt{3}d_{\triangle }}{2}n)^{2}}
\label{psi_hex_II} \\
&&+i\sum\limits_{n\text{ even}}e^{i\frac{2\pi }{d_{\triangle }}nx-\frac{b}{2}%
(y-\frac{\sqrt{3}d_{\triangle }}{2}n)^{2}}{\Large \}}  \notag
\end{eqnarray}

Naive nonmagnetic translation in the "diagonal" direction, see Fig. \ref%
{figII2}, now gives
\begin{equation}
\varphi _{\triangle }\left( x+\frac{d_{\triangle }}{2},y+\frac{\sqrt{3}%
d_{\triangle }}{2}\right) =ie^{i\frac{2\pi }{d_{\triangle }}x}\varphi
_{\triangle }\left( x,y\right)  \label{hex_trans1_II}
\end{equation}%
This is again a \textquotedblleft regauging\textquotedblright ,\ which
generally accompanies a symmetry transformation. The "magnetic translation"
now will be

\begin{equation}
T_{\mathbf{d}_{2}}=e^{-i(\frac{2\pi }{d_{\triangle }}x+\frac{\pi }{2})}e^{i(%
\frac{d_{\triangle }}{2}p_{x}+\frac{\sqrt{3}d_{\triangle }}{2}p_{y})}.
\label{magn_trans_hex_II}
\end{equation}%
The normalization is
\begin{equation}
\frac{1}{vol}\int_{cell}\left\vert \varphi _{\Diamond }\left( x,y\right)
\right\vert ^{2}=1.  \label{norm_II}
\end{equation}
which gives: $\left\vert c\right\vert ^{2}=3^{1/4}\pi ^{1/2}b$. Combining
the even and the odd parts, the normalized function also can be written in a
form%
\begin{eqnarray}
\varphi _{\triangle }\left( \mathbf{r}\right) &=&\varphi \left( b^{1/2}%
\mathbf{r}\right) ;  \label{fi_final_II} \\
\varphi \left( \mathbf{r}\right) &=&3^{1/8}\sum\limits_{l=-\infty }^{\infty
}e^{i\left[ \frac{\pi }{2}l^{2}+3^{1/4}\pi ^{1/2}lx\right] -\frac{1}{2}%
(y-3^{1/4}\pi ^{1/2}l)^{2}}.  \notag
\end{eqnarray}%
This form will be used extensively in the following sections.

\emph{\textbf{General rhombic lattice}}

All the rhombic lattices with magnetic field $b$ are obtained from the
Ansatz $C_{n+2}=C_{n}$ by assuming the phase $C_{1}=iC_{0}$:

\begin{equation}
\varphi (x,y,b)\equiv \sqrt{\frac{2}{d_{\theta }}\frac{\sqrt{\pi }}{\sqrt{b}}%
}\sum\limits_{l=-\infty }^{\infty }e^{i(\frac{2\pi }{d_{\theta }}xl+\frac{%
\pi }{2}l^{2})-\frac{b}{2}(y-\frac{2\pi }{d_{\theta }b}l)^{2}}\text{.}
\end{equation}%
The hexagonal lattice corresponds to $\theta =60^{o}$, see Fig. \ref{figII2}%
. One can check that a rhombic lattice indeed is invariant under magnetic
translations by $\mathbf{d}_{1}$ and $\mathbf{d}_{2}$. The flux quantization
takes a form
\begin{equation}
\frac{1}{2}d_{\theta }^{2}\tan \theta =\frac{2\pi }{b}\text{.}
\label{rhombic_a_II}
\end{equation}%
One notices $d_{\theta }=d_{\theta }(b)=d_{\theta }(1)/\sqrt{b}$,and that
generally we have a following relation,
\begin{equation}
\varphi (x,y,b)\equiv \varphi (b^{1/2}x,b^{1/2}y)
\end{equation}%
where the right side equation $\varphi (x,y)$ is the solution in the case of
$b=1$ and we replace $x,y$ by $\sqrt{b}x,\sqrt{b}y$. There are of course
infinitely many invariant functions differing by a "fractional" translation
as well as by rotation of the lattice. These symmetries are "broken
spontaneously" by the lattice. According to Goldstone theorem, they lead to
existence of soft phonon modes in the crystalline phase and will be studied
in section III.

\emph{\textbf{General magnetic translations and their algebra}}

Let us generalize the discussion by considering an arbitrary Landau gauge.
Using the experience with regauging of the two nontrivial translations in
our gauge, which generally is defined a matrix

\begin{equation}
\mathcal{B}=%
\begin{tabular}{|l|l|}
\hline
$0$ & $b$ \\ \hline
$0$ & $0$ \\ \hline
\end{tabular}%
,\text{ \ \ }A_{i}=\mathcal{B}_{ij}r_{j}.  \label{gauge_matrix_II}
\end{equation}%
Magnetic translation operator for a general vector $\mathbf{d}$ should be
defined as

\begin{equation}
T_{\mathbf{d}}=e^{-i\left( \frac{1}{2}d_{i}\mathcal{B}_{ij}+r_{i}\mathcal{B}%
_{ij}\right) d_{j}}e^{i\mathbf{d}\cdot \widehat{\mathbf{p}}}=e^{i\mathbf{d}%
\cdot \widehat{\mathbf{P}}}\text{,}  \label{magtrans_general_II}
\end{equation}%
with a generator $\widehat{\mathbf{P}}$ defined by
\begin{equation}
\widehat{P}_{i}=-i\partial _{i}-\mathcal{B}_{ji}r_{j}=\widehat{p}_{i}-%
\mathcal{B}_{ji}r_{j}\text{.}  \label{generator_II}
\end{equation}%
This can be derived using the general formula
\begin{equation}
e^{K}e^{L}=e^{K+L+\frac{1}{2}[K,L]}\text{,}  \label{matrix_formula_II}
\end{equation}%
valid when commutator $[K,L]$ is proportional to the identity operator.
Applying the formula to the case of the expression eq.(\ref%
{magtrans_general_II}) with $K=-i\left( \frac{1}{2}d_{i}\mathcal{B}%
_{ij}+r_{i}\mathcal{B}_{ij}\right) d_{j},$ $L=i\widehat{\mathbf{p}}\cdot
\mathbf{d}$\textbf{,} and using the basic algebra $\left[ r_{i},\widehat{p}%
_{j}\right] =i\delta _{ij},$ one indeed obtains a number
\begin{equation}
\lbrack K,L]=\left[ r_{i}\mathcal{B}_{ij}d_{j},\widehat{\mathbf{p}}\cdot
\mathbf{d}\right] =id_{i}\mathcal{B}_{ij}d_{j}\text{.}
\label{commutator_AB_II}
\end{equation}

The expression for magnetic translations can also be derived from a
requirement that they commute with "Hamiltonian" $\widehat{H}$ defined in
eq.(\ref{H_II}). In fact they commute with both covariant derivatives $%
D_{i}, $%
\begin{equation}
D_{i}=\partial _{i}+i\mathcal{B}_{ij}r_{j},  \label{cov_der_II}
\end{equation}%
as well. However, using the same basic algebra, one also observes that
magnetic translations generally do not commute: $T_{\mathbf{d}_{1}}T_{%
\mathbf{d}_{2}}$ differs from $T_{\mathbf{d}_{2}}T_{\mathbf{d}_{1}}$ by a
phase. This is a consequence of the Campbell-Baker-Hausdorff formula $%
e^{K}e^{L}=e^{L}e^{K}e^{[K,L]}$, which follows immediately from eqs.(\ref%
{matrix_formula_II}) and (\ref{commutator_AB_II})%
\begin{equation}
e^{i\mathbf{d}_{1}\cdot \widehat{\mathbf{P}}}e^{i\mathbf{d}_{2}\cdot
\widehat{\mathbf{P}}}=e^{i\mathbf{d}_{2}\cdot \widehat{\mathbf{P}}}e^{i%
\mathbf{d}_{1}\cdot \widehat{\mathbf{P}}}e^{-[\mathbf{d}_{1}\cdot \widehat{%
\mathbf{P}},\mathbf{d}_{2}\cdot \widehat{\mathbf{P}}]}
\label{magntrans_commutator_II}
\end{equation}%
with the constant commutator given by $[\mathbf{d}_{1}\cdot \widehat{\mathbf{%
P}},\mathbf{d}_{2}\cdot \widehat{\mathbf{P}}]=ib\mathbf{d}_{1}\times \mathbf{%
d}_{2}$. The group property therefore is
\begin{equation}
T_{\mathbf{d}_{1}}T_{\mathbf{d}_{2}}=e^{-ib\mathbf{d}_{1}\times \mathbf{d}%
_{2}}T_{\mathbf{d}_{2}}T_{\mathbf{d}_{1}},  \label{magn_trans_commutator_II}
\end{equation}%
from which the requirement to have an integer number of fluxons per unit
cell of a lattice follows:
\begin{equation}
b\mathbf{d}_{1}\times \mathbf{d}_{2}=2\pi \times integer.
\label{quantization_II}
\end{equation}%
Note that the generator of magnetic translations is not proportional to
covariant derivative $D_{i}=\partial _{i}-i\mathcal{B}_{ij}r_{j}$. The
relation is nonlocal,
\begin{equation}
P_{i}=-iD_{i}+\varepsilon _{ij}r_{j},  \label{P-D_connection_II}
\end{equation}%
where $\varepsilon _{ij}$ is the antisymmetric tensor.

\subsubsection{The Abrikosov lattice solution: choice of the lattice
structure based on minimization of the quartic contribution to energy}

\emph{\textbf{The Abrikosov $\beta $ constant of a lattice structure}}

To lift the degeneracy between all the possible "wave functions" with
arbitrary normalization on the ground Landau level, one can try to minimize
the quartic term in free energy eq.(\ref{F_II}). It is reasonable to assume
that more "symmetric" configurations will have an advantage. In particular
lattices will be preferred over "chaotic" inhomogeneous ones. Moreover
hexagonal lattice should be perhaps the leading candidate due to its
relative isotropy and high symmetry. This configuration is preferred the in
London limit \cite{Tinkham} since vortices repel each other and try to self
assemble into the most homogeneous configuration. A simpler square lattice
was considered in fact as the best candidate by Abrikosov and we start from
this lattice to try to fix the variational parameter $C$. The energy
constrained to the LLL is%
\begin{equation}
G=\int d\mathbf{r}\left[ -a_{H}|\Psi |^{2}+\frac{1}{2}|\Psi |^{4}+\frac{%
\kappa ^{2}}{4}\left( \mathbf{b}-\mathbf{h}\right) ^{2}\right] \text{.}
\label{G_LLL_II}
\end{equation}%
The quartic contribution to energy density is proportional to the space
average of $\left\vert \varphi \right\vert ^{4}$ which is called the
Abrikosov $\beta _{\Diamond }$:

\begin{eqnarray}
\beta _{\Diamond } &=&\frac{1}{d_{\Diamond }^{2}}\int_{-d_{\Diamond
}/2}^{d_{\Diamond }/2}dx\int_{-d_{\Diamond }/2}^{d_{\Diamond
}/2}dy\left\vert \varphi _{\Diamond }\left( x,y\right) \right\vert ^{4}
\label{betaA_square_II} \\
&=&\frac{1}{d_{\Diamond }^{2}}\int_{-d_{\Diamond }/2}^{d_{\Diamond
}/2}dx\int_{-d_{\Diamond }/2}^{d_{\Diamond }/2}dy\sum_{n_{i}}e^{i\frac{2\pi
}{d_{\Diamond }}\left( n_{1}-n_{2}+n_{3}-n_{4}\right) x}  \notag \\
&&\times e^{-\frac{b}{2}\left[ \left( y-d_{\Diamond }n_{1}\right)
^{2}+\left( y-d_{\Diamond }n_{2}\right) ^{2}+\left( y-d_{\Diamond
}n_{3}\right) ^{2}+\left( y-d_{\Diamond }n_{4}\right) ^{2}\right] }  \notag
\end{eqnarray}%
In principle one can slightly generalize the method we used to calculate
analytically both integrals and sums \cite{Saint}, however will refrain from
doing so here, since in Appendix A a more efficient method will be
presented. The result is $\beta _{\Diamond }=1.18$. More generally one it is
shown there that for any lattice this constant is given by
\begin{equation}
\beta _{\theta }=\sum_{n_{1},n_{2}\mathbf{=}-\infty }^{\infty }e^{-\frac{b%
\mathbf{X}^{2}\left( n_{1},n_{2}\right) }{2}},  \label{betaA_rhombic}
\end{equation}%
where the summation is over the lattice sites $\mathbf{X}\left(
n_{1},n_{2}\right) \mathbf{=}n_{1}\mathbf{d}_{1}+n_{2}\mathbf{d}_{2}$. For
example for $\beta _{\triangle }=1.16$ for the hexagonal lattice.

\emph{\textbf{Energy, entropy and specific heat}}

Free energy density of the leading order solution is indeed negative.
Substituting a variational solution, one has
\begin{gather}
\frac{1}{vol}\int_{\mathbf{r}}\overline{F}=\frac{1}{vol}\int_{\mathbf{r}%
}\left\vert C\right\vert ^{2}[\varphi ^{\ast }\widehat{H}\varphi -\frac{1-t-b%
}{2}\left\vert \varphi \right\vert ^{2}  \label{F_square_II} \\
+\frac{1}{2}\left\vert C\right\vert ^{2}\left\vert \varphi \right\vert
^{4}]=-\left\vert C\right\vert ^{2}a_{H}+\frac{1}{2}\left\vert C\right\vert
^{4}\beta _{\Diamond }\text{.}  \notag
\end{gather}%
The FLL and the transition to the normal state can therefore described well
by a "dimensionally reduced" $D=0$ $U(1)$ symmetric model with the complex
"order parameter" $C$. It is similar to the Meissner state in the absence of
magnetic field but in $D=0$ with the only difference being that between $%
\beta _{A}$ and $1$ (which is just about 10\%). One minimizes it with
respect to $C$:%
\begin{equation}
\left\vert C\right\vert ^{2}=a_{H}/\beta _{\Diamond }\text{.}
\label{norm_square_min_II}
\end{equation}%
The average energy density at minimum (still on the subspace of square
lattices) is given by%
\begin{equation}
\frac{1}{vol}\int_{\mathbf{r}}\overline{F}=-\frac{a_{H}^{2}}{2\beta
_{\Diamond }}=-\frac{\left( 1-b-t\right) ^{2}}{8\beta _{\Diamond }}
\label{F_square2_II}
\end{equation}%
or, returning to the unscaled units, the energy density is
\begin{equation}
\frac{F}{vol}=-\frac{H_{c2}^{2}}{4\pi \kappa ^{2}}\frac{a_{H}^{2}}{\beta
_{\Diamond }}.  \label{F_square_regular_units_II}
\end{equation}

The first derivative with respect to temperature $T$, the entropy density%
\begin{equation}
S=-\frac{H_{c2}^{2}}{4\pi \kappa ^{2}\beta _{A}T_{c}}a_{H},
\label{entropy_square_II}
\end{equation}%
smoothly vanishes at transition to the normal phase. On the other hand the
second derivative, the specific heat divided by temperature, jumps to a
constant%
\begin{equation}
\frac{C_{v}}{T}=\frac{H_{c2}^{2}}{8\pi \kappa ^{2}\beta _{A}T_{c}^{2}}
\label{spec_heat_square_II}
\end{equation}%
from zero in the normal phase. Note that in this section we use a simple GL
model which neglects the normal state contribution to free energy, eq.(\ref%
{GL_F_I}), retaining only terms depending on the order parameter. The
additional term is a smooth "background", also referred to as a
"contribution of normal electrons".

Of course a similar argument is valid for any lattice symmetry with
corresponding Abrikosov parameter $\beta _{A}$. What is the correct shape of
the vortex lattice? To minimize the energy in this approximation is
equivalent to the minimization of $\beta _{\theta }$ with respect to shape
of the lattice. This is achieved for the hexagonal lattice, although
differences are not large. The square lattice incidentally has the largest
energy among all the rhombic structures, some $2\%$ higher than that of the
hexagonal lattice. This sounds rather small, but for a comparison, the
typical latent heat at melting (difference in internal energies between
lattice and homogeneous liquid) is of the same order of magnitude.

\emph{\textbf{Magnetization to leading order in $1/\kappa ^{2}$}}

Magnetization can be obtained via minimization of the Gibbs free energy with
respect to magnetic induction $B.$ In our units and within LLL approximation
one can differentiate eq.(\ref{G_LLL_II}) and the Maxwell term with respect
to $b$:
\begin{equation}
\kappa ^{2}\left[ h-b\left( \mathbf{r}\right) \right] =4\pi \kappa
^{2}m\left( \mathbf{r}\right) =|\Psi \left( \mathbf{r}\right) |^{2}.
\label{minimization_II}
\end{equation}%
The magnetization $m\left( \mathbf{r}\right) $ is therefore proportional to
the superfluid density $|\Psi \left( \mathbf{r}\right) |^{2}$ and is thus
highly inhomogeneous. Its space average is%
\begin{equation}
m\equiv \frac{\int_{\mathbf{r}}m\left( \mathbf{r}\right) }{4\pi vol}=\frac{%
C^{2}|\varphi _{\Diamond }\left( \mathbf{r}\right) |^{2}}{4\pi \kappa ^{2}}=-%
\frac{1-t-b}{8\pi \kappa ^{2}\beta _{\Diamond }}.  \label{average_magn_II}
\end{equation}%
For large $\kappa $ (typical value for high $T_{c}$ superconductors is $%
\kappa =100$) the magnetization is of order $1/\kappa ^{2}$ compared to $H$
and therefore negligible. This justifies an assumption of constant magnetic
induction, which can be slightly corrected:
\begin{equation}
b=\frac{-\frac{1-t}{2\beta _{\Diamond }}+\kappa ^{2}h}{\kappa ^{2}-\frac{1}{%
2\beta _{\Diamond }}}\simeq h-\frac{1-t-h}{2\kappa ^{2}\beta _{\Diamond }}%
\text{.}  \label{b_vs_h_II}
\end{equation}%
Rescaling back to regular units, one has

\begin{equation}
M=\frac{1}{4\pi }\left( B-H\right) \simeq -\frac{H_{c2}}{4\pi \kappa ^{2}}%
\frac{a_{H}}{\beta _{\Diamond }},  \label{magnetization_II}
\end{equation}%
with
\begin{equation}
a_{H}=\frac{1}{2}\left( 1-\frac{T}{T_{c}}-\frac{B}{H_{c2}}\right) \simeq
\frac{1}{2}\left( 1-\frac{T}{T_{c}}-\frac{H}{H_{c2}}\right) ,
\label{a_H_natural_II}
\end{equation}%
valid up to corrections of order $\kappa ^{-2}$.

\emph{\textbf{A general relation between the current density and the
superfluid density on LLL}}

The pattern of \ supercurrent flow around vortex cores can be readily
obtained by substituting the Abrikosov vortex approximation into the
expression for the supercurrent density eq.(\ref{Js_I}). We derive here a
general relation between an arbitrary static LLL function eq.(\ref%
{varlattice_II}) and the supercurrent. It will be helpful for understanding
of the mechanism behind the flux flow, occurring in dynamical situations,
when electric field is able to penetrate a superconductor. The covariant
derivatives acting on the LLL basis elements give:%
\begin{eqnarray}
D_{x}\phi _{k_{x}} &=&\left( \partial _{x}-iby\right) \left\{ \pi
^{-1/4}b^{1/2}e^{ik_{x}x-\frac{b}{2}\left( y-k_{x}/b\right) ^{2}}\right\}
\notag \\
&=&-i\pi ^{-1/4}b^{1/2}\left( by-k_{x}\right) e^{ik_{x}x-\frac{b}{2}\left(
y-k_{x}/b\right) ^{2}}  \notag \\
&=&i\left( b/2\right) ^{1/2}\phi _{N=1,k_{x}}
\label{covariant_derivatives_II} \\
D_{y}\phi _{k_{x}} &=&\partial _{y}\left\{ \pi ^{-1/4}b^{1/2}e^{ik_{x}x\left[
-\frac{b}{2}\left( y-k_{x}/b\right) ^{2}\right] }\right\}  \notag \\
&=&-\pi ^{-1/4}b^{1/2}\left( by-k_{x}\right) e^{ik_{x}x-\frac{b}{2}\left(
y-k_{x}/b\right) ^{2}}  \notag \\
&=&\left( b/2\right) ^{1/2}\phi _{N=1,k_{x}}\text{.}  \notag
\end{eqnarray}

\begin{figure}[t]
\centering \rotatebox{270}{\includegraphics[width=0.3%
\textwidth,height=8cm]{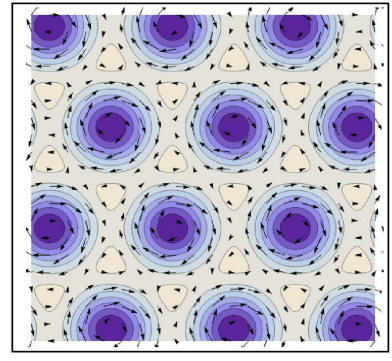}}
\caption{Superflow around the vortex centers in the hexagonal lattice}
\label{figII3}
\end{figure}

The covariant derivatives, which are linear combinations of "raising" and
"lowering" Landau level operators,
\begin{eqnarray}
D_{x} &=&i\sqrt{\frac{b}{2}}\left( \widehat{a}^{\dagger }+\widehat{a}\right)
;\text{ }D_{y}=\sqrt{\frac{b}{2}}\left( \widehat{a}^{\dagger }-\widehat{a}%
\right) ;  \label{creation_anihillation_II} \\
\widehat{a}^{\dagger } &=&\frac{-i\partial _{x}+\partial _{y}-by}{\sqrt{2b}};%
\text{\ }\widehat{a}=-\frac{-i\partial _{x}+\partial _{y}+by}{\sqrt{2b}}.
\notag
\end{eqnarray}%
therefore raise an LLL function to the first LL. One can check that the
following relation is valid:%
\begin{equation}
i\Psi ^{\ast }\left( \mathbf{r}\right) D_{i}\Psi \left( \mathbf{r}\right)
+c.c.=\varepsilon _{ij}\partial _{j}\left( \left\vert \Psi \left( \mathbf{r}%
\right) \right\vert ^{2}\right) ,  \label{vector_relation_II}
\end{equation}%
where $\varepsilon _{ij}$ is the antisymmetric tensor. We therefore have
established an exact relation between the current density and (scaled with $%
J_{GL}=\frac{c\Phi _{0}}{2\pi ^{2}\kappa ^{2}\xi ^{3}}$ $J_{GL}=\frac{c\Phi
_{0}}{4\pi ^{2}\kappa ^{2}\xi ^{3}}$ according to eq.(\ref{J_GL_AppA}))
superfluid density,
\begin{equation}
\overline{J}_{i}\left( \mathbf{r}\right) =\frac{J_{i}\left( \mathbf{r}%
\right) }{J_{GL}}=-\frac{1}{2}\varepsilon _{ij}\overline{\partial }%
_{j}\left( \left\vert \overline{\Psi }\left( \overline{\mathbf{r}}\right)
\right\vert ^{2}\right) \text{,}  \label{J-psi2_relation_II}
\end{equation}%
valid, however, on LLL states only. In regular units the current density is
related to (unscaled) order parameter field by
\begin{equation}
J_{i}\left( \mathbf{r}\right) =-\frac{e^{\ast }{\hbar }}{2m^{\ast }}%
\varepsilon _{ij}\partial _{j}\left( \left\vert \Psi \left( \mathbf{r}%
\right) \right\vert ^{2}\right) \text{,}
\end{equation}%
The supercurrent indeed creates a vortex around a dip in the superfluid
density, Fig.\ref{figII3}. The overall current is of course zero, since the
bulk integral is transformed into a surface one.
An approximate solution described in this subsection is perhaps valid "near"
the $H_{c2}\left( T\right) $ line, however to estimate the range of validity
and to obtain a better approximation, one would prefer a systematic
perturbative scheme over an uncontrollable variational principle. This is
provided by the $a_{H}$ expansion.

\subsection{Systematic expansion around the bifurcation point.}

\subsubsection{Expansion and the leading order}

We have defined the operator $\widehat{H}$ in eq.(\ref{H_II}) in such a way
that its spectrum will start from zero. This allows the development of the
bifurcation point perturbation theory for the GL equation eq.(\ref{GLeq_II}%
). This type of the perturbation theory is quite different from the
one used in linear equations like Schr\"{o}dinger equation.

One develops a perturbation theory in small $a_{H}$ around the $H_{c2}\left(
T\right) $ line :
\begin{equation}
\Psi =\sqrt{a_{H}}\left( \Psi ^{\left( 0\right) }+a_{H}\Psi ^{\left(
1\right) }+a_{H}^{2}\Psi ^{\left( 2\right) }+...\right)
\label{psi_expansion_II}
\end{equation}%
Note the fractional power of the expansion parameter in front of the
"regular" series. This is related to the mean field critical exponent for a $%
\phi ^{4}$ type equation being $1/2$, so that all the terms in the free
energy have the same power $a_{H}^{2}$ and are "relevant", as we mentioned
in Introduction. Substituting this series into eq.(\ref{GLeq_II}) one
observes that the leading ($a_{H}^{1/2}$) order equation gives the lowest
LLL restriction already motivated in the heuristic approach of the previous
subsection:
\begin{equation}
\widehat{H}\Psi ^{\left( 0\right) }=0,  \label{leading_II}
\end{equation}%
resulting in $\Psi ^{\left( 0\right) }=C^{\left( 0\right) }\varphi $ with
normalization undetermined. It will be determined by the next order. The
next to leading ($a_{H}^{3/2}$) order equation is:%
\begin{equation}
\widehat{H}\Psi ^{\left( 1\right) }-C^{\left( 0\right) }\varphi +C^{\left(
0\right) }\left\vert C^{\left( 0\right) }\right\vert ^{2}\varphi \left\vert
\varphi \right\vert ^{2}=0.  \label{next_II}
\end{equation}%
Multiplying it with $\varphi ^{\ast }$ and integrating over coordinates, one
obtains%
\begin{equation}
\int_{\mathbf{r}}\varphi ^{\ast }\left[ \widehat{H}\Psi ^{\left( 1\right)
}-C^{\left( 0\right) }\varphi +C^{\left( 0\right) }\left\vert C^{\left(
0\right) }\right\vert ^{2}\varphi \left\vert \varphi \right\vert ^{2}\right]
=0.  \label{scalar_product_II}
\end{equation}%
The first term vanishes since Hermitian operator $\widehat{H}$ in the scalar
product, defined as

\begin{equation}
\frac{1}{L_{x}L_{y}}\int_{\mathbf{r}}f^{\ast }\left( r\right) g\left(
r\right) \equiv \left[ f|g\right] \text{,}  \label{scalar
product_II}
\end{equation}%
can be applied on it and vanished by virtue of eq.(\ref{leading_II}). This
way one recovers the "naive" result of eq.(\ref{norm_square_min_II}):%
\begin{equation}
-1+\left\vert C^{\left( 0\right) }\right\vert ^{2}\frac{1}{L_{x}L_{y}}\int_{%
\mathbf{r}}\left\vert \varphi \right\vert ^{4}=-1+\left\vert C^{\left(
0\right) }\right\vert ^{2}\beta _{\Delta }=0\text{.}  \label{C0_II}
\end{equation}%
Note that to this order different lattices or in fact any LLL functions are
"approximate solutions".

\subsubsection{Higher orders corrections to the solution}

\emph{\textbf{Next to leading order}}

Higher order corrections would in principle contain higher Landau level
eigenfunctions in the basis of solutions of the linearized GL equation eq.(%
\ref{linearizedGL_II}) for eigenvalues $E_{N}$, eq.(\ref{HLL_functions_II}).
As on the LLL for higher Landau levels one can combine them into a lattice
with a certain (here hexagonal) symmetry:

\begin{eqnarray}
\varphi _{\Delta N}\left( \mathbf{r}\right) &=&\int_{k_{x}}C_{k_{x}}\phi
_{Nk_{x}}(\mathbf{r})=\varphi _{N}\left( b^{1/2}\mathbf{r}\right) ;
\label{HLL_lat_functions_II} \\
\varphi _{N}\left( \mathbf{r}\right) &=&\frac{3^{1/8}}{\sqrt{2^{N}N!}}%
\sum\limits_{l=-\infty }^{\infty }e^{il\left( \frac{\pi l}{2}+3^{1/4}\pi
^{1/2}x\right) -\frac{1}{2}(y-3^{1/4}\pi ^{1/2}l)^{2}}\text{.}  \notag
\end{eqnarray}%
The coefficients are the same as given in the previous subsection, eq.(\ref%
{psi_hex_II}).

\begin{figure}[t]
\centering \rotatebox{270}{\includegraphics[width=0.3%
\textwidth,height=8cm]{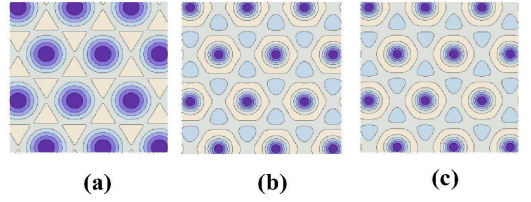}}
\caption{Convergence of the bifurcation perturbation theory}
\label{figII4}
\end{figure}

The order $\left( a_{H}\right) ^{i+1/2}$ correction can be expanded in the
Landau levels basis, eq.(\ref{HLL_lat_functions_II}) as%
\begin{equation}
\Psi ^{i}\left( \mathbf{r}\right) =C^{\left( i\right) }\varphi \left( b^{1/2}%
\mathbf{r}\right) +\sum_{N=1}^{\infty }C_{N}^{\left( i\right) }\varphi
_{N}\left( b^{1/2}\mathbf{r}\right)  \label{psi_i_II}
\end{equation}%
(to simplify notations the LLL coefficient is denoted simply $C^{\left(
i\right) }$ rather than $C_{0}^{\left( i\right) }$, suppressing $N=0$, the
convention we have been using already for $\varphi \equiv \varphi _{0}$).
Inserting this into eq.(\ref{next_II}), one obtains to order $a_{H}^{3/2}$:
\begin{equation}
\sum_{N=1}^{\infty }NbC_{N}^{\left( 1\right) }\varphi _{N}=C^{\left(
0\right) }\varphi -C^{\left( 0\right) }|C^{\left( 0\right) }|^{2}\varphi
|\varphi |^{2}.  \label{next1_II}
\end{equation}%
The scalar product with $\varphi ^{N}$ determines $C_{N}^{\left( 1\right) }$%
:
\begin{equation}
C_{N}^{\left( 1\right) }=-\frac{\beta _{N}}{Nb\beta _{\Delta }^{3/2}},
\label{gn_II}
\end{equation}%
where
\begin{equation}
\beta _{N}\equiv \frac{1}{vol}\int_{\mathbf{r}}|\varphi |^{2}\varphi
_{N}\varphi ^{\ast }.  \label{betaN_II}
\end{equation}%
To find $C^{\left( 1\right) }$ we need in addition also the order $%
a_{H}^{5/2}$ equation:
\begin{equation}
\widehat{H}\Psi _{2}=\sum_{N=1}^{\infty }NbC_{N}^{\left( 2\right) }\varphi
_{N}=\Psi _{1}-(C^{\left( 0\right) })^{2}(2\Psi _{1}|\varphi |^{2}+\Psi
_{1}^{\ast }\varphi ^{2})\text{.}  \label{next_next_II}
\end{equation}%
Inner product with $\varphi $ gives:
\begin{equation}
C^{\left( 1\right) }=\frac{3}{2}\sum_{N=1}^{\infty }\frac{(\beta _{N})^{2}}{%
Nb\beta _{\Delta }^{5/2}}.  \label{g1_II}
\end{equation}%
The expansion can be relatively easily continued. Fig. \ref{figII4} shows
three successive approximation. The convergence is quite fast even as far
from the $H_{c2}\left( T\right) $ line at for $b=0.1,t=0.5$.

\emph{\textbf{Orders $a_{H}^{2}$ and $a_{H}^{3}$ in the expansion of free
energy.}}

The mean field expression for the free energy to order $a_{H}^{2}$ was
obtained already using heuristic approach, eq.(\ref{F_square2_II})..
Inserting the next correction eqs.(\ref{g1_II}) and (\ref{gn_II}) into eq.(%
\ref{F_II}) one obtains the free energy density:

\begin{eqnarray}
\frac{F^{\left( 2\right) }+F^{\left( 3\right) }}{4\text{ }vol\text{ }\Delta }
&=&-\frac{a_{H}^{2}}{2\beta _{\Delta }}-\frac{a_{H}^{3}}{\beta _{\Delta
}^{3}b}\sum_{N=1}^{\infty }\frac{(\beta _{N}){}^{2}}{N}  \label{F_corr_II} \\
&=&-0.43a_{H}^{2}-0.0078\frac{a_{H}^{3}}{b}\text{,}  \notag
\end{eqnarray}%
where $\Delta =\frac{H_{c2}^{2}}{8\pi \kappa ^{2}}$ is the unit of energy
density.

It is interesting to note that $\beta _{N}\neq 0$ only when $n=6j$, where $j$
is an integer. This is due to hexagonal symmetry of the vortex lattice \cite%
{Lascher65}. For $n=6j$ it decreases very fast with $j$: $\beta
_{6}=-0.2787,\beta _{12}=0.0249.$ Because of this the coefficient of the
next to leading order is very small (additional factor of $6$ in the
denominator). We might preliminarily conclude therefore that the
perturbation theory in $a_{H}$ works much better that might be naively
anticipated and can be used very far from transition line. If we demand that
the correction is smaller then the main contribution the corresponding line
on the phase diagram will be $b=0.015\cdot (1-t)$. For example the LLL
melting line corresponds to $a_{H}\sim 1$. This overly optimistic conclusion
is however incorrect as calculation of the following term shows.\bigskip

\emph{\textbf{How precise is LLL?}}

Now we discuss in what region of the parameter space the expansion outlined
above can be applied. First of all note that all the contributions to $\Psi
_{1}$ are proportional to $1/b$. This is a general feature: the actual
expansion parameter is $a_{H}/b$. One can ask whether the expansion is
convergent and, if yes, what is its radius of convergence. Looking just at
the leading correction and comparing it to the LLL one gets a very
optimistic estimate. For this purpose higher orders coefficients were
calculated \cite{Li99}. The results for the $\Psi _{2}$ are following:
\begin{gather}
C_{N}^{\left( 2\right) }=\frac{1}{Nb}{\large [}C_{N}^{\left( 1\right) }-%
\frac{1}{\beta _{\Delta }}\times  \label{g2N_II} \\
\sum_{M=0}^{\infty }C_{M}^{\left( 1\right) }\left( 2\left[ N,0|M,0\right] +%
\left[ 0,0|M,N\right] \right) {\large ]}  \notag
\end{gather}%
and
\begin{gather}
C^{\left( 2\right) }=-\frac{3}{2\beta _{\Delta }}\beta _{N}C_{N}^{\left(
2\right) }-\frac{1}{2\sqrt{\beta _{\Delta }}}\times  \label{g2_II} \\
\sum_{L,M=0}^{\infty }C_{L}^{\left( 1\right) }C_{M}^{\left( 1\right) }\left(
\left[ 0,0|L,M\right] +2\left[ M,0|L,0\right] \right) ,  \notag
\end{gather}%
where
\begin{equation}
\left[ K,L|M,N\right] \equiv \frac{1}{vol}\int_{\mathbf{r}}\varphi
_{K}^{\ast }\varphi _{L}^{\ast }\varphi _{M}\varphi _{N}.  \label{[KL|MN]_II}
\end{equation}

We already can see that $C_{N}^{\left( 2\right) }$ and $C^{\left( 2\right) }$
are proportional to $C_{N}^{\left( 1\right) }$ and in addition there is a
factor of $1/N$. Since, due to hexagonal lattice symmetry all the $%
C_{N}^{\left( 1\right) }$, $N\neq 6j$ vanish, so do $C_{N}^{\left( 2\right)
} $. We have checked that there is no more small parameters, so we conclude
that the leading order coefficient is much larger than first (factor $6\cdot
5$), but the second is only $6$ times larger than the third. The correction
to free energy density is
\begin{equation}
\frac{F_{.}^{\left( 4\right) }}{4\text{ }vol\text{ }\Delta }=\frac{0.056}{%
6^{2}}\frac{a_{H}^{4}}{b^{2}}\text{.}  \label{F4_II}
\end{equation}%
Accidental smallness by factor $1/6$ of the coefficients in the $a_{H}/b$
expansion due to symmetry means that the range of validity of this expansion
is roughly $a_{H}<6b$ or $B<\frac{H_{c2}\left( T\right) }{13}$. Moreover
additional smallness of all the HLL corrections compared to the LLL means
that they constitute just several percent of the correct result inside the
region of applicability. To illustrate this point we plot on Fig. \ref%
{figII4} the perturbatively calculated solution for $b=0.1,t=0.5$. One can
see that although the leading LLL function has very thick vortices (Fig. \ref%
{figII4}a), the first nonzero correction makes them of order of the
coherence length (Fig. \ref{figII4}b). Following correction of the order $%
\left( a_{H}/b\right) ^{2}$ makes it practically indistinguishable from the
numerical solution. Amazingly the order parameter between the vortices
approaches its vacuum value. Paradoxically starting from the region close to
$H_{c2}$ the perturbation theory knows how to correct the order parameter so
that it looks very similar to the London approximation (valid only close to $%
H_{c1}$) result \ of well separated vortices.

We conclude therefore that the expansion in $a_{H}/b$ works in the mean
field better that one can naively expect.

\section{Thermal fluctuations and melting of the vortex solid into a liquid}

In this section a theory of thermal fluctuations and of melting of the
vortex lattice in type II superconductors in the framework of Ginzburg -
Landau approach is presented. Far from $H_{c1}(T)$ the lowest Landau level
approximation can be used. Within this approximation the model simplifies
and results depend just on one parameter: the LLL reduced temperature. To
obtain an accurate description of both the vortex lattice and the vortex
liquid different methods are applied. In the crystalline phase basic
excitations are phonons. Their spectrum and interactions are rather unusual
and the low temperature perturbation theory requires to develop a certain
technique. Generally perturbation theory to the two loop order is
sufficient, but for certain purposes (like finding a spinodal in which
metastable crystalline state becomes unstable) a self consistent "gaussian"
approximation is required. In the liquid state both the perturbation theory
and gaussian approximations are insufficient to get a precision required to
describe the first order melting transition and one utilizes more
sophisticated methods. Already gaussian approximation shows that the
metastable liquid state persists (within LLL) till zero temperature. The
high temperature renormalized series (around the gaussian variational state)
supplemented by interpolation to a $T=0$ metastable \textquotedblright
perfect liquid\textquotedblright\ state are sufficient. The melting line
location is determined and magnetization and specific heat jumps along it
are calculated. The magnetization of liquid is larger than that of solid by
1.8\% irrespective of the melting temperature, while the specific heat jump
is about 6\% and decreases slowly with temperature.

\subsection{The LLL scaling and the quasi - momentum basis}

\subsubsection{The LLL scaling}

\emph{\textbf{Units and the LLL scaled temperature}}

If the magnetic field is sufficiently high,\ we can keep only the $N=0$ LLL
modes. This is achieved by enforcing the following constraint,
\begin{equation}
-\frac{{\hbar }^{2}}{2m^{\ast }}\mathbf{D}^{2}\Psi =\frac{{\hbar }e^{\ast }}{%
2m^{\ast }c}B\Psi \text{,}  \label{LLL_III}
\end{equation}%
where covariant derivatives were defined in eq.(\ref{covariant_derivative_I}%
). Using it the free energy eq.(\ref{G_I}) simplifies:
\begin{eqnarray}
G\left[ \Psi ,\mathbf{A}\right] &=&\int dr{\large \{}\frac{{\hbar }^{2}}{%
2m_{c}^{\ast }}\left\vert \partial _{z}\Psi \right\vert ^{2}\mathbf{+}\alpha
T_{c}\left( 1-t-b\right) |\Psi |^{2}  \notag \\
&&+\frac{\beta }{2}|\Psi |^{4}+\frac{\left( \mathbf{B}-\mathbf{H}\right) ^{2}%
}{8\pi }{\large \}}  \label{G_III}
\end{eqnarray}

Originally the Ginzburg - Landau statistical sum, eq.(\ref{Z_I}), had five
dimensionless parameters, three material parameters $\kappa =\lambda /\xi
,\gamma _{a}=\left( m_{c}^{\ast }/m^{\ast }\right) ^{1/2},$ and the Ginzburg
number, defined by
\begin{equation}
Gi\equiv \left( \frac{e^{\ast 2}\kappa ^{2}\xi T_{c}\gamma _{a}}{2\pi c^{2}{%
\hbar }^{2}}\right) ^{2}  \label{Gi_III}
\end{equation}%
and two external parameters $t=T/T_{c}$ and $b=B/H_{c2}$. However, since
there is now no gradient term in directions perpendicular to the field, it
is missing one independent parameter. The Gibbs energy,%
\begin{equation}
\mathcal{G}=-T\log \left\{ \int_{\Psi ,B}\exp \left[ -\frac{1}{T}\int G\left[
\Psi ,\mathbf{B}\right] \right] \right\} \text{,}  \label{Geff_III}
\end{equation}%
thus possesses the \textquotedblright LLL scaling\textquotedblright\ \cite%
{Thouless75,Ruggeri76,Ruggeri78,Shenoy72}. To exhibit these scaling
relations, it is useful to use units of coordinates and fields, which are
dependent not just on material parameters (as those used in section II), but
also on external parameters, magnetic field and temperature. One uses the
magnetic length rather than coherence length as a unit of length in
directions perpendicular to magnetic field, $x=\frac{\xi }{\sqrt{b}}%
\overline{x},$ $y=\frac{\xi }{\sqrt{b}}\overline{y}$, while in the field
direction different factor is used, $z=\frac{\xi }{\gamma _{a}}\left( \frac{%
\sqrt{Gi}tb}{4}\right) ^{-1/3}\overline{z}$. Magnetic field is rescaled as
before with $H_{c2}$, while the order parameter field has an additional
factor: $\Psi ^{2}=2\Psi _{0}^{2}\left( \frac{\sqrt{Gi}tb}{4}\right)
^{2/3}\psi ^{2}$. The usefulness of the fractional powers additional factors
will become clear later.

The dimensionless Boltzmann factor becomes
\begin{eqnarray}
g\left[ \psi ,b\right] &\equiv &\frac{G\left[ \Psi ,\mathbf{A}\right] }{T}=%
\frac{1}{2^{5/2}\pi }f\left[ \psi \right]  \label{g[psi,b]_III} \\
&&+\frac{\kappa ^{2}}{2^{5/2}\pi }\left( \frac{\sqrt{Gi}tb}{4}\right)
^{-4/3}\int d\overline{r}\frac{\left( \mathbf{b}-\mathbf{h}\right) ^{2}}{4};
\notag \\
f\left[ \psi \right] &=&\int d\overline{r}\left[ \frac{1}{2}|\partial _{%
\overline{z}}\psi |^{2}+a_{T}|\psi |^{2}+\frac{1}{2}|\psi |^{4}\right] \text{%
,}  \label{f_definition_III}
\end{eqnarray}%
where the LLL scale "temperature" is
\begin{equation}
a_{T}=-\left( \frac{\sqrt{Gi}tb}{4}\right) ^{-\frac{2}{3}}a_{H}=-\frac{1-t-b%
}{2}\left( \frac{\sqrt{Gi}tb}{4}\right) ^{-\frac{2}{3}}.  \label{aT_III}
\end{equation}%
The constant $a_{H}$ was defined in eq.(\ref{ah_II}) and extensively used in
the previous section. The scaled temperature therefore is the only remaining
dimensionless parameter in eq.(\ref{g[psi,b]_III}) in addition to the
coefficient of the last term. Factors of $2^{5/2}\pi $ in definition of
"dimensionless free energy" $f$ in eq.(\ref{g[psi,b]_III}) are traditionally
kept and will appear frequently in what follows. Assuming nonfluctuating
constant magnetic field, one can disregard the last term in eq.(\ref%
{g[psi,b]_III}), and consider the thermal fluctuations of the order
parameter only. This assumption is typically valid in almost all
applications and will be discussed in subsection E. Certain physical
quantities, the "LLL scaled" ones, are functions of this parameter only. We
list the most important such quantities below.

\emph{\textbf{Scaled quantities}}

The scaled free energy density is:
\begin{equation}
\emph{f}_{d}\left( a_{T}\right) =-\frac{2^{5/2}\pi }{V^{\prime }}\log \int
\mathcal{D}\psi \mathcal{D}\psi ^{\ast }e^{-\frac{1}{2^{5/2}\pi }f\left[
\psi \right] },  \label{scaled_free_energy_III}
\end{equation}%
where $V^{\prime }$ is the rescaled volume and $\emph{f}\left( a_{T}\right) $
is related to the free energy density in unscaled units by
\begin{equation}
\mathcal{F}_{d}=\left( \frac{\sqrt{Gi}tb}{4}\right) ^{4/3}\frac{H_{c2}^{2}}{%
2\pi \kappa ^{2}}\emph{f}_{d}\left( a_{T}\right) \text{.}
\label{F_physunits_III}
\end{equation}%
Turning to magnetization, let us return to conventional units eq.(\ref{G_III}%
) and neglect fluctuations of magnetic field (considered in \cite%
{Halperin74,Dasgupta81,Herbut96,Herbut07,Lobb87}).
Within LLL magnetization in the presence of thermal fluctuations is
determined from%
\begin{equation}
\frac{\delta }{\delta B}\mathcal{G}=Z^{-1}\int_{\Psi }\frac{\delta }{\delta B%
}G\left[ \Psi ,B\right] e^{-\frac{1}{T}\int G\left[ \Psi ,B\right] }=0\text{.%
}  \label{magnetization_eq_III}
\end{equation}%
Taking the derivative, one obtains%
\begin{eqnarray}
&&-Z^{-1}\int_{\Psi }\left[ \int_{r}\frac{\alpha T_{c}}{H_{c2}}|\Psi |^{2}+%
\frac{B-H}{4\pi }\right] e^{-\frac{1}{T}\int G[\Psi ,B]}  \notag \\
&=&-\frac{\alpha T_{c}}{H_{c2}}\left\langle |\Psi |^{2}\right\rangle -\frac{%
B-H}{4\pi }=0\text{,}  \label{magnetization_eq2_III}
\end{eqnarray}%
where from now on $\left\langle ...\right\rangle $ denotes the thermal
average. The magnetization on LLL is therefore proportional to the
superfluid density%
\begin{equation}
M=-\frac{\alpha T_{c}}{H_{c2}}\left\langle |\Psi |^{2}\right\rangle \text{.}
\label{M_psi^2_relation_III}
\end{equation}%
This motivates the definition of the LLL scaled magnetization proportional
to $\left\langle |\Psi |^{2}\right\rangle $,
\begin{equation}
m\left( a_{T}\right) =-\left\langle |\psi |^{2}\right\rangle =-\frac{%
\partial }{\partial a_{T}}f_{d}\left( a_{T}\right)
\label{scaled_magnetization_III}
\end{equation}%
which is related to magnetization by%
\begin{equation}
\frac{M}{H_{c2}}=\frac{1}{4\pi \kappa ^{2}}\left( \frac{\sqrt{Gi}}{4}%
tb\right) ^{2/3}m\left( a_{T}\right) .  \label{magnetization_t_scaled_III}
\end{equation}%
Consequently $\frac{M}{\left( TB\right) ^{3/2}}$ depends on $a_{T}$ only,
the statement called "the LLL scaling" proposed in \cite%
{Thouless75,Ruggeri76,Ruggeri78,Tesanovic92,Tesanovic94}. It has been
experimentally demonstrated in numerous experiments.

The specific heat contribution due to the vortex matter is generally defined
by $C=-T\frac{\partial ^{2}}{\partial T^{2}}G(T,H)$. Usually, since the GL
approach is applied near $T_{c}$, one can replace $T$ by $T_{c}$ in the
Boltzmann factor, leaving the temperature dependence just inside the
coefficient of $|\Psi |^{2}$ in eq.(\ref{G_III}). In this case the
normalized specific heat is defined as
\begin{equation}
c=\frac{C}{C_{mf}}\text{,}  \label{spec_heat_relative_III}
\end{equation}%
where $C_{mf}=\frac{H_{c2}^{2}T}{8\pi \kappa ^{2}\beta _{\Delta }T_{c}^{2}}$
is the mean field specific heat of solid calculated in the previous section.
Substituting $\emph{G}(T,H)$, if very near phase transition temperature, we
can put $t=1$ in the scaling factor $\frac{\sqrt{Gi}}{4}tb,$in this case, we
obtain:%
\begin{equation}
c=-\beta _{\Delta }\frac{\partial ^{2}}{\partial a_{T}^{2}}\emph{f}%
_{d}\left( a_{T}\right) \text{.}  \label{scaled_c_III}
\end{equation}%
Since the range of applicability of LLL can extend beyond vicinity of $T_{c}$%
, especially at strong fields (since they depress order parameter), one
should use a more complicated formula which does not utilize $T\simeq T_{c}$%
:
\begin{eqnarray}
c &=&-\beta _{A}{\Huge [}\frac{16}{9t^{2}}\left( bt\right) ^{4/3}\emph{f}%
_{d}(a_{T})-\frac{4\left( b-1-t\right) }{3t^{2}}\left( bt\right) ^{2/3}
\notag \\
&&\times \emph{f}^{\prime }(a_{T})+\frac{\left( 2-2b+t\right) }{9t^{2}}^{2}%
\emph{f}_{d}^{\prime \prime }(a_{T}){\Huge ].}  \label{c_precise_III}
\end{eqnarray}%
It no longer possesses the LLL scaling.

\subsubsection{Magnetic translations and the quasi - momentum basis}

It is necessary to use the representations of translational symmetry in
order to classify various excitations of both the Abrikosov lattice and a
homogeneous state created when thermal fluctuations become strong enough. As
we have seen in subsection IIB, presence of magnetic field makes the use of
the translational symmetry a nontrivial task, due to the need to "regauge".
Here we use an algebraic approach to construct the quasi - momenta basis and
then to determine the excitation spectrum of the lattice and the liquid,
which in turn determines its elastic and thermal properties.

\emph{\textbf{\smallskip The quasi - momentum basis}}

We motivated the definition of the magnetic translation symmetries eq.(\ref%
{magtrans_general_II}) by the property that they transform various lattices
onto themselves. More formally the $x-y$ plane translation operators $T_{d}$%
, eq.(\ref{magtrans_general_II}), represent symmetries since they commute
with "Hamiltonian" $\widehat{H}$ of eq.(\ref{H_II}). Excitations of the
lattice are no longer invariant under the symmetry transformations. This in
particular means that we cannot longer consider the problem as two
dimensional. However, as in the solid state physics, it is convenient to
expand them in the basis of eigenfunctions of the generators of the magnetic
translations operators defined in eq.(\ref{magtrans_general_II}) and simple
translations in the field, $z$, direction:%
\begin{eqnarray}
\widehat{\mathbf{P}}\varphi _{Nk} &=&\mathbf{k}\varphi _{Nk};\text{ \ \ \ \ }%
T_{\mathbf{d}}\varphi _{Nk}=e^{i\mathbf{k\cdot d}}\varphi _{Nk};
\label{eigenvector_III} \\
\widehat{p}_{z}\varphi _{Nk} &=&k_{z}\varphi _{Nk};\text{ \ \ \ \ }%
T_{d_{z}}\varphi _{Nk}=e^{ik_{z}d_{z}}\varphi _{Nk}\text{.}  \notag
\end{eqnarray}%
with commutation relation eq.(\ref{magn_trans_commutator_II}): $T_{\mathbf{d}%
_{1}}T_{\mathbf{d}_{2}}=e^{-ib\mathbf{d}_{1}\times \mathbf{d}_{2}}T_{\mathbf{%
d}_{2}}T_{\mathbf{d}_{1}}$. The tree dimensional quasi - momentum vector is
denoted by $k\equiv \left( \mathbf{k},k_{z}\right) $. It is easy to
construct these functions explicitly. On the $N^{th}$ Landau level the 2D
quasi - momentum $\mathbf{k}$ function is given by:%
\begin{equation}
\varphi _{N\mathbf{k}}\left( \mathbf{r}\right) =T_{\widetilde{\mathbf{k}}%
}\varphi _{N}\left( \mathbf{r}\right) ,
\label{quasimomentum_construction_III}
\end{equation}%
where $\widetilde{k}_{i}=\varepsilon _{ij}k_{j}$ for $i=x,y$ and $\varphi
_{N}\left( \mathbf{r}\right) $ for a\ given lattice symmetry was constructed
in IIA. Here we will take the hexagonal lattice functions of eq.(\ref%
{HLL_lat_functions_II}). Indeed

\begin{equation}
T_{\mathbf{d}}\varphi _{N\mathbf{k}}=T_{\mathbf{d}}T_{\widetilde{\mathbf{k}}%
}\varphi _{N}=e^{-i\mathbf{d\times }\widetilde{\mathbf{k}}}T_{\widetilde{%
\mathbf{k}}}T_{\mathbf{d}}\varphi _{N}=e^{i\mathbf{k\cdot d}}\varphi _{N%
\mathbf{k}}\text{.}  \label{proof_construction_III}
\end{equation}%
To write it explicitly, the most convenient form of the magnetic translation
is that of eq.(\ref{magtrans_general_II}), which gives
\begin{equation}
\varphi _{N\mathbf{k}}=e^{-i\left( \frac{1}{2}\widetilde{k}_{i}\mathcal{B}%
_{ij}\widetilde{k}_{j}+x_{i}\mathcal{B}_{ij}\widetilde{k}_{j}\right) }e^{i%
\widetilde{\mathbf{k}}\cdot \mathbf{p}}\varphi _{N}.  \label{fi_N_III}
\end{equation}%
Since $T_{\mathbf{d}}$ is unitary, the normalization is the same as that of $%
\varphi _{N}$. On LLL in our gauge one has:%
\begin{eqnarray}
\varphi _{\mathbf{k}} &=&e^{ixk_{x}}\varphi _{\mathbf{0}}\left( \mathbf{r}+%
\widetilde{\mathbf{k}}\right) =3^{1/8}\sum\limits_{l=-\infty }^{\infty }
\label{2D_shift_function_III} \\
&&e^{i[\frac{\pi l^{2}}{2}+3^{1/4}\pi ^{1/2}\left( x+k_{y}\right) l+xk_{x}]-%
\frac{1}{2}[y-k_{x}-3^{1/4}\pi ^{1/2}l]^{2}}.  \notag
\end{eqnarray}%
In the direction along the field one uses the usual momentum:%
\begin{equation}
\varphi _{k}\left( r\right) =e^{ik_{z}z}\varphi _{\mathbf{k}}\left( \mathbf{r%
}\right) ,  \label{fi_k_3D_III}
\end{equation}%
where, as before, we use the notation $r\equiv \left( \mathbf{r},z\right) $.

The values of the quasi - momentum cover a Brillouin zone in the $x-y$
plane. As usual, it is convenient to work in basis vectors of the reciprocal
lattice, $\mathbf{k}=k_{1}\widetilde{\mathbf{d}}_{1}+k_{2}\widetilde{\mathbf{%
d}}_{2}$, with the basis vectors%
\begin{equation}
\widetilde{\mathbf{d}}_{1}=\frac{\sqrt{3^{\frac{1}{2}}}}{\sqrt{\pi }}\left(
1,-\frac{1}{\sqrt{3}}\right) ;\widetilde{\mathbf{d}}_{2}=\frac{\sqrt{3^{%
\frac{1}{2}}}}{\sqrt{\pi }}\left( 0,\frac{2}{\sqrt{3}}\right) \text{.}
\label{reciprocal_III}
\end{equation}%
The measure is
\begin{eqnarray}
\int_{B.z.}dk_{x}dk_{y} &\equiv &2\pi \int_{0}^{1}dk_{1}\int_{0}^{1}dk_{2};
\label{measure_III} \\
\int_{k}d^{3}k &\equiv &\int_{-\infty }^{\infty }dk_{z}\int_{B.z.}d\mathbf{k}%
\text{\textbf{.}}  \notag
\end{eqnarray}

Beyond LLL the quasi - momentum basis consists of $\varphi _{\mathbf{k}}^{N}(%
\mathbf{r}),$ $N^{th}$ Landau level \textquotedblright wave
functions\textquotedblright\ with quasi-momentum $\mathbf{k}$:
\begin{eqnarray}
\varphi _{\mathbf{k}}^{N}(\mathbf{r}) &=&\sqrt{\frac{3^{1/4}}{2^{N}N!}}%
\sum\limits_{l=-\infty }^{\infty }H_{n}(y-k_{x}-3^{1/4}\pi ^{1/2}l)
\label{HLL_functions_III} \\
&&\times e^{i[\frac{\pi l^{2}}{2}+3^{1/4}\pi ^{1/2}\left( x+k_{y}\right)
l+xk_{x}]-\frac{1}{2}[y-k_{x}-3^{1/4}\pi ^{1/2}l]^{2}}.  \notag
\end{eqnarray}%
The construction is identical to LLL. Even in the homogeneous liquid state,
which is obviously more symmetric than the hexagonal lattice, we find it
convenient to use this basis:
\begin{equation}
\psi (r)=\frac{1}{\left( 2\pi \right) ^{3/2}}\int_{k}\sum_{N=0}^{\infty
}\varphi _{k}^{N}(r)\psi _{k}^{N}\text{.}  \label{psi_via_quasimomenta_III}
\end{equation}

\emph{\textbf{Energy in the quasi - momentum basis}}

As was discussed in section II, the lowest energy configurations belong to
LLL. There is an energy gap to any $N>0$ configuration, so it is reasonable
that, for temperatures small enough, their contribution is small.
Restricting the set of states over which we integrate to LLL

\begin{equation}
\psi \left( r\right) =\frac{1}{\left( 2\pi \right) ^{3/2}}\int_{k}\varphi
_{k}\left( r\right) \psi _{k},  \label{LLL_restriction_III}
\end{equation}%
one has the Boltzmann factor $\frac{1}{2^{5/2}\pi }f\left[ \psi \right] $ ,
eq.(\ref{f_definition_III}), and other physical quantities via new variables
$\psi _{k}$. The first two terms in eq.(\ref{f_definition_III},) are simple%
\begin{eqnarray}
f_{0}\left[ \psi \right] &=&\frac{1}{\left( 2\pi \right) ^{3}}\int_{k}\left(
k_{z}^{2}/2+a_{T}\right) \int_{r}\varphi _{k}^{\ast }\left( r\right) \varphi
_{l}\left( r\right) \psi _{k}^{\ast }\psi _{l}  \notag \\
&=&\int_{k}\left( k_{z}^{2}/2+a_{T}\right) \psi _{k}^{\ast }\psi _{k}.
\label{f0_III}
\end{eqnarray}

The quartic term is

\begin{eqnarray}
f_{int}\left[ \psi \right] &=&\frac{L_{x}L_{y}}{2\left( 2\pi \right) ^{5}}%
\int_{k,l,k^{\prime },l^{\prime }}\delta \left( k_{z}+l_{z}-k_{z}^{\prime
}-l_{z}^{\prime }\right)  \label{f_int_III} \\
&&\times \left[ \mathbf{k},\mathbf{l}|\mathbf{k}^{\prime },\mathbf{l}%
^{\prime }\right] \psi _{k}^{\ast }\psi _{l}^{\ast }\psi _{k^{\prime }}\psi
_{l^{\prime }},  \notag
\end{eqnarray}%
with
\begin{equation}
\left[ \mathbf{k},\mathbf{l}|\mathbf{k}^{\prime },\mathbf{l}^{\prime }\right]
\equiv \frac{1}{L_{x}L_{y}}\int_{\mathbf{r}}\varphi _{\mathbf{k}}^{\ast }(%
\mathbf{r})\varphi _{\mathbf{l}}^{\ast }(\mathbf{r})\varphi _{\mathbf{k}%
^{\prime }}(\mathbf{r})\varphi _{\mathbf{l}^{\prime }}(\mathbf{r})
\label{Lambda_III}
\end{equation}%
calculated in Appendix A. Generally the expression is not very simple due to
the so called "Umklapp" processes since when four quasi - momenta involved
We turn now to the first application of this basis: calculation of harmonic
excitations spectrum of the vortex lattice.

\subsection{Excitations of the vortex lattice and perturbations around it.}

\subsubsection{Shift of the field and the excitation spectrum}

\emph{\textbf{Shift of the field and diagonalization of the quadratic part}}

For negative $a_{T}$ and neglecting thermal fluctuations the minimum of
energy is achieved by choosing one of the degenerate lattice solutions, the
hexagonal lattice $\varphi _{\Delta }$ in our case. This was the main
subject of the previous section. When thermal fluctuations are weak, one can
expand in temperature around the mean field solution. The zero quasimomentum
field is then shifted by the mean field solutions. In our new LLL units we
therefore express the complex fields $\psi _{k}$ via two "shifted" real
fields ($O_{k}=O_{-k}^{\ast }$, $A_{k}=A_{-k}^{\ast }$):%
\begin{equation}
\psi _{k}=v_{0}\left( 2\pi \right) ^{3/2}\delta _{k}+\frac{c_{\mathbf{k}}}{%
\sqrt{2}}\left( O_{k}+iA_{k}\right)  \label{shift_III}
\end{equation}%
with value of the field found in section II in the LLL units being%
\begin{equation}
v_{0}=\sqrt{-\frac{a_{T}}{\beta _{\Delta }}}.  \label{v0_III}
\end{equation}%
Notations "$O$" and "$A$" indicate an analogy to optical and acoustic
phonons in atomic crystals. The constants $c_{\mathbf{k}}$ will be chosen
later and will help to diagonalize the quadratic part of the free energy.
Substituting this into the energy eqs.(\ref{f0_III}) and (\ref{f_int_III}),
one obtains a constant "mean field" energy density of section II,

\begin{equation}
\frac{f_{mf}}{vol}=-\frac{a_{T}^{2}}{2\beta _{A}},  \label{f_const_III}
\end{equation}%
while the quadratic part is
\begin{gather}
f_{2}=\frac{1}{2}\int_{k}\left[ k_{z}^{2}/2-a_{T}+2v_{0}^{2}\left\vert c_{%
\mathbf{k}}\right\vert ^{2}\beta _{\mathbf{k}}\right] \left( O_{k}^{\ast
}-iA_{k}^{\ast }\right) \times  \label{f_quad_III} \\
\left( O_{k}+iA_{k}\right) +\frac{v_{0}^{2}}{4}\int_{k}\left[ \gamma _{%
\mathbf{k}}c_{\mathbf{k}}^{2}\left( O_{k}^{\ast }+iA_{k}^{\ast }\right)
\left( O_{k}+iA_{k}\right) +c.c\right] ,  \notag
\end{gather}%
where functions,

\begin{eqnarray}
\beta _{\mathbf{k}} &=&\frac{1}{vol}\int_{\mathbf{r}}\left\vert \varphi
\left( \mathbf{r}\right) \right\vert ^{2}\left\vert \varphi _{\mathbf{k}%
}\left( \mathbf{r}\right) \right\vert ^{2}=\left[ \mathbf{0},\mathbf{k|0,k}%
\right] ,  \label{beta_gamma_III} \\
\text{ \ \ \ }\gamma _{\mathbf{k}} &=&\frac{1}{vol}\int_{\mathbf{r}}\varphi
^{\ast 2}\left( \mathbf{r}\right) \varphi _{\mathbf{k}}\left( \mathbf{r}%
\right) \varphi _{-\mathbf{k}}\left( \mathbf{r}\right) =\left[ \mathbf{0},%
\mathbf{0|k,-k}\right] ,  \notag
\end{eqnarray}%
are calculated and given explicitly in Appendix A. There is no linear
term,since we shifted by the mean field solution.

The choice
\begin{equation}
c_{\mathbf{k}}=\sqrt{\frac{\gamma _{\mathbf{k}}^{\ast }}{\left\vert \gamma _{%
\mathbf{k}}\right\vert }}  \label{ck_III}
\end{equation}%
eliminates the $OA$ terms, diagonalizing $f_{2}$:%
\begin{equation}
f_{2}=\frac{1}{2}\int_{k}\varepsilon _{k}^{O}O_{k}^{\ast }O_{k}+\varepsilon
_{k}^{A}A_{k}^{\ast }A_{k}\text{.}  \label{f2_diag_III}
\end{equation}%
The resulting spectrum is:epsilon\_III%
\begin{eqnarray}
\varepsilon _{k}^{O} &=&\mu _{O\mathbf{k}}^{2}+k_{z}^{2}/2;\text{ \ \ }%
\varepsilon _{k}^{A}=\mu _{A\mathbf{k}}^{2}+k_{z}^{2}/2;  \label{epsilon_III}
\\
\mu _{O\mathbf{k}}^{2} &=&a_{T}+v_{0}^{2}\left( 2\beta _{\mathbf{k}%
}+\left\vert \gamma _{\mathbf{k}}\right\vert \right) =-\frac{a_{T}}{\beta
_{\Delta }}\left( 2\beta _{\mathbf{k}}+\left\vert \gamma _{\mathbf{k}%
}\right\vert -\beta _{\Delta }\right) ;  \notag \\
\text{ \ \ \ }\mu _{A\mathbf{k}}^{2} &=&a_{T}+v_{0}^{2}\left( 2\beta _{%
\mathbf{k}}-\left\vert \gamma _{\mathbf{k}}\right\vert \right) =-\frac{a_{T}%
}{\beta _{\Delta }}\left( 2\beta _{\mathbf{k}}-\left\vert \gamma _{\mathbf{k}%
}\right\vert -\beta _{\Delta }\right) \text{.}  \notag
\end{eqnarray}%
The cubic and quartic terms describing the anharmonicities or interactions
of the excitations (phonons) are%
\begin{eqnarray}
f_{3} &=&\int_{k,l,m}\delta \left( k_{z}-l_{z}-m_{z}\right) \Lambda
_{3}\left( \mathbf{k},\mathbf{l},\mathbf{m}\right) \times \\
&&\left[ \left( O_{k}^{\ast }-iA_{k}^{\ast }\right) \left(
O_{l}+iA_{l}\right) \left( O_{m}+iA_{m}\right) +c.c.\right] ;  \notag
\end{eqnarray}%
\begin{gather}
f_{4}=\int_{k,l,k^{\prime },l^{\prime }}\delta \left(
k_{z}-l_{z}+k_{z}^{\prime }-l_{z}^{\prime }\right) \Lambda _{4}\left(
\mathbf{k},\mathbf{l},\mathbf{k}^{\prime },\mathbf{l}^{\prime }\right)
\label{f4_III} \\
\times \left( O_{k}^{\ast }-iA_{k}^{\ast }\right) \left( O_{l}+iA_{l}\right)
\left( O_{k^{\prime }}^{\ast }-iA_{k^{\prime }}^{\ast }\right) \left(
O_{l^{\prime }}+iA_{l^{\prime }}\right) ,  \notag
\end{gather}%
where%
\begin{equation}
\Lambda _{3}\left( \mathbf{k},\mathbf{l},\mathbf{m}\right) \equiv v_{0}\frac{%
L_{x}L_{y}}{2^{5}\pi ^{7/2}}\left[ \mathbf{k},\mathbf{0}|\mathbf{l},\mathbf{m%
}\right] c_{\mathbf{k}}^{\ast }c_{\mathbf{l}}c_{\mathbf{m}}
\end{equation}%
\begin{equation}
\Lambda _{4}\left( \mathbf{k},\mathbf{l},\mathbf{k}^{\prime },\mathbf{l}%
^{\prime }\right) \equiv \frac{L_{x}L_{y}}{2^{8}\pi ^{5}}\ \left[ \mathbf{k},%
\mathbf{k}^{\prime }|\mathbf{l},\mathbf{l}^{\prime }\right] c_{\mathbf{k}%
}^{\ast }c_{\mathbf{l}}c_{\mathbf{k}^{\prime }}^{\ast }c_{\mathbf{l}^{\prime
}},  \label{Lambda1_III}
\end{equation}%
with $\left[ \mathbf{k},\mathbf{k}^{\prime }|\mathbf{l},\mathbf{l}^{\prime }%
\right] $ defined in eq.(\ref{Lambda_III}).

\emph{\textbf{Supersoft Goldstone (shear) modes}}

While the O mode is "massive" even for small quasi - momenta, the A mode is
a Goldstone boson resulting from spontaneous breaking of several continuous
symmetries and is therefore "massless". The broken symmetries include the
electric charge $U\left( 1\right) ,$ (magnetic) translations and rotations.
Spectrum of Goldstone modes is typically "soft" and quadratic in momentum.
This is indeed the case, as far as the field direction $z$ is concerned, eq.(%
\ref{epsilon_III}), but the situation in the perpendicular directions is
different \cite{Shenoy72,Eilenberger67}.

We use expansion of the functions $\beta _{\mathbf{k}}$ and\ $\gamma _{%
\mathbf{k}}$, eq.(\ref{beta_gamma_III}), derived in Appendix A:

\begin{eqnarray}
\beta _{\mathbf{k}} &=&\beta _{\Delta }-\frac{\beta _{\Delta }}{4}\mathbf{k}%
^{2}+\beta _{4\Delta }\mathbf{k}^{4};\text{ \ \ }
\label{beta_gamma_expansion_III} \\
\text{\ }\gamma _{\mathbf{k}} &=&\beta _{\Delta }-\frac{\beta _{\Delta }}{2}%
\mathbf{k}^{2}-i\beta _{\Delta }k_{x}k_{y}  \notag \\
&&+i\frac{\beta _{\Delta }k_{x}k_{y}\mathbf{k}^{2}}{2}+\frac{\beta _{\Delta }%
}{8}\left( \mathbf{k}^{4}-4k_{x}^{2}k_{y}^{2}\right)  \notag
\end{eqnarray}%
\bigskip with constants given in Appendix A, $\beta _{4\Delta }=0.132$. The
acoustic spectrum consequently has the following expansion at small momenta:
\begin{equation}
\mu _{A\mathbf{k}}^{2}=\left( 2\beta _{4\Delta }-\beta _{\Delta }/8\right)
v_{0}^{2}\left\vert \mathbf{k}\right\vert ^{4}+...=0.1a_{H}\left\vert
\mathbf{k}\right\vert ^{4}  \label{supersoft_III}
\end{equation}%
All the quadratic term cancel and the Goldstone bosons are "supersoft".

One can further investigate the structure of these supersoft modes and
identify them with "shear modes" \cite%
{Moore89,Moore92,Zhuravlev99,Zhuravlev02}. To conclude, there are many
broken continuous symmetries (translations in two directions, rotations and
the $U\left( 1\right) $ phase transformations, forming a rather uncommon in
physics Lie group) leading to a single Goldstone mode. The commutators of
the magnetic translations generators $\widehat{\mathbf{P}}$ and the $U\left(
1\right) $ generator $\widehat{Q}=1$ are (using the explicit form eq.(\ref%
{generator_II})):%
\begin{equation}
\lbrack \widehat{P}_{x},\widehat{P}_{y}]=i\widehat{Q};[\widehat{P}_{x},%
\widehat{Q}]=0;[\widehat{P}_{y},\widehat{Q}]=0,  \label{Lie algebra_III}
\end{equation}%
and form the so called Heisenberg - Weyl algebra. However the Goldstone mode
is much softer than the regular one: $\left\vert \mathbf{k}\right\vert ^{4}$
instead of $\left\vert \mathbf{k}\right\vert ^{2}$. The situation is not
entirely unique, since ferromagnetic spin waves, Tkachenko modes in
superfluid and excitations in 2D electron gas within LLL share this
property. A rigorous general derivation of the modification of the Goldstone
theorem in this case is still not available. Note also that, when the
magnetic part is not neglected, the modes become massive via a kind of
Anderson - Higgs mechanism, which gives them a small "mass" of order $%
1/\kappa ^{2}$ in our units.

This exceptional "softness" apparently should lead to an instability of the
vortex lattice against thermal fluctuations. Indeed naive calculation of the
correlator in perturbation theory shows that certain quantities including
superfluid density $\left\vert \psi \right\vert ^{2}$ are infrared (IR)
divergent \cite{Maki71}. This was even considered an indication that the
vortex lattice does not exist \cite{Nikulov95,Nikulov95b,Moore97}, despite
large body of experimental evidence, even at that time. As a result, the
perturbation theory around the Abrikosov solution was not developed beyond
the one - loop order for a long time. One could argue \cite{Brandt} that
real physics is dominated by the small mass $1/\kappa ^{2} $ of the shear
mode, acting as a cutoff that prevents IR divergencies, but basic physical
properties related to thermal fluctuations near $H_{c2}\left( T\right) $
seemed to be independent of the cutoff, especially for high $T_{c} $
superconductors. In \cite{Rosenstein99} the IR divergencies were
reconsidered and it was found that they all cancel exactly at each order in
physical quantities like free energy, magnetization etc. We therefore
systematically consider the (renormalized) perturbation theory for free
energy up to two loops and then turn to other physical quantities.

\subsubsection{Feynman diagrams. Perturbation theory to one loop.}

\emph{\textbf{Feynman diagrams for the loop expansion}}

To develop a perturbation theory, the coefficient in front of the Boltzmann
factor, eq.(\ref{f_definition_III}) is considered large%
\begin{equation}
f=\frac{1}{\alpha }\left[ f_{0}+f_{2}+f_{3}+f_{4}\right] .
\label{B.factor_pert_III}
\end{equation}%
The "small parameter" $\alpha $ is actually $1$, but will be useful to
organize the perturbation theory before the actual expansion parameter is
uncovered in the process of assembling the series. One does not have to
consider a linear in fields term $f_{1}$ since it involves only the $k=0$
Goldstone excitations and does not contribute to bulk energy density \cite%
{Jevicki77}. The free energy is calculated from eq.(\ref%
{scaled_free_energy_III}) by expanding exponent of "vertices" $f_{3}\left[
\psi \right] $ and $f_{4}\left[ \psi \right] $, so that all the integrals
become gaussian:
\begin{gather}
\emph{f}\left( a_{T}\right) =\frac{1}{\alpha }f_{mf}-2^{5/2}\pi \log {\large %
\{}\int_{\psi }e^{-\frac{1}{\alpha }\left( f_{2}+f_{3}+f_{4}\right) }{\large %
\}}  \label{free_energy_expansion_III} \\
=\frac{1}{\alpha }f_{0}-2^{5/2}\pi \log {\large \{}\int_{\psi }e^{-\frac{1}{%
\alpha }f_{2}}[1-\frac{1}{\alpha }\left( f_{3}+f_{4}\right) +...]{\large \}}
\notag \\
=\frac{1}{\alpha }f_{0}-2^{5/2}\pi \log {\large \{}\int_{\psi }e^{-\frac{1}{%
\alpha }f_{2}}{\large \}}+connected\text{ }diagrams.  \notag
\end{gather}

\begin{figure}[t]
\centering \rotatebox{0}{\includegraphics[width=0.3%
\textwidth,height=8cm]{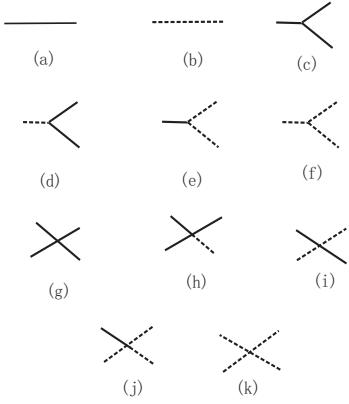}}
\caption{Feynman rules for vortex lattice}
\label{figIII1}
\end{figure}

The propagators entering Feynman diagrams (Fig. \ref{figIII1}a,b) are read
from the quadratic part, eq.(\ref{f0_III}):
\begin{equation}
G_{0}^{O,A}(k)=\alpha \frac{2^{5/2}\pi }{\epsilon _{k}^{O,A}}\text{.}
\label{G0_III}
\end{equation}%
The leading order propagators are denoted by dashed and solid lines for the $%
A$ and the\ $O$ modes respectively. Nonquadratic parts of the free energy
are the three - leg and the four - leg \textquotedblright
vertices\textquotedblright , Fig. \ref{figIII1}c-f and Fig. \ref{figIII1}g-k
respectively. It is important for disappearance of "spurious" IR
divergencies (to be discussed later) to realize that vertices involving the $%
A$ field are "soft", namely at small momentum they behave like powers of $k$%
. For example, the $A_{k}A_{l}A_{m}$ vertex, Fig. \ref{figIII1}f, is very
\textquotedblright soft\textquotedblright . At small momenta it is
proportional to the fourth power of momenta

The power of $\alpha ^{L-1}$, $L=\frac{1}{2}\left( 3N_{3}+4N_{4}\right)
-N_{3}-N_{4}+1$, where $N_{3},N_{4}$ are numbers of the three - leg and the
four - leg vertices, in front of a contribution means that topologically the
number of "loops" is $L$ \cite{Itzykson}. The leading term, the mean field
energy is of order $\alpha ^{-1}$.

\emph{\textbf{Energy to the one loop order}}

Important point to note is that in the "ordered" phase, despite the fact
that we are talking about perturbation theory, the shift or, in other words,
definition of the "physical" excitation fields $O_{k}$ and $A_{k}$ in terms
of the original fields $\psi _{k}$ can change from order to order \cite%
{Itzykson}. The shift $v$ in eq.(\ref{shift_III}) is therefore renormalized,
that is,
\begin{equation}
v^{2}=v_{0}^{2}+\alpha v_{1}^{2}+....  \label{v_renorm_III}
\end{equation}%
One finds $v_{1}$ in the same way $v_{0}$ was found, namely, by minimizing
the effective the free energy at the minimal order in which it appears. Let
us therefore explicitly write several leading contributions to the energy%
\begin{equation}
\emph{f}=\frac{1}{\alpha }\emph{f}_{0}+\emph{f}_{1}+\alpha \emph{f}_{2}+....
\label{free_energy_alpha_exp_III}
\end{equation}

We us start from the "mean field" part in eq.(\ref{free_energy_expansion_III}%
):%
\begin{gather}
\frac{f_{mf}}{vol}=\frac{1}{\alpha }\left[ -a_{T}v^{2}+\frac{1}{2}\beta
_{\Delta }v^{4}\right]  \label{mean_field_exp_III} \\
=\frac{1}{\alpha }\left[ -a_{T}v_{0}^{2}+\frac{1}{2}\beta _{\Delta }v_{0}^{4}%
\right]  \notag \\
+\left[ -a_{T}v_{1}^{2}+\beta _{\Delta }v_{0}^{2}v_{1}^{2}\right] +\alpha %
\left[ \frac{1}{2}\beta _{\Delta }v_{1}^{4}\right] +O\left( \alpha
^{2}\right) \text{.}  \notag
\end{gather}%
$\ $ The leading order is $\alpha ^{-1}$ and comes solely from the mean
field contribution, which is therefore the leading contribution in eq.(\ref%
{free_energy_expansion_III}) and coincides with eq.(\ref{f_const_III}):%
\begin{equation}
\frac{\emph{f}_{0}}{vol}=-\frac{a_{T}^{2}}{\beta _{\Delta }}.
\end{equation}%
This part of energy can also be viewed as an equation determining $v_{0}$.

Substituting $v_{0}$ into the expression in the second square bracket in eq.(%
\ref{mean_field_exp_III}) makes it zero. The only contribution to the order $%
\alpha ^{0}$ comes from the second term in eq.(\ref%
{free_energy_expansion_III}), the "trace log",$-2^{5/2}\pi \log \{\int_{\psi
}e^{-\frac{1}{\alpha }f_{2}}\}$ equal to:%
\begin{align}
& =\frac{1}{2^{3/2}\pi ^{2}}\int_{k}\left\{ \log \left[ G_{0}^{O}(k)\right]
+\log \left[ G_{0}^{A}(k)\right] \right\}  \label{trlog_III} \\
& =\frac{1}{2^{3/2}\pi ^{2}}\int_{k}\left[ \log \left( \mu _{O\mathbf{k}%
}^{2}+k_{z}^{2}/2\right) +\log \left( \mu _{A\mathbf{k}}^{2}+k_{z}^{2}/2%
\right) \right] .  \notag
\end{align}%
When we take the leading order in the expansion of the excitation spectrum
in powers of $\alpha $
\begin{eqnarray}
\left( \mu _{\mathbf{k}}^{O,A}\right) ^{2} &=&a_{T}+v^{2}\left( 2\beta _{%
\mathbf{k}}\pm \left\vert \gamma _{\mathbf{k}}\right\vert \right)
\label{spectrum expansion_III} \\
&=&a_{T}+v_{0}^{2}\left( 2\beta _{\mathbf{k}}\pm \left\vert \gamma _{\mathbf{%
k}}\right\vert \right) +\alpha v_{1}^{2}\left( 2\beta _{\mathbf{k}}\pm
\left\vert \gamma _{\mathbf{k}}\right\vert \right) +...  \notag \\
&=&\left( \mu _{0\mathbf{k}}^{O,A}\right) ^{2}+\alpha v_{1}^{2}\left( 2\beta
_{\mathbf{k}}\pm \left\vert \gamma _{\mathbf{k}}\right\vert \right) +...,
\notag
\end{eqnarray}%
the one loop energy becomes:%
\begin{eqnarray}
\frac{\emph{f}_{1}}{vol} &=&\frac{1}{2^{3/2}\pi ^{2}}\int_{k}\{\log [\left(
\mu _{0\mathbf{k}}^{O}\right) ^{2}+k_{z}^{2}/2]+\log [  \notag \\
\left( \mu _{0\mathbf{k}}^{A}\right) ^{2}+k_{z}^{2}/2]\} &=&\frac{1}{\pi }%
\int_{\mathbf{k}.}\left( \mu _{0\mathbf{k}}^{O}+\mu _{0\mathbf{k}%
}^{A}\right) =2.848\left\vert a_{T}\right\vert ^{1/2}.  \label{f1_III}
\end{eqnarray}

\subsubsection{Renormalization of the field shift and spurious infrared
divergencies.}

\emph{\textbf{Energy to two loops. Infrared divergent renormalization of the
shift}}

\begin{figure}[t]
\centering \rotatebox{0}{\includegraphics[width=0.3%
\textwidth,height=8cm]{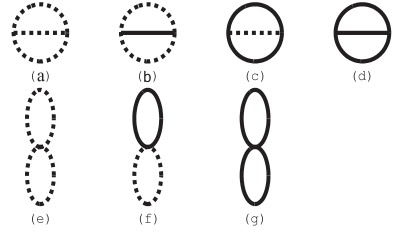}}
\caption{Two loops connected diagrams contributing to free energy.}
\label{figIII2}
\end{figure}

To order $\alpha $, corresponding to two loops, one has the first
contribution from the mean field part, which contains $v_{1}$, namely, the
third square bracket in eq.(\ref{mean_field_exp_III}). The "trace log" term,
eq.(\ref{trlog_III}), contributes due to leading correction to the
excitation spectrum eq.(\ref{spectrum expansion_III}):%
\begin{eqnarray}
&&\frac{1}{2^{3/2}\pi ^{2}}\alpha v_{1}^{2}\int_{k}[\frac{2\beta _{\mathbf{k}%
}+|\gamma _{\mathbf{k}}|}{\left( \mu _{0\mathbf{k}}^{O}\right)
^{2}+k_{z}^{2}/2}+\frac{2\beta _{\mathbf{k}}-|\gamma _{\mathbf{k}}|}{\left(
\mu _{0\mathbf{k}}^{A}\right) ^{2}+k_{z}^{2}/2}]  \notag \\
&=&\alpha v_{1}^{2}\frac{1}{2\pi }\int_{\mathbf{k}}[\frac{2\beta _{\mathbf{k}%
}+|\gamma _{\mathbf{k}}|}{\mu _{0\mathbf{k}}^{O}}+\frac{2\beta _{\mathbf{k}%
}-|\gamma _{\mathbf{k}}|}{\mu _{0\mathbf{k}}^{A}}],  \label{trlog_1_III}
\end{eqnarray}%
while the rest of the contributions in eq.(\ref{free_energy_expansion_III})
are drawn as Feynman two - loop diagrams in fig.Fig. \ref{figIII2} and
cannot contain $v_{1}$, since propagators and vertices already provide one
factor $\alpha $.\ The minimization with respect to $v_{1}^{2}$ results in:
\begin{eqnarray}
v_{1}^{2} &=&-\frac{1}{2\pi \beta _{\Delta }}\int_{\mathbf{k}}[\frac{2\beta
_{\mathbf{k}}+|\gamma _{\mathbf{k}}|}{\mu _{0\mathbf{k}}^{O}}+\frac{2\beta _{%
\mathbf{k}}-|\gamma _{\mathbf{k}}|}{\mu _{0\mathbf{k}}^{A}}]
\label{v1square_III} \\
&=&-\frac{1}{2\pi }\int_{\mathbf{k}}\frac{1}{\mu _{0\mathbf{k}}^{A}}-\frac{1%
}{2\pi \beta _{\Delta }}\int_{\mathbf{k}}[\frac{2\beta _{\mathbf{k}}+|\gamma
_{\mathbf{k}}|}{\mu _{0\mathbf{k}}^{O}}-\frac{\beta _{\Delta }}{a_{T}}\mu _{0%
\mathbf{k}}^{A}].  \notag
\end{eqnarray}

Due to additional softness of the $A$ mode $e_{0\mathbf{k}}^{A}\propto
\left\vert \mathbf{k}\right\vert ^{4}$, the first (\textquotedblright
bubble\textquotedblright )\ integral diverges logarithmically near $\mathbf{%
k\rightarrow 0}$\textbf{:}
\begin{equation}
\int d^{2}\mathbf{k}\left( \mu _{0\mathbf{k}}^{A}\right) ^{-1}\simeq \lbrack
-\frac{1}{a_{T}\left( 2\beta _{4\Delta }/\beta _{\Delta }-1/8\right) }%
]^{1/2}\int \frac{d^{2}\mathbf{k}}{\left\vert \mathbf{k}\right\vert ^{2}}%
\propto \log L.  \label{logdiv_III}
\end{equation}%
This means apparently that for the infinite infrared cutoff fluctuations
destroy the inhomogeneous ground state, namely the state with lowest energy
is a homogeneous liquid. It is plausible that since the divergence is
logarithmic, we might be at lower critical dimensionality in which an analog
of Mermin - Wagner theorem \cite{Mermin66,Itzykson} is applicable. Even this
does not necessarily means that perturbation theory starting from ordered
ground state is useless\cite{Jevicki77}. A rigorous way to proceed in these
situations have been found while considering simpler models like the "$%
\varphi ^{4}$ model", $F=\frac{1}{2}(\triangledown \varphi
^{a})^{2}+V(\varphi ^{a2})$, in $D=2$ with number of components larger then
one, say $a=1,2$. Considering the corresponding statistical sum, one first
integrates exactly zero modes, existing due to spontaneous breaking of a
continuous symmetry ($U\left( 1\right) $ in our case, field rotations in the
$\varphi ^{4}$ model) and then develops a perturbation theory via steepest
descent method for the rest of the variables. When the zero mode (the above
mentioned Goldstone boson with $k=0$) is taken out, there appears a single
configuration with lowest energy and the steepest descent is well defined.
For invariant quantities like energy this procedure simplifies: one actually
can forget for a moment about integration over zero mode and proceed with
the calculation, as if it is done in the ordered phase. The invariance of
the quantities ensures that the zero mode integration trivially factorizes.
This is no longer true for noninvariant quantities for which the machinery
of \textquotedblright collective coordinates method\textquotedblright\
should be used \cite{Rajaraman}.
In our case, we first note that the shift of the field is not a $U\left(
1\right) $ or translation invariant quantity, so invariant quantities like
energy might be still calculable. Moreover the sign of the divergence is
negative and a physically reasonable possibility that the shift decreases as
a power of cutoff:%
\begin{eqnarray}
v^{2} &\approx &v_{0}^{2}\left[ 1-\alpha c\log L+\frac{1}{2}\left( \alpha
c\log L\right) ^{2}+...\right]  \label{shift_divergence_III} \\
&\approx &v_{0}^{2}e^{-\alpha c\log L}=\frac{v_{0}^{2}}{L^{\alpha c}}.
\notag
\end{eqnarray}

\emph{\textbf{IR divergences in energy. The "nondiagrammatic" mean field and
Trlog contributions.}}

Substituting the IR divergent correction $v_{1}$, eq.(\ref{v1square_III}),
back into the free energy, eqs.(\ref{mean_field_exp_III}) and (\ref%
{trlog_1_III}), one obtains a divergent contribution for the
"nondiagrammatic" terms in eq.(free\_energy\_expansion\_III):%
\begin{eqnarray}
&&\frac{1}{2}\beta _{\Delta }v_{1}^{4}+v_{1}^{2}\frac{1}{2\pi }\int_{\mathbf{%
k}}[\frac{2\beta _{\mathbf{k}}+|\gamma _{\mathbf{k}}|}{\mu _{0\mathbf{k}}^{O}%
}+\frac{2\beta _{\mathbf{k}}-|\gamma _{\mathbf{k}}|}{\mu _{0\mathbf{k}}^{A}}]
\label{divergences_renorm_III} \\
&=&-\frac{1}{2\left( 2\pi \right) ^{2}\beta _{\Delta }}\{\int_{\mathbf{k}}%
\frac{\beta _{\Delta }}{\mu _{0\mathbf{k}}^{A}}+\int_{\mathbf{k}}(\frac{%
2\beta _{\mathbf{k}}+|\gamma _{\mathbf{k}}|}{\mu _{0\mathbf{k}}^{O}}-\frac{%
\beta _{\Delta }}{a_{T}}\mu _{0\mathbf{k}}^{A})\}^{2}  \notag
\end{eqnarray}%
containing both the $\left( \log L\right) ^{2}$%
\begin{equation}
\frac{\emph{f}_{div}^{1}}{vol}=-\frac{\beta _{\Delta }}{2\left( 2\pi \right)
^{2}}\int_{\mathbf{k,l}}\frac{1}{\mu _{0\mathbf{k}}^{A}\mu _{0\mathbf{l}}^{A}%
}  \label{nondiagram_leading_III}
\end{equation}%
and the sub leading $\log L$ divergences. However we haven't finished yet
with the order $\alpha $. They also likely to have divergences, naively even
worse than logarithmic. We therefore return to the rest of contributions to
the two loop order.

\emph{\textbf{"Setting sun" diagrams.}}

One gets several classes of diagrams on Fig. \ref{figIII2}, some of them IR
divergent. The naively most divergent diagram fig.2a actually converges. It
contains however two $AAA$ vertices, each one of them is proportional to the
fourth power of momenta. The integrals over$\ k_{z}$ and $l_{z}$ can be
explicitly performed using a formula
\begin{gather}
\frac{1}{2\pi }\int_{k_{z}}\int_{l_{z}}\frac{1}{k_{z}^{2}/2+\mu _{\mathbf{k}%
}^{2}}\frac{1}{l_{z}^{2}/2+\mu _{\mathbf{l}}^{2}}\frac{1}{%
(k_{z}+l_{z})^{2}/2+\mu _{\mathbf{k+l}}^{2}}  \notag \\
=\frac{\pi }{\mu _{\mathbf{k}}\mu _{\mathbf{l}}\mu _{\mathbf{k+l}}\left( \mu
_{\mathbf{k}}+\mu _{\mathbf{l}}+\mu _{\mathbf{k+l}}\right) }\text{.}
\label{QM_formula1_III}
\end{gather}%
The divergences appear, when one or more factors in denominator belong to
the $A$ mode for which $\mu _{\mathbf{k}}\propto \left\vert \mathbf{k}%
\right\vert ^{2}$ for small $\mathbf{k}$. However, if the numerator vanishes
at these momenta, the diagram is finite. The numerators contains vertices
involving the same "supersoft" field $A$ and typically vertices in theories
with spontaneous symmetry breaking are also soft (this fact is known in
field theoretical literature as "soft pions" theorem due to their appearance
in particle physics). In the present case they are "super soft". The $AAA$
vertex function, eq.(\ref{Lambda1_III}), is
\begin{gather}
\Lambda _{3}^{AAA}\left( \mathbf{k},\mathbf{l},\mathbf{m}\right) =\frac{%
iL_{x}L_{y}}{2^{5}\pi ^{\frac{7}{2}}}\frac{v_{0}}{3}\{c_{\mathbf{k}}^{\ast
}c_{\mathbf{l}}c_{\mathbf{m}}\left[ 0,-\mathbf{k}|\mathbf{l,m}\right]
\label{3vertex_III} \\
+c_{\mathbf{l}}^{\ast }c_{\mathbf{k}}c_{\mathbf{m}}\left[ 0,-\mathbf{l}|%
\mathbf{k,m}\right] +c_{\mathbf{m}}^{\ast }c_{\mathbf{l}}c_{\mathbf{k}}\left[
0,-\mathbf{m}|\mathbf{l,k}\right] -c_{\mathbf{k}}c_{\mathbf{l}}^{\ast }c_{%
\mathbf{m}}^{\ast }\times  \notag \\
\left[ -\mathbf{l,-m}|0,\mathbf{k}\right] -c_{\mathbf{l}}c_{\mathbf{k}%
}^{\ast }c_{\mathbf{m}}^{\ast }\left[ -\mathbf{k,-m}|0,\mathbf{l}\right] -c_{%
\mathbf{m}}c_{\mathbf{l}}^{\ast }c_{\mathbf{k}}^{\ast }\left[ -\mathbf{l,-k}%
|0,\mathbf{m}\right] \}.  \notag
\end{gather}%
One easily sees that for each of the "dangerous" momenta $\mathbf{k=0,l=0}$
or $\mathbf{m=0}$ each one of two vertices vanishes. For example when $%
\mathbf{k=0}$%
\begin{gather}
\Lambda _{3}^{AAA}\left( \mathbf{0},\mathbf{l},\mathbf{m}\right) =i\frac{%
L_{x}L_{y}}{2^{5}\pi ^{7/2}}\frac{v_{0}}{3}\{c_{\mathbf{l}}c_{\mathbf{m}}%
\left[ 0,0|\mathbf{l,m}\right] +  \notag \\
c_{\mathbf{l}}^{\ast }c_{\mathbf{m}}\left[ 0,-\mathbf{l}|\mathbf{0,m}\right]
+c_{\mathbf{m}}^{\ast }c_{\mathbf{l}}\left[ 0,-\mathbf{m}|\mathbf{l,0}\right]
-c_{\mathbf{l}}^{\ast }c_{\mathbf{m}}^{\ast }\left[ -\mathbf{l,-m}|0,\mathbf{%
0}\right]  \notag \\
-c_{\mathbf{l}}c_{\mathbf{m}}^{\ast }\left[ 0\mathbf{,-m}|0,\mathbf{l}\right]
-c_{\mathbf{m}}c_{\mathbf{l}}^{\ast }\left[ -\mathbf{l,0}|0,\mathbf{m}\right]
\}  \label{3vertex_div_III} \\
=v_{0}i\frac{1}{2^{3}\pi ^{3/2}}\frac{1}{3}\delta _{\mathbf{l}+\mathbf{m}%
}\left\{ \left\vert \gamma _{\mathbf{l}}\right\vert +2\beta _{\mathbf{l}%
}-\left\vert \gamma _{\mathbf{l}}\right\vert -2\beta _{\mathbf{l}}\right\}
=0.  \notag
\end{gather}%
This means that there are at least two powers of $k$ in the numerator and
the integral converges. There are no other power-wise divergencies left to
the two loop order. Analogous analysis of the $OOA$ vertex shows that the $%
OOA$ setting sun diagram, Fig. \ref{figIII2}c is also convergent.

Naively logarithmically divergent $AAO$ setting sun diagram, Fig. \ref%
{figIII2}b actually has both the $\log ^{2}L$ and the $\log L$ divergences.
The $AAO$ vertex function is

\begin{gather}
\Lambda _{3}^{AAO}\left( \mathbf{k},\mathbf{l},\mathbf{m}\right) =\frac{%
v_{0}L_{x}L_{y}}{2^{5}\pi ^{7/2}}\{c_{\mathbf{l}}c_{\mathbf{m}}^{\ast }c_{%
\mathbf{k}}^{\ast }\left[ -\mathbf{k,-m}|0,\mathbf{l}\right]  \label{OAA_III}
\\
+c_{\mathbf{k}}^{\ast }c_{\mathbf{l}}c_{\mathbf{m}}\left[ 0,-\mathbf{k}|%
\mathbf{l,m}\right] +c_{\mathbf{l}}^{\ast }c_{\mathbf{k}}c_{\mathbf{m}}\left[
0,-\mathbf{l}|\mathbf{k,m}\right] -c_{\mathbf{m}}^{\ast }c_{\mathbf{l}}c_{%
\mathbf{k}}\times  \notag \\
\left[ 0,-\mathbf{m}|\mathbf{l,k}\right] -c_{\mathbf{l}}^{\ast }c_{\mathbf{m}%
}c_{\mathbf{k}}^{\ast }\left[ -\mathbf{l,-k}|0,\mathbf{m}\right] +c_{\mathbf{%
l}}c_{\mathbf{m}}^{\ast }c_{\mathbf{k}}^{\ast }\left[ -\mathbf{l,-m}|0,%
\mathbf{k}\right] \}.  \notag
\end{gather}%
We will need its asymptotic when one of the momenta of the soft excitation $%
A $ is small

\begin{equation}
\Lambda _{3}^{AAO}\left( \mathbf{0},\mathbf{l},\mathbf{m}\right) =\frac{v_{0}%
}{2^{2}\pi ^{3/2}}\delta _{\mathbf{l+m}}\left\vert \gamma _{\mathbf{l}%
}\right\vert .  \label{LambdaAAO_asympt_III}
\end{equation}%
The diagram of Fig.\ref{figIII2}b, after integration over the field
direction momenta $k_{z},l_{z}$, is:
\begin{equation}
-2^{5}\pi ^{3}L_{z}\int_{\mathbf{k},\mathbf{l,m}}\frac{\Lambda
_{3}^{AAO}\left( \mathbf{k},\mathbf{l},\mathbf{m}\right) \Lambda
_{3}^{AAO}\left( -\mathbf{k},-\mathbf{l},-\mathbf{m}\right) }{\mu _{\mathbf{k%
}}^{A}\mu _{\mathbf{l}}^{A}\mu _{\mathbf{m}}^{O}\left( \mu _{\mathbf{k}%
}^{A}+\mu _{\mathbf{l}}^{A}+\mu _{\mathbf{m}}^{O}\right) }
\label{OAA_setting_sun_III}
\end{equation}

The leading divergence is determined by the asymptotics of the integrand as
both $\mathbf{k}$ and $\mathbf{l}$ approach zero. Consequently it is given
by the integral when the two vertex functions replaced with their values
taken at $\mathbf{k}=\mathbf{l=0}$ and momenta of \ $\mu _{\mathbf{k}}^{O}$
and $\left( \mu _{\mathbf{k}}^{O}+\mu _{\mathbf{l}}^{A}+\mu _{\mathbf{m}%
}^{A}\right) $ in the denominator also taken to zero. The $\log ^{2}L$
divergent part near $\mathbf{k}=\mathbf{l=0}$ is therefore
\begin{eqnarray}
\emph{f}_{div}^{2} &=&-2^{5}\pi ^{3}L_{z}\int_{\mathbf{k},\mathbf{l,m}}\frac{%
\Lambda _{3}^{AAO}\left( \mathbf{0},\mathbf{l},\mathbf{m}\right) \Lambda
_{3}^{AAO}\left( 0,-\mathbf{l},-\mathbf{m}\right) }{\mu _{\mathbf{k}}^{A}\mu
_{\mathbf{l}}^{A}\mu _{\mathbf{l}}^{O}\left( \mu _{\mathbf{0}}^{A}+\mu _{%
\mathbf{l}}^{A}+\mu _{\mathbf{l}}^{O}\right) }  \notag \\
&=&-\frac{L_{z}L_{x}L_{y}}{2^{2}\pi ^{2}}v_{0}^{2}\int_{\mathbf{k},\mathbf{l}%
}\frac{2\left\vert \gamma _{\mathbf{0}}\right\vert ^{2}}{\mu _{\mathbf{k}%
}^{A}\mu _{\mathbf{l}}^{A}\mu _{\mathbf{0}}^{O}\mu _{\mathbf{0}}^{O}}  \notag
\\
&=&-\frac{vol}{\left( 2\pi \right) ^{2}}\int_{\mathbf{k},\mathbf{l}}\frac{%
\beta _{\Delta }}{\mu _{\mathbf{k}}^{A}\mu _{\mathbf{l}}^{A}}.
\label{setsun_leading_III}
\end{eqnarray}

\emph{\textbf{The "bubble" diagrams and cancellation of the leading
divergences}}

Diagrams given in Figs .\ref{figIII2}e,f,g, can be easily evaluated:

\begin{eqnarray}
\frac{\emph{f}_{\left( e,f,g\right) }}{vol} &=&\frac{1}{\left( 2\pi \right)
^{2}}\int_{\mathbf{k,l}}\beta _{\mathbf{k-l}}\left( \frac{1}{\mu _{\mathbf{k}%
}^{O}}+\frac{1}{\mu _{\mathbf{k}}^{A}}\right) \left( \frac{1}{\mu _{\mathbf{l%
}}^{O}}+\frac{1}{\mu _{\mathbf{l}}^{A}}\right)  \notag \\
&&+\frac{1}{2\beta _{A}}\frac{1}{\left( 2\pi \right) ^{2}}\left\{ \int
\left\vert \gamma _{\mathbf{k}}\right\vert \left( \frac{1}{\mu _{\mathbf{k}%
}^{O}}-\frac{1}{\mu _{\mathbf{k}}^{A}}\right) \right\} ^{2}.
\label{diag_efg_III}
\end{eqnarray}%
The leading divergence is%
\begin{equation}
\frac{\emph{f}_{div}^{3}}{vol}=\frac{3}{2}\frac{1}{\left( 2\pi \right) ^{2}}%
\int_{\mathbf{k,l}}\beta _{\mathbf{\Delta }}\frac{1}{\mu _{\mathbf{k}%
}^{A}\mu _{\mathbf{l}}^{A}}.  \label{bubble_leading_III}
\end{equation}%
One observes that sum of three leading $\left( \log L\right) ^{2}$
divergences given in eqs. (\ref{nondiagram_leading_III}), (\ref%
{setsun_leading_III}) and (\ref{bubble_leading_III}) cancel. There are still
sub leading $\log L$ divergences. They require more care, since "not
dangerous" momenta cannot be put to zero, and are treated next.

\emph{\textbf{Cancellation of the IR divergencies}}

The two - loop contribution to energy in a "standard" form:%
\begin{equation}
\emph{f}=\frac{V}{\left( 2\pi \right) ^{2}}\int_{\mathbf{k,l}}\frac{F\left(
\mathbf{k,l}\right) }{\mu _{\mathbf{k}}^{A}\mu _{\mathbf{l}}^{A}}
\label{standard_form_III}
\end{equation}%
In order to demonstrate cancelation of the IR divergences we investigate the
value of the numerator $F\left( \mathbf{k,l}\right) $ at $\mathbf{k}=0$ and $%
\mathbf{l}=0$ and show that $F\left( \mathbf{k=0,l}\right) =0$ and $F\left(
\mathbf{k,l=0}\right) =0$. The part due to nondiagrammatic terms eq.(\ref%
{divergences_renorm_III}) can be written as:%
\begin{gather}
F^{1}\left( \mathbf{k,l}\right) =-\frac{1}{2\beta _{\mathbf{\Delta }}}[\frac{%
2\beta _{\mathbf{k}}+|\gamma _{\mathbf{k}}|}{\mu _{\mathbf{k}}^{O}}\mu _{%
\mathbf{k}}^{A}+2\beta _{\mathbf{k}}-|\gamma _{\mathbf{k}}|]  \notag \\
\times \lbrack \frac{2\beta _{\mathbf{l}}+|\gamma _{\mathbf{l}}|}{\mu _{%
\mathbf{l}}^{O}}\mu _{\mathbf{l}}^{A}+2\beta _{\mathbf{l}}-|\gamma _{\mathbf{%
l}}|]  \label{canvcel_III} \\
F^{1}\left( 0\mathbf{,l}\right) =-\frac{\mu _{\mathbf{l}}^{A}}{2}[\frac{%
2\beta _{\mathbf{l}}+|\gamma _{\mathbf{l}}|}{\mu _{\mathbf{l}}^{O}}+\frac{%
2\beta _{\mathbf{l}}-|\gamma _{\mathbf{l}}|}{\mu _{\mathbf{l}}^{A}}]  \notag
\end{gather}%
similarly, the setting sun diagram,%
\begin{gather}
F^{2}\left( \mathbf{0,l}\right) =-\frac{v_{0}^{2}2\left\vert \gamma _{%
\mathbf{l}}\right\vert ^{2}}{\mu _{\mathbf{l}}^{O}\left( \mu _{\mathbf{l}%
}^{A}+\mu _{\mathbf{l}}^{O}\right) }=-\frac{\beta _{\mathbf{\Delta }%
}v_{0}^{2}\left\vert \gamma _{\mathbf{l}}\right\vert }{\mu _{\mathbf{l}%
}^{O}\left( \mu _{\mathbf{l}}^{A}+\mu _{\mathbf{l}}^{O}\right) }\times
\notag \\
\frac{(\mu _{\mathbf{l}}^{O})^{2}-(\mu _{\mathbf{l}}^{A})^{2}}{\left\vert
a_{T}\right\vert }=-\mu _{\mathbf{l}}^{A}\left\vert \gamma _{\mathbf{l}%
}\right\vert (\frac{1}{\mu _{\mathbf{l}}^{A}}-\frac{1}{\mu _{\mathbf{l}}^{O}}%
),  \label{cancel2_III}
\end{gather}%
according to eq.(\ref{setsun_leading_III}). The divergent part of the bubble
diagrams can be written as:%
\begin{equation}
F^{3}\left( \mathbf{0,l}\right) =\frac{\mu _{\mathbf{l}}^{A}}{2}[\left\vert
\gamma _{\mathbf{l}}\right\vert (\frac{1}{\mu _{\mathbf{l}}^{A}}-\frac{1}{%
\mu _{\mathbf{l}}^{O}})+2\beta _{\mathbf{l}}(\frac{1}{\mu _{\mathbf{l}}^{A}}-%
\frac{1}{\mu _{\mathbf{l}}^{O}})].  \label{camcel3_III}
\end{equation}%
One explicitly observes that $F^{1}\left( \mathbf{0,l}\right) +F^{2}\left(
\mathbf{0,l}\right) +F^{3}\left( \mathbf{0,l}\right) =0$ . The same happens $%
F^{1}\left( \mathbf{k,0}\right) +F^{2}\left( \mathbf{k,0}\right)
+F^{3}\left( \mathbf{k,0}\right) =0$ . Therefore all the IR divergences, e.g.%
\textit{\ }the $\log L$ and $\left( \log L\right) ^{2},$ cancelled. Similar
cancellations of all the logarithmic IR divergencies occur in scalar models
with Goldstone bosons in $D=2$ and $D=3$ (where the divergencies are known
as \textquotedblright spurious\textquotedblright \cite{Jevicki77,David81}).

\emph{\textbf{Vortex lattice energy}}

The finite result for the Gibbs free energy to two loops (finite parts of
the integrals were calculated numerically). Up to two loops the calculation%
\cite{Li02,Li02a,Li02b} (extending the one carried in ref.\cite{Rosenstein99}
to Umklapp processes) gives:
\begin{equation}
f_{d}\left( a_{T}\right) =\frac{\emph{f}}{vol}=-\frac{a_{T}^{2}}{2\beta
_{\Delta }}+2.848\left\vert a_{T}\right\vert ^{1/2}+\frac{2.4}{a_{T}}\text{.}
\label{scaled_two_loop_III}
\end{equation}%
In regular units the free energy density is%
\begin{equation}
\frac{\mathcal{F}}{vol}=\frac{H_{c2}^{2}}{2\pi \kappa ^{2}}\left( \frac{%
\sqrt{Gi}bt}{4\pi }\right) ^{4/3}\left( -\frac{a_{T}^{2}}{2\beta _{\Delta }}%
+2.848\left\vert a_{T}\right\vert ^{1/2}+\frac{2.4}{a_{T}}\right) \text{.}
\label{energy_two_loop_III}
\end{equation}%
Below we use this expression to determine the melting line and various
thermal and magnetic properties of the vortex solid: magnetization, entropy,
specific heat. Near the melting point $a_{T}\approx -9.5$ the precision
becomes of order 0.1\% allowing comparison with the free energy of vortex
liquid, which is much harder to get. Eventually the (asymptotic) expansion
becomes inapplicable near the spinodal point at which the crystal is
unstable due to thermal "softening". This is considered using gaussian
approximation in subsection D.

\subsubsection{Correlators of the $U\left( 1\right) $ phase and the
structure function}

\emph{\textbf{Correlator of the $U\left( 1\right) $ phase and helicity
modulus}}

The correlator of the order parameter is finite at all distances in the
absence of thermal fluctuations%
\begin{equation}
C_{mf}\left( r,r^{\prime }\right) =\psi ^{\ast }\left( r\right) \psi \left(
r^{\prime }\right) =v_{0}^{2}\varphi ^{\ast }\left( r\right) \varphi \left(
r^{\prime }\right) ,  \label{phase_corr_mf_III}
\end{equation}%
where $v_{0}^{2}=\frac{\left\vert a_{T}\right\vert }{\beta _{\Delta }}$ is
finite exhibiting the phase long range order in the vortex lattice (despite
periodic modulation). However the order is expected to be disturbed by
thermal fluctuations. Leading order perturbation theory gives an early
indication of the loss of the order in directions perpendicular to the
field. The leading correction consists of the $\alpha $ correction to the
shift $v$, eq.(\ref{v1square_III}) and sum of two propagators%
\begin{gather}
C\left( r,r^{\prime }\right) =\int_{k,l}\varphi _{k}^{\ast }\left( r\right)
\varphi _{l}\left( r^{\prime }\right) \times  \label{corr_phase_III} \\
<[v\delta _{k}+\frac{c_{\mathbf{k}}^{\ast }}{\sqrt{2}\left( 2\pi \right)
^{3/2}}\left( O_{k}-iA_{k}\right) ][v\delta _{l}+\frac{c_{\mathbf{l}}}{\sqrt{%
2}\left( 2\pi \right) ^{3/2}}  \notag \\
\times \left( O_{l}+iA_{l}\right) ]>=C^{\left( 0\right) }\left( r,r^{\prime
}\right) +\alpha \{v_{1}^{2}\varphi ^{\ast }\left( r\right) \varphi \left(
r^{\prime }\right) +  \notag \\
\frac{1}{2\left( 2\pi \right) ^{3}}\int_{k}\left[ G_{0}^{A}\left( k\right)
+G_{0}^{O}\left( k\right) \right] \varphi _{k}^{\ast }\left( r\right)
\varphi _{k}\left( r^{\prime }\right) \}+O\left( \alpha ^{2}\right) .  \notag
\end{gather}%
One observes that the logarithmic divergences of the second and the third
terms cancel, but that the correlator in the $x-y$ plane depends on the
large distance $\mathbf{r}-\mathbf{r}^{\prime }$ as a log:%
\begin{gather}
C\left( \mathbf{r},z=0,\mathbf{r}^{\prime },z^{\prime }=0\right) \sim
v_{0}^{2}\varphi ^{\ast }\left( \mathbf{r}\right) \varphi \left( \mathbf{r}%
^{\prime }\right)  \label{convergent_corr_III} \\
\times \{1+\frac{\alpha }{2\pi }\int_{k}\left[ e^{i\mathbf{k\cdot }\left(
\mathbf{r}-\mathbf{r}^{\prime }\right) }-1\right] \frac{1}{\mu _{0\mathbf{k}%
}^{A}}\}  \notag \\
=v_{0}^{2}\varphi ^{\ast }\left( \mathbf{r}\right) \varphi \left( \mathbf{r}%
^{\prime }\right) \left[ 1+\alpha c\log \left\vert \mathbf{r}-\mathbf{r}%
^{\prime }\right\vert \right] .  \notag
\end{gather}%
It is expected, that exactly as the expectation value $v$ dependence on IR
cutoff, the actual correlator is not growing logarithmically, by rather
decreasing as a power $\left\vert \mathbf{r}-\mathbf{r}^{\prime }\right\vert
^{-c}$. This is an example of the Berezinsky - Kosterlitz - Thouless
phenomenon \cite{Itzykson}. It appears however at rather high dimensionality
$D=3$. Note however that the LLL constraint (large magnetic fields)
effectively reduce dimensionality, enhancing the role of thermal
fluctuations.

In the direction parallel to the field the correlations are still long
range. Indeed the helicity modulus is%
\begin{gather}
C\left( \mathbf{r}=\mathbf{0},z,\mathbf{r}^{\prime }=\mathbf{0},z^{\prime
}\right) \sim v_{0}^{2}\times  \label{Helicity_III} \\
\{1+\frac{\alpha 4\pi \sqrt{2}}{2\left( 2\pi \right) ^{3}}\int_{k}\left[
e^{ik_{z}\left( z-z^{\prime }\right) }-1\right] \frac{1}{\left( \mu _{0%
\mathbf{k}}^{A}\right) ^{2}+k_{z}^{2}/2}\}\sim v_{0}^{2}.  \notag
\end{gather}

\emph{\textbf{Structure function. Definitions}}

The superfluid density correlator, defined by
\begin{equation}
\widetilde{S}(r)=\frac{1}{vol}\int_{r^{\prime }}\left\langle \left\vert \psi
\left( r^{\prime }\right) \right\vert ^{2}\left\vert \psi \left( r^{\prime
}+r\right) \right\vert ^{2}\right\rangle ,  \label{S(r)_III}
\end{equation}%
quantifies spontaneous breaking of the translational and rotational
symmetries only as in both locations the superfluid density is invariant
under the $U\left( 1\right) $ gauge transformations. This is different from
the phase correlator $\left\langle \psi ^{\ast }\left( r\right) \psi \left(
r^{\prime }\right) \right\rangle $ discussed in the previous subsection,
which decays as a power as indicated by the IR divergences. As in the case
of the $U\left( 1\right) $ phase correlations, it is easier to consider the
Fourier transform of the correlator, the structure function. Since
translational symmetry is not broken along the field direction, one can
restrict the discussion to the lateral correlations and consider partial
Fourier transform:%
\begin{equation}
S(\mathbf{q},0)=\int d\mathbf{r}e^{i\mathbf{q}\cdot \mathbf{r}}\widetilde{S}(%
\mathbf{r},r_{z}=0)  \label{S(q)_III}
\end{equation}%
In this subsection the structure function is calculated to leading order in
thermal fluctuations strength (harmonic approximation) within the LLL,
namely neglecting higher $a_{H}$ corrections. We discuss these corrections
later.

\emph{\textbf{Structure function of the vortex crystal without thermal
fluctuations.}}

Substituting the LLL mean field solution eq.(\ref{fi_final_II}) into the
definition of structure function one obtains:
\begin{eqnarray}
S_{mf}(\mathbf{q},0) &\equiv &\frac{1}{L_{x}L_{y}}\left( \frac{a_{T}}{\beta
_{A}}\right) ^{2}\int_{\mathbf{r}}e^{i\mathbf{q}\cdot \mathbf{r}}\int_{%
\mathbf{r}^{\prime }}|\varphi (\mathbf{r}^{\prime })|^{2}|\varphi (\mathbf{r}%
^{\prime }\mathbf{+r})|^{2}  \notag \\
&=&\left( \frac{a_{T}}{\beta _{A}}\right) ^{2}\frac{1}{2\pi }%
\int_{cell}|\varphi (\mathbf{r}^{\prime })|^{2}e^{-i\mathbf{q}\cdot \mathbf{r%
}^{\prime }}\int_{\mathbf{r}}|\varphi (\mathbf{r})|^{2}e^{i\mathbf{q}\cdot
\mathbf{r}}  \notag \\
&=&\left( \frac{a_{T}}{\beta _{A}}\right) ^{2}4\pi ^{2}\sum_{\mathbf{K}%
}\delta (\mathbf{q-K})e^{-\frac{\mathbf{q}^{2}}{2}},
\label{mf_structuref_III}
\end{eqnarray}%
where we made use of formulas in Appendix A and delta function peaks are
located at vectors of the reciprocal lattice. The height of the peak
decreases rapidly beyond the reciprocal magnetic length (which is our unit).
When mesoscopic thermal fluctuations are significant, they might broaden the
peaks far below the temperature at which the lattice becomes unstable (the
spinodal point).

\emph{\textbf{Leading order corrections to thermal broadening of Bragg peaks}%
}

The calculation of the structure function closely follows that of free
energy. The correlator is calculated using the Wick contractions:
\begin{gather}
\widetilde{S}(\mathbf{r},z=0)=\frac{1}{L_{x}L_{y}}\int_{k,l,k^{\prime
},l^{\prime },\mathbf{r}^{\prime }}\varphi _{\mathbf{k}}^{\ast }(\mathbf{r}%
^{\prime })\varphi _{\mathbf{l}}(\mathbf{r}^{\prime })\varphi _{\mathbf{k}%
^{\prime }}^{\ast }(\mathbf{r}^{\prime }+\mathbf{r})  \notag \\
\times \varphi _{\mathbf{l}^{\prime }}(\mathbf{r}^{\prime }+\mathbf{r}%
)[v\delta _{k}+\frac{c_{\mathbf{k}}^{\ast }}{\sqrt{2}\left( 2\pi \right)
^{3/2}}\left( O_{k}-iA_{k}\right) ][v\delta _{l}  \notag \\
+\frac{c_{\mathbf{l}}}{\sqrt{2}\left( 2\pi \right) ^{3/2}}\left(
O_{l}+iA_{l}\right) ]\times \lbrack v\delta _{k^{\prime }}+\frac{c_{\mathbf{k%
}^{\prime }}^{\ast }}{\sqrt{2}\left( 2\pi \right) ^{3/2}}\times  \notag \\
\left( O_{k^{\prime }}-iA_{k^{\prime }}\right) ][v\delta _{l^{\prime }}+%
\frac{c_{\mathbf{l}^{\prime }}}{\sqrt{2}\left( 2\pi \right) ^{3/2}}\left(
O_{l^{\prime }}+iA_{l^{\prime }}\right) ].  \label{therm_structuref_III}
\end{gather}%
The leading order ($\alpha ^{0}$) term is the mean field part, eq.(\ref%
{mf_structuref_III}), while the first order term is the harmonic fluctuation
part.\

\ The fluctuation part contains $v_{1}$ corrections term $S_{4}(\mathbf{q}%
,0) $, same in structure as the leading order but with (IR diverging)
coefficient $2\left( \frac{a_{T}}{\beta _{A}}\right) v_{1}^{2}4\pi ^{2}$
instead of $\left( \frac{a_{T}}{\beta _{A}}\right) ^{2}4\pi ^{2}$ and four
contractions (diagrams):

\begin{eqnarray}
&&\frac{v_{0}^{2}}{L_{x}L_{y}[\sqrt{2}\left( 2\pi \right) ^{3/2}]^{2}}%
\int_{k,l,k^{\prime },l^{\prime },\mathbf{r}^{\prime }}\varphi _{\mathbf{k}%
}^{\ast }(\mathbf{r}^{\prime })\varphi _{\mathbf{l}}(\mathbf{r}^{\prime
})\varphi _{\mathbf{k}^{\prime }}^{\ast }(\mathbf{r}^{\prime }+\mathbf{r})
\notag \\
&&\times \varphi _{\mathbf{l}^{\prime }}(\mathbf{r}^{\prime }+\mathbf{r}%
)\{c_{\mathbf{k}}^{\ast 2}\left( G_{0k}^{O}-G_{0k}^{A}\right) \delta
_{l}\delta _{l^{\prime }}\delta _{k+k^{\prime }}+
\label{strucf_contractions_III} \\
&&c_{\mathbf{l}}^{2}\left( G_{0l}^{O}-G_{0l}^{A}\right) \delta _{k}\delta
_{k^{\prime }}\delta _{l+l^{\prime }}+2\left( G_{0k}^{O}+G_{0k}^{A}\right)
\delta _{k^{\prime }}\delta _{l^{\prime }}\delta _{k+l}  \notag \\
&&+2\left( G_{0k}^{O}+G_{0k}^{A}\right) \delta _{l}\delta _{k^{\prime
}}\delta _{k+l^{\prime }}\}\text{.}  \notag
\end{eqnarray}%
Performing integrations and Fourier transforms using methods described in
Appendix A the first two contribution are:
\begin{eqnarray}
S_{1}(\mathbf{q},0) &=&\frac{4\pi a_{T}}{\beta _{A}}\cos \left( k_{x}k_{y}+%
\mathbf{k}\times \mathbf{Q}+\theta _{\mathbf{k}}\right)  \label{S1_III} \\
&&\times e^{-\frac{\mathbf{q}^{2}}{2}}\left[ \left( \mu _{0\mathbf{k}%
}^{O}\right) ^{-1}-\left( \mu _{0\mathbf{k}}^{A}\right) ^{-1}\right]  \notag
\end{eqnarray}%
where $\mathbf{Q}$ and $\mathbf{k}$ are the "integer" and the "fractional"
parts of $\mathbf{q}$, \ in a sense $\mathbf{q}=\mathbf{k}+$\ $\mathbf{Q}$
for $\mathbf{k}$ inside Brillouin zone and\ $\mathbf{Q}$ on the reciprocal
lattice. The third term is
\begin{equation}
S_{2}(\mathbf{q},0)=\frac{4\pi a_{T}}{\beta _{A}}e^{-\frac{\mathbf{q}^{2}}{2}%
}\left[ \left( \mu _{0\mathbf{k}}^{O}\right) ^{-1}+\left( \mu _{0\mathbf{k}%
}^{A}\right) ^{-1}\right] ,  \label{S2_III}
\end{equation}%
while the last is:
\begin{eqnarray}
S_{3}(\mathbf{q},0) &=&\frac{4\pi a_{T}}{\beta _{A}}\delta _{n}(\mathbf{q}%
)e^{-\frac{\mathbf{q}^{2}}{2}}\int_{\mathbf{k}}\cos (\mathbf{k}\times
\mathbf{Q})  \label{S3_III} \\
&&\times \left[ \left( \mu _{0\mathbf{k}}^{O}\right) ^{-1}+\left( \mu _{0%
\mathbf{k}}^{A}\right) ^{-1}\right] .  \notag
\end{eqnarray}

\emph{\textbf{Cancellation of the infrared divergences}}

Although all of the four terms $S_{1}$, $S_{2},$ $S_{3}$ and $S_{4}$ are
divergent as any of the peaks is approached, $\mathbf{k}\rightarrow \mathbf{0%
}$, the sums $S_{1},S_{2}$ and $S_{3},S_{4}$ are not. We start with the
first two:
\begin{equation}
S_{1}+S_{2}=\frac{4\pi a_{h}}{\beta _{A}}e^{-\frac{\mathbf{q}^{2}}{2}}f_{1}(%
\mathbf{q}),  \label{S12_III}
\end{equation}%
where%
\begin{gather}
f_{1}(\mathbf{q})=\left[ 1+\cos \left( k_{x}k_{y}+\mathbf{k}\times \mathbf{Q}%
+c_{k}\right) \right] \left( \mu _{0\mathbf{k}}^{O}\right) ^{-1}  \notag \\
+\left[ 1-\cos (k_{x}k_{y}+\mathbf{k}\times \mathbf{Q}+c_{\mathbf{k}})\right]
\left( \mu _{0\mathbf{k}}^{A}\right) ^{-1}  \label{f1function_III}
\end{gather}%
When $\mathbf{k}\rightarrow \mathbf{0,}$ it can be shown that $%
k_{x}k_{y}+c_{k}=O\left( \mathbf{k}^{2\cdot 2}\right) $ , thus $(k_{x}k_{y}+%
\mathbf{k}\times \mathbf{Q}+c_{k}$ $\rightarrow \mathbf{k}\times \mathbf{Q}$
, and \ $1-\cos (k_{x}k_{y}+\mathbf{k}\times \mathbf{Q}+c_{k})\rightarrow
\left( \mathbf{k}\times \mathbf{Q}\right) ^{2}$. Hence it will cancel the $%
1/k^{2}$ singularity coming from $1/\mu _{0\mathbf{k}}^{A}$. Thus $f_{1}(%
\mathbf{q})$ approaches $const.+const.\cdot \frac{\left( \mathbf{k}\times
\mathbf{Q}\right) ^{2}}{k^{2}}$ when $\mathbf{Q}\neq \mathbf{0}$, and
approaches $const.+const.\cdot k^{6}$ when $\mathbf{Q}=\mathbf{0}$.
Similarly the sum of $S_{4}(\mathbf{q},0)$ and $S_{3}(\mathbf{q},0)$ is not
divergent, although separately they are. Their sum is.
\begin{equation}
S_{3}(\mathbf{q},0)+S_{4}(\mathbf{q},0)=\frac{4\pi a_{T}^{1/2}}{\beta _{A}}%
\delta _{n}(\mathbf{q})e^{-\frac{\mathbf{q}^{2}}{2}}\left[ f_{2}(\mathbf{Q}%
)+f_{3}\right] ,  \label{S34_III}
\end{equation}%
with
\begin{eqnarray}
f_{2}(\mathbf{Q}) &=&\int_{\mathbf{k}}\left[ -1+\cos \left( \mathbf{k}\times
\mathbf{Q}\right) \right] \left[ \left( \mu _{0\mathbf{k}}^{O}\right)
^{-1}+\left( \mu _{0\mathbf{k}}^{A}\right) ^{-1}\right] ;  \notag \\
f_{3} &=&-\int_{\mathbf{k}}\left( \mu _{0\mathbf{k}}^{O}+\mu _{0\mathbf{k}%
}^{A}\right) =-8.96  \label{f2_function_III}
\end{eqnarray}

\emph{\textbf{Supersoft phonons and the "halo" shape of the Bragg peaks}}

The sum of all the four terms can be cast in the following form:
\begin{gather}
S(\mathbf{q},0)=\frac{4\pi ^{2}}{\beta _{A}}\frac{a_{T}^{2}}{\beta _{A}}%
\delta _{n}(\mathbf{q})e^{-\frac{\mathbf{q}^{2}}{2}}+\frac{4\pi a_{T}^{1/2}}{%
\beta _{A}}\times  \label{fullS_III} \\
e^{-\frac{\mathbf{q}^{2}}{2}}\left[ f_{1}(\mathbf{q})+\delta _{n}(\mathbf{q}%
)f_{2}(\mathbf{Q})+\delta _{n}(\mathbf{q})f_{3}\right] ;  \notag
\end{gather}%
The results were compared\cite{Li02} with numerical simulation of the LLL
system in \cite{Sasik95}. For reciprocal lattice vectors close to origin the
value of $f_{2}(\mathbf{Q})$ are:

\begin{center}
\textbf{Table 1}

Values of $f_{2}(\mathbf{Q})$ \ with small $n_{1,}n_{2}$.

\begin{tabular}{|c|c|c|c|}
\hline
$n_{1,}n_{2}$ & $(0,1),(1,0),(1,1)$ & $(1,2),(0,2),(2,2)$ & $1,3$ \\ \hline
$f_{2}(\mathbf{Q})/\left( 2\pi \right) $ & $-5.20$ & $-7.11$ & $-8.31$ \\
\hline
\end{tabular}
\end{center}

The correction to the height of the peak at $\mathbf{Q}$, $\frac{c_{1}\Delta
(\mathbf{q})}{(1+c_{1}f_{3})}f_{2}(\mathbf{Q})$, is quite small. The
theoretical prediction has roughly the same characteristic saddle shape
\textquotedblright halos\textquotedblright\ around the peaks as in MC
simulation ref. \cite{Sasik95} and experiment \cite{Lieber99}. Conversely,
MC simulation result provides the nonperturbative evidence $\mu _{0\mathbf{k}%
}^{A}\rightarrow \left\vert k\right\vert ^{2}$ for small $k_{x},k_{y}$. In
eq.(\ref{fullS_III}), \ if $\mu _{0\mathbf{k}}^{A}\rightarrow \left\vert
k\right\vert ,$ we would get a contribution from the most singular term $%
const.+const.\cdot \frac{\left( \mathbf{k}\times \mathbf{Q}\right) ^{2}}{k}$%
. This term will become constant when $k\rightarrow 0$, and we will not get
the same characteristic saddle shape \textquotedblright
halos\textquotedblright\ around the peaks as in ref. \cite{Sasik95}.
Consequently the $\mu _{0\mathbf{k}}^{A}\rightarrow \left\vert k\right\vert
^{2}$ asymptotics\ for $k\rightarrow 0$ is crucial for such characteristic
shape and thus the MC simulation result provides a nonperturbative evidence
for it.

\emph{\textbf{Magnetization profile}}

Another quantity which can be measured is the magnetic field distribution.
In addition to constant magnetic field background there are $1/\kappa ^{2}$
magnetization corrections due to field produced by supercurrent. To leading
order in $1/\kappa ^{2}$ magnetization is given by eq.(\ref%
{M_psi^2_relation_III}). The superfluid density\ $\left\langle \left\vert
\psi (r)\right\vert ^{2}\right\rangle $ is calculated as in eq.(\ref%
{corr_phase_III}):
\begin{gather}
\left\langle \left\vert \psi (\mathbf{r},z=0)\right\vert ^{2}\right\rangle =%
\frac{a_{T}}{\beta _{A}}\left\vert \varphi (\mathbf{r})\right\vert ^{2}+%
\frac{1}{2\pi }\int_{\mathbf{k}}\left\vert \varphi _{\mathbf{k}%
}(x)\right\vert ^{2}(\frac{1}{\mu _{0\mathbf{k}}^{A}}+\frac{1}{\mu _{0%
\mathbf{k}}^{O}})  \notag \\
-\;\frac{1}{2\pi }\int_{\mathbf{k}}(\frac{2\beta _{\mathbf{k}}+|\gamma _{%
\mathbf{k}}|}{\beta _{\Delta }\mu _{0\mathbf{k}}^{O}}+\frac{2\beta
_{k}-|\gamma _{k}|}{\beta _{\Delta }\mu _{0\mathbf{k}}^{A}})\left\vert
\varphi (\mathbf{r})\right\vert ^{2}.  \label{magn(r)_III}
\end{gather}%
Its Fourier transform $\rho (\mathbf{q})\equiv \int d\mathbf{r}e^{i\mathbf{q}%
\cdot \mathbf{r}}\left\langle \left\vert \psi (\mathbf{r},z=0)\right\vert
^{2}\right\rangle $ can be easily calculated:
\begin{eqnarray}
\rho (\mathbf{q}) &=&4\pi ^{2}\delta _{n}(\mathbf{q})\{\frac{a_{T}}{\beta
_{A}}+\frac{1}{2\pi }\int_{\mathbf{k}}[(e^{i\mathbf{k\times q}}-\frac{2\beta
_{\mathbf{k}}+|\gamma _{\mathbf{k}}|}{\beta _{\Delta }})  \notag \\
&&\times \left( \mu _{0\mathbf{k}}^{O}\right) ^{-1}+(e^{i\mathbf{k\times q}}-%
\frac{2\beta _{\mathbf{k}}-|\gamma _{\mathbf{k}}|}{\beta _{\Delta }})\left(
\mu _{0\mathbf{k}}^{A}\right) ^{-1}]\}  \notag \\
&&\times e^{-\frac{\mathbf{q}^{2}}{4}+i\pi n_{1}(n_{2}+1)}  \label{rho_III}
\end{eqnarray}%
Performing integrals, one obtains:
\begin{eqnarray}
\rho (\mathbf{q}) &=&4\pi ^{2}\delta _{n}(\mathbf{q})e^{-\frac{\mathbf{q}^{2}%
}{4}+i\pi n_{1}(n_{2}+1)}\times  \label{rho(q)_final_III} \\
&&\{\frac{a_{T}}{\beta _{A}}+\frac{1}{2\pi }\left[ f_{3}+f_{2}(\mathbf{Q)}%
\right] a_{T}^{-1/2}\}  \notag
\end{eqnarray}%
The function $f_{2}(\mathbf{Q)}$ and constant $f_{3}$ appeared in eq.(\ref%
{f2_function_III}).

\subsection{Basic properties of the vortex liquid. Gaussian approximation.}

\subsubsection{The high temperature perturbation theory and its shortcomings}

\emph{\textbf{The loop expansion}}

\begin{figure}[t]
\centering \rotatebox{270}{\includegraphics[width=0.3%
\textwidth,height=8cm]{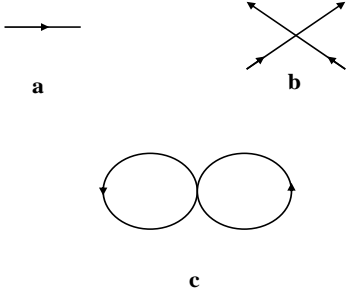}}
\caption{(a),(b)Feynman rules in the homogeneous phase;(c), the two loop
correction to energy}
\label{figIII4}
\end{figure}

Unlike the perturbation theory in the crystalline state, in which various
translational, rotational and gauge $U\left( 1\right) $ symmetries are
spontaneously broken, the perturbation theory at high temperature is quite
straightforward. One directly uses the quadratic and the quartic terms in
the Boltzmann factor, eq.(\ref{f_definition_III}) as a "large" part $K$ and
a "perturbation" $V:$%
\begin{equation}
K=\frac{1}{2^{5/2}\pi }f_{0};\text{ \ \ }V=\frac{1}{2^{5/2}\pi }f_{int}\text{%
.}  \label{K_V_III}
\end{equation}%
Again the "parameter" $\alpha $ is actually $1$, but is regarded as small
and the actual expansion parameter will become apparent shortly. The Feynman
rules for a field $\psi $, namely the propagator
\begin{equation}
G_{0}(k)=\frac{2^{5/2}\pi }{k_{z}^{2}/2+a_{T}}  \label{pert_prop_III}
\end{equation}%
and the four - point vertex are given on Fig. \ref{figIII4}a,b respectively
is defined in eq.(\ref{Lambda_III}). Since $\psi $ is a complex field, we
use an arrow to indicate the "orientation" of the propagator.

The leading contribution to the LLL scaled free energy, that of the
quadratic theory, is
\begin{eqnarray}
\emph{f}_{1} &=&2^{5/2}\pi Tr\log D^{-1}(k)=\frac{vol}{2^{1/2}\pi ^{2}}%
\int_{k\,}\log G(k)  \label{one loop_III} \\
&=&\frac{vol}{2^{1/2}\pi ^{2}}\int_{k_{z}}\log \frac{k_{z}^{2}/2+a_{T}}{%
2^{5/2}\pi }=vol\text{ }4a_{T}^{1/2}+const  \notag
\end{eqnarray}%
The $const$ in the last line is ultraviolet divergent, but unimportant for
our purposes and will be generally suppressed. Corrections can be
conveniently presented as Feynman diagrams.

The next, "two - loop", correction is the diagram in fig. \ref{figIII4}c,
which reads%
\begin{eqnarray}
\emph{f}_{2} &=&\frac{1}{\left( 2\pi \right) ^{6}}\int_{k,l\,}\Lambda \left(
k,l,k,l\right) G_{0}(k)G_{0}(l)  \label{two_loop_III} \\
&=&vol\text{ }4a_{T}^{-1}  \notag
\end{eqnarray}%
Observe that the $k_{z}$ integrations can be reduced to corresponding
integrations in quantum mechanics of the anharmonic oscillator \cite%
{Thouless75,Ruggeri76,Ruggeri78}, so that the series resemble a
dimensionally reduced $D-2=1$ field theory or quantum mechanics.

\emph{\textbf{Actual expansion parameter and the applicability range}}

This expansion can be carried to a high order after several simple tricks
are learned \cite{Thouless75,Ruggeri76,Ruggeri78,Hu-MacDonald93,Hu94}. The
result to four loops is:
\begin{equation}
f_{d}=4a_{T}^{1/2}+\frac{4}{a_{T}}-\frac{17}{2a_{T}^{5/2}}+\frac{907}{%
24a_{T}^{4}}\text{.}  \label{f_four_loops_III}
\end{equation}%
One observes that the small parameter is $a_{T}^{-3/2}$ although
coefficients grow and series are asymptotic. The difference with analogous
expansion in the crystalline phase is that the sign of $a_{T}$ is opposite
and the leading order is $\sqrt{a_{T}}$ rather than $a_{T}^{2}$.
Phenomenologically the region of positive large $a_{T}$ is not very
interesting since at that point, for example, magnetization is already very
small. Also higher Landau levels effects become significant as will be
discussed in subsection E, where HLL effects \cite{Prange69} are considered.

Therefore attempts were made to extend the series to smaller temperatures.
One of the simplest methods is to perform a Hartree - Fock type resummation
order by order. Let us first describe in some detail a certain variant of
this type of approximation called generally gaussian, since it will be
extensively used to treat thermal fluctuations as well as disorder effects
in the following sections. It will be shown in subsection D that the
approximation, is not just a variational scheme, but constitutes a first
approximant in a convergent series of approximants (which however are not
series in an external parameter) called "optimized perturbation theory"
(OPT).

\subsubsection{General gaussian approximation}

\emph{\textbf{Variational principle}}

We start from the simplest, one parameter version of the gaussian
approximation which is quite sufficient to describe the basic properties of
the vortex liquid well below the mean field transition point $a_{T}=0$.
Within this approximation one introduces a variational parameter $\mu $
(which is physically an excitation energy of the vortex liquid) adding and
subtracting a simple quadratic expression $\mu ^{2}|\psi |^{2}$ from the
Boltzmann factor:
\begin{equation}
f(\mu )=K+\alpha V  \label{K_V_division_III}
\end{equation}%
\begin{eqnarray}
K &=&\frac{1}{2^{5/2}\pi }\int_{r}\left( \mu ^{2}|\psi |^{2}+\frac{1}{2}%
|\partial _{z}\psi |^{2}\right) =\int_{k}\psi _{k}^{\ast }G^{-1}\psi _{k}
\label{K,V_expressions_III} \\
V &=&\frac{1}{2^{5/2}\pi }[a_{T}\int_{k}|\psi _{k}|^{2}+\frac{1}{2}%
\int_{k}\Lambda \left( k,l,k^{\prime },l^{\prime }\right) \psi _{k}^{\ast
}\psi _{l}\psi _{k^{\prime }}^{\ast }\psi _{l^{\prime }}],  \notag
\end{eqnarray}%
where the constant $a$ was defined by$\ $%
\begin{equation}
a\equiv a_{T}-\mu ^{2}.  \label{a_definition_III}
\end{equation}%
Now one considers $K$ as an \textquotedblright
unperturbed\textquotedblright\ part and $\alpha V$ as a small perturbation.
This is a different partition than the one we used previously to develop a
perturbation theory. Despite the fact that $\alpha =1$, we develop
perturbation theory as before. To first order in $\alpha ,$ the scaled free
energy is:
\begin{gather}
\emph{f}_{gauss}=-2^{5/2}\pi \log \left\{ \int_{\psi }e^{-K+\alpha V}\right\}
\label{f_gauss_III} \\
=-2^{5/2}\pi \log \{\int_{\psi }e^{-K+\alpha V}\}\approx -2^{5/2}\pi \log
\{\int_{\psi _{k}}e^{-K}\left[ 1+\alpha V\right] \}  \notag \\
=-2^{5/2}\pi \log \{Z_{\mu }+\alpha \int_{\psi _{k}}e^{-K}V\}\approx
-2^{5/2}\pi \lbrack \log Z_{\mu }-\alpha \left\langle V\right\rangle _{\mu }]
\notag
\end{gather}%
where $Z_{\mu }$ is the gaussian partition function $Z_{\mu }=\int_{\psi
_{k}}e^{-K}$ and thermal averages denoted by $\left\langle ..\right\rangle $
are made in this quadratic theory.

Collecting terms, one obtains
\begin{equation}
\frac{\emph{f}_{gauss}}{vol}=2\mu +\alpha \left[ 2\mu +\frac{2a}{\mu }+\frac{%
4}{\mu ^{2}}\right] .  \label{gauss_en_liquid_final_III}
\end{equation}%
Now comes the improvement. One optimizes the solvable quadratic large part
by minimizing energy for $\alpha =1$ with respect to $\mu $. The
optimization condition is called \textquotedblright gap
equation\textquotedblright ,
\begin{equation}
\mu ^{3}-a_{T}\mu -4=0,  \label{gap_eq_liq_III}
\end{equation}%
since the BCS approximation is one of the famous applications of the general
gaussian approximation.

\emph{\textbf{Existence of a metastable homogeneous state down to zero
temperature. Pseudocritical fixed point.}}

It is clear that the overheated solid becomes unstable at some finite
temperature. It not clear however whether over - cooled liquid becomes
unstable at some finite temperature (like water) or exists all the way down
to $T=0$ as a metastable state. It was shown by variety of methods that
liquid (gas) phase of the classical one component Coulomb plasma exists as a
metastable state down to zero fluctuation temperature with energy gradually
approaching that of the Madelung solid and excitation energy diminishing
\cite{Carvalho99}. It seems plausible that the same would happen with any
system of particles repelling each other with sufficiently long range
forces. In fact the vortex system in the London approximation becomes a sort
of \ repelling particles with the force even more long range than Coulombic.

\begin{figure}[t]
\centering \rotatebox{270}{\includegraphics[width=0.36%
\textwidth,height=9.6cm]{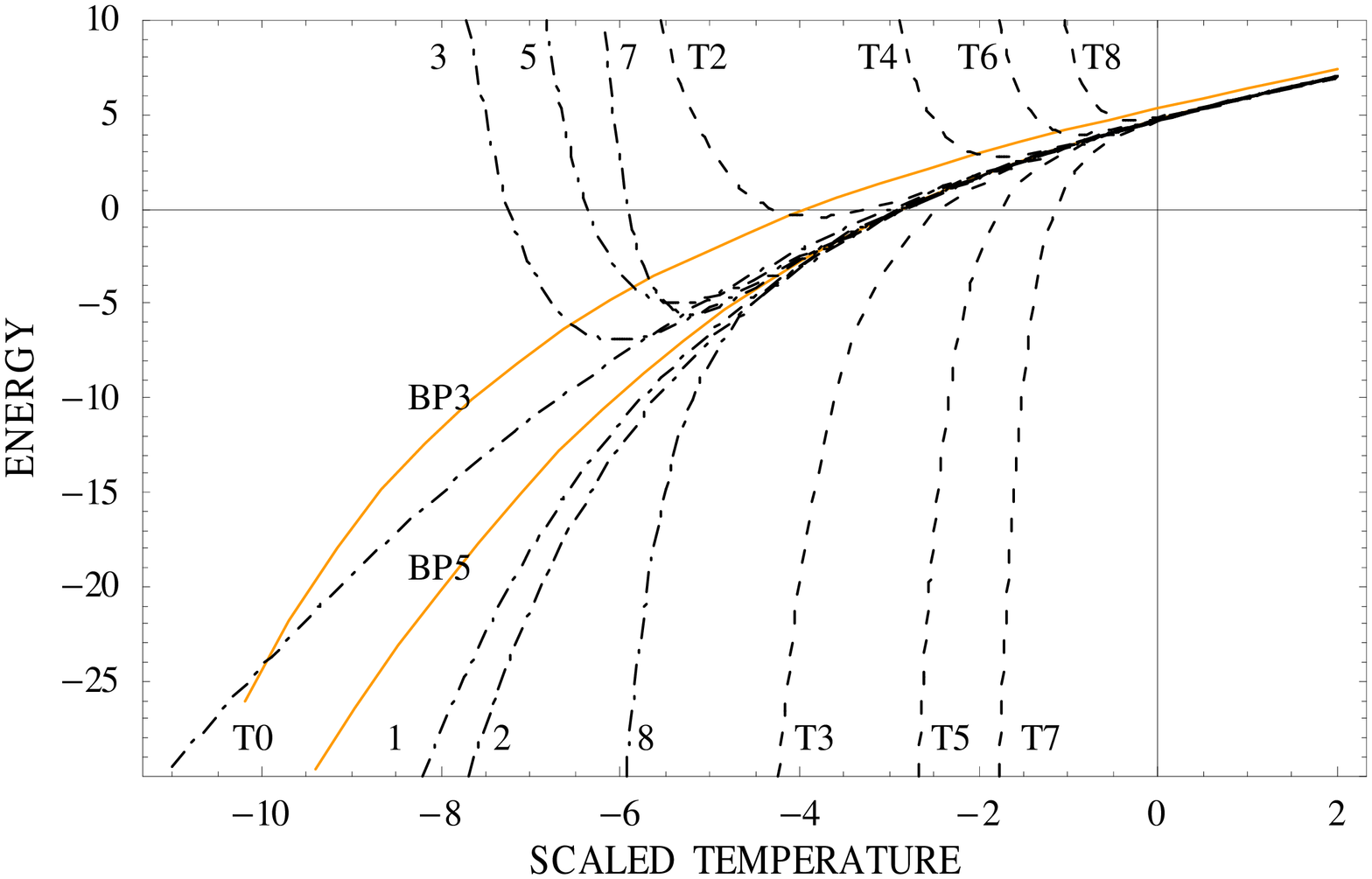}}
\caption{Free energy in liquid. The curve T0 is the gaussian approximation,
while T1,... are higher order renormalized perturbation theory results.
Optimized perturbation theory gives curves 1, 2, ... and finally BP lines
are the Borel - Pade results. }
\label{figIII6}
\end{figure}

Note that there always exists one solution of this cubic equation for
positive $\mu $ for all values of $a_{T}$, negative as well as positive. The
excitation energy in the liquid decreases asymptotically as
\begin{equation}
\mu _{a_{T}\rightarrow -\infty }\sim -\frac{4}{a_{T}}
\label{pseudocritical_excitations_III}
\end{equation}%
at temperatures approaching zero. Importantly it becomes small at the
melting point located at $a_{T}=-9.5$, see below\textbf{.} The gaussian
energy is plotted on Fig. \ref{figIII6} (marked as the $T0$ line).
The existence of the solution means that the homogeneous phase exists all
the way down to $T=0$ albeit as a metastable state below the melting point
at which the free energy of the solid is smaller, see Fig. \ref{figIII9}.
Physically this rather surprising fact is intimately related to repulsion of
the vortex lines. It is well known that if in addition to repulsion there
exists an attraction like a long range attractive forces between atoms and
molecules, they will lead to a spinodal point of the liquid \cite{Lovett77}.
However, if the attractive part is absent like in, for instance, electron
liquid, one component plasma etc., the spinodal point is pushed down to zero
temperature. It becomes a "pseudocritical" point, namely, exhibits
criticality, but globally unstable due to existence of a lower energy state
\cite{Compagner74}. Scaled LLL free energy density diverges as a power as
well
\begin{equation}
f(a_{T}\rightarrow -\infty ){\sim }-\frac{a_{T}^{2}}{2},
\label{liq_energy_asympt_III}
\end{equation}%
see Fig. \ref{figIII6}.

Assuming absence of singularities on the liquid branch allows to develop an
essentially precise theory of the LLL GL model in vortex liquid (even
including overcooled liquid) using the Borel - Pade (BP) \cite{Baker} method
at any temperature. This calculation is carried out in subsection D4. The
gaussian liquid state can be used as a starting point of "renormalized"
perturbation theory around it. Such an expansion was first developed by
Ruggeri and Thouless \cite{Thouless75,Ruggeri76,Ruggeri78} for the GL model.

\subsection{More sophisticated theories of vortex liquid.}

\subsubsection{Perturbation theory around the gaussian state}

After the variational spectrum $\mu $ was fixed, one can expand in
presumably small terms in eq.(\ref{K_V_division_III}) multiplied by $\alpha $
up to a certain order. Here we summarize the Feynman rules.

\emph{\textbf{Feynman diagrams}}

\begin{figure}[t]
\centering \rotatebox{270}{\includegraphics[width=0.3%
\textwidth,height=8cm]{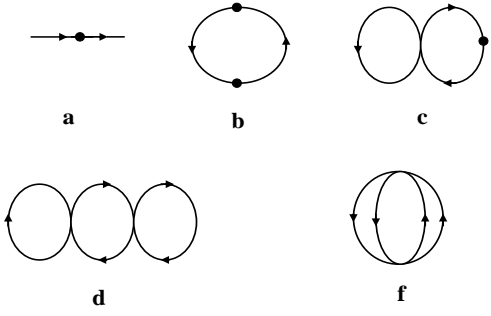}}
\caption{ Additional diagram of the renormalized perturbation theory shown
in (a). Bubbles or cacti diagrams summed by the optimized expansion are
shown in (b) - (d). A diagram which is not of that type is shown in (f).}
\label{figIII8}
\end{figure}

The propagator in the quasi - momentum space,
\begin{equation}
G_{k}=\frac{2^{5/2}\pi }{\mu ^{2}+k_{z}^{2}/2},  \label{Gk_III}
\end{equation}%
is the same as in usual perturbation theory, Fig. \ref{figIII4}a, but with
gaussian mass $\mu $. The\ four - leg interaction vertex is also the same as
Fig. \ref{figIII4}b, but there is an additional two - leg term. It has a
factor $\alpha $ and treated as a vertex and can be represented by a dot on
a line, Fig. \ref{figIII8}a, with a value of $\frac{\alpha }{2^{5/2}\pi }a$.
The second term is a four line vertex, Fig. \ref{figIII4}b, with a value of $%
\frac{\alpha }{2^{7/2}\pi }$.

To calculate the effective energy density $\ \emph{f}=-2^{5/2}\pi \ln Z$,
one draws all the connected vacuum diagrams. To the three loop order one
has:
\begin{eqnarray}
\frac{\emph{f}_{1}}{vol} &=&-\frac{1}{2\mu ^{5}}\left( 17+8a\mu +a^{2}\mu
^{2}\right) ;  \label{f1,f2_III} \\
\frac{\emph{f}_{2}}{vol} &=&\frac{1}{24\mu ^{8}}\left( 907+510a\mu
+96a^{2}\mu ^{2}+6a^{3}\mu ^{3}\right) .  \notag
\end{eqnarray}

The liquid LLL (scaled) free energy is generally written as (using the gap
equation)
\begin{equation}
\frac{\emph{f}}{vol}=4\mu \lbrack 1+g\left( x\right) ]\text{.}
\label{f_via_g_III}
\end{equation}%
The function $g$ can be expanded as
\begin{equation}
g\left( x\right) =\sum_{n=1}c_{n}x^{n},  \label{expansion_g_III}
\end{equation}%
where the high temperature small parameter is $x=\frac{1}{2}\mu ^{-3}$.\ The
coefficients $c_{n}$ which were calculated to $6^{th}$ order in \cite%
{Thouless75,Ruggeri76,Ruggeri78} and extended to $9^{th}$ order in \cite%
{Hikami90,Hikami91}. The consecutive approximants are plotted on fig.6 ($T1$
to $T9$).

\emph{\textbf{Applicability range and ways to improve it}}

One clearly sees that the series are asymptotic and can be used only at $%
a_{T}>-2$. Therefore the great effort invested in these high order
evaluations still falls short of a required values to describe the melting
of the vortex lattice. One can improve on this by optimizing the variational
parameter $\mu $ at each order instead of fixing it at the first order
calculation. This will lead in the following subsections to a convergent
series instead of the asymptotic one. The radius of convergence happens to
be around $a_{T}=-5$ short of melting and roughly at the spinodal point of
the vortex solid (see next subsection).

Another direction is to capitalize on the "pseudocritical fixed point" at
zero temperature. Indeed, the excitation energy, for example, behaves as a
power $\mu \propto a_{T}^{-1},$ other physical quantities are also
"critical", at least according to gaussian approximation. It is therefore
possible to consider supercooled liquid or liquid above the melting line but
at low enough temperature as being in the neighborhood of a pseudocritical
point. To this end the experience with critical phenomena is helpful. One
generally develops an expansion around a weak coupling unstable fixed point
(high temperature in our case) and "flows" towards a strong coupling stable
fixed point (zero temperature in our case) \cite{Itzykson}. Practically,
when higher order expansions are involved, one makes use of the
renormalization group methods in a form of the Pade - Borel resummation \cite%
{Baker}. This route will be followed in subsection 3 and will lead to a
theory valid for arbitrarily low temperature. The OPE will serve as a
consistency check on the upper range of applicability of the resummation,
which is generally hard to predict.

\subsubsection{Optimized perturbation theory.}

\emph{\textbf{General idea of the optimized gaussian perturbation theory}}

We will use a variant of OPT, the optimized gaussian series \cite{Kleinert}
to study the vortex liquid\cite{Li01,Li02,Li02a,Li02b}. It is based on the
\textquotedblright principle of minimal sensitivity\textquotedblright\ idea
\cite{Stevenson81,Okopinska87}, first introduced in quantum mechanics. Any
perturbation theory starts from dividing the Hamiltonian into a solvable
\textquotedblright large\textquotedblright\ part and a perturbation. Since
we can solve any quadratic Hamiltonian we have a freedom to choose
\textquotedblright the best\textquotedblright\ such quadratic part. Quite
generally such an optimization converts an asymptotic series into a
convergent one (see a comprehensive discussion, references and a proof in
\cite{Kleinert}). The free energy is divided into the \textquotedblright
large\textquotedblright\ quadratic part and a perturbation introducing
variational parameter $\mu \ $like for gaussian approximation, eq.(\ref%
{K_V_division_III}), although now the minimization will be made on orders of
$\alpha $ higher than the first.

Expanding the logarithm of the statistical sum to order $\alpha ^{n+1}$
\begin{gather}
Z=\int_{\psi }\exp (-K)\exp (-\alpha V)=\int_{\psi }\sum_{j=0}\frac{1}{j!}%
\left( \alpha V\right) ^{j}\exp (-K),  \notag \\
\widetilde{\emph{f}}_{n}[\mu ]=-2^{5/2}\pi \log Z=-2^{5/2}\pi
\label{OPT_def_III} \\
\times \{\log [\int_{\psi }e^{-K}]-\sum_{j=1}^{n+1}\frac{\left( -\alpha
\right) ^{j}}{j!}\left\langle V^{i}\right\rangle _{K}\},  \notag
\end{gather}%
where $\left\langle {}\right\rangle _{K}$ denotes the sum of all the
connected Feynman diagrams with $G$ as a propagator and then taking $\alpha
\rightarrow 1,$ we obtain a functional of $G$. To define the $n^{th}$ order
OPT approximant $f_{n}$ one minimizes $\widetilde{f}_{n}[G]$ with respect to
$G$:
\begin{equation}
\emph{f}_{n}=\min_{\mu }\widetilde{\emph{f}}_{n}[\mu ].  \label{min_III}
\end{equation}%
Till now the method has been applied and comprehensively investigated in
quantum mechanics only (\cite{Kleinert} and references therein) although
attempts in field theory have been made \cite%
{Duncan93,Guida95,Guida96,Bender94,Bellet96a,Bellet96b}.

\emph{\textbf{Implementation and the convergence radius in GL}}

We can obtain all the OPT diagrams which do not appear in the gaussian
theory by insertions of bubbles and the additional vertex fig1c.insertions
from the diagrams contributing to the non - optimized theory. Bubbles or
\textquotedblright cacti\textquotedblright\ diagrams, see Fig.8, are
effectively inserted into energy by a technique known in field theory \cite%
{Kleinert}. One writes $\emph{f}$ in the following form:
\begin{equation}
\emph{f}=4\mu _{1}+4\mu _{1}f\left( x\right) ,  \label{f_opt_III}
\end{equation}%
where $x=\frac{\alpha }{2\varepsilon _{1}^{3/2}}$ and $\mu _{1}$ is given by
a solution of cubic equation
\begin{equation}
\mu _{1}^{3}-\mu _{2}^{2}\mu _{1}-4\alpha =0.  \label{cubic_III}
\end{equation}%
Summing up all the insertions of the mass vertex, which now has a value of $%
\frac{\alpha }{2^{5/2}\pi }a,$ is achieved by

\begin{equation}
\varepsilon _{2}=\varepsilon +\alpha a.  \label{epsilon2_III}
\end{equation}%
\ We then expand $\emph{f}$ to order $\alpha ^{n+1},$ and then taking $%
\alpha =1$, to obtain $\emph{f}_{n}$. The solution of eq.(\ref{cubic_III})
can be obtained perturbatively in $\alpha $:
\begin{equation}
\mu _{1}=\mu _{2}+\frac{2\alpha }{\mu _{2}^{2}}-\frac{6\alpha ^{2}}{\mu
_{2}^{5}}+\frac{32\alpha ^{3}}{\mu _{2}^{8}}-\frac{210\alpha ^{4}}{\mu
_{2}^{11}}+...  \label{mu1_mu2_III}
\end{equation}

The nth OPT approximant $\emph{f}_{n}$\ is obtained by minimization of $%
\widetilde{\emph{f}}_{n}(\mu )$ with respect to $\mu $:
\begin{equation}
\left[ \frac{\partial }{\partial \left( \mu ^{2}\right) }-\frac{\partial }{%
\partial a}\right] \widetilde{\emph{f}}_{n}\left( \mu ,a\right) =0.
\label{minimization_III}
\end{equation}%
The above equation is equal to $\mu ^{-\left( 3n+4\right) }$ times a
polynomial $g_{n}\left( z\right) $ of order $n$ in $z\equiv a\mu $. That eq.(%
\ref{minimization_III}) is of this type can be seen by noting that\ the
function $\emph{f}$ depends on combination $\frac{\alpha }{\left( \mu
^{2}+\alpha a_{H}\right) ^{2}}$ only. We were unable to prove this
rigorously, but have checked it to the $40^{th}$ order in $\alpha .$ This
property simplifies greatly the task: one has to find roots of polynomials
rather than solving transcendental equations. There are $n$ (real or
complex) solutions for $g_{n}\left( z\right) =0$. However (as in the case of
anharmonic oscillator \cite{Kleinert}) \ the best root is the real root with
the smallest absolute value,.

We then obtain $\mu $ solving the cubic equation, $z_{n}=a\mu =\left(
a_{T}-\mu ^{2}\right) \mu $, explicitly:%
\begin{eqnarray}
\mu &=&2^{1/3}a_{T}\left( -27z+\sqrt{-108a_{T}^{3}+729z^{2}}\right) ^{-1/3}
\label{mu(z)_III} \\
&&+\frac{1}{32^{1/3}}\left( -27z+\sqrt{-108a_{T}^{3}+729z^{2}}\right) ^{1/3}.
\notag
\end{eqnarray}%
For $z_{0}=-4,$we obtain the gaussian result, dashed line marked
\textquotedblright T0\textquotedblright\ on Fig. \ref{figIII6}. \

Feynman rules undergo minor modifications. The mass insertion vertex, now
has a value of $\frac{\alpha }{2^{5/2}\pi }a_{H}$, while the four line
vertex is $\frac{\alpha }{2^{5/2}\pi }$. However since the propagator in the
field direction $z$ and perpendicular factorizes,\ the $k_{z}$ integrations
can be reduced to corresponding integrations in quantum mechanics of the
anharmonic oscillator, as we explained in subsection B. Expanding $\emph{f}$%
\ in $\alpha $ to order $n+1$, then one then sets $\alpha =1$ to obtain $%
\widetilde{\emph{f}}_{n}$. We list here first few OPT approximants $%
\widetilde{\emph{f}}_{n}$:%
\begin{eqnarray}
\widetilde{f}_{0} &=&4\mu +\frac{2a_{H}}{\mu }+\frac{4}{\mu ^{2}};
\label{f_tilde_III} \\
\widetilde{f}_{1} &=&\widetilde{f_{0}}-\frac{1}{2\mu ^{5}}\left(
17+8a_{H}\mu +a_{H}^{2}\mu ^{2}\right) ;  \notag \\
\widetilde{f_{2}} &=&\widetilde{f_{1}}+\frac{1}{24\mu ^{8}}\left(
907+510a_{H}\mu +96a_{H}^{2}\mu ^{2}+6a_{H}^{3}\mu ^{3}\right) ,  \notag
\end{eqnarray}%
with higher orders given in ref.\cite{Li02a}.

\emph{\textbf{Rate of convergence of OPT}}

The remarkable convergence of OPE in simple models was investigated in
numerous works \cite{Duncan93,Guida95,Guida96,Bender94,Bellet96a,Bellet96b}.
It was found that at high orders the convergence of partition function of
simple integrals (similar to the \textquotedblright zero dimensional
GL\textquotedblright\ studied in \cite{Wilkin93} ),

\begin{equation}
Z=\int_{-\infty }^{\infty }d\varphi e^{-(a\varphi ^{2}+\varphi ^{4})}
\label{toyZ_III}
\end{equation}%
is exponentially fast. The reminder in bound by \cite%
{Duncan93,Guida95,Guida96,Bender94,Bellet96a,Bellet96b}
\begin{equation}
r_{N}=|Z-Z_{N}|<c_{1}\exp [-c_{2}N].  \label{rN_III}
\end{equation}%
For quantum mechanical anharmonic oscillator (both positive and negative
quadratic term) it is just a bit slower:
\begin{equation}
R_{N}=|E-E_{N}|<c_{1}\exp [-c_{2}N^{1/3}],  \label{R_N_III}
\end{equation}%
where $E$ is the ground state energy. We follow here the convergence proof
of \cite{Duncan93,Guida95,Guida96,Bender94,Bellet96a,Bellet96b}. The basic
idea is to construct a conformal map from the original coupling $g$ to a
coupling of bounded range and isolate a nonanalytic prefactor. Suppose we
have a perturbative expansion (usually asymptotic, sometimes non Borel
summable)
\begin{equation}
E(g)=\sum_{n=0}^{\infty }c_{n}g^{n}.  \label{E_Borel_III}
\end{equation}%
One defines a set of conformal maps dependent on parameter $\rho $ of
coupling $g$ onto new coupling $\ \beta :$
\begin{equation}
\overline{g}(\beta ,\rho )=\rho \frac{\beta }{(1-\beta )^{\kappa }}.
\label{conformal_map_III}
\end{equation}%
While range of $g$ is the cut complex plane the range of $\beta $ is
compact. The value of parameter $\rho $ for each approximant will be defined
later. Then one defines a "scaled" energy
\begin{equation}
\Omega (\beta ,\rho )=(1-\beta )^{\sigma }E(\overline{g}(\beta ,\rho )),
\label{omega_III}
\end{equation}%
where the prefactor $(1-\beta )^{\sigma }$ is determined by strong coupling
limit so that $\Omega (\beta ,\rho )$ is bounded everywhere. Approximants to
$\Omega $ are expansion to $N$th order in $\beta ,$
\begin{equation}
\Omega _{N}(\beta ,\overline{\rho })=\sum_{n=0}^{N}\frac{1}{n!}\frac{%
\partial ^{n}}{\partial \beta ^{n}}\left[ (1-\beta )^{\alpha }E(\overline{g}%
(\beta ,\overline{\rho }))\right] ,  \label{OmegaN_III}
\end{equation}%
with parameter $\overline{\rho }$ substituted by $\overline{\rho }=\frac{g}{%
\beta }(1-\beta )^{\kappa }$. The energy approximant becomes
\begin{equation}
E_{N}(\beta )=\frac{\Omega _{N}(\beta )}{(1-\beta )^{\sigma }}.
\label{E_N_beta_III}
\end{equation}%
Two exponents $\ \sigma =\frac{1}{2}$ and $\kappa =\frac{3}{2}$ , for
example, anharmonic oscillator and 3D GL model. OPE is equivalent to
choosing $\beta $ which minimizes $E_{N}(\beta )$. It can be shown quite
generally (see Appendix C of paper in \cite{Bender94} and \cite{Kleinert})
that the minimization equation is a polynomial one in $\rho $ . This is in
line with our observation in the previous subsection that minimization
equations are polynomial in $z$ with $\rho $ identified as $-\frac{1}{z}.$

The remainder $R_{N}=|E-E_{N}|$ using dispersion relation is bounded by
\begin{equation}
R_{N}<c_{1}g^{\sigma /\kappa }(\overline{\rho }N^{b})^{N}+c_{2}\exp
[-N\left( \frac{\overline{\rho }}{g}\right) ^{1/\kappa }],
\label{R_N_final_III}
\end{equation}%
where exponent $b$ is determined by discontinuity of $E(g)$ at small
negative $g$:
\begin{equation}
Disc\text{ }E(g)\sim \exp [-\frac{const}{(-g)^{1/b}}].
\label{discontinuity_III}
\end{equation}%
The constants are $b=1$ for anharmonic oscillator and $b=3/4$ for 3D GL
model \cite{Thouless75,Ruggeri76,Ruggeri78}. \ For 3D GL model, we found
that $R_{N}<c_{1}\exp \left( -c_{2}N^{1/3}\right) ,$ as in anharmonic
oscillator.

\bigskip

\subsubsection{Overcooled liquid and the Borel - Pade interpolation}

\emph{\textbf{Borel - Pade resummation\bigskip}}

We have already observed using the gaussian approximation that there exists
a pseudo - critical fixed point at zero fluctuation temperature $\alpha
_{T}\rightarrow -\infty $. One can therefore attempt to use the RG "flow"
from the weak coupling point, the perturbation at high temperature to this
strongly couple fixed point. This procedure always have an element of
interpolation. It should be consistent with the perturbation theory, but
goes far beyond it. Technically it is achieved by the Borel - Pade (BP)
approximants. We will not attempt to describe the method in detail, see
textbooks \cite{Baker}, and concentrate on application.

The procedure is not unique. One starts from the renormalized perturbation
series of $g\left( x\right) $, calculated in subsection B, eq.(\ref%
{expansion_g_III}), $g\left( x\right) =\sum c_{n}x^{n}$. We will denote by $%
g_{k}\left( x\right) $ the $[k,k-1]$ BP transform of $g(x)$ (other BP
approximants clearly violate the correct low temperature asymptotics and are
not considered).\ The BP transform is defined as
\begin{equation}
\int_{0}^{\infty }g_{k}^{\prime }\left( x\text{ }t\right) \exp \left(
-t\right) dt  \label{BP_III}
\end{equation}%
where $g_{k}^{\prime }$ is the $[k,k-1]$ Pade transform of the better
convergent series
\begin{equation}
\sum_{n=1}^{2k-1}\frac{c_{n}x^{n}}{n!}.  \label{Borel_series_III}
\end{equation}%
The $[k,k-1]$ Pade transform of a function is defined as a rational function
of the form
\begin{equation}
\frac{\sum_{i=0}^{k}h_{i}x^{i}}{\sum_{i=0}^{k-1}d_{i}x^{i}},
\label{rational_III}
\end{equation}%
whose expansion up to order $2k-1$ coincides with that of the function the
series eq.(\ref{Borel_series_III}).

The results are plotted on Fig. \ref{figIII6} as solid lines for $k=3,4$ and
$5$. The lines for $k=4,5$ are practically indistinguishable on the plot.
The energy converges therefore, even at low temperatures below melting. It
describes therefore the metastable liquid up to zero temperature. Due to
inherent non - unique choice of the BP approximants it is crucial to compare
the results with convergent series (within the range of convergence). This
is achieved by comparison with the OPT results of the previous subsection.

\emph{\textbf{Comparison with other results}}

As is shown on Fig. \ref{figIII6}, the two highest available BP approximants
are consistent with the converging OPT series described above practically in
the whole range of $\alpha _{T}$. One can compare the results with existing
(not very extensive) Monte Carlo simulation and agreement is well within the
MC precision. Moreover similar method was applied to the 2D GL model which
was simulated extensively \cite%
{Kato93,Tesanovic-Xing91,ONeill93,Hu-MacDonald93,Hu94} and for which longer
series are available \cite{Hikami90,Hikami91} and agreement is still
perfect. We conclude that the method is precise enough to study the melting
problem.

Now let \ us mention several issues, which prevented its use and acceptance
early on. Ruggeri and Thouless \cite{Thouless75,Ruggeri76,Ruggeri78} tried
to use BP to calculate the specific heat without much success because their
series were too short \cite{Wilkin93}. In addition they tried to force it to
conform to the solid expression at low temperatures, which is impossible.
Attempts to use BP for calculation of melting also ran into problems.
Hikami, Fujita and Larkin \cite{Hikami90,Hikami91} tried to find the melting
point by comparing the BP energy with the one loop solid energy and obtained
$a_{T}=-7$. However their one loop solid energy was incorrect (by factor $%
\sqrt{2}$) and in any case it was not precise enough, since the two loop
contribution is essential.

To conclude the BP method and the OPT are precise enough to quantitatively
determine thermodynamic properties of the vortex liquid, including the
supercooled one. The precision is good enough in order to determine the
melting line. We therefore turn to the physical consequences of the
analytical methods for both the crystalline and the melted liquid states.

\emph{\textbf{Magnetization and specific heat in vortex liquids\bigskip}}

As long as the free energy is known, one differentiates it to calculated
other physical quantities like entropy, magnetization and specific heat
using general LLL formulas. Since the BP formulas, although analytical, are
quite bulky (and can be found in Mathematica file) we will not provide them.
The magnetization curves were compared to those in fully oxidized $YBCO$ in
\cite{Li02} to data of ref. \cite{Nishizaki00} and with $Nb$ (after
correction to a rather small $\kappa $) to data of ref. \cite{Salem02},
while the specific heat data were compared with experimental in $SnNb_{3}$
in ref. \cite{Lortz06} and in $Nb$ (also after the finite $\kappa $
correction).

\subsection{First order melting and metastable states}

\subsubsection{The melting line and discontinuity at melt}

\emph{\textbf{Location of the melting line}}

\begin{figure}[t]
\centering \rotatebox{270}{\includegraphics[width=0.36%
\textwidth,height=9.6cm]{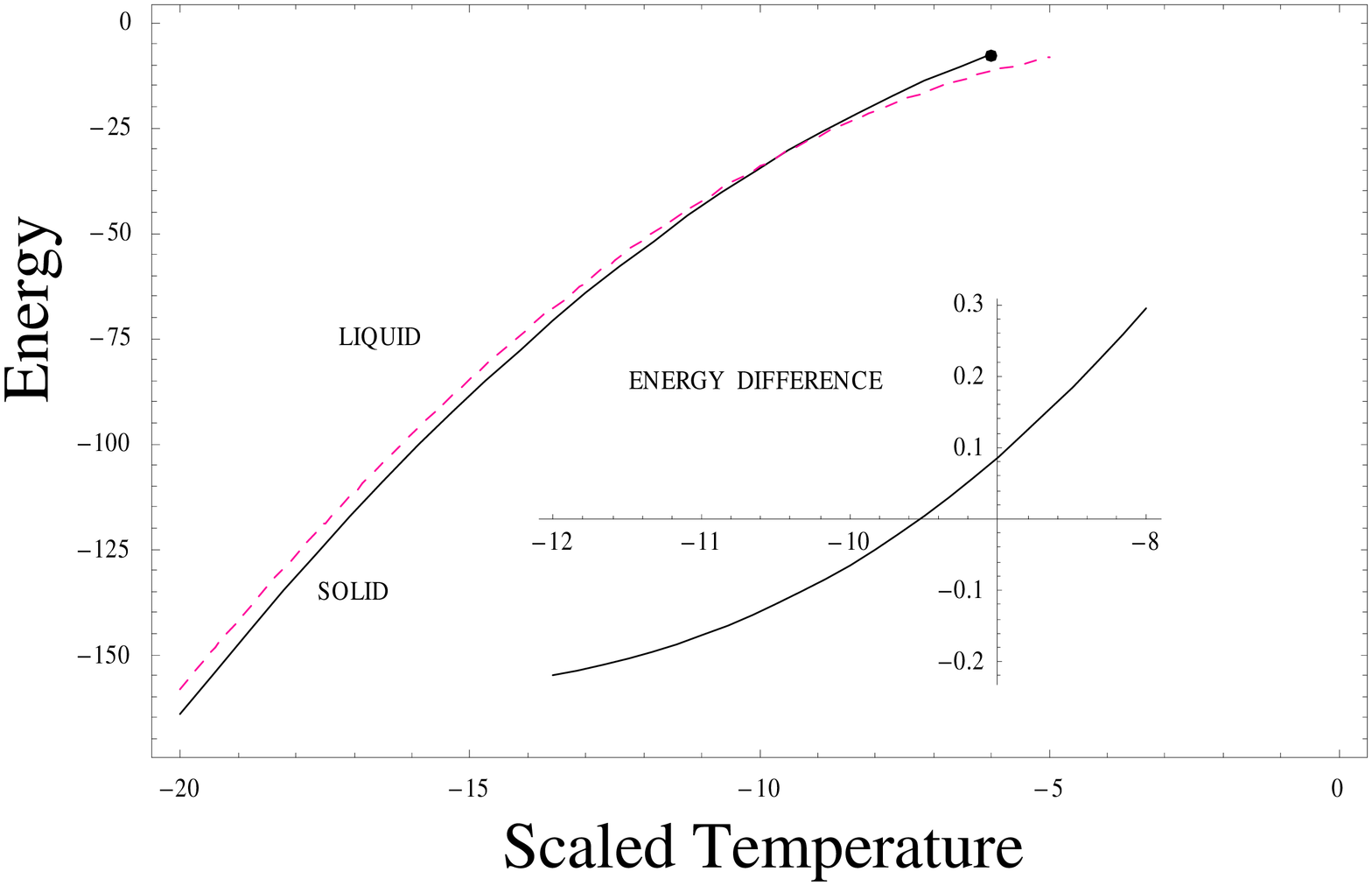}}
\caption{The melting point and the spinodal point of the crystal. Free
energy of the crystalline and the liquid states are equal at melt, while
metastable crystal becomes unstable at spinodal point.}
\label{figIII9}
\end{figure}

Comparing solid two - loop free energy given by eq.(\ref{energy_two_loop_III}%
) and liquid BP energy, Fig. \ref{figIII9}, we find that they intersect at $%
a_{T}^{m}=-9.5$ (see insert for the difference). The available 3D Monte
Carlo simulations \cite{Sasik95} unfortunately are not precise enough to
provide an accurate melting point since the LLL scaling is violated and one
gets values $a_{T}^{m}=-14.5$,$-13.2$,$-10.9$ at magnetic fields $1T$,$2T$,$%
5T$ respectively. This is perhaps due to small sample size ($\sim $100
vortices). The situation in 2D is better since the sample size is much
larger. We performed similar calculation to that in 3D for the 2D LLL GL
liquid free energy, combined it with the earlier solid energy calculation
\cite{Rosenstein99,Li02}
\begin{equation}
\frac{\emph{f}_{sol}}{vol}=-\frac{a_{T}^{2}}{2\beta _{A}}+2\log \frac{%
\left\vert a_{T}\right\vert }{4\pi ^{2}}-\frac{19.9}{a_{T}^{2}}-2.92.
\label{2D_solid_III}
\end{equation}%
and find that the melting point obtained $a_{T}^{m}=-13.2$. It is \ in good
agreement with numerous MC simulations \cite%
{Hu-MacDonald93,Hu94,Li-Nattermann03,Kato93}.

\emph{\textbf{Comparison with phenomenological Lindemann criterion and
experiments}}

Phenomenologically melting line can be located using Lindemann criterion or
its more refined version using Debye - Waller factor. The more refined
definition is required since vortices are not point - like. It was found
numerically for Yukawa gas \cite{Stevens93} that the Debye - Waller factor $%
e^{-2W}$ (ratio of the structure function at the second Bragg peak at
melting to its value at $T=0$) is about $60\%$. To one loop order one gets
using methods of \cite{Li99b} to calculate the Debye - Waller factor at the
melting line obtained here by using the non-perturbative method
\begin{equation}
e^{-2W}=0.50.  \label{Lindemann_III}
\end{equation}

The higher loop correction to this factor is supposed to be positive
and the total value might be equal to a value around $0.6$ (we did
not undertake this calculation due to the complexity). However we
apply a "one loop" criterion (Debye - Waller factor is $0.5$
calculated to one loop), and this method was applied to the layered
superconductor based on Lawrence- Doniach-
Ginzburg Landau model and the rotating Bose Einstein condensate in \cite%
{Feng09,Wu07}, and the results were both in surprisingly good agreements
with numerical calculations in \cite{Hu-MacDonald97,Cooper01}.

The melting line in accord with numerous experiments in both clean low $%
T_{c} $ materials like $NbSe_{2}$\cite%
{Kokubo04,Kokubo05,Xiao04,Adesso,Thakur,Kokubo07} and $Nb_{3}Sn$\cite%
{Lortz06}, in which the line can be inferred from the peak effect (see
below) and various dynamical effects or high $T_{c}$ like the fully oxidized
$YBa_{2}Cu_{3}O_{7}$ \cite{Nishizaki00}, see fit in \cite{Li02}. The fully
oxidized $YBCO$ is best suited for the application of the present theory,
since pinning on the mesoscopic scale is negligible. For example the melting
line is extended beyond $30T$ as shown in \cite{Li02}.

Melting lines of optimally doped untwined \cite%
{Schilling96,Schilling97,Bouquet01,Welp91,Welp96,Willemin98} $%
YBa_{2}Cu_{3}O_{7-\delta }$ and $DyBa_{2}Cu_{3}O_{7}$ \cite%
{Roulin96,Roulin96b,Revaz98} are also fitted extremely well\cite{Li03}. More
recently both $NbSe_{2}$ and thick films of $Nb_{3}Ge$ were fitted in ref.
\cite{Kokubo07} in which disorder is significant, but the pristine melting
line is believed to be clearly seen in dynamics via peak effect. In Table 2
parameters inferred from these fits are given where the data for $%
YBCO_{7-\delta }$, $YBCO_{7}$, $DyBCO_{6.7}$ are taken from \cite%
{Schilling96}, \cite{Nishizaki00}, \cite{Roulin96} respectively. Parameters
like $Gi$ characterizing the strength of thermal fluctuations differ a bit
from the often mentioned \cite{Blatter}. Similar fits were made in 2D for
organic superconductor \cite{Fruchter}. Unlike the Lindemann criterion, the
quantitative calculation allows determination of various discontinuities
across the melting line (since we have energies of both phases) to which we
turn next.

\begin{center}
\textbf{Table 2} Parameters of high $T_{c}$ superconductors deduced from the
melting line \vskip1cm

\begin{tabular}{|c|c|c|c|c|c|}
\hline
$material$ & $T_{c}$ & $H_{c2}$ & $Gi$ & $\kappa $ & $\gamma _{a}$ \\ \hline
$YBCO_{7-\delta }$ & $93.07$ & $167.53$ & $1.910^{-4}$ & $48.5$ & $7.76$ \\
\hline
$YBCO_{7}$ & $88.16$ & $175.9$ & $7.010^{-5}$ & $50$ & $4$ \\ \hline
$DyBCO_{6.7}$ & $90.14$ & $163$ & $3.210^{-5}$ & $33.77$ & $5.3$ \\ \hline
\end{tabular}
\end{center}

\subsubsection{Discontinuities at melting}

\emph{\textbf{Magnetization jump}}

The scaled magnetization is defined by $m\left( a_{T}\right) =-\frac{d}{%
da_{T}}\emph{f}\left( a_{T}\right) $ can be calculated in both phases and
the difference $\Delta m=m_{s}-m_{l}$ at the melting point $a_{T}^{m}=-9.5$
is
\begin{equation}
\frac{\Delta M}{M_{s}}=\frac{\Delta m}{m_{s}}=0.018  \label{deltaM_III}
\end{equation}%
This was compared in ref.\cite{Li03} with experimental results on fully
oxidized $YBa_{2}Cu_{3}O_{7}$ \cite{Nishizaki00} and optimally doped
untwined $YBa_{2}Cu_{3}O_{7-\delta }$ \cite{Welp91,Welp96}. These samples
probably have the lowest degree of disorder not included in calculations.

\emph{\textbf{Specific heat jump}}

In addition to the delta function like spike at melting following from the
magnetization jump discussed above experiment shows also a specific heat
jump \cite{Schilling96,Schilling97,Bouquet01,Lortz06}. The theory allows to
quantitatively estimate it. The specific heat jump is:
\begin{gather}
\Delta c=0.0075\left( \frac{2-2b+t}{t}\right) ^{2}  \label{delta_c_III} \\
-0.20Gi^{1/3}\left( b-1-t\right) \left( \frac{b}{t^{2}}\right) ^{2/3}  \notag
\end{gather}%
It was compared in ref. \cite{Li03}, with the experimental values of \cite%
{Willemin98}. See also comparison with specific heat in $NbSn_{3}$ of ref.
\cite{Lortz06}.

In addition\ the value of the specific heat jump in the 2D GL model is in
good agreement with MC simulations \cite{Hu-MacDonald93,Hu94,Kato93}, while
the 3D MC result is still unavailable.

\subsubsection{Gaussian approximation in the crystalline phase and the
spinodal line}

\emph{\textbf{Gaussian Variational Approach with shift of the field}}

Gaussian variational approach in the phase exhibiting spontaneously broken
symmetry is quite a straightforward, albeit more cumbersome, extension of
the method to include "shift" $v\left( r\right) $. In our case of one
complex field one should consider the most general quadratic form
\begin{gather}
K=\int_{r,r^{\prime }}\left[ \psi ^{\ast }(r)-v^{\ast }(r)\right]
G^{-1}(r,r^{\prime })\left[ \psi (r^{\prime })-v(r^{\prime })\right]  \notag
\\
+\left[ \psi (r)-v(r)\right] H^{\ast }(r,r^{\prime })\left[ \psi (r^{\prime
})-v(r^{\prime })\right] +c.c  \label{K_shifted_III}
\end{gather}

To obtain \textquotedblright shift\textquotedblright\ $v$ and
\textquotedblright width of the gaussian\textquotedblright\ which is a
matrix containing $G$ and $H$, one minimizes the gaussian effective free
energy \cite{Cornwall74}, which is an upper bound on the energy. Assuming
hexagonal symmetry (a safe assumption for the present purpose), the shift
should be proportional to the zero quasi - momentum function, $v(r)=v\varphi
(r),$ with a constant $v$ taken real thanks to the global $U(1)$ gauge
symmetry. On LLL, as in perturbation theory, we will use the phonon
variables $O_{k}$ and $A_{k}$ defined in quasimomentum basis eqs. (\ref%
{LLL_restriction_III}),(\ref{shift_III}) instead of $\psi (r)$%
\begin{equation}
\psi (r)=v\varphi (\mathbf{r})+\frac{1}{\sqrt{2}\left( 2\pi \right) ^{\frac{3%
}{2}}}\int_{k}e^{ik_{z}z}c_{\mathbf{k}}\varphi _{\mathbf{k}}(\mathbf{r}%
)\left( O_{k}+iA_{k}\right) .
\end{equation}%
The phase defined after eq. (\ref{ck_III}) is quite important for
simplification of the problem and was introduced for future convenience. The
most general quadratic form in these variables is
\begin{eqnarray}
K &=&\int_{k}O_{k}G_{OO}^{-1}(k)O_{-k}+A_{k}G_{AA}^{-1}(k)A_{-k}
\label{General_K_III} \\
&&+O_{k}G_{OA}^{-1}(k)A_{-k}+A_{k}G_{OA}^{-1}(k)O_{-k},  \notag
\end{eqnarray}%
with matrix of functions to be determined together with the constant$\ v$ by
the variational principle. The gaussian free energy is
\begin{gather}
\frac{\emph{f}_{gauss}}{vol}=\left( a_{T}v^{2}+\frac{\beta _{A}}{2}%
v^{4}\right) +\int_{k}2^{5/2}\pi \frac{1}{2\left( 2\pi \right) ^{3}}\times
\notag \\
\log \left[ \det (G^{-1})\right] +\frac{1}{2\left( 2\pi \right) ^{3}}%
\{\left( k_{z}^{2}/2+a_{T}\right)  \notag \\
\times \left[ G_{OO}\left( k\right) +G_{AA}\left( k\right) \right]
+v^{2}[\left( 2\beta _{\mathbf{k}}+\left\vert \gamma _{\mathbf{k}%
}\right\vert \right) G_{OO}\left( k\right)  \notag \\
+\left( 2\beta _{\mathbf{k}}-\left\vert \gamma _{\mathbf{k}}\right\vert
\right) G_{AA}\left( k\right) ]\}+\frac{1}{4\left( 2\pi \right) ^{6}}\times
\label{f_gauss_shift_III} \\
\{\frac{1}{2\beta _{\Delta }}[\int_{k}\left\vert \gamma _{\mathbf{k}%
}\right\vert \left( G_{OO}\left( k\right) -G_{AA}\left( k\right) \right)
]^{2}+4[\int_{k}\left\vert \gamma _{\mathbf{k}}\right\vert G_{OA}\left(
k\right) ]^{2}]  \notag \\
+\int_{k,l}\beta _{\mathbf{k-l}}\left[ G_{OO}\left( k\right) +G_{AA}\left(
k\right) \right] \left[ G_{OO}\left( l\right) +G_{AA}\left( l\right) \right]
\},  \notag
\end{gather}%
leading to the following minimization equations are:
\begin{gather}
v^{2}+\frac{a_{T}}{\beta _{A}}=-\frac{1}{2\left( 2\pi \right) ^{3}\beta
_{\Delta }}\int_{k}\left( 2\beta _{\mathbf{k}}+\left\vert \gamma _{\mathbf{k}%
}\right\vert \right) G_{OO}\left( k\right)  \notag \\
+\left( 2\beta _{\mathbf{k}}-\left\vert \gamma _{\mathbf{k}}\right\vert
\right) G_{AA}\left( k\right)  \label{OA_III} \\
2^{5/2}\pi \left[ G(k)^{-1}\right] _{OO}=k_{z}^{2}/2+a_{T}+v^{2}\left(
2\beta _{\mathbf{k}}+\left\vert \gamma _{\mathbf{k}}\right\vert \right) +%
\frac{1}{2\left( 2\pi \right) ^{3}}  \notag \\
\int_{l}\left( 2\beta _{\mathbf{k-l}}+\frac{\left\vert \gamma _{\mathbf{k}%
}\right\vert \left\vert \gamma _{\mathbf{l}}\right\vert }{\beta _{\Delta }}%
\right) G_{OO}\left( l\right) +\left( 2\beta _{\mathbf{k-l}}-\frac{%
\left\vert \gamma _{\mathbf{k}}\right\vert \left\vert \gamma _{\mathbf{l}%
}\right\vert }{\beta _{\Delta }}\right) G_{AA}\left( l\right)  \notag
\end{gather}

\bigskip and%
\begin{gather}
2^{5/2}\pi \left[ G(k)^{-1}\right] _{AA}=\frac{k_{z}^{2}}{2}%
+a_{T}+v^{2}\left( 2\beta _{k}-\left\vert \gamma _{\mathbf{k}}\right\vert
\right) +\frac{1}{2\left( 2\pi \right) ^{3}}\int_{l}  \notag \\
\left( 2\beta _{\mathbf{k-l}}+\frac{\left\vert \gamma _{\mathbf{k}%
}\right\vert \left\vert \gamma _{\mathbf{l}}\right\vert }{\beta _{\Delta }}%
\right) G_{AA}\left( l\right) +\left( 2\beta _{\mathbf{k-l}}-\frac{%
\left\vert \gamma _{\mathbf{k}}\right\vert \left\vert \gamma _{\mathbf{l}%
}\right\vert }{\beta _{\Delta }}\right) G_{OO}\left( l\right)  \notag \\
2^{5/2}\pi \left[ G(k)^{-1}\right] _{OA}=-\frac{2^{5/2}\pi G_{OA}(k)}{%
G_{OO}(k)G_{AA}(k)-G_{OA}(k)^{2}}  \notag \\
=4\frac{\left\vert \gamma _{\mathbf{k}}\right\vert }{\beta _{\Delta }}\frac{1%
}{2\left( 2\pi \right) ^{3}}\int_{l}\left\vert \gamma _{\mathbf{l}%
}\right\vert G_{OA}\left( l\right)
\end{gather}

These equations look quite intractable, however they can be simplified.

\emph{\textbf{How to eliminate the off - diagonal terms}}

The crucial observation is that after we have inserted the phase $c_{\mathbf{%
k}}=\sqrt{\gamma _{\mathbf{k}}/\left\vert \gamma _{\mathbf{k}}\right\vert }$
in eq. (\ref{f_gauss_shift_III}) using our experience with perturbation
theory, $G_{AO}$ appears explicitly only on the right hand side of the last
equation. It also implicitly appears on the left hand side due to a need to
invert the matrix $G$. Obviously $G_{OA}(k)=0$ is a solution and in this
case the matrix diagonalizes. However general solution can be shown to
differ from this simple one just by a global gauge transformation.
Subtracting the $OO$ equation from the $AA$ equation above, eq. (\ref{OA_III}%
) and using the $OA$ equation, we observe that matrix $G^{-1}$ has a form:
\begin{eqnarray}
G^{-1} &\equiv &\left(
\begin{array}{cc}
G_{OO}^{-1}(k) & G_{AO}^{-1}(k) \\
G_{AO}^{-1}(k) & G_{AA}^{-1}(k)%
\end{array}%
\right)  \label{inverse_prop_guess_III} \\
&=&\frac{1}{2^{5/2}\pi }\left(
\begin{array}{cc}
k_{z}^{2}/2+\mu _{O\mathbf{k}}^{2} & \mu _{AO\mathbf{k}}^{2} \\
\mu _{AO\mathbf{k}}^{2} & k_{z}^{2}/2+\mu _{A\mathbf{k}}^{2}%
\end{array}%
\right) ,  \notag
\end{eqnarray}%
with
\begin{eqnarray}
\mu _{O\mathbf{k}}^{2} &=&E_{\mathbf{k}}+\Delta _{1}\left\vert \gamma _{%
\mathbf{k}}\right\vert ;\mu _{A\mathbf{k}}^{2}=E(k)-\Delta _{1}\left\vert
\gamma _{\mathbf{k}}\right\vert ;\text{ \ }  \notag \\
\text{\ }\mu _{AO\mathbf{k}}^{2} &=&\Delta _{2}\left\vert \gamma _{\mathbf{k}%
}\right\vert  \label{Ansatz_III}
\end{eqnarray}%
where $\Delta _{1},\Delta _{2}$ are constants. Substituting this into the
gaussian energy one finds that it depends on $\Delta _{1},\Delta _{2}$ via
the combination $\Delta =\sqrt{\Delta _{1}^{2}+\Delta _{2}^{2}}$ only.
Therefore without loss of generality we can set $\Delta _{2}=0$, thereby
returning to the $G_{OA}=0$ case.

Using this observation, the gap equations significantly simplify. The
function $E_{\mathbf{k}}$ and the constant $\Delta $ satisfy:
\begin{eqnarray}
E_{\mathbf{k}} &=&a_{T}+2v^{2}\beta _{k}+2\times  \label{mode_III} \\
&<&\beta _{\mathbf{k-l}}\left( \frac{1}{\mu _{O\mathbf{l}}}+\frac{1}{\mu _{A%
\mathbf{l}}}\right) >_{\mathbf{l}};  \notag \\
\beta _{\Delta }\Delta &=&-a_{T}-2<\beta _{\mathbf{l}}\left( \frac{1}{\mu _{O%
\mathbf{l}}}+\frac{1}{\mu _{A\mathbf{l}}}\right) >_{\mathbf{l}}.
\label{delta_III}
\end{eqnarray}

\bigskip and shift equation%
\begin{equation}
v^{2}+\frac{a_{T}}{\beta _{A}}=-<\frac{2\beta _{\mathbf{k}}+\left\vert
\gamma _{\mathbf{k}}\right\vert }{\mu _{O\mathbf{k}}}+\frac{2\beta _{\mathbf{%
k}}-\left\vert \gamma _{\mathbf{k}}\right\vert }{\mu _{A\mathbf{k}}}>
\end{equation}

The gaussian energy (after integration over $k_{z}$) becomes:
\begin{gather}
\frac{\emph{f}}{vol}=v^{2}a_{T}+\frac{\beta _{A}}{2}v^{4}+\emph{f}_{1}+\emph{%
f}_{2}+\emph{f}_{3};  \notag \\
\emph{f}_{1}=<\mu _{O\mathbf{k}}+\mu _{A\mathbf{k}}>_{\mathbf{k}};  \notag \\
\emph{f}_{2}=a_{T}<\left( \mu _{O\mathbf{k}}^{-1}+\mu _{A\mathbf{k}%
}^{-1}\right) +v^{2}[\left( 2\beta _{k}+\left\vert \gamma _{k}\right\vert
\right) \mu _{O\mathbf{k}}^{-1}  \notag \\
+\left( 2\beta _{k}-\left\vert \gamma _{k}\right\vert \right) \mu _{A\mathbf{%
k}}^{-1}]>_{\mathbf{k}};  \label{Ansatz_energy_III} \\
\emph{f}_{3}=<\beta _{\mathbf{k-l}}\left( \mu _{O\mathbf{k}}^{-1}+\mu _{A%
\mathbf{k}}^{-1}\right) \left( \mu _{O\mathbf{l}}^{-1}+\mu _{A\mathbf{l}%
}^{-1}\right) >_{\mathbf{k,l}}  \notag \\
+\frac{1}{2\beta _{\Delta }}\left[ <\left\vert \gamma _{k}\right\vert \left(
\mu _{O\mathbf{l}}^{-1}-\mu _{A\mathbf{l}}^{-1}\right) >_{\mathbf{k}}\right]
^{2}.  \notag
\end{gather}%
The problem becomes quite manageable numerically after one spots an
unexpected small parameter.

\emph{\textbf{The mode expansion}}

Using a formula eq.(\ref{betaN_A})
\begin{eqnarray}
\beta _{\mathbf{k}} &=&\sum_{n=0}^{\infty }\chi ^{n}\beta _{n}(\mathbf{k})
\label{beta_mode_III} \\
\beta _{n}(\mathbf{k}) &\equiv &\sum_{\left\vert \mathbf{X}\right\vert
^{2}=na_{\Delta }^{2}}\exp [i\mathbf{k\bullet X}],  \notag
\end{eqnarray}%
derived in Appendix A and the hexagonal symmetry of the spectrum, one
deduces that $E_{\mathbf{k}}$ can be expanded in \textquotedblright
modes\textquotedblright\
\begin{equation}
E_{\mathbf{k}}=\sum E_{n}\beta _{n}(\mathbf{k})
\end{equation}%
The integer $n$ determines the distance of a points on reciprocal lattice
from the origin, and $\chi \equiv \exp [-a_{\Delta }^{2}/2]=\exp [-2\pi /%
\sqrt{3}]=0.0265$. One estimates that $E_{n}\simeq \chi ^{n}a_{T},$
therefore the coefficients decrease exponentially with $n$. Note that for
some integers, for example $n=2,5,6$, $\beta _{n}=0$. Retaining only first $%
s $ modes will be called \textquotedblright the $s$ mode
approximation\textquotedblright . We minimized numerically the gaussian
energy by varying $v,\Delta $ and first few modes of $E_{\mathbf{k}}$.

\begin{center}
\textbf{Table 3.}

Mode expansion.

\begin{tabular}{|c|c|c|c|c|}
\hline
$a_{T}$ & $-30$ & $-20$ & $-10$ & $-5.5$ \\ \hline
$\emph{f}$ & $-372.2690$ & $-159.5392$ & $-33.9873$ & $-6.5103$ \\ \hline
\end{tabular}
\end{center}

The sample results of free energy density for various $a_{T}$ with $3$ modes
are given in Table 3. In practice two modes are also quite enough. We see
that in the interesting region of not very low temperatures the energy
converges extremely fast. In practice two modes are quite enough.

\emph{\textbf{Spinodal point\bigskip}}

One can show that above
\begin{equation}
a_{T}^{spinodal}=-5.5  \label{spinodal_III}
\end{equation}%
there is no solution for the gap equations. The corresponding value in 2D is
$a_{T}^{spinodal}=-7$ and is consistent with the relaxation time measured in
Monte Carlo simulations ref.\cite{Kato93}. The spinodal point was observed
in $NbSe_{2}$ \cite{Xiao04,Thakur,Adesso} at the position consistent with
the theoretical estimate.

\emph{\textbf{Corrections to the gaussian approximation}}

The lowest order correction to the gaussian approximation (that is sometime
called the post - gaussian correction) was calculated in ref. \cite%
{Li02,Li02a,Li02b} to determine the precision of the gaussian approximation.
This is necessary in order to fit experiments and compare with low
temperature perturbation theory and other nonperturbative methods.

A general idea behind calculating systematic corrections to the gaussian
approximation was already described for liquid in subsection C and
modifications are quite analogous to those done for the gaussian
approximation. Results for the specific heat were compared in ref.\cite%
{Li02b}. Generally the post - gaussian result is valid till $a_{T}=-7 $ and
rules out earlier approximations, as the one in ref.\cite%
{Tesanovic92,Tesanovic94} (dotted line).

\section{Quenched disorder and the vortex glass.}

In any superconductor there are impurities either present naturally or
systematically produced using the proton or electron irradiation. The
inhomogeneities both on the microscopic and the mesoscopic scale greatly
affect thermodynamic and especially dynamic properties of type II
superconductors in magnetic field. The field penetrates the sample in a form
of Abrikosov vortices, which can be pinned by disorder. In addition, in high
$T_{c}$ superconductors, thermal fluctuations also greatly influence the
vortex matter, for example in some cases thermal fluctuations will
effectively reduce the effects of disorder. As a result the $T-H$ phase
diagram of the high $T_{c}$ superconductors is very complex due to the
competition between thermal fluctuations and disorder, and it is still far
from being reliably determined, even in the best studied superconductor, the
optimally doped $YBCO$ superconductor.

It is the purpose of this section to describe the glass transition and
static and thermodynamic properties of both the disordered reversible and
the irreversible glassy phase. The disorder is represented by the random
component of the coefficients of the GL free energy, eq.(\ref{disorder_I}),
and the main technique used is the replica formalism. The most general so
called hierarchical homogeneous (liquid) Ansatz \cite{Mezard91} and its
stability is considered to obtain the glass transition line and to determine
the nature of the transition for various values of the disorder strength of
the GL coefficients. In most cases the glassy phase exhibits the phenomenon
of "replica symmetry breaking, when ergodicity is lost due to trapping of
the system in multiple metastable states. In this case physical quantities
do not possess a unique value, but rather have a distribution. We start with
the case of negligible thermal fluctuations.

\subsection{Quenched disorder as a perturbation of the vortex lattice}

\subsubsection{The free energy density in the presence of pinning potential}

\emph{\textbf{GL model with $\delta T_{c}$ disorder}}

We start with space variations of the coefficient of $\left\vert \Psi
\right\vert ^{2}$, eq.(\ref{disorder_I}) distributed as a white noise, eq.(%
\ref{average_I}). It can be regarded as a local variation of $T_{c}$. As was
mentioned in section I other types of disorder are present and might be
important, however, as will be shown later are more complicated.

Since a point - like disorder breaks the translational symmetry in all
directions including that of the magnetic field $z$, one has to consider
configurations dependent on all three coordinates and take into account
anisotropy, discussed in subsection IE. We restrict to the case $m_{a}^{\ast
}=m_{b}^{\ast }\equiv m^{\ast }:$

\begin{eqnarray}
F\left[ W\right] &=&\int_{r}\frac{{\hbar }^{2}}{2m^{\ast }}\left\vert
\mathbf{D}\Psi \right\vert ^{2}+\frac{{\hbar }^{2}}{2m_{c}^{\ast }}%
\left\vert \partial _{z}\Psi \right\vert ^{2}+\alpha \left( T-T_{c}\right)
\notag \\
&&\times \left[ 1+W\left( r\right) \right] \left\vert \Psi \right\vert ^{2}+%
\frac{\beta }{2}\left\vert \Psi \right\vert ^{4},  \label{F_IV}
\end{eqnarray}%
where $W(r)$ is the $\delta T_{c}$ random disorder (real) field, which we
assume to be a white noise with variance that can be written in the
following form:%
\begin{equation}
\overline{W(r)W(r^{\prime })}=n\xi ^{2}\xi _{c}\delta ^{3}\left( r-r^{\prime
}\right) .  \label{variance_IV}
\end{equation}%
The dimensionless parameter $n$ is proportional to the density of pinning
centers and a single pin's "strength", while $\xi _{c}\equiv \xi \left(
m^{\ast }/m_{c}^{\ast }\right) ^{1/2}$ is the coherence length in the field
direction. The units we use here are the same as before with the addition of
$\xi _{c}$ as the unit of length in the $z$ direction. As in previous
sections, we will confined ourselves mainly to the region in parameter space
described well by the lowest Landau level approximation (LLL) defined next.

\emph{\textbf{The disordered LLL GL free energy in the quasi-momentum basis}}

In the units and the field normalization described in IIA the LLL energy
becomes:
\begin{gather}
F\left[ W\right] =\int_{r}[\frac{1}{2}\left\vert \partial _{z}\Psi
\right\vert ^{2}-a_{H}\left\vert \Psi \right\vert ^{2}  \label{F_LLL_IV} \\
+\frac{1-t}{2}W(r)\left\vert \Psi \right\vert ^{2}+\frac{1}{2}\left\vert
\Psi \right\vert ^{4}],  \notag
\end{gather}%
where $a_{H}=\frac{1}{2}\left( 1-b-t\right) $ and
\begin{equation}
\overline{W(r)W(r^{\prime })}=n\delta ^{3}\left( r-r^{\prime }\right)
\end{equation}%
in the new length unit. The order parameter field on LLL can be expanded in
the quasi - momentum basis defined in IIIA as%
\begin{equation}
\Psi \left( r\right) =\frac{1}{\left( 2\pi \right) ^{3/2}}\int_{k}\varphi
_{k}\left( r\right) \Psi _{_{k}},  \label{quasimom_expansion_IV}
\end{equation}%
where $k\equiv (\mathbf{k},k_{z})$, functions are defined in eqs.(\ref%
{fi_k_3D_III}),(\ref{HLL_functions_III}) and$\ $the integration measure was
defined in section IIIA to be the Brillouin zone in the $x-y$ plane and the
full range of momenta in the $z$ direction. We consider the hexagonal
lattice, although modifications required to consider a different lattice
symmetry are minor. Using the quasimomentum LLL functions of eq.(\ref%
{fi_k_3D_III}), the disorder term becomes

\begin{equation}
F_{dis}=\frac{1-t}{2}\int_{r}W\left( r\right) \left\vert \Psi \left(
r\right) \right\vert ^{2}=\int_{k,l}w_{k,l}\Psi _{_{k}}^{\ast }\Psi _{_{l}}
\label{Fdis_IV}
\end{equation}%
with%
\begin{equation}
w_{k,l}=\frac{1-t}{2\text{ }\left( 2\pi \right) ^{3}}\int_{r}W\left(
r\right) \varphi _{k}^{\ast }\left( r\right) \varphi _{l}\left( r\right)
\text{.}  \label{w_IV}
\end{equation}%
The rest of the terms can be written as
\begin{gather}
F_{clean}=\int_{k}\left( k_{z}^{2}/2-a_{H}\right) \Psi _{_{k}}^{\ast }\Psi
_{_{k}}+\frac{1}{2\left( 2\pi \right) ^{3}}\times  \label{Fclean_IV} \\
\int_{k,k^{\prime },l,l^{\prime }}\left[ k,k^{\prime }|l,l^{\prime }\right]
\Psi _{_{k}}^{\ast }\Psi _{_{k^{\prime }}}^{\ast }\Psi _{_{l}}\Psi
_{_{l^{\prime }}}\sum\limits_{Q}\delta \left( k+k^{\prime }-l-l^{\prime
}-Q\right)  \notag
\end{gather}%
with $\left[ k,l|k^{\prime }l^{\prime }\right] =\frac{1}{vol}\int_{r}\varphi
_{k}^{\ast }\left( r\right) \varphi _{l}\left( r\right) \varphi _{k^{\prime
}}^{\ast }\left( r\right) \varphi _{l^{\prime }}\left( r\right) $ and where $%
Q=\left( \overrightarrow{\mathbf{Q}},0\right) $ and $\overrightarrow{\mathbf{%
Q}}$ is the reciprocal lattice vectors as $k,l,k^{\prime },l^{\prime }$
satisfy the momentum conservation up to a reciprocal lattice vector. $\left[
k,l|k^{\prime }l^{\prime }\right] $ will be equal to zero if $k+k^{\prime
}-l-l^{\prime }\neq Q$.

\subsubsection{Perturbative expansion in disorder strength.}

\emph{\textbf{Expansion around the Abrikosov solution}}

The GL equations derived from the free energy in the quasimomentum basis are%
\begin{gather}
\left( k_{z}^{2}/2-a_{H}\right) \Psi _{_{k}}+\alpha \int_{l}w_{k,l}\Psi
_{_{l}}+\int_{k^{\prime }l,l^{\prime }}  \label{equation_exact_IV} \\
\sum\limits_{Q}\delta \left( k+k^{\prime }-l-l^{\prime }-Q\right) \frac{%
\left[ k,k^{\prime }|l,l^{\prime }\right] }{\left( 2\pi \right) ^{3}}\Psi
_{_{k^{\prime }}}^{\ast }\Psi _{_{l}}\Psi _{_{l^{\prime }}}=0.  \notag
\end{gather}%
The parameter $\alpha =1$ inserted there will help with counting orders. The
expansion in orders of the disorder strength $\alpha $ reads:
\begin{equation}
\Psi =\Psi ^{\left( 0\right) }+\alpha \Psi ^{\left( 1\right) }+\alpha
^{2}\Psi ^{\left( 2\right) }+...\text{.}  \label{dis_expansion_IV}
\end{equation}%
The clean case Abrikosov solution of section II is defined as the
quasimomentum zero. Therefore

\begin{equation}
\Psi ^{\left( 0\right) }=\left( 2\pi \right) ^{3/2}\sqrt{\frac{a_{H}}{\beta
_{\Delta }}}\delta _{k}\text{.}  \label{psi0_IV}
\end{equation}%
The delta function appears due to its long - range translational order. Now
the equation eq.(\ref{equation_exact_IV}) can be solved order by order in $%
\alpha $. Since contributions linear in disorder potential will average to
zero, in order to get the leading contribution of disorder one should
calculate the free energy to the second order in $\alpha $. Multiplying
exact equation eq.(\ref{equation_exact_IV}) by $\Psi _{_{k}}^{\ast }$ and
integrating over $k$, one can express the order four in $\Psi $ term via
simpler quadratic ones:%
\begin{equation}
F=\frac{1}{2}\int_{k}\left( k_{z}^{2}/2-a_{H}\right) \left\vert \Psi
_{_{k}}\right\vert ^{2}+\frac{\alpha }{2}\int_{k,l}\Psi _{_{k}}^{\ast
}w_{k,l}\Psi _{_{l}}\text{.}  \label{F_simplified_IV}
\end{equation}%
Substituting the expansion eq.(\ref{dis_expansion_IV}) and using delta
functions of $\Psi ^{\left( 0\right) }$ of eq.(\ref{psi0_IV}) one gets the
following $\alpha ^{2}$ terms%
\begin{gather}
F^{\left( 2\right) }=-\frac{a_{H}^{3/2}\left( 2\pi \right) ^{3/2}}{2\beta
_{\Delta }^{1/2}}[\Psi _{0}^{\left( 2\right) \ast }+\Psi _{0}^{\left(
2\right) }]+\frac{1}{2}\int_{k}\left( k_{z}^{2}/2-a_{H}\right)  \notag \\
\times \left\vert \Psi _{_{k}}^{\left( 1\right) }\right\vert ^{2}+\frac{%
a_{H}^{1/2}\left( 2\pi \right) ^{3/2}}{2\beta _{\Delta }^{1/2}}%
\int_{k}[w_{0,k}\Psi _{k}^{\left( 1\right) }+\Psi _{_{k}}^{\left( 1\right)
\ast }w_{k,0}].  \label{F_simplified_2_IV}
\end{gather}%
Therefore the second order correction to $\Psi $ is needed only for zero
quasi - momentum.

\emph{\textbf{First order elastic response of the vortex lattice}}

To order $\alpha $ one obtains the following equation%
\begin{gather}
\left( k_{z}^{2}/2-a_{H}\right) \Psi _{_{k}}^{\left( 1\right)
}+w_{k,0}\left( 2\pi \right) ^{3/2}\sqrt{\frac{a_{H}}{\beta _{\Delta }}}+
\label{order1_eq_IV} \\
2\frac{a_{H}}{\beta _{\Delta }}\beta _{\mathbf{k}}\Psi _{_{k}}^{\left(
1\right) }+\frac{a_{H}}{\beta _{\Delta }}\gamma _{\mathbf{k}}\Psi
_{_{-k}}^{\left( 1\right) \ast }=0  \notag
\end{gather}%
as $Q=0$ because of the conservation of quasimomentum in this case. This
equation and its complex conjugate lead to a system of two linear equations
for two variables $\Psi _{_{k}}^{\left( 1\right) }$ and $\Psi
_{_{-k}}^{\left( 1\right) \ast }$.Solution, not surprisingly, involves the
spectrum of harmonic excitations of the vortex lattice already familiar from
the perturbative corrections due to thermal fluctuations, IIIA:

\begin{eqnarray}
\Psi _{k}^{\left( 1\right) } &=&-\frac{\left( 2\pi \right) ^{3/2}}{%
\varepsilon _{k}^{A}\varepsilon _{k}^{O}}\frac{a_{H}^{1/2}}{\beta _{\Delta
}^{1/2}}[(k_{z}^{2}/2-a_{H}+2\frac{a_{H}}{\beta _{\Delta }}\beta _{\mathbf{k}%
})w_{k,0}  \label{psi1_IV} \\
&&-\frac{a_{H}}{\beta _{\Delta }}\gamma _{\mathbf{k}}w_{-k,0}^{\ast }],
\notag
\end{eqnarray}%
where $\varepsilon _{k}^{A}$, $\varepsilon _{k}^{O}$ are defined in eq.(\ref%
{epsilon_III}).

\emph{\textbf{Disorder average of the pinning energy to leading order}}

The relevant equation (zero quasi - momentum) at the second order in $\alpha
$ is:
\begin{gather}
-a_{H}\Psi _{_{0}}^{\left( 2\right) }+\int_{k}w_{0,k}\Psi _{k}^{\left(
1\right) }+\frac{a_{H}}{\beta _{\Delta }}\left[ 2\beta _{\Delta }\Psi
_{_{0}}^{\left( 2\right) }+\beta _{\Delta }\Psi _{_{0}}^{\left( 2\right)
\ast }\right]  \notag \\
+\frac{a_{H}^{1/2}}{\beta _{\Delta }^{1/2}\left( 2\pi \right) ^{3/2}}\int_{l}%
\left[ 2\beta _{\mathbf{k}}\Psi _{_{k}}^{\left( 1\right) \ast }+\gamma _{%
\mathbf{k}}^{\ast }\Psi _{k}^{\left( 1\right) }\right] \Psi _{_{k}}^{\left(
1\right) }=0  \label{psi2_eq_IV}
\end{gather}%
leading to%
\begin{gather}
a_{H}\left[ \Psi _{_{0}}^{\left( 2\right) }+\Psi _{_{0}}^{\left( 2\right)
\ast }\right] =-\int_{k}\Psi _{_{k}}^{\left( 1\right) }\{w_{0,k}+
\label{psi2_solution_IV} \\
\frac{a_{H}^{1/2}}{\beta _{\Delta }^{1/2}\left( 2\pi \right) ^{3/2}}\left[
2\beta _{\mathbf{k}}\Psi _{_{k}}^{\left( 1\right) \ast }+\gamma _{\mathbf{k}%
}^{\ast }\Psi _{_{-k}}^{\left( 1\right) }\right] \}.  \notag
\end{gather}

Substituting this into eq.(\ref{F_simplified_2_IV}) and simplify the
equation using eq.(\ref{order1_eq_IV}), we obtain the energy expressed via $%
\Psi _{_{k}}^{\left( 1\right) }$
\begin{equation}
F^{\left( 2\right) }=\frac{a_{H}^{1/2}\left( 2\pi \right) ^{3/2}}{2\beta
_{\Delta }^{1/2}}\int_{k}\left\{ w_{0,k}\Psi _{k}^{\left( 1\right) }+\Psi
_{_{k}}^{\left( 1\right) \ast }w_{k,0}\right\}  \label{psi2_IV}
\end{equation}%
and using the expression for $\Psi _{_{k}}^{\left( 1\right) }$ eq.(\ref%
{psi1_IV}) one obtains various terms quadratic in disorder $w$. The disorder
averages are

\begin{eqnarray}
\overline{w_{k,l}w_{k^{\prime },l^{\prime }}} &=&\frac{\left( 1-t\right)
^{2}nV}{4}\text{ }\left[ k,k^{\prime }|l,l^{\prime }\right] ;  \notag \\
\overline{w_{k,l}w_{k^{\prime },l^{\prime }}^{\ast }} &=&\frac{\left(
1-t\right) ^{2}nV}{4\text{ }}\left[ k,l^{\prime }|l,k^{\prime }\right] ;%
\text{ \ \ } \\
\overline{w_{k,l}^{\ast }w_{k^{\prime },l^{\prime }}^{\ast }} &=&\frac{%
\left( 1-t\right) ^{2}nV}{4\text{ }}\left[ l,l^{\prime }|k,k^{\prime }\right]
,  \notag
\end{eqnarray}%
and so that the pinning energy becomes after some algebra

\begin{equation}
\frac{\overline{F^{\left( 2\right) }}}{vol}=-\frac{\left( 1-t\right)
^{2}na_{H}}{4\beta _{A}2\left( 2\pi \right) ^{3}}\int_{k}(\frac{\beta _{%
\mathbf{k}}+\left\vert \gamma _{_{k}}\right\vert }{\epsilon _{k}^{O}}+\frac{%
\beta _{\mathbf{k}}-\left\vert \gamma _{_{\mathbf{k}}}\right\vert }{\epsilon
_{k}^{A}}).  \label{average_F2_IV}
\end{equation}%
Integrating over $k_{z},$ one obtains finally,

\begin{equation}
\frac{\overline{F^{\left( 2\right) }}}{vol}=-\frac{\left( 1-t\right)
^{2}na_{H}}{16\sqrt{2}\text{ }\pi \beta _{\Delta }}<\frac{\beta _{\mathbf{k}%
}+\left\vert \gamma _{_{k}}\right\vert }{\mu _{O\mathbf{k}}}+\frac{\beta _{%
\mathbf{k}}-\left\vert \gamma _{_{\mathbf{k}}}\right\vert }{\mu _{A\mathbf{k}%
}}>  \label{average_F2_1_IV}
\end{equation}%
where $\mu _{\mathbf{k}}^{A,O}$ are given in eq.(\ref{epsilon_III}).

Using expansion for small $\mathbf{k}$ of the functions $\beta _{\mathbf{k}}$
and $\left\vert \gamma _{_{\mathbf{k}}}\right\vert $ derived in Appendix A,
one can see that the second term is finite
\begin{equation}
\approx \int \frac{\beta _{02}\left\vert \mathbf{k}\right\vert ^{2}}{%
\left\vert \mathbf{k}\right\vert ^{2}}d^{2}\mathbf{k}.  \label{div3D_IV}
\end{equation}%
Numerically
\begin{equation}
\frac{\overline{F}}{vol}=-\frac{a_{H}^{2}}{2\beta _{\Delta }}-\frac{\left(
1-t\right) ^{2}n}{2^{5/2}\text{ }\pi \beta _{\Delta }}a_{H}^{1/2}.
\label{final_F_IV}
\end{equation}

\emph{\textbf{Stronger disorder: 2D GL and columnar defects}}

The same calculation can be performed in 2D with the result%
\begin{equation}
\frac{\overline{F^{\left( 2\right) }}}{vol}=-\frac{\left( 1-t\right)
^{2}na_{H}}{4\text{ }\beta _{A}\left( 2\pi \right) ^{2}}\int_{k}(\frac{\beta
_{\mathbf{k}}+\left\vert \gamma _{_{k}}\right\vert }{\mu _{O\mathbf{k}}^{2}}+%
\frac{\beta _{\mathbf{k}}-\left\vert \gamma _{_{\mathbf{k}}}\right\vert }{%
\mu _{A\mathbf{k}}^{2}}).  \label{F_2D_IV}
\end{equation}%
This is logarithmically IR divergent at any value of the disorder strength%
\begin{equation}
\approx \int \frac{\beta _{\Delta 2}\left\vert \mathbf{k}\right\vert ^{2}}{%
\left\vert \mathbf{k}\right\vert ^{4}}d^{2}\mathbf{k}.  \label{div_2D_IV}
\end{equation}%
Therefore either the dependence is not analytic or (more probably) disorder
significantly modifies the structure of the solution. Generalization in
another direction, that of long - rage correlated disorder can also be
easily performed. One just replaces the white noise variance by a general one%
\begin{equation}
\overline{W(r)W(r^{\prime })}=K\left( r-r^{\prime }\right) .
\label{general_variance_IV}
\end{equation}

For columnar defects the variance is independent of $z,$
\begin{equation}
K\left( r-r^{\prime }\right) =n\delta \left( \mathbf{r}-\mathbf{r}^{\prime
}\right) ,
\end{equation}%
and one again obtains a logarithmic divergence.%
\begin{equation}
\frac{\overline{F^{\left( 2\right) }}}{vol}\propto <\frac{\beta _{\mathbf{k}%
}+\left\vert \gamma _{_{k}}\right\vert }{\mu _{O\mathbf{k}}^{2}}+\frac{\beta
_{\mathbf{k}}-\left\vert \gamma _{_{\mathbf{k}}}\right\vert }{\mu _{A\mathbf{%
k}}^{2}}>.  \label{columnar_IV}
\end{equation}

\subsubsection{Disorder influence on the vortex liquid and crystal. Shift of
the melting line}

\emph{\textbf{Disorder correction to free energy}}

Thermal fluctuations in the presence of quenched are still described by
partition function
\begin{equation}
Z=-\frac{1}{\omega _{t}}\left[ F[\Psi ]+\frac{1-t}{2}\int_{r}W\left(
r\right) \left\vert \Psi \left( r\right) \right\vert ^{2}\right] .
\end{equation}%
If $W$ is small, we can calculate $Z$ by perturbation theory in $W$. To the
second order\ free energy $-T\ln Z$ is
\begin{gather}
G=G_{clean}+\frac{1-t}{2}\int_{r}W(r)\left\langle \left\vert {\Psi (r)}%
\right\vert ^{2}\right\rangle -\frac{1}{8\omega _{t}}(1-t)^{2}\times \\
\int_{r,r^{\prime }}W(r)W(r^{\prime })[\left\langle \left\vert {\Psi (r)}%
\right\vert ^{2}\left\vert {\Psi (r}^{\prime }{)}\right\vert
^{2}\right\rangle -\left\langle \left\vert {\Psi (r)}\right\vert
^{2}\right\rangle \left\langle \left\vert \Psi {(r}^{\prime }{)}\right\vert
^{2}\right\rangle ],  \notag
\end{gather}%
where $\left\langle {}\right\rangle $ and $f_{clean}$ denote the thermal
average and free energy of \ the clean system. Averaging now over disorder
one obtains
\begin{gather}
\Delta G_{dis}=G-G_{clean}=  \label{deltaG_IV} \\
-\frac{n}{8\omega _{t}}(1-t)^{2}\int_{r}\left[ {\left\langle {\left\vert
\Psi {(r)}\right\vert ^{4}}\right\rangle -\left\langle {\left\vert \Psi {(r)}%
\right\vert ^{2}}\right\rangle }^{2}\right] .  \notag
\end{gather}%
Therefore one has to calculate the superfluid density thermal
correlator. In LLL approximation and LLL units,

\begin{gather}
\Delta G_{dis}/V=-r\left( t\right) \triangle f;  \notag \\
\triangle f=\frac{1}{2}\left[ {\left\langle {\left\vert \Psi _{LLL}{(r)}%
\right\vert ^{4}}\right\rangle }_{r}{-\left\langle {\left\vert \Psi _{LLL}{%
(r)}\right\vert ^{2}}\right\rangle }_{r}^{2}\right] ; \\
r\left( t\right) =\frac{n}{4\omega _{t}}(1-t)^{2}=\frac{n_{0}(1-t)^{2}}{t}%
;n_{0}=\frac{n}{4\sqrt{2Gi\pi }}  \notag
\end{gather}

Calculations of this kind in both solid and liquid were the subject of the
previous section.

\emph{\textbf{Correlators in the crystalline and the liquid states}}

Within LLL (and using the LLL units introduced in section III) the one loop
disorder correction to the crystal's energy is:%
\begin{equation}
\Delta \emph{f}_{crystal}=2.14\left\vert {a_{T}}\right\vert ^{1/2}.
\label{delta_crystal_IV}
\end{equation}
An explicit expression for $\emph{f}_{liq}(a_{T})$, obtained using the
Borel-Pade resummation of the renormalized high temperature series
(confirmed by optimized Gaussian series and Monte Carlo simulation) is
rather bulky and can be found in ref. \cite{Li02}. One can derived an
expression for the disorder correction in liquid by differentiating the
"clean" partition function with respect to parameters:
\begin{equation}
\Delta \emph{f}_{liq}=\frac{1}{3}(\emph{f}_{liq}-2a_{T}^{\prime }\emph{f}%
_{liq}^{\prime })/3-\frac{1}{2}(\emph{f}_{liq}^{\prime })^{2}.
\end{equation}
These two results enable us to find the location of the transition line and,
in addition, to calculate discontinuities of various physical quantities
across the transition line.
\begin{figure}[t]
\centering \rotatebox{0}{\includegraphics[width=0.45%
\textwidth,height=9cm]{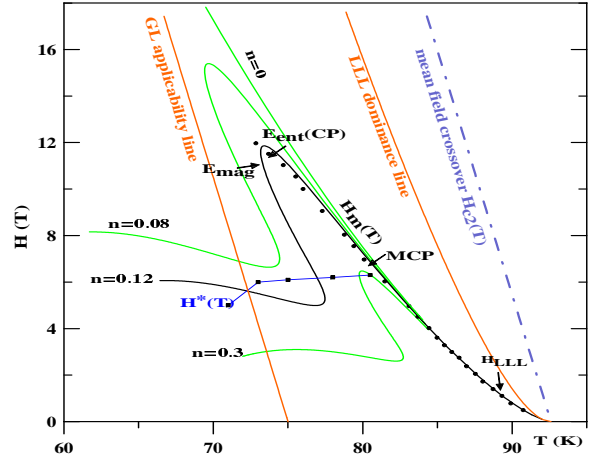}}
\caption{Phase diagram for YBCO.}
\label{figIV1}
\end{figure}

\emph{\textbf{The "downward shift" of the first order transition line in the
$T-H$ plane}}

\begin{figure}[t]
\centering \rotatebox{270}{\includegraphics[width=0.3%
\textwidth,height=8cm]{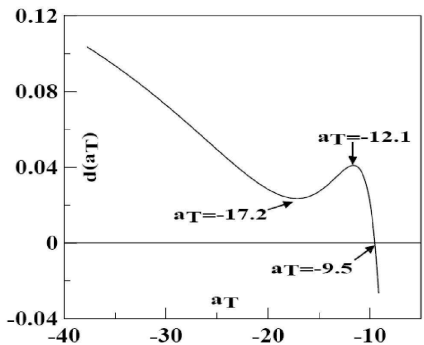}}
\caption{Universal function $d\left( a_{T}\right) $ determining the shift of
the melting line due to disorder}
\label{figIV2}
\end{figure}

It was noted in section III that in a clean system a homogeneous state
exists as a metastable overcooled liquid state all the way down to zero
temperature (not just below the melting temperature corresponding to $%
a_{T}=-9.5$, see the $n=0$ line in Fig.\ref{figIV1}) . This is of importance
since interaction with disorder can convert the metastable state into a
stable one. Indeed generally a homogeneous state gains more than a
crystalline state from pinning, since it can easier adjust itself to the
topography of the pinning centers. At large $\left\vert {a_{T}}\right\vert $
in particular $\Delta \emph{f}_{liq}\propto a_{T}^{2}$ compared to just $%
\Delta \emph{f}_{sol}\propto \left\vert {a_{T}}\right\vert ^{1/2}$. As a
result in the presence of disorder the transition line shifts to lower
fields. The equation for the melting line is
\begin{equation}
d(a_{T})\equiv (f_{liq}-f_{sol})/(\Delta f_{liq}-\Delta f_{sol})=n(t).
\end{equation}%
The universal function $d(a_{T})$, plotted in Fig.\ref{figIV2} turns out to
be non-monotonic. This is an important fact. Since $n(t)$ is a monotonic
function of $t$, one obtains the transition lines for various $n$ in Fig.\ref%
{figIV1} by \textquotedblleft sweeping\textquotedblright\ the Fig.\ref%
{figIV2}. A peculiar feature of $d(a_{T})$ is that it has a local minimum at
$a_{T}\approx -17.2$ and a local maximum at $a_{T}\approx -12.1$ (before
crossing zero at $\ a_{T}\approx -9.5)$. Therefore between these two points
there are three solutions to the melting line equation. As a result,
starting from the zero field at $T_{c},$ the transition field $H(T)$ reaches
a maximum at $E_{ent}$ beyond which the curve sharply turns down (this
feature was called \textquotedblleft inverse melting\textquotedblright\ in
\cite{Avraham01}) and at $E_{mag}$ backwards. Then it reaches a minimum and
continues as the Bragg glass -- vortex glass line roughly parallel to the $T$
axis.

The temperature dependence of the disorder strength $n(t)$, as of any
parameter in the GL approach should be derived from a microscopic theory or
fitted to experiment. General dependence near $T_{c}$ is: $n(t)=n(1-t)^{2}/t$
. The extra factor $(1-t)^{2}$, not appearing in a phenomenological
derivation \cite{Blatter}, is due to the fact that near $H_{c2}$ order
parameter is small $\left\vert \psi \right\vert ^{2}\propto (1-t)$ and
disorder (oxygen deficiencies) locally destroys superconductivity rather
than perturbatively modifies the order parameter. The curves in Fig.\ref%
{figIV1} correspond to the disorder strength $n_{0}=0.08,0.12,0.3$. The best
fit for the low field part of the experimental melting line $H_{m}(T)$ of
the optimally doped \textit{YBCO} (data taken from \cite{Schilling96}, $%
T_{c} $ $=92.6$, $\ \gamma =8.3$) gives $Gi=2.0\times 10^{-4}$, $H_{c2}=190\
\ T$, \ $\kappa =\lambda /\xi =50$ (it is consistent with other experiments,
for example, \cite{Nishizaki02,Deligiannis00}). This part is essentially
independent of disorder. The upper part of the melting curve is very
sensitive to disorder: both the length of the \textquotedblleft
finger\textquotedblright\ and its slope depend on $n_{0}$. The best fit is $%
n_{0}=0.12.$ This value is of the same order of magnitude as the one
obtained phenomenologically using eq.(3.82) in ref. \cite{Blatter}. We
speculate that the low temperature part of the \textquotedblleft
unified\textquotedblright\ line corresponds to the solid -- vortex glass
transition $H^{\ast }(T)$ observed in numerous experiments \cite%
{Nishizaki02,Schilling96,Schilling97,Bouquet01,Kokkaliaris00,Radzyner02,Pal01,Pal02}%
, see data (squares in Fig.\ref{figIV1}) taken from \cite{Nishizaki02}. A
complicated shape of the \textquotedblleft wiggling\textquotedblright\ line
has been recently observed \cite{Pal01,Pal02}. Now we turn to a more
detailed characteristics of the phase transition.

\emph{\textbf{Discontinuities across the transition and the Kauzmann point.
Absence of a second order transition.}}

\textbf{\textit{\ }}Magnetization and specific heat of both solid and liquid
can be calculated from the above expressions for free energy. Magnetization
of liquid along the melting line $H_{m}(T)$ is larger than that of solid.
The magnetization jump is compared in \cite{Li03}a with the SQUID
experiments \cite{Schilling97}\ in the range $80-90K$ (triangles) and of the
torque experiments (stars \cite{Willemin98} and circles \cite{Nishizaki02}
). One observes that the results of the torque experiments compare
surprisingly well above $83K$, but those of \cite{Nishizaki02} vanish
abruptly below $83K$ unlike the theory and are inconsistent with the
specific heat experiments \cite{Schilling96,Deligiannis00} discussed below.
The SQUID data are lower than theoretical (same order of magnitude though).
We predict that at lower temperatures (somewhat beyond the range
investigated experimentally so far) magnetization reaches its maximum and
changes sign at the point $E_{mag}$ (at which magnetization of liquid and
solid are equal).

In ref.\cite{Li03}, entropy jump was calculated using the Clausius --
Clapeyron relation $\partial H_{m}(T)/\partial T=-\Delta S/\Delta M$ and
compared with an experimental one deduced from the spike of the specific
heat (\cite{Schilling96}, and an indirect measurement from the magnetization
jumps (in ref.\cite{Nishizaki02}). At high temperatures the theoretical
values are a bit lower than the experimental and both seem to approach a
constant at $T_{c}$. The theoretical entropy jump and the experimental one
of \cite{Schilling96} vanish at $E_{ent}$ (Fig. \ref{figIV1}) near $75K$.
Such points are called Kauzmann points. Below this temperature entropy of
the liquid becomes smaller than that of the solid. Note that the equal
magnetization point $E_{mag}$ is located at a slightly lower field than the
equal entropy point $E_{ent}$. Experimentally a Kauzmann point was
established in \textit{BSCCO} as a point at which the \textquotedblleft
inverse melting\textquotedblright\ appears \cite{Avraham01}. The Kauzmann
point observed at a lower temperature in \textit{YBCO} in ref. \cite%
{Radzyner02} is different from $E_{ent}$ since it is a minimum rather than a
maximum of magnetic field. It is also located slightly outside the region of
applicability of our solution. The point $E_{ent}$ is observed in \cite%
{Pal01,Pal02} in which the universal line is continuous.

In addition to the spike, the specific heat jump has also been observed
along the melting line $H_{m}(T)$ \cite%
{Schilling97,Schilling96,Deligiannis00}. Theoretically the jump does not
vanish either at $E_{ent}$ or $E_{mag}$, but is rather flat in a wide
temperature range. Our results are larger than experimental jumps of \cite%
{Schilling96} (which are also rather insensitive to temperature) by a factor
of 1.4 to 2 \cite{Li03}. In many experimental papers there appears a segment
of the second order phase transition continuing the first order melting line
beyond a certain point. In \cite{Bouquet01} it was shown that at that point
the specific heat profile shows \textquotedblright
rounding\textquotedblright . We calculated the specific heat profile above
the universal first order transition line. It exhibits a \textquotedblright
rounding\textquotedblright\ feature similar to that displayed by the data of
\cite{Schilling96,Schilling97,Schilling02,Bouquet01} with no sign of the
criticality. The height of the peak is roughly of the size of the specific
heat jump. We therefore propose not to interpret this feature as an evidence
for a second order transition above the first order line.

\emph{\textbf{Limitations of the perturbative approach}}

Of course the perturbative approach is limited to small couplings only. In
fact, when the correction is compared to the main part of the lattice energy
the range become too narrow for practical applications in low temperature
superconductors. For high $T_{c}$ superconductors thermal fluctuations
cannot be neglected at higher temperatures since it "melts" the lattice and
even at low temperatures provides thermal depinning. On the conceptual side,
it is clear that disorder contributes to destruction of the translational
and rotational order. Therefore at certain disorder strength, vortex matter
might restore the translation and rotation symmetries, even without help of
thermal fluctuations. It is possible to use the perturbation theory in
disorder with the liquid state as a starting point in the case of large
thermal fluctuations, however it fails to describe the most interesting
phenomenon of the vortex glass introduced by Fisher \cite{Fisher89,Fisher91}%
. Therefore one should try to develop non - perturbative methods to describe
disorder This is the subject the following sections.

\subsection{The vortex glass}

When thermal fluctuations are significant the efficiency of imperfections to
pin the vortex matter is generally diminished. This phenomenon is known as
"thermal depinning". In addition, as we have learned in section III, the
vortex lattice becomes softer and eventually melts via first order
transition into the vortex liquid. The inter - dependence of pinning,
interactions and thermal fluctuations is very complex and one needs an
effective nonperturbative method to evaluate the disorder averages. Such a
method, using the replica trick was developed initially in the theory of
spin glasses. It is more difficult to apply it in a crystalline phase, so we
start from a simpler homogeneous phase (the homogeneity might be achieved by
both the thermal fluctuations and disorder) and return to the crystalline
phase in the following subsection.

\subsubsection{Replica approach to disorder}

\emph{\textbf{The replica trick}}

The replica method is widely used to study disordered electrons in metals
and semiconductors, spin glasses and other areas of condensed matter physics
and far beyond it \cite{Itzykson}. It was applied to vortex matter in the
elastic medium approximation \cite%
{Nattermann90,Bogner01,Korshunov93,Giamarchi94,Giamarchi95a,Giamarchi95b,Giamarchi96,Giamarchi97}%
. In the following we describe the method in some detail.

The main problem in calculation of disorder averages is that one typically
has to take the average of non - polynomial functions of the statistical sum
eq.(\ref{Z_I}):
\begin{equation}
Z=\int \mathcal{D}\Psi \mathcal{D}\Psi ^{\ast }e^{-\frac{1}{\omega _{t}}%
\left[ F[\Psi ]+\frac{1-t}{2}\int_{r}W\left( r\right) \left\vert \Psi \left(
r\right) \right\vert ^{2}\right] }.  \label{stat_sum_IV}
\end{equation}%
Most interesting physical quantities are calculated by taking derivatives of
the free energy which is a logarithm of $Z$. Applying a simple mathematical
identity to represent the logarithm as a small power, $\log \left( z\right) =%
{\lim_{n\rightarrow 0}}\frac{1}{n}(z^{n}-1)$, the average over the free
energy is written as:
\begin{equation}
\overline{\mathcal{F}}=-\omega _{t}{\lim_{n\rightarrow 0}}\frac{1}{n}%
\overline{(Z^{n}-1)}.  \label{F_bar_IV}
\end{equation}%
The quantity $Z^{n}$ can be looked upon as a statistical sum over $n$
identical "replica" fields $\Psi _{a}$ , $a=1,...,n$:
\begin{equation}
Z^{n}\left[ W\right] =\prod_{b}\int_{\Psi _{b}}e^{-\sum_{a=1}^{n}[\frac{%
F[\Psi _{a}]}{\omega _{t}}+\frac{1-t}{2\omega _{t}}\int_{r}W\left( r\right)
\left\vert \Psi _{a}\left( r\right) \right\vert ^{2}]}  \label{Zn_IV}
\end{equation}%
where $F[\Psi _{a}]$ is the free energy (in physical units meantime) without
disorder. Note that the disorder potential enters in the exponent. The
disorder measure, consistent with variance in eq.(\ref{variance_IV}) is a
gaussian. Therefore disorder average is a gaussian integral which can be
readily performed:

\begin{eqnarray}
\overline{Z^{n}} &=&\frac{1}{norm}\int \mathcal{D}We^{-\frac{1}{2n}%
\int_{r}W^{2}\left( r\right) }Z^{n}\left[ W\right]  \label{Zn_averaged_IV} \\
&=&\int_{\Psi _{a}}e^{-\frac{1}{\omega _{t}}F_{n}},  \notag
\end{eqnarray}%
where%
\begin{equation}
F_{n}\equiv \sum_{a}F[\Psi _{a}]+\frac{1}{2}r\left( t\right)
\sum_{a,b}\int_{r}\left\vert \Psi _{a}\right\vert ^{2}\left\vert \Psi
_{b}\right\vert ^{2}.  \label{Fn_IV}
\end{equation}%
After the disorder average different replicas are no longer independent. In
LLL limit and units,

\begin{eqnarray}
\overline{Z^{n}} &=&\int_{\Psi _{a}}e^{-\frac{1}{4\pi \sqrt{2}}F_{n}}, \\
F_{n} &\equiv &\sum_{a}F[\Psi _{a}]+\frac{r\left( t\right) }{2}%
\sum_{a,b}\int_{r}\left\vert \Psi _{a}\right\vert ^{2}\left\vert \Psi
_{b}\right\vert ^{2}  \notag
\end{eqnarray}

This statistical physics model is a type of scalar field theory and the
simplest nonperturbative scheme commonly used to treat such a model is
gaussian approximation already introduced in IIIB. Its validity and
precision can be checked only by calculating corrections.

\emph{\textbf{Correlators and distributions}}

Correlators averaged over both the thermal fluctuations and disorder can be
generated by the usual trick of introducing an external "source" into
statistical sum eq.(\ref{stat_sum_IV}):%
\begin{gather}
Z\left[ W,S^{\ast },S\right] =\int_{\Psi ,\Psi ^{\ast }}exp\{-\frac{1}{%
\omega _{t}}[F[\Psi ]+\frac{1}{2}\left( 1-t\right)  \label{Z[W,S]_IV} \\
\times \int_{r}W\left( r\right) \left\vert \Psi \left( r\right) \right\vert
^{2}+\int_{r}\Psi \left( r\right) S^{\ast }\left( r\right) +\Psi ^{\ast
}\left( r\right) S\left( r\right) ]\}  \notag \\
=\int_{\Psi ,\Psi ^{\ast }}e^{-\frac{1}{\omega _{t}}F[\Psi ,W\left( r\right)
,S\left( r\right) ,S^{\ast }\left( r\right) ]}.  \notag
\end{gather}%
and taking functional derivatives of the free energy in the presence of
sources%
\begin{equation}
\mathcal{F}\left[ W,S^{\ast },S\right] =-\omega _{t}\log Z\left[ W,S^{\ast
},S\right] \text{.}
\end{equation}%
The first two thermal correlators are%
\begin{gather}
\left\langle \Psi \left( r\right) \right\rangle =\frac{1}{Z\left[ W,S^{\ast
},S\right] }\int_{\Psi ,\Psi ^{\ast }}\Psi \left( r\right) e^{-\frac{1}{%
\omega _{t}}F[\Psi ,W\left( r\right) ,S\left( r\right) ,S^{\ast }\left(
r\right) ]}  \notag \\
-\frac{\omega _{t}}{Z\left[ W,0,0\right] }\frac{\delta }{\delta S^{\ast
}\left( r\right) }Z\left[ W,S^{\ast },S\right] |_{S,S^{\ast }=0}  \notag \\
=\frac{\delta }{\delta S^{\ast }\left( r\right) }\mathcal{F}\left[ W,S^{\ast
},S\right] |_{S,S^{\ast }=0};  \label{psi_corr_IV} \\
\left\langle \Psi ^{\ast }\left( r\right) \Psi \left( r^{\prime }\right)
\right\rangle _{c}=\left\langle \Psi ^{\ast }\left( r\right) \Psi \left(
r^{\prime }\right) \right\rangle -\left\langle \Psi ^{\ast }\left( r\right)
\right\rangle \left\langle \Psi \left( r^{\prime }\right) \right\rangle
\notag \\
=\frac{\delta ^{2}}{\delta S\left( r\right) \delta S^{\ast }\left( r^{\prime
}\right) }\mathcal{F}\left[ W,S^{\ast },S\right] |_{S,S^{\ast }=0}.  \notag
\end{gather}%
Now the disorder averages of these quantities are made using the replica
trick%
\begin{gather}
\overline{\left\langle \Psi \left( r\right) \right\rangle }=-\omega _{t}%
\frac{\delta }{\delta S^{\ast }\left( r\right) }{\lim_{n\rightarrow 0}}\frac{%
1}{n}\overline{(Z\left[ S,S^{\ast }\right] ^{n}-1)}
\label{psi_dis_average_IV} \\
=-\omega _{t}\lim_{n\rightarrow 0}\frac{1}{n}\frac{\delta }{\delta S^{\ast
}\left( r\right) }\int_{\Psi _{a}}e^{-\frac{1}{\omega _{t}}[F_{n}[\Psi
_{a}]+S^{\ast }\left( r\right) \sum_{a}\Psi ^{a}\left( r\right) ]}|_{S^{\ast
}=0}  \notag \\
=\lim_{n\rightarrow 0}\frac{1}{n}\int_{\Psi _{a}}\sum_{a}\Psi ^{a}\left(
r\right) e^{-\frac{1}{\omega _{t}}F_{n}[\Psi _{a}]}=\frac{1}{n}%
\sum_{a}\left\langle \Psi ^{a}\left( r\right) \right\rangle  \notag
\end{gather}%
Similar calculation for the two field correlator result in%
\begin{equation}
\overline{\left\langle \Psi ^{\ast }\left( r\right) \Psi \left( r^{\prime
}\right) \right\rangle _{c}}=\lim_{n\rightarrow 0}\frac{1}{n}%
\sum_{a,b}\left\langle \Psi ^{\ast a}\left( r\right) \Psi ^{b}\left(
r^{\prime }\right) \right\rangle .  \label{two_field_average_IV}
\end{equation}%
In disorder physics it is of interest to know the disorder distribution of
physical quantities like magnetization (which within LLL is closely related
to the correlator, see IIIB). The simplest example is the second moment of
the order parameter distribution $\overline{\left\langle \Psi ^{\ast }\left(
r\right) \right\rangle \left\langle \Psi \left( r^{\prime }\right)
\right\rangle }$. This is harder to evaluate due to two thermal averages.
One still uses eq.(\ref{psi_corr_IV}) twice:
\begin{gather}
\left\langle \Psi ^{\ast }\left( r\right) \right\rangle \left\langle \Psi
\left( r^{\prime }\right) \right\rangle =\frac{1}{Z^{2}\left[ W\right] }%
\int_{\Psi _{1}\Psi _{2}}\Psi _{1}\left( r\right) \Psi _{2}\left( r^{\prime
}\right)  \notag \\
\times e^{-\frac{1}{\omega _{t}}F[\Psi _{1},W\left( r\right) ]-\frac{1}{%
\omega _{t}}F[\Psi _{2},W\left( r\right) ]} \\
=\lim_{n\rightarrow 0}\int_{\Psi _{1}\Psi _{2}}\Psi _{1}\left( r\right) \Psi
_{2}\left( r^{\prime }\right) e^{-\frac{1}{\omega _{t}}F[\Psi _{1},W\left(
r\right) ]-\frac{1}{\omega _{t}}F[\Psi _{2},W\left( r\right) ]}\times  \notag
\\
Z^{n-2}\left[ W\right] =\lim_{n\rightarrow 0}\int_{\Psi _{1}\Psi _{2}}\Psi
_{1}\left( r\right) \Psi _{2}\left( r^{\prime }\right) e^{-\frac{1}{\omega
_{t}}\sum\limits_{i=1}^{n}F[\Psi _{i},W\left( r\right) ]}  \notag
\end{gather}

\bigskip The disorder average leads to \cite{Mezard87}:%
\begin{gather}
\overline{\left\langle \Psi ^{\ast }\left( r\right) \right\rangle
\left\langle \Psi \left( r^{\prime }\right) \right\rangle }%
=\lim_{n\rightarrow 0}\int_{\Psi _{1}\Psi _{2}}\Psi _{1}\left( r\right) \Psi
_{2}\left( r^{\prime }\right) e^{-\frac{1}{\omega _{t}}F_{n}}  \notag \\
=\lim_{n\rightarrow 0,a\neq b}\int_{\Psi _{a}\Psi _{b}}\Psi _{a}\left(
r\right) \Psi _{b}\left( r^{\prime }\right) e^{-\frac{1}{\omega _{t}}%
F_{n}}=Q_{a,b}  \label{double_thermal_IV}
\end{gather}

In case of replica symmetry breaking, the formula above shall be
written as

\begin{equation}
\overline{\left\langle \Psi ^{\ast }\left( r\right) \right\rangle
\left\langle \Psi \left( r^{\prime }\right) \right\rangle }%
=\lim_{n\rightarrow 0,a\neq b}\frac{1}{n\left( n-1\right) }%
\sum\limits_{a\neq b}Q_{a,b}.
\end{equation}

Therefor%
\begin{eqnarray}
\overline{\left\langle \Psi ^{\ast }\left( r\right) \Psi \left( r^{\prime
}\right) \right\rangle } &=&\overline{\left\langle \Psi ^{\ast }\left(
r\right) \Psi \left( r^{\prime }\right) \right\rangle _{c}}+\overline{%
\left\langle \Psi ^{\ast }\left( r\right) \right\rangle \left\langle \Psi
\left( r^{\prime }\right) \right\rangle }  \notag \\
&=&\lim_{n\rightarrow 0}\frac{1}{n}\sum_{a}\left\langle \Psi ^{\ast a}\left(
r\right) \Psi ^{a}\left( r^{\prime }\right) \right\rangle
\end{eqnarray}

\emph{\textbf{Disordered LLL theory}}

Restricting the order parameter to LLL (eq.(\ref{LLL_III})) by expanding it
in quasi - momentum LLL functions eq.(\ref{quasimom_expansion_IV}), one
obtains the disordered LLL theory. Let us also rewrite the model in the same
units we have used in section III. The resulting Boltzmann factor is $\
\frac{1}{2^{5/2}\pi }f$
\begin{equation}
f=\sum_{a}\left\{ \psi _{k}^{a\ast }\left( k_{z}^{2}/2+a_{T}\right) \psi
_{k}^{a}+f_{int}\left[ \psi _{a}\right] \right\} +f_{dis},
\label{B._factor_IV}
\end{equation}%
with the disorder term

\begin{eqnarray}
f_{dis} &=&\sum_{a,b}\frac{r\left( t\right) L_{x}L_{y}}{2\left( 2\pi \right)
^{5}}\int_{k,l,k^{\prime },l^{\prime }}\delta \left(
k_{z}-l_{z}+k_{z}^{\prime }-l_{z}^{\prime }\right)  \notag \\
&&\times \left[ \mathbf{k},\mathbf{k}^{\prime }|\mathbf{l},\mathbf{l}%
^{\prime }\right] \psi _{k}^{a\ast }\psi _{l}^{a}\psi _{l^{\prime }}^{b\ast
}\psi _{k^{\prime }}^{c},  \label{F_dis_IV}
\end{eqnarray}%
in which $\left[ \mathbf{k},\mathbf{k}^{\prime }|\mathbf{l},\mathbf{l}%
^{\prime }\right] $ was defined in eq.(\ref{Lambda_III}).

\subsubsection{Gaussian approximation}

\emph{\textbf{Gaussian energy in homogeneous (amorphous) phase}}

One can recover the perturbative results of the previous subsection and even
generalize them to finite temperatures, by expanding in $r$, however the
replica method's advantage is more profound when nonperturbative effects are
involved. We now apply the gaussian approximation, which has been already
used in vortex physics in the framework of the elastic medium approach, \cite%
{Korshunov93,Korshunov90,Giamarchi94,Giamarchi95a,Giamarchi95b,Giamarchi96,Giamarchi97}
following its use in polymer and disordered magnets' physics \cite{Mezard91}%
. As usual, homogeneous phases are simpler than the crystalline phase
considered in the previous subsection, so we start from the case in which
both the translational and the $U\left( 1\right) $ symmetries are respected
by the variational correlator:%
\begin{equation}
\overline{\left\langle \psi _{k}^{a\ast }\psi _{k}^{b}\right\rangle }%
=G_{ab}(k_{z})=[\frac{2^{5/2}\pi }{\frac{k_{z}^{2}}{2}I+\mu ^{2}}]_{a,b}%
\text{.}  \label{replica_propagator_IV}
\end{equation}%
Since the gaussian approximation in the vortex liquid within the GL approach
was described in detail in section III, we just have to generalize various
expressions to the case of $n$ replicas. The gaussian effective free energy
is expressed via variational parameter \cite{Mezard91,Li02,Li02a,Li02b} $\mu
_{ab}$ which in the present case is a matrix in the replica space. The
bubble and the trace log integrals appearing in the free energy are very
simple:%
\begin{eqnarray}
\frac{1}{\left( 2\pi \right) ^{3}}\int_{k}[\frac{2^{5/2}\pi }{\frac{k_{z}^{2}%
}{2}I+\mu ^{2}}]_{a,b} &=&2\left[ \mu ^{-1}\right] _{ab}\equiv 2u_{ab},
\notag \\
\frac{2^{5/2}\pi }{\left( 2\pi \right) ^{3}}\int_{k}\left[ LogG^{-1}(k_{z})%
\right] _{aa} &=&4\mu _{aa}+const,  \label{bubble_trlog_IV}
\end{eqnarray}%
where the "inverse mass" matrix $u_{ab}$ was defined. As a result the
gaussian effective free energy density can be written in a form:%
\begin{gather}
n\text{ }\emph{f}_{gauss}=\sum_{a}-\frac{2^{5/2}\pi }{\left( 2\pi \right)
^{3}}\int_{k}[LogG^{-1}(k_{z})]_{aa}+  \label{nf_gauss_IV} \\
\frac{1}{\left( 2\pi \right) ^{3}}\int_{k}[\left( k_{z}^{2}/2+a_{T}\right)
G(k_{z})-I]_{_{aa}}+4\left( u_{aa}\right) ^{2}-2r\sum_{a,b}\left\vert
u_{ab}\right\vert ^{2}  \notag \\
=2\sum_{a}\left\{ \mu _{aa}+a_{T}u_{aa}+2\left( u_{aa}\right) ^{2}\right\}
-2r\sum_{a,b}\left\vert u_{ab}\right\vert ^{2}  \notag
\end{gather}%
where we discarded an (ultraviolet divergent) constant and higher order in $%
n $, and for simplicity, $r\left( t\right) $ is denoted by $r$.

\emph{\textbf{Minimization equations}}

It is convenient to introduce a real (not necessarily symmetric) matrix $%
Q_{ab}$, which is in one - to - one linear correspondence with Hermitian
(generally complex) matrix $u_{ab}$ via%
\begin{equation}
Q_{ab}=\text{Re}[u_{ab}]+\text{Im}[u_{ab}].  \label{Qdef_IV}
\end{equation}%
Unlike $u_{ab}$, all the matrix elements of $Q_{ab}$ are independent. In
terms of this matrix the free energy can be written as

\begin{equation}
\frac{n}{2}\emph{f}_{gauss}=\sum_{a}\left( u^{-1}\right)
_{aa}+a_{T}Q_{aa}+2\left( Q_{aa}\right) ^{2}-r\sum_{a,b}Q_{ab}^{2}.
\label{f_replica_IV}
\end{equation}%
Taking derivative with respect to independent variables $Q_{ab},$ gives the
saddle point equation for this matrix element:

\begin{gather}
\frac{n}{2}\frac{\delta \emph{f}}{\delta Q_{ab}}=-\frac{1}{2}\left[ \left(
1-i\right) \left( u^{-2}\right) _{ab}+c.c.\right] +  \label{first_der_IV} \\
a_{T}\delta _{ab}+4Q_{aa}\delta _{ab}-2rQ_{ab}=0.  \notag
\end{gather}%
Since the electric charge (or the superconducting phase) $U(1)$ symmetry is
assumed, we consider only solutions with real $u_{ab}$. In this case $%
u_{ab}=Q_{ab}$ is a symmetric real matrix.

\emph{\textbf{The replica symmetric matrices Ansatz and the Edwards -
Anderson order parameter}}

Experience with very similar models in the theory of disordered magnets
indicates that solutions of these minimization equations are most likely to
belong to the class of hierarchical matrices, which will be described in the
next subsection, We limit ourselves here to most obvious of those, namely to
matrices which respect the $Z_{n}$ replica permutation symmetry
\begin{equation}
Q_{ab}\rightarrow Q_{p\left( a\right) p\left( b\right) }
\label{replica_symmetry_IV}
\end{equation}%
for any of $n!$ permutations $a\rightarrow p\left( a\right) .$ If we
include also disorder in $\left\vert \psi \left( r\right)
\right\vert ^{4}$ term, one will find in the low temperature region,
replica symmetry is spontaneously breaking as soon as the Edwards
-Anderson (EA) order parameter is non zero. However we will limit
our discussion to replica symmetric solution (not considering
disorder in $\left\vert \psi \left( r\right) \right\vert ^{4}$ ) and
think that the glass transition appears when the EA order parameter
is non zero. This transition line from zero EA to non zero EA
obtained in the following is very near to the replica symmetry
breaking transition line considering weak disorder in $\left\vert
\psi \left( r\right) \right\vert ^{4}$ \cite{Vinokur06}. We also
believe that even without disorder in $\left\vert \psi \left(
r\right) \right\vert ^{4}$ term, the replica symmetry is breaking if
we can solve the model non perturbatively.

The most general matrix of the replica symmetric solution has a form%
\begin{equation}
Q_{ab}=u_{ab}=\delta _{ab}\widetilde{u}+\left( 1-\delta _{ab}\right) \lambda
\label{RS_matrix_IV}
\end{equation}%
The off diagonal elements are equal to the Edwards - Anderson (EA) order
parameter $\lambda $. A nonzero value for this order parameter signals that
the annealed and the quenched averages are generally different. Let us
calculate $\overline{\left\langle \psi ^{\ast }\left( r\right) \right\rangle
\left\langle \psi \left( r\right) \right\rangle }$ starting from the eq.(\ref%
{double_thermal_IV}). Using the eq.(\ref{replica_propagator_IV}) , one
obtains in gaussian approximation
\begin{equation}
\overline{\left\langle \psi ^{\ast }\left( r\right) \right\rangle
\left\langle \psi \left( r\right) \right\rangle _{r}}=2\lambda
\label{EA_calculation_IV}
\end{equation}%
where $\left\langle {}\right\rangle _{r}$ contains also space average.

One can visualize this phase as a phase with locally broken $U\left(
1\right) $ symmetry with various directions of the phase at different
locations with zero average $\overline{\left\langle \psi \left( r\right)
\right\rangle }=0$ but a distribution of non zero value of characteristic
width $\lambda $. Distribution of more complicated quantities will be
discussed in the last subsection. Here we will refer to this state as glass,
although in the subsection E it will be referred to as the "ergodic pinned
liquid" (EPL) distinguished from the "nonergodic pinned liquid" (NPL) in
which, in addition, the ergodicity is broken. Broken ergodicity is related
to "replica symmetry breaking", however, as we show there, in the present
model of the $\delta T_{c}$ disorder and within gaussian approximation RSB
does not occur. If the EA order parameter is zero, disorder doesn't have a
profound effect on the properties of the vortex matter. We refer to this
state just as "liquid".

The dynamic properties of such phase are generally quite different from
those of the non -glassy $\lambda $\ (zero EA order parameter) phase. In
particular it is expected to exhibit infinite conductivity \cite%
{Fisher89,Fisher91,Dorsey92}. However if $u_{ab}$ is replica symmetric,\
pinning does not results in the multitude of time scales. Certain time scale
sensitive phenomena like various memory effects \cite%
{Paltiel00,Paltiel00b,Xiao02} and the responses to \textquotedblleft
shaking\textquotedblright\ \cite{Zeldov05} are expected to be different from
the case when $u_{ab}$ breaks the replica permutation symmetry.

We show in the following subsection that within the gaussian approximation
and the limited disorder model that we consider (the $\delta T_{c}$
inhomogeneity only) RSB does not occur. After that is shown, we can consider
the remaining problem without using the machinery of hierarchical matrices.

\emph{\textbf{Properties of the replica symmetric matrices}}

It is easy to work with the RS matrices like $u_{ab}$ in eq.(\ref%
{RS_matrix_IV}). It has two eigenvalues. A replica symmetric eigenvector
\begin{equation}
u%
\begin{pmatrix}
1 \\
1 \\
.. \\
1%
\end{pmatrix}%
=\Delta ^{\prime }%
\begin{pmatrix}
1 \\
1 \\
.. \\
1%
\end{pmatrix}%
;\text{ \ \ }\Delta ^{\prime }\equiv \widetilde{u}+\left( n-1\right) \lambda
\simeq \widetilde{u}-\lambda ,\text{ }  \label{RS_eigenvalue_IV}
\end{equation}%
where sub leading terms at small $n$, were omitted in the last line, and the
rest of the space (replica asymmetric vectors) which is $n-1$ times
degenerate. For example%
\begin{equation}
u%
\begin{pmatrix}
1 \\
-1 \\
0 \\
..%
\end{pmatrix}%
=\Delta
\begin{pmatrix}
1 \\
-1 \\
0 \\
..%
\end{pmatrix}%
;\text{ \ \ }\Delta \equiv \widetilde{u}-\lambda \simeq \widetilde{u}%
-\lambda ,  \label{RA_eigenvalue_IV}
\end{equation}%
The counting seems strange, but mathematically can be defined and works.
Numerous attempts to discredit replica calculations on these grounds were
proven baseless. Note that the two eigenvalues differ by order $n$ terms
only. Projectors on these spaces are%
\begin{eqnarray}
P_{S} &=&\frac{1}{n}\left(
\begin{array}{cccc}
1 & 1 & .. & 1 \\
1 & 1 & .. & 1 \\
.. & .. & .. & .. \\
1 & 1 & .. & 1 \\
1 & 1 &  & 1%
\end{array}%
\right) ;\text{ \ \ \ }P_{A}=\mathcal{I}-P_{s};  \label{projectors_IV} \\
u &=&\Delta ^{\prime }P_{S}+\Delta P_{A}.  \notag
\end{eqnarray}%
Here $\mathcal{I}$ is the unit matrix $\delta _{ab}.$ It is easy to invert
RS matrices and multiply them using this form. For example
\begin{equation}
u^{-1}=\Delta ^{\prime -1}P_{S}+\Delta ^{-1}P_{A}.  \label{inverse_IV}
\end{equation}

\subsubsection{The glass transition between the two replica symmetric
solutions}

\emph{\textbf{The unpinned liquid and the \textquotedblleft ergodic
glass\textquotedblright\ replica symmetric solutions of the minimization
equations}}

The minimization equation eq.(\ref{first_der_IV}) for RS matrices takes a
form%
\begin{gather}
-\Delta ^{\prime -2}P_{S}-\Delta ^{-2}P_{A}+\left( a_{T}+4\widetilde{u}%
\right) \mathcal{I}  \label{minim_derivation1_IV} \\
-2r\left( \Delta ^{\prime }P_{S}+\Delta P_{A}\right) =0  \notag
\end{gather}%
Expressing it via independent matrices $I$ and $P_{s}$ one obtains:%
\begin{gather}
\left[ \Delta ^{-2}-\Delta ^{\prime -2}-2r\left( \Delta ^{\prime }-\Delta
\right) \right] P_{S}+ \\
\left( -\Delta ^{-2}+a_{T}+4\widetilde{u}-2r\Delta \right) \mathcal{I}=0.
\notag
\end{gather}%
To leading order in $n$ (first) the $P_{s}$ equation is%
\begin{equation}
\lambda \left( \Delta ^{-3}-r\right) =0\text{.}  \label{RS_first_IV}
\end{equation}%
This means that there exists a RS symmetric solution $\lambda =0$. In
addition there is a non diagonal one. It turns out that there is a third
order transition between them.

The second equation,%
\begin{equation}
-\Delta ^{-2}+a_{T}+4\widetilde{u}-2r\Delta =0\text{,}  \label{RS_second_IV}
\end{equation}
for the diagonal ("liquid") solution, $\Delta _{l}=\widetilde{u}_{l},$ is
just a cubic equation:

\begin{equation}
-\widetilde{u}_{l}^{-2}+a_{T}+4\widetilde{u}_{l}-2r\widetilde{u}_{l}=0\text{.%
}  \label{diageq_IV}
\end{equation}%
For the non - diagonal solution the first equation eq.(\ref{RS_first_IV})
gives $\Delta =r^{1/3}$, which, when plugged into the first equation, gives:%
\begin{equation}
\widetilde{u}_{g}=\frac{1}{4}\left( 3r^{2/3}-a_{T}\right) ;\lambda =\frac{1}{%
4}\left( 3r^{2/3}-a_{T}\right) -r^{-1/3}.  \label{u_tillde_IV}
\end{equation}%
The matrix $u$ therefore is
\begin{equation}
u_{ab}^{g}=r^{-1/3}\delta _{ab}+\lambda .  \label{uab_IV}
\end{equation}%
The two solutions coincide when $\lambda _{g}=0$ leading to the glass line
equation%
\begin{equation}
a_{T}^{g}=r^{-1/3}\left( 3r-4\right) .  \label{glass_line_eq_IV}
\end{equation}

\emph{\textbf{Free energy and its derivatives. The third order glass line.}}

Let us now calculate energies of the solutions. Energy of such a RS matrix
is given, using eq.(\ref{inverse_IV}), by%
\begin{gather}
\frac{n}{2}\emph{f}_{gauss}=\sum_{a}\{\left( \Delta ^{\prime -1}P_{S}+\Delta
^{-1}P_{A}\right) _{aa}+a_{T}\widetilde{u}  \label{RS_free_en_derivation_IV}
\\
+2\widetilde{u}^{2}-r\left( \Delta ^{\prime 2}P_{S}+\Delta ^{2}P_{A}\right)
_{aa}\}  \notag \\
\simeq n\left\{ 2\Delta ^{-1}-\Delta ^{-2}\lambda +a_{T}\widetilde{u}+2%
\widetilde{u}^{2}-r\left( \Delta ^{2}+2\Delta \lambda \right) \right\} ,
\notag
\end{gather}%
where leading order in small $n$ was kept. The RS energy is
\begin{equation}
\emph{f}_{l}=2\widetilde{u}_{l}^{-1}+2a_{T}\widetilde{u}_{l}+2\left(
4-2r\right) \widetilde{u}_{l}^{2}\text{,}  \label{RS_free_energy_IV}
\end{equation}%
which can be further simplified by using eq.(\ref{diageq_IV}),%
\begin{equation}
\emph{f}_{l}=4\widetilde{u}_{l}^{-1}\text{.}
\end{equation}%
The glass free energy is even simpler:
\begin{equation}
\emph{f}_{g}=6r^{1/3}-\frac{1}{4}\left( 3r^{2/3}-a_{T}\right) ^{2}.
\label{uab2_IV}
\end{equation}

Since in addition to the energy, the first derivative of the scaled energy,
the scaled entropy
\begin{equation}
\frac{d\emph{f}}{da_{T}}=2r^{-1/3}  \label{entropy_IV}
\end{equation}%
and the second derivative, specific heat,
\begin{equation}
\frac{d^{2}\emph{f}}{da_{T}^{2}}=-\frac{1}{2}  \label{specific_heat_IV}
\end{equation}%
respectively, coincide for both solution on the transition line defined by
eq.(\ref{glass_line_eq_IV}). The third derivatives are different so that the
transition is a third order one.

\emph{\textbf{Hessian and the stability domain of a solution}}

Up to now we have found two homogeneous solutions of the minimization
equations. There might be more and the solutions might not be stable, when
considered on the wider set of gaussian states. In order to prove that a
solution is stable beyond the set of replica symmetric matrices $u$, one has
to calculate the second derivative of free energy (so called Hessian) with
respect to arbitrary real matrix $Q_{ab}$ defined in eq.(\ref{Qdef_IV}):

\begin{gather}
H_{(ab)(cd)}\equiv \frac{n}{2}\frac{\delta ^{2}f_{eff}}{\delta Q_{ab}\delta
Q_{cd}}  \label{Hessian_IV} \\
=\{\frac{1}{2}\left[ \left( u^{-2}\right) _{ac}\left( u^{-1}\right)
_{db}-i\left( u^{-2}\right) _{ad}\left( u^{-1}\right) _{cb}\right] +\frac{1}{%
2}[\left( u^{-1}\right) _{ac}\times  \notag \\
\left( u^{-2}\right) _{db}-i\left( u^{-1}\right) _{ad}\left( u^{-2}\right)
_{cb}]+c.c.\}+4\delta _{ac}\delta _{bd}\delta _{ab}-2r\delta _{ac}\delta
_{bd}.  \notag
\end{gather}%
The Hessian should be considered as a matrix in a space, which itself is a
space of matrices, so that Hessian's index contains two pairs of indices of $%
u$. We will use a simplified notation for\ the product of the Kronecker
delta functions with more than two indices: $\delta _{ac}\delta _{bd}\delta
_{ab}\equiv \delta _{abcd}$. It is not trivial to define what is meant by
"positive definite" when the number of components approaches zero. It turns
out that the correct definition consists in finding all the eigenvalues of
the Hessian "super" matrix.

\emph{\textbf{Stability of the liquid solution}}

For the diagonal solution the Hessian is a very simple operator on the space
of real symmetric matrices:

\begin{equation}
H_{(ab)(cd)}=c_{I}I_{abcd}+c_{J}J_{abcd},  \label{Hessian1_IV}
\end{equation}%
where the operators $I$ (the identity in this space) and $J$ are defined as
\begin{equation}
I\equiv \delta _{ac}\delta _{bd};\text{ \ \ \ \ }J=\delta _{abcd}
\label{I&J_IV}
\end{equation}%
and their coefficients in the liquid phase are:

\begin{equation}
c_{I}=2\left( \widetilde{u}_{l}^{-3}-r\right) ,\text{ \ \ }c_{J}=4,
\label{coefficients_Hessian_IV}
\end{equation}%
with $\widetilde{u}_{l}$ being a solution of eq.(\ref{diageq_IV}). The
corresponding eigenvectors in the space of symmetric matrices are
\begin{equation}
v_{(cd)}\equiv A\delta _{cd}+B.
\end{equation}%
To find eigenvalues $h$ of $H$ we apply the Hessian on a vector $v$. The
result is (dropping terms vanishing in the limit $n\rightarrow 0$):%
\begin{eqnarray}
H_{(ab)(cd)}v_{cd} &=&A\left( c_{I}+c_{J}\right) \delta _{ab}+B\left(
c_{I}+c_{J}\delta _{ab}\right)  \notag \\
&=&h\left( A\delta _{ab}+B\right)  \label{Hessian_II}
\end{eqnarray}%
There two eigenvalues are therefore: $h^{(1)}=c_{I}$ and $%
h^{(2)}=c_{I}+c_{J} $. Since $c_{J}=4>0,$ the sufficient condition for
stability is:
\begin{equation}
c_{I}=2\left( \widetilde{u}^{-3}-r\right) >0.  \label{criterion_IV}
\end{equation}%
It is satisfied everywhere below the transition line of eq.(\ref%
{glass_line_eq_IV}).

\emph{\textbf{The stability of the glass solution}}

The analysis of stability of the non - diagonal solution is slightly more
complicated. The Hessian for the non - diagonal solution is:

\begin{equation}
H_{(ab)(cd)}=c_{V}V+c_{U}U+c_{J}J,  \label{Hessian3_IV}
\end{equation}%
where new operators are%
\begin{equation}
V_{(ab)(cd)}=\delta _{ac}+\delta _{bd};U_{(ab)(cd)}=1  \label{V&U_IV}
\end{equation}%
and coefficients are

\begin{equation}
c_{V}=-3\lambda r^{2/3};\text{ \ \ }c_{U}=4\lambda ^{2}r^{1/3};c_{J}=4
\label{coef2_IV}
\end{equation}%
In the present case, one obtains three different eigenvalues, \cite%
{Alameida78,Fischer,Dotsenko}
\begin{equation}
h^{(1.2)}=2\left( 1\pm \sqrt{1-4\lambda r^{2/3}}\right)  \label{h12_IV}
\end{equation}%
and $h^{(3)}=0.$ Note that the eigenvalue of Hessian on the antisymmetric
matrices are degenerate with eigenvalue $h^{(1)}$ in this case (we will come
back later on this eigenvalue). For $\lambda <0$ the solution is unstable
due to negative $h^{(2)}.$ For $\lambda >0,$ both eigenvalues are positive
and the solution is stable. The line $\lambda =0$ coincides with the third
order transition line, hence the non diagonal solution is stable when the
diagonal is unstable and \textit{vise versa}. We conclude that one of the
two RS solutions is stable for any value of external parameters (here
represented by $a_{T}$ and $r$). There still might be a replica asymmetric
solution with first order transition to it, but this possibility will be
ruled out within the gaussian approximation and in the homogeneous phase
without the $\left\vert \psi \right\vert ^{4}$ disorder term in the next
subsection. Therefore the transition does not correspond to RSB. Despite
this in the phase with nonzero EA order parameter there are Goldstone bosons
corresponding to $h^{(3)}$ in the replica limit of $n\rightarrow 0$. The
criticality and the zero modes due to disorder (pinning) in this phase might
lead to great variety of interesting phenomena both in statics and dynamics.

\emph{\textbf{Generalizations and comparison with experimental
irreversibility line}}

\begin{figure}[t]
\centering \rotatebox{270}{\includegraphics[width=0.45%
\textwidth,height=9cm]{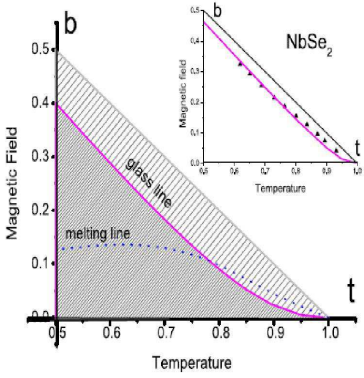}} \caption{$NbSe_{2}$ phase
diagram} \label{figIV3}
\end{figure}

\begin{figure}[t]
\centering \rotatebox{360}{\includegraphics[width=0.45%
\textwidth,height=9cm]{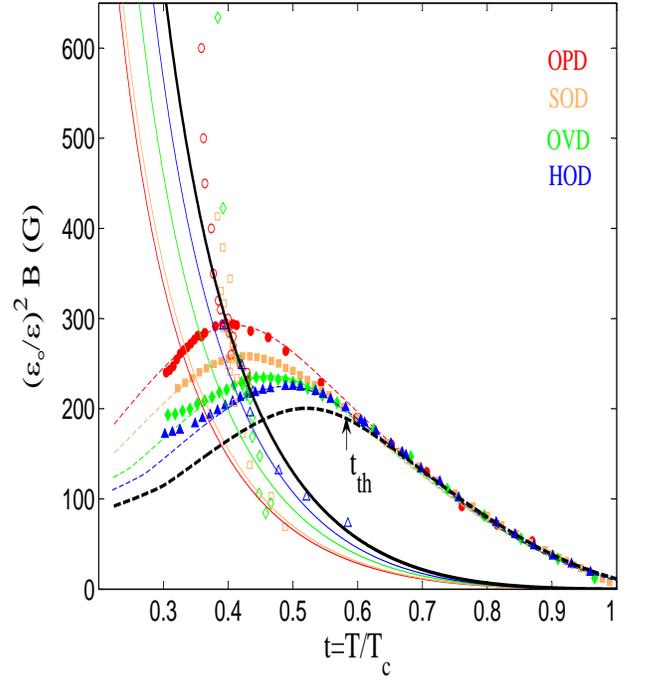}} \caption{Melting line
and order - disorder lines in layered superconductor BSCCO
\cite{Zeldov07}} \label{figIV4}
\end{figure}

The glass line resembles typical irreversibility lines in both low $T_{c}$
and high $T_{c}$ materials, see Fig. \ref{figIV3}, where the irreversibility
line of $NbSe_{2}$ is fitted.
The theory can be generalized to 2D GL model describing thin films or very
anisotropic layered superconductors. The glass line is given in 2D%
\begin{equation}
a_{T}^{g}=2\sqrt{2}\frac{R-1}{\sqrt{R}}  \label{2D_glass_IV}
\end{equation}%
An examples of \ organic superconductor \cite{Shibauchi98} was given in \cite%
{Vinokur06}. The data of BSCCO \cite{Zeldov05} are compared to the
theoretical results on Fig. \ref{figIV4}.

\subsubsection{The disorder distribution moments of the LLL magnetization\ }

As was discussed in III, the magnetization within LLL is proportional to the
superfluid density, whose average is

\begin{gather}
\overline{\left\langle \psi ^{\ast }\left( r\right) \psi \left( r\right)
\right\rangle _{r}}=\lim_{n\rightarrow 0}\frac{1}{n}\sum_{a}\left\langle
\Psi ^{\ast a}\left( r\right) \Psi ^{a}\left( r\right) \right\rangle _{r}
\label{correlator_IV} \\
\frac{1}{n}\sum\limits_{a}\frac{4\pi \sqrt{2}}{\left( 2\pi \right) ^{3}}%
\int_{k}[(\frac{k_{z}^{2}}{2}\emph{1}+\mu ^{2-1})]_{aa}=\frac{2}{n}%
\sum\limits_{a}u_{aa}=2\widetilde{u}.  \notag
\end{gather}%
The variance of the distribution is determined from the two thermal averages
disorder average:
\begin{equation}
\overline{\lbrack \left\langle \psi ^{\ast }\psi \right\rangle \left\langle
\psi ^{\ast }\psi \right\rangle ]_{r}}=\frac{1}{n\left( n-1\right) }%
\sum_{a\neq b}\left\langle \left\vert \psi ^{\ast a}\left( r\right)
\right\vert ^{2}\left\vert \psi ^{b}\left( r\right) \right\vert
^{2}\right\rangle _{r}  \label{variance_deriv_IV}
\end{equation}%
Within gaussian approximation (Wick contractions) the correlators are%
\begin{equation}
\overline{\left[ \left\langle \psi ^{\ast }\psi \right\rangle \left\langle
\psi ^{\ast }\psi \right\rangle \right] _{r}}=4(\widetilde{u}^{2}+\lambda
^{2})  \label{psi_thermal_square_IV}
\end{equation}

Therefore variance of the distribution is given by $\lambda $. This variance
determines the width of the magnetization loop. In turn, according to the
phenomenological Bean model \cite{Tinkham} , the width of the magnetization
loop is proportional to the critical current. The distribution of
magnetization is not symmetric, as the third moment shows:

\begin{equation}
\overline{\left[ \left\langle \psi ^{\ast }\psi \right\rangle ^{3}\right]
_{r}}=8\left( \widetilde{u}^{3}+3\widetilde{u}\lambda ^{2}+\lambda
^{3}\right) \text{.}  \label{psi_thermal_cube_IV}
\end{equation}%
It's calculation is more involved. The third irreducible cumulant is
therefore nonzero:

\begin{equation}
\overline{\left[ \left\langle \psi ^{\ast }\psi \right\rangle ^{3}\right]
_{r}}-3\cdot 2\widetilde{u}\overline{\left[ \left\langle \psi ^{\ast }\psi
\right\rangle ^{2}\right] _{r}}+2\left( 2\widetilde{u}\right) ^{3}=8\lambda
^{3}.  \label{three_variance_IV}
\end{equation}%
In analogy to ref. \cite{Mezard91} one can define "glass susceptibility"%
\begin{equation}
\chi =\left\langle \psi ^{\ast }\psi \right\rangle -\left\langle \psi ^{\ast
}\right\rangle \left\langle \psi \right\rangle =2\left( \widetilde{u}%
-\lambda \right)  \label{glass_suss_IV}
\end{equation}%
useful in description of the "glassy" state. Its variance of susceptibility
vanishes without RSB:

\begin{equation}
\overline{\chi ^{2}}=\overline{\chi }^{2}\text{.}
\label{glass_suss_square_IV}
\end{equation}%
We return to the replica symmetry breaking after considering the crystalline
phase.\bigskip

\subsection{Gaussian theory of a disordered crystal}

\subsubsection{Replica symmetric Ansatz in Abrikosov crystal}

In subsection A we used perturbation theory in disorder to assess the basic
properties of the vortex crystal. However we learned in the previous
subsection that certain properties like the glass related phenomena cannot
be captured by perturbation theory and one has to resort to simplest
nonperturbative methods available. In the homogeneous phase gaussian
approximation in the replica symmetric subspace was developed and we now
generalize to a more complicated crystalline case. This is quite analogous
to what we did with thermal fluctuations, so the description contains less
details.

\emph{\textbf{Replica symmetric shift of the free energy}}

Within the gaussian approximation the expectation values of the fields as
well as their propagators serve as variational parameters. To implement it,
it is convenient to shift and \textquotedblright rotate\textquotedblright\
the fields according to eq.(\ref{shift_III}):

\begin{equation}
\psi _{a}\left( r\right) =v_{a}\varphi \left( r\right) +\frac{1}{4\pi ^{3/2}}%
\int_{k}c_{\mathbf{k}}\exp \left( ik_{z}\right) \varphi _{\mathbf{k}}\left(
\mathbf{r}\right) (O_{k}^{a}+iA_{k}^{a}),  \label{psi_shift_IV}
\end{equation}%
where factor $c_{\mathbf{k}}\equiv \frac{\gamma _{k}}{\left\vert \gamma
_{k}\right\vert }$ was introduced for convenience and $\varphi _{\mathbf{k}%
}(x)$ are the quasimomentum functions. In principle the shift as well as
fields are replica index dependence. However assumption of the unbroken
replica symmetry means that

\begin{equation}
v_{a}=v,  \label{shift_IV}
\end{equation}%
is the only variational shift parameter.

To evaluate gaussian energy we first substitute this into free energy and
write quadratic, cubic and quartic parts in fields $A$ and $O$. The
quadratic terms originating from the interaction and disorder term are
listed below. The $\ OO$ terms coming from the interaction part are: the $%
O_{a}^{\ast }O_{b}$\ term

\begin{gather}
\frac{v^{2}}{2^{4}\pi ^{3}}\int_{r}\left\vert \varphi \left( r\right)
\right\vert ^{2}\int_{k,l}\varphi _{\mathbf{k}}^{\ast }(x)c_{-\mathbf{k}}c_{%
\mathbf{k}}\varphi _{\mathbf{l}}\left( \mathbf{r}\right) O_{-k}^{a}O_{l}^{b}
\label{O_terms_IV} \\
=\frac{v^{2}}{2}\int_{k}\beta _{\mathbf{k}}O_{k}^{a}O_{-k}^{b},  \notag
\end{gather}%
the $O_{a}^{\ast }O_{b}^{\ast }$ term $\frac{v^{2}}{4}\int_{k}\left\vert
\gamma _{\mathbf{k}}\right\vert O_{k}^{a}O_{-k}^{b}$, and finally the $%
O_{a}O_{b}$ term $\frac{v^{2}}{4}\int_{k}\left\vert \gamma _{\mathbf{k}%
}\right\vert O_{k}^{a}O_{-k}^{b}$. Sum over all four $OO$ terms is therefore

\begin{equation}
\int_{k}v^{2}(\beta _{\mathbf{k}}+\left\vert \gamma _{k}\right\vert
)O_{k}^{a}O_{-k}^{b}.  \label{O_terms_res_IV}
\end{equation}%
Similarly the $AA$ terms sum up to:
\begin{equation}
\int_{k}v^{2}(\beta _{\mathbf{k}}-\left\vert \gamma _{\mathbf{k}}\right\vert
)A_{k}^{a}A_{-k}^{b},  \label{A_terms_res_IV}
\end{equation}%
while the$\ OA$ terms cancel. The disorder term contributes (leading order
in $n$, as usual for replica method):
\begin{eqnarray}
&&\frac{r}{2}v^{2}[n\int_{k}\beta _{\mathbf{k}}{\sum_{a}}%
A_{k}^{a}A_{-k}^{a}+\int_{k}(\beta _{\mathbf{k}}-\left\vert \gamma _{\mathbf{%
k}}\right\vert ){\sum_{a,b}}A_{k}^{a}A_{-k}^{b}]  \label{f2A_IV} \\
&=&\frac{r}{2}v^{2}\int_{k}[{\sum_{a}}\left( \beta _{\mathbf{k}}-\left\vert
\gamma _{\mathbf{k}}\right\vert \right) A_{k}^{a}A_{-k}^{a}+{\sum_{a\neq b}}%
(\beta _{\mathbf{k}}-\left\vert \gamma _{\mathbf{k}}\right\vert
)A_{k}^{a}A_{-k}^{b}]  \notag
\end{eqnarray}%
to the $A$ part and
\begin{equation}
\frac{r}{2}v^{2}[\int_{k}{\sum_{a}}\left( \beta _{\mathbf{k}}+\left\vert
\gamma _{\mathbf{k}}\right\vert \right) O_{k}^{a}O_{-k}^{a}+{\sum_{a\neq b}}%
\int_{k}(\beta _{\mathbf{k}}+\left\vert \gamma _{\mathbf{k}}\right\vert
)O_{k}^{a}O_{-k}^{b}]  \label{f2O_IV}
\end{equation}%
to the $O$ part. The quadratic part of the free energy therefore is:

\begin{gather}
f_{2}=\frac{1}{2}\int_{k}\sum_{a}[a_{T}+v^{2}\left( 2\beta _{\mathbf{k}%
}-\left\vert \gamma _{\mathbf{k}}\right\vert \right) +rv^{2}\left( \beta _{%
\mathbf{k}}-\left\vert \gamma _{\mathbf{k}}\right\vert \right)
]A_{k}^{a}A_{-k}^{a}  \notag \\
+rv^{2}{\sum_{a\neq b}}\left( \beta _{\mathbf{k}}-\left\vert \gamma _{%
\mathbf{k}}\right\vert \right) A_{k}^{a}A_{-k}^{b}+[a_{T}+v^{2}\left( 2\beta
_{\mathbf{k}}+\left\vert \gamma _{\mathbf{k}}\right\vert \right)
+rv^{2}\times  \notag \\
\left( \beta _{\mathbf{k}}+\left\vert \gamma _{\mathbf{k}}\right\vert
\right) ]O_{k}^{a}O_{-k}^{a}+rv^{2}{\sum_{a\neq b}}\left( \beta _{\mathbf{k}%
}-\left\vert \gamma _{\mathbf{k}}\right\vert \right) O_{k}^{a}O_{-k}^{b}.
\label{f2_all_IV}
\end{gather}%
There is no linear term and cubic term is not needed since its contraction
vanishes. The quartic term will be taken into account later. We will not
need cubic terms within the gaussian approximation, while the quartic terms
are not affected by the shift of fields. We are ready therefore to write
down the gaussian variational energy.

\emph{\textbf{Gaussian energy}}

Now we describe briefly the contributions to the gaussian energy. The mean
field terms (namely containing the shift only with no pairings) are
\begin{equation}
f_{mf}=\int_{x}[nf(\psi _{a})-\frac{r}{2}n^{2}\left\vert \psi
_{a}\right\vert ^{4}].  \label{f_mf_IV}
\end{equation}%
which using eq.(\ref{psi_shift_IV}) takes a form:
\begin{equation}
\emph{f}_{mf}=na_{T}v^{2}+\frac{n}{2}\beta _{A}v^{4}-\frac{r}{2}n^{2}\beta
_{A}v^{4}.  \label{mf_term_IV}
\end{equation}%
The last term can be omitted since the power of $n$ exceeds $1$. Gaussian
effective energy in addition to $f_{mf}$ contains the Trlog term and the
\textquotedblright bubble diagrams\textquotedblright . The Trlog term comes
from free gaussian part, see section III. The reference \textquotedblright
best gaussian (or quadratic) energy\textquotedblright\ is defined
variationally as a quadratic form
\begin{eqnarray}
&&\frac{1}{2}\int_{k}\left( \varepsilon
_{k}^{A}A_{k}^{a}A_{-k}^{a}+\varepsilon _{k}^{O}O_{k}^{a}O_{-k}^{a}\right) +
\\
&&{\sum_{a\neq b}}\left[ \left( \overline{\mu }_{\mathbf{k}}^{A}\right)
^{2}A_{k}^{a}A_{-k}^{b}+\left( \overline{\mu }_{\mathbf{k}}^{O}\right)
^{2}O_{k}^{a}O_{-k}^{b}\right] ;  \notag
\end{eqnarray}%
and $\varepsilon $%
\begin{equation}
_{k}^{A}=k_{z}^{2}/2+\left( \mu _{\mathbf{k}}^{A}\right) ^{2};\varepsilon
_{k}^{O}=k_{z}^{2}/2+\left( \mu _{\mathbf{k}}^{O}\right) ^{2}
\end{equation}%
where $\mu _{\mathbf{k}}^{A},\mu _{\mathbf{k}}^{O},\overline{\mu }_{\mathbf{k%
}}^{A},\overline{\mu }_{\mathbf{k}}^{O},$ are all variational parameters. We
assumed no mixing of the $A$ and the $O$ modes, following the experience in
the clean case and the structure of the quadratic part determined in the
previous subsection.

In the following we keep sub leading terms in $n$ since they contribute to
order $n$ in energy. The Trlog (divided by volume) is sum of logarithms of
all the eigenvalues.

\begin{gather}
\emph{f}_{tr\log }=\frac{1}{2^{3/2}\pi ^{2}}\int_{k}(n-1)\log [\varepsilon
_{k}^{A}-\left( \overline{\mu }_{\mathbf{k}}^{A}\right) ^{2}]+\log
[\varepsilon _{k}^{A}+(n-1)  \notag \\
\times \left( \overline{\mu }_{\mathbf{k}}^{A}\right) ^{2}]+(n-1)\log
[\varepsilon _{k}^{O}-\left( \overline{\mu }_{\mathbf{k}}^{O}\right)
^{2}]+\log [\varepsilon _{k}^{O}+(n-1)\left( \overline{\mu }_{\mathbf{k}%
}^{O}\right) ^{2}]  \notag \\
=\frac{1}{\pi }\int_{\mathbf{k}}(n-1)[\left( \mu _{\mathbf{k}}^{A}\right)
^{2}-\left( \overline{\mu }_{\mathbf{k}}^{A}\right) ^{2}]^{1/2}+[\left( \mu
_{\mathbf{k}}^{A}\right) ^{2}+(n-1)\left( \overline{\mu }_{\mathbf{k}%
}^{A}\right) ^{2}]^{\frac{1}{2}}  \notag \\
+(n-1)[\left( \mu _{\mathbf{k}}^{O}\right) ^{2}-\left( \overline{\mu }_{%
\mathbf{k}}^{O}\right) ^{2}]^{1/2}+[\left( \mu _{\mathbf{k}}^{O}\right)
^{2}+(n-1)\left( \overline{\mu }_{\mathbf{k}}^{O}\right) ^{2}]^{\frac{1}{2}}
\label{f_trlog_IV}
\end{gather}%
where in the last line integration is over Brillouin zone. One observes that
the order $O\left( n\right) $ terms cancel, while the relevant order is:%
\begin{align}
\emph{f}_{tr\log }& =\frac{n}{2\pi }\int_{\mathbf{k}}2[\left( \mu _{\mathbf{k%
}}^{A}\right) ^{2}-\left( \overline{\mu }_{\mathbf{k}}^{A}\right) ^{2}]^{%
\frac{1}{2}}+\left( \overline{\mu }_{\mathbf{k}}^{A}\right) ^{2}[\left( \mu
_{\mathbf{k}}^{A}\right) ^{2}-\left( \overline{\mu }_{\mathbf{k}}^{A}\right)
^{2}]^{-\frac{1}{2}}  \notag \\
& +2[\left( \mu _{\mathbf{k}}^{O}\right) ^{2}-\left( \overline{\mu }_{%
\mathbf{k}}^{O}\right) ^{2}]^{\frac{1}{2}}+\left( \overline{\mu }_{\mathbf{k}%
}^{O}\right) ^{2}[\left( \mu _{\mathbf{k}}^{O}\right) ^{2}-\left( \overline{%
\mu }_{\mathbf{k}}^{O}\right) ^{2}]^{-\frac{1}{2}}.  \label{trlog_IV}
\end{align}%
The diagrams are of two kinds. Those including one propagator and ones which
have two propagators from the part quartic in fields. The propagators are
expectation values of pair of fluctuating fields obtained by inverting the
replica symmetric matrix like in the previous subsection. For example for
the acoustic mode one gets:
\begin{eqnarray}
4\pi \sqrt{2}p_{k}^{A} &=&\langle A_{k}^{a}A_{-k}^{a}\rangle =4\pi \sqrt{2}%
\frac{\varepsilon _{k}^{A}-2\left( \overline{\mu }_{\mathbf{k}}^{A}\right)
^{2}}{[\varepsilon _{k}^{A}-\left( \overline{\mu }_{\mathbf{k}}^{A}\right)
^{2}]^{2}},  \notag \\
4\pi \sqrt{2}\overline{p}_{k}^{A} &=&\langle A_{k}^{a}A_{-k}^{b}\rangle =-%
\frac{4\pi \sqrt{2}\left( \overline{\mu }_{\mathbf{k}}^{A}\right) ^{2}}{%
[\varepsilon _{k}^{A}-\left( \overline{\mu }_{\mathbf{k}}^{A}\right)
^{2}]^{2}}.  \label{p_def_IV}
\end{eqnarray}%
The integrals of the propagators over $k_{z}$ give:

\begin{gather}
\frac{1}{2\left( 2\pi \right) ^{3}}\int_{k_{z}}4\pi \sqrt{2}p_{k}^{A}=p_{%
\mathbf{k}}^{A}=\frac{1}{2\pi }[  \label{p_res_IV} \\
(\left( \mu _{\mathbf{k}}^{A}\right) ^{2}-\left( \overline{\mu }_{\mathbf{k}%
}^{A}\right) ^{2})^{-1/2}-\frac{\left( \overline{\mu }_{\mathbf{k}%
}^{A}\right) ^{2}}{2}(\left( \mu _{\mathbf{k}}^{A}\right) ^{2}-\left(
\overline{\mu }_{\mathbf{k}}^{A}\right) ^{2})^{-3/2}],  \notag \\
\frac{1}{2\left( 2\pi \right) ^{3}}\int_{k_{z}}4\pi \sqrt{2}\overline{p}%
_{k}^{A}=\overline{p}_{\mathbf{k}}^{A}  \notag \\
=\frac{1}{2\pi }[-\frac{\left( \overline{\mu }_{\mathbf{k}}^{A}\right) ^{2}}{%
2}(\left( \mu _{\mathbf{k}}^{A}\right) ^{2}-\left( \overline{\mu }_{\mathbf{k%
}}^{A}\right) ^{2})^{-3/2}],  \notag
\end{gather}%
and similarly for $O$.

The contraction of the quadratic parts, after the integration and expanded
to order $n$ results in:%
\begin{gather}
\emph{f}_{2}=-\frac{1}{2}f_{tr\log }+n\int_{\mathbf{k}}\{p_{\mathbf{k}%
}^{A}[a_{T}+v^{2}(2\beta _{\mathbf{k}}-\left\vert \gamma _{\mathbf{k}%
}\right\vert )-rv^{2}(\beta _{\mathbf{k}}-  \notag \\
\left\vert \gamma _{\mathbf{k}}\right\vert )]+rv\overline{p}_{k}^{A}(\beta _{%
\mathbf{k}}-\left\vert \gamma _{\mathbf{k}}\right\vert )\}+n\int_{\mathbf{k}%
}\{p_{\mathbf{k}}^{O}[a_{T}+v^{2}(2\beta _{\mathbf{k}}+\left\vert \gamma _{%
\mathbf{k}}\right\vert )  \notag \\
-rv^{2}(\beta _{\mathbf{k}}+\left\vert \gamma _{\mathbf{k}}\right\vert )]+rv%
\overline{p}_{k}^{O}(\beta _{\mathbf{k}}+\left\vert \gamma _{\mathbf{k}%
}\right\vert )\}  \label{f2_term_IV}
\end{gather}%
For quartic terms coming from two contractions of the interaction and the
disorder part one obtains,
\begin{eqnarray}
\emph{f}_{int} &=&n\int_{\mathbf{k,l}}\left( p_{\mathbf{k}}^{A}+p_{\mathbf{k}%
}^{O}\right) \beta _{\mathbf{k-l}}\left( p_{\mathbf{l}}^{A}+p_{\mathbf{l}%
}^{O}\right)  \label{int_term_IV} \\
&&+\frac{\gamma _{\mathbf{k}}\gamma _{\mathbf{l}}}{2\beta _{\Delta }}\left(
p_{\mathbf{k}}^{A}-p_{\mathbf{k}}^{O}\right) \left( p_{\mathbf{l}}^{A}-p_{%
\mathbf{l}}^{O}\right)  \notag
\end{eqnarray}%
and%
\begin{gather}
\emph{f}_{dis}=-\frac{r}{2}\int_{\mathbf{k,l}}[\left( p_{\mathbf{k}}^{A}+p_{%
\mathbf{k}}^{O}\right) \beta _{\mathbf{k-l}}\left( p_{\mathbf{l}}^{A}+p_{%
\mathbf{l}}^{O}\right) +  \notag \\
\frac{\gamma _{\mathbf{k}}\gamma _{\mathbf{l}}}{\beta _{\Delta }}\left( p_{%
\mathbf{k}}^{A}-p_{\mathbf{k}}^{O}\right) \left( p_{\mathbf{l}}^{A}-p_{%
\mathbf{l}}^{O}\right) ]-\left( \overline{p}_{\mathbf{k}}^{A}+\overline{p}_{%
\mathbf{k}}^{O}\right) \beta _{\mathbf{k-l}}  \label{f_dis_IV} \\
\left( \overline{p}_{\mathbf{l}}^{A}+\overline{p}_{\mathbf{l}}^{O}\right) -%
\frac{\gamma _{\mathbf{k}}\gamma _{\mathbf{l}}}{\beta _{\Delta }}\left(
\overline{p}_{\mathbf{k}}^{A}-\overline{p}_{\mathbf{k}}^{O}\right) \beta _{%
\mathbf{k-l}}\left( \overline{p}_{\mathbf{l}}^{A}-\overline{p}_{\mathbf{l}%
}^{O}\right)  \notag
\end{gather}%
respectively. Finally we get
\begin{equation}
\emph{f}_{gauss}=\emph{f}_{mf}+\emph{f}_{2}+\emph{f}_{tr\log }+\emph{f}%
_{int}+\emph{f}_{dis}.  \label{fgauss_IV}
\end{equation}

\bigskip

\subsubsection{Solution of the gap equations}

\emph{\textbf{Gap and shift equations}}

The gaussian energy is minimized with respect to variational parameters.
Differentiating with respect to $v^{2}$ one gets the "shift" equation:%
\begin{gather}
0=a_{T}+\beta _{\Delta }v^{2}+\int_{\mathbf{k}}(2\beta _{\mathbf{k}%
}-\left\vert \gamma _{\mathbf{k}}\right\vert )p_{\mathbf{k}}^{A}-r\left(
\beta _{\mathbf{k}}-\left\vert \gamma _{\mathbf{k}}\right\vert \right)
\label{shift_eq_IV} \\
\left( p_{\mathbf{k}}^{A}-\overline{p}_{\mathbf{k}}^{A}\right) +\int_{%
\mathbf{k}}(2\beta _{\mathbf{k}}+\left\vert \gamma _{\mathbf{k}}\right\vert
)p_{\mathbf{k}}^{O}-r\left( \beta _{\mathbf{k}}+\left\vert \gamma _{\mathbf{k%
}}\right\vert \right) \left( p_{\mathbf{k}}^{O}-\overline{p}_{\mathbf{k}%
}^{O}\right)  \notag
\end{gather}%
while differentiating with respect to four variational parameters in the
propagator matrix gap equations are obtained%
\begin{eqnarray}
E_{\mathbf{k}} &=&a_{T}+2\beta _{\mathbf{k}}v^{2}-rv^{2}\beta _{\mathbf{k}%
}+\left( 2-r\right) \int_{\mathbf{l}}\beta _{\mathbf{k-l}}\left( p_{\mathbf{l%
}}^{A}-p_{\mathbf{l}}^{O}\right)  \notag \\
\Delta _{\mathbf{k}} &=&\left( r-1\right) v^{2}\eta _{\mathbf{k}}+\left(
1-r\right) \eta _{\mathbf{k}}\int_{\mathbf{l}}\frac{\eta _{\mathbf{l}}}{%
\beta _{A}}\left( p_{\mathbf{l}}^{A}-p_{\mathbf{l}}^{O}\right)
\label{E_gap_IV} \\
\overline{E}_{\mathbf{k}} &=&-rv^{2}\beta _{\mathbf{k}}-r\int_{\mathbf{l}%
}\beta _{\mathbf{k-l}}\left( \overline{p}_{\mathbf{l}}^{A}+\overline{p}_{%
\mathbf{l}}^{O}\right)  \notag \\
\text{\ }\overline{\Delta }_{\mathbf{k}} &=&rv^{2}\eta _{\mathbf{k}}-r\eta _{%
\mathbf{k}}\int_{\mathbf{l}}\frac{\eta _{\mathbf{l}}}{\beta _{\Delta }}%
\left( \overline{p}_{\mathbf{l}}^{A}-\overline{p}_{\mathbf{l}}^{O}\right) ,
\label{Ebar_gap_IV}
\end{eqnarray}%
where
\begin{eqnarray*}
E_{\mathbf{k}} &=&\frac{1}{2}\left[ \left( \mu _{\mathbf{k}}^{O}\right)
^{2}+\left( \mu _{\mathbf{k}}^{A}\right) ^{2}\right] ,\text{ \ }\overline{E}%
_{\mathbf{k}}=\frac{1}{2}\left[ \left( \overline{\mu }_{\mathbf{k}%
}^{O}\right) ^{2}+\left( \overline{\mu }_{\mathbf{k}}^{A}\right) ^{2}\right]
\\
\Delta _{\mathbf{k}} &=&\frac{1}{2}\left[ \left( \mu _{\mathbf{k}%
}^{O}\right) ^{2}-\left( \mu _{\mathbf{k}}^{A}\right) ^{2}\right] ,\text{ \ }%
\overline{\Delta }_{\mathbf{k}}=\frac{1}{2}\left[ \left( \overline{\mu }_{%
\mathbf{k}}^{O}\right) ^{2}-\left( \overline{\mu }_{\mathbf{k}}^{A}\right)
^{2}\right] \text{\ }
\end{eqnarray*}%
\bigskip

\emph{\textbf{\ Solution by the mode expansion}}

\bigskip

One can observe that the Ansatz
\begin{equation}
\left( \mu _{\mathbf{k}}^{O}\right) ^{2}-\left( \mu _{\mathbf{k}}^{A}\right)
^{2}=\eta _{\mathbf{k}}\Delta ;\text{ \ \ \ }\left( \overline{\mu }_{\mathbf{%
k}}^{O}\right) ^{2}-\left( \overline{\mu }_{\mathbf{k}}^{A}\right) ^{2}=\eta
_{\mathbf{k}}\overline{\Delta }  \label{delta_Ansatz_IV}
\end{equation}%
satisfies the gap equations, leading to simpler set for two unknown
functions and three unknown parameters satisfying eqs.(\ref{shift_eq_IV}),(%
\ref{E_gap_IV}),(\ref{Ebar_gap_IV}) and:%
\begin{eqnarray}
\Delta &=&\left( 1-r\right) \left[ v^{2}+\int_{\mathbf{l}}\frac{\eta _{%
\mathbf{l}}}{\beta _{A}}\left( p_{\mathbf{l}}^{A}-p_{\mathbf{l}}^{O}\right) %
\right]  \label{delta_eqs_IV} \\
\overline{\Delta } &=&-r\left[ v^{2}-\int_{\mathbf{l}}\frac{\eta _{\mathbf{l}%
}}{\beta _{\Delta }}\left( \overline{p}_{\mathbf{l}}^{A}-\overline{p}_{%
\mathbf{l}}^{O}\right) \right]  \notag
\end{eqnarray}%
The equation can be solved by using mode expansion:%
\begin{equation}
\beta _{k}=\sum_{n=0}^{\infty }\chi ^{n}\beta _{n}(k);\beta _{n}(k)\equiv
\sum_{\left\vert \mathbf{X}\right\vert ^{2}=na_{\Delta }^{2}}\exp [i\mathbf{%
k\bullet X}]  \label{mode_expansion_IV}
\end{equation}%
As in IIIC, The integer $n$ determines the distance of a points on
reciprocal lattice from the origin. and $\chi \equiv \exp [-a_{\Delta
}^{2}/2]=\exp [-2\pi /\sqrt{3}]=0.0265$. One estimates that $E_{n}\simeq
\chi ^{n}a_{T},$ therefore the coefficients decrease exponentially with $n$.
Note that for some integers, for example $n=2,5,6$, $\beta _{n}=0$.
Retaining only first $s$ modes will be called \textquotedblright the $s$
mode approximation\textquotedblright .
\begin{eqnarray}
E\left( \mathbf{k}\right) &=&E_{0}+E_{1}\chi \beta _{1}(\mathbf{k}%
)+...E_{n}\chi ^{n}\beta _{n}(\mathbf{k})...+  \label{E_expansion_IV} \\
\overline{E\left( \mathbf{k}\right) } &=&\overline{E_{0}}+\overline{E_{1}}%
\chi \beta _{1}(\mathbf{k})+...\overline{E_{n}}\chi ^{n}\beta _{n}(\mathbf{k}%
)...+.  \notag
\end{eqnarray}%
The expression deviates significantly from the perturbative one, especially
at low temperatures and when the 2D case is considered.

\bigskip

\emph{\textbf{Generalizations and comparison to experiments}}

\bigskip

As we noted already in 2D disorder leads, at least perturbatively, to more
profound restructuring of the vortex lattice than in 3D. In fact
perturbation theory becomes invalid. The gaussian methods described above
remove the difficulty and allow calculation of the order - disorder lime. In
this case one does not encounters the "wiggle" but rather a smooth decrease
of the order - disorder field, when temperature becomes lower. In Fig. \ref%
{figIV4} the ODO line of strongly anisotropic high $T_{c}$ superconductor $%
BSCCO$ is shown.

One generally observes that there is always off diagonal component in the
correlator of the "optical" phonon field $O$. However the off diagonal
Edwards - Anderson parameter part for a more important low energy excitation
"acoustic" branch appears only below a line quite similar to the glass line
in the homogeneous phase.

\bigskip

\subsection{Replica symmetry breaking}

When thermal fluctuations are significant the efficiency of imperfections to
pin the vortex matter is generally diminished. This phenomenon is known as
"thermal depinning". In addition, as we have learned in section III, the
vortex lattice becomes softer and eventually melts via first order
transition into the vortex liquid. The inter - dependence of pinning,
interactions and thermal fluctuations is very complex and one needs an
effective nonperturbative method to evaluate the disorder averages. Such a
method, using the replica trick was developed initially in the theory of
spin glasses. It is more difficult to apply it in a crystalline phase, so we
start from a simpler homogeneous phase (the homogeneity might be achieved by
both the thermal fluctuations and disorder) and return to the crystalline
phase in the following subsection.

\subsubsection{Hierarchical matrices and absence of RSB for the $\protect%
\delta T_{c}$ disorder in gaussian approximation}

\emph{\textbf{The hierarchical matrices and their Parisi's parametrization}}

Experience with very similar models in the theory of disordered magnets
indicates that solutions of these minimization equations are most likely to
belong to the class of hierarchical matrices, which are comprehensively
described, for example in \cite{Fischer,Dotsenko,Mezard91}, We limit
ourselves here to operational knowledge of working with such matrices
contained in Appendix of ref. \cite{Mezard91} and collect several rules of
using the Parisi's representation in Appendix B. General hierarchical
matrices $u$ are parametrized using the diagonal elements $\widetilde{u}$
and the Parisi's (monotonically increasing) function $u_{x}$ specifying the
off diagonal elements with $0<x<1$ \cite{Parisi80}. Physically different $x$
represent time scales in the glass phase. In particular the Edwards -
Anderson (EA) order parameter is $u_{x=1}=\lambda >0$ .

A nonzero value for this order parameter signals that the annealed and the
quenched averages are different. The dynamic properties of such phase are
generally quite different from those of the non glassy $\lambda =0$\ phase.
In particular it is expected to exhibit infinite conductivity \cite%
{Fisher89,Fisher91,Dorsey92}. We will refer to this phase as the "ergodic
pinned liquid" (EPL) distinguished from the "nonergodic pinned liquid" (NPL)
in which, in addition, the ergodicity is broken. Broken ergodicity is
related to "replica symmetry breaking" discussed below, however, as we show
shortly, in the present model of the $\delta T_{c}$ disorder and within
gaussian approximation RSB does not occur.

In terms of $\ $Parisi parameter\ \ $\widetilde{u}$ and $u_{x}$ the matrix
equation eq.(\ref{first_der_IV}) takes a form:

\begin{equation}
-\widetilde{u^{-2}}+a_{T}+\left( 4-2r\right) \widetilde{u}=0
\label{gap_eq_Parisi_IV}
\end{equation}%
\begin{equation}
\left( u^{-2}\right) _{x}+2ru_{x}=0.
\end{equation}%
Dynamically (see next section), if $u_{x}$ is a constant, \ pinning does not
results in the multitude of time scales. Certain time scale sensitive
phenomena like various memory effects \cite{Paltiel00,Paltiel00b,Xiao02} and
the responses to \textquotedblleft shaking\textquotedblright\ \cite{Zeldov05}
are expected to be different from the case when $u_{x}$ takes multiple
values. If $u_{x}$ takes a finite different number of $n$ values, we call $%
n-1$ step RSB. On the other hand, if $u_{x}$ is continuous, the continuous
replica symmetry breaking (RSB) occurs. We show below that within the
gaussian approximation and the limited disorder model that we consider (the $%
\delta T_{c}$ inhomogeneity only) RSB does not occur. After that is shown,
we can consider the remaining problem without using the machinery of
hierarchical matrices.

\emph{\textbf{Absence of replica symmetry breaking}}

In order to show that $u_{x}$ is a constant, it is convenient to rewrite the
second equation via the matrix $\mu ,$ the matrix inverse to $u$:

\begin{equation}
\left( \mu ^{2}\right) _{x}+2r(\mu ^{-1})_{x}=0.  \label{proof1_IV}
\end{equation}%
Differentiating this equation with respect to $x$ one obtains;

\begin{equation}
2\left[ \left\{ \mu \right\} _{x}-r\left( \left\{ \mu \right\} _{x}\right)
^{-2}\right] x\frac{d\mu _{x}}{dx}=0,  \label{proof2_IV}
\end{equation}%
where we used a set of standard notations in the spin glass theory: \cite%
{Mezard91}
\begin{eqnarray}
\left\{ \mu \right\} _{x} &\equiv &\widetilde{\mu }-\left\langle \mu
_{x}\right\rangle -[\mu ]_{x};\text{ \ \ }\left\langle \mu _{x}\right\rangle
\equiv \int_{0}^{1}dx\mu _{x};  \label{proof3_IV} \\
\text{ \ \ }[\mu ]_{x} &=&\int_{0}^{x}dy\left( \mu _{x}-\mu _{y}\right) .
\notag
\end{eqnarray}%
If one is interested in a continuous monotonic part $\frac{d\mu _{x}}{dx}%
\neq 0,$ the only solution of eq.(\ref{proof1_IV}) is%
\begin{equation}
\left\{ \mu \right\} _{x}=r^{1/3}  \label{proof4_IV}
\end{equation}%
Differentiating this again and dropping the nonzero derivative $\frac{d\mu
_{x}}{dx}$ again, one further gets a contradiction: $\frac{d\mu _{x}}{dx}=0$
. This proves that there are no such monotonically increasing continuous
segments. One can therefore generally have either the replica symmetric
solutions, namely $u_{x}=\lambda $ or look for a several step - like RSB
solutions \cite{Fischer,Dotsenko}. One can show that the constant $u_{x}$
solution is stable. Therefore, if a step - like RSB solution exists, it
might be only an additional local minimum. We explicitly looked for a one
step solution and found that there is none.

\section{Summary and perspective}

In this section we summarize and provide references to original papers,
point out further applications and generalizations of results presented
here. The bibliography of works on the GL theory of the vortex matter is so
extensive that there, no doubt, many important papers and even directions
are missed in this brief outline. Some of them however can be found in books
\cite{Kopnin,Larkin-Varlamov,Saint,Ketterson,Tinkham} and reviews \cite%
{Blatter,Nattermann,Brandt,Giamarchi-Bhattacharya}.

\subsection{GL equations.}

\emph{\textbf{Microscopic derivations of the GL equations}}

Phenomenological Ginzburg - Landau equations \cite{Ginzburg50} preceded a
microscopic theory of superconductivity. Soon after the BCS theory appeared
Gorkov and others derived from it the GL equations. Derivations and the
relations of the GL parameters to the microscopic parameters in the BCS
theory are reviewed in the book by Larkin and Varlamov \cite{Larkin-Varlamov}
(where extensive bibliography can be found). The dynamical versions of the
theory were derived using several methods and the parameter $\gamma $ in the
time dependent GL equation related no the normal state conductivity \cite%
{Larkin-Varlamov}. Most of the methods described here can be generalized to
the case, when the non - dissipative imaginary part of $\gamma $ is also
present. In particular this leads to the Hall current \cite%
{Ullah90,Ullah91,Dorsey93} and was used to explain the "Hall anomaly" in
both low $T_{c}$ and high $T_{c}$ superconductors.

The $\delta T$ disorder was introduced phenomenologically in statics in \cite%
{Larkin70}. Other coefficients of the GL free energy may also have random
components \cite{Blatter}. How these new random variables influence the LLL
model was discussed in \cite{Vinokur06}.

\emph{\textbf{Anisotropy}}

High $T_{c}$ cuprates are layered superconductors which can be better
described by the Lawrence - Doniach (LD) model \cite{Lawrence-Doniach71}
than the 3D GL model discussed in the present review. The LD model, is a
version of the GL model with a discretized $z$ coordinate. However in many
cases one can use two simpler limiting cases. If anisotropy is not very
large one can use anisotropic 3D GL, eq.(\ref{F0_I}). The requirement, that
the GL can be effectively used, therefore limits us to optimally doped $%
YBCO_{7-\delta }$ and similar materials for which the anisotropy parameter
is not very large: $\gamma _{a}=\sqrt{m_{c}^{\ast }/m_{a,b}^{\ast }}=4-8$.
Effects of layered structure are dominant in $BSCCO$ or $Tl$ based compounds
($\gamma _{a}>80$) and noticeable for cuprates with anisotropy of order $%
\gamma _{a}=50,$ like $LaBaCuO$ or $Hg1223$. Anisotropy effectively reduces
dimensionality leading to stronger thermal fluctuations according to eq,(\ref%
{Gi_I}). Very anisotropic layered superconductors can be described by 2D GL
model%
\begin{equation}
F=L_{z}\int d^{2}\mathbf{r}\left[ \frac{{\hbar }^{2}}{2m^{\ast }}\left\vert
\mathbf{D}\psi \right\vert ^{2}+a^{\prime }|\psi |^{2}+\frac{b^{\prime }}{2}%
|\psi |^{4}\right] ,  \label{2DGL_VI}
\end{equation}%
which can be approached by the methods presented here. For LD model
analytical methods become significantly more complicated. The gaussian
approximation study of thermal fluctuations was however performed \cite%
{Ikeda95,Larkin-Varlamov} and used to explained the, so called crossing
point of the magnetization curves, as well as crossover between the $3D$ to
the $2D$ behavior\cite%
{Baraduc94,SalemSugui94,Junod98,Rosenstein01,Huh02,Tesanovic92,Lin05}
In many simulations this model rather than GL was adopted \cite%
{Ryu96,Wilkin97}. 
The GL model can be extended also in direction of introducing anisotropy in
the $a-b$ plane, like the four - fold symmetric anisotropy leading to
transition from the rhombic lattice to the square lattice \cite%
{Chang98,Chang98b,Klironomos03,Park98,Rosenstein99b} observed in many high $%
T_{c}$ and low $T_{c}$ type II superconductors alike \cite{Eskildsen01,Li06a}
. 

\emph{\textbf{Dynamics}}

Dynamics of vortex matter can be described by a time dependent
generalization of the GL equations \cite{Larkin-Varlamov}. The bifurcation
method presented here can be extended to moving vortex lattice \cite%
{Thompson-Hu71,Li04}. The extension is nontrivial since the linear operator
appearing in the equations is non Hermitian.

One also can consider thermal transport \cite{Ullah90,Ullah91}, for example
the Nernst effect\cite{Ussishkin02,Ussishkin03,Mukerjee04,Tinh09},
measured recently experiments \cite{Wang02,Wang06,Pourret06}.

\subsection{Theory of thermal fluctuations in GL model}

Here we briefly list various alternative approaches to those described in
the present review. It is important to mention an unorthodox
opinion concerning the very nature of the crystalline state and melting
transition. Although a great variety of recent experiments indicate that the
transition is first order (for alternative interpretation see \cite%
{Nikulov95,Nikulov95b}), some authors doubt the existence of a stable solid
phase \cite{Moore89,Moore92,Kienappel97} and therefore of the transition all
together.

\emph{\textbf{Functional renormalization group for the LLL theory}}

While applying the renormalization group (RG) on the one loop level,
Br\'{e}zin, Nelson and Thiaville \cite{Brezin85} found no fixed
points of the (functional) RG equations and thus concluded that the
transition to the solid phase, if it exists, is not continuous. The
RG method therefore cannot provide a quantitative theory of the
melting transition. It is widely believed however that the finite
temperature transition exists and is first order (see however
\cite{Newman96}, who found another solution of functional RG
equations).

\emph{\textbf{Large number of components limit}}

The GL theory can be generalized (in several different ways) to an $N$
component order parameter field. The large $N$ limit of this theory can be
computed in a way similar to that in the $N$ component scalar models widely
used in theory of critical phenomena \cite{Itzykson}. The most
straightforward generalization has been studied in \cite{Affleck85} in the
homogeneous phase leading to a conclusion that there is no instability of
this state in the 3D GL. However since there are other ways to extend the
theory to the $N$ component case, it was shown in \cite{Moore98} that the
state in which only one component has a nonzero expectation value (similar
to the one component Abrikosov lattice) is not the ground state of the most
straightforward generalization. Subsequently it was found \cite%
{Lopatin99,Li04a} that there exists a generalization for which this is in
fact the case . 

\emph{\textbf{Diagrams resummation}}

In many body theories one can resum various types of diagrams. In
fact one can consider Hartree-Fock, $1/N$ and even one loop RG as
kinds of the diagrams resummation. Moore and Yeo
\cite{Yeo96,Yeo96b,Yeo01} and more recently Yeo with his coworkers
\cite{Yeo06,Park08} followed a strategy used in strongly coupled
electron systems to resum all the parquet diagrams.
The thermal fluctuation in GL model had been studied using various
analytic methods \cite{Koshelev94}, but in the vortex liquid
region near the melting point,or overcooled liquid,
non-perturbative method must be used, for example, Borel-Pade
resummation method to obtain density density correlation was
carried out in ref.\cite{Hu94}.

\emph{\textbf{Numerical simulations}}

The LLL GL model was studied numerically in both 3D \cite{Sasik95} and 2D
\cite{Tesanovic-Xing91,ONeill93,Kato93,Li-Nattermann03}. The melting was
found to be first order. The results serve as an important check on the
analytic theory described in this review. In many simulations the XY model
is employed \cite{Ryu96,Hu97,Wilkin97}.
It is believed that results are closely related to that of the Ginzburg -
Landau model. The methods allows consideration of disorder and dynamics \cite%
{Hu01,Olsson01,Chen03,Olsson07} and fluctuations of the magnetic field \cite%
{Sudbo98,Sudbo99}.

\emph{\textbf{Density functional}}

The density functional theory is a general method to tackle a strongly
coupled system. The method crucially depends on the choice of the
functional. It was applied to the GL model by Herbut and Te\v{s}anovi\'{c}
\cite{Herbut94} and was employed in \cite{Menon94,Menon99,Menon02} to study
the melting and in \cite{Hu05} to the layered systems.

\emph{\textbf{Vortex matter theory}}

Elastic moduli were first calculated from the GL model by Brandt \cite%
{Brandt77,Brandt77b,Brandt86} by considering an infinitesimal shift
of zeroes of the order parameter. He found that the compression and
the shear moduli are dispersive. This feature is important in
phenomenological applications like the Lindemann criterion for both
the melting and the order
- disorder lines (considered different) \cite%
{Houghton89,Ertas96,Kierfeld00,Mikitik01,Mikitik03}, as well to estimates of
the critical current and the collective pinning theory (see reviews \cite%
{Brandt,Blatter} and references therein). The dispersion however is ignored
in most advanced applications of the elasticity theory to statics\cite%
{Giamarchi94,Giamarchi95a,Giamarchi95b,Giamarchi96,Giamarchi97,Nattermann}
or dynamics\cite{Giamarchi96,Chauve00,Giamarchi98,Giamarchi-Bhattacharya}.
Recently a phase diagram of strongly type II superconductors was discussed
using a modified elasticity theory taking into account dislocations of the
vortex lattice \cite{Dietel06,Dietel07,Dietel09}

Te\v{s}anovi\'{c} and coworkers noted \cite{Tesanovic92,Tesanovic94} a
remarkable fact that most of the fluctuations effects are just due to
condensation energy. The lateral correlations part are just around 2\% and
therefore can be neglected. The theory explains an approximate intersection
of the magnetization curves and is used to analyze data \cite%
{Zhou93,Pierson95,Pierson96,Pierson98,Pierson98b}.

\emph{\textbf{Beyond LLL}}

To quantitatively describe vortex matter higher Landau levels (HLL)
corrections are sometimes required. For example in optimally doped $YBCO$
superconductor one can establish the LLL scaling for fields above $3T$ and
temperature above $70K$ (see, for example, \cite{Sok95}). A glance at the
data however shows that above $T_{c}$ the scaling is generally unconvincing:
the LLL magnetization is much larger that the experimental one above $T_{c}$%
. Naively, on the solid side, when the distance from the mean field
transition line is smaller than the inter - Landau level gap, one expects
that the higher Landau modes can be neglected. More careful examination of
the mean field solution presented in subsection IIB reveals that a weaker
condition should be used for a validity test of the LLL approximation.
However the fluctuation corrections involving HLL in strongly fluctuating
superconductors might be important. Ikeda and collaborators calculated the
fluctuation spectrum in solid including HLL \cite{Ikeda90,Ikeda95}. In the
vortex liquid the HLL contribution has been studied by Lawrie \cite{Lawrie94}
in the framework of the gaussian (Hartree - Fock) approximation. He found
the region of validity of LLL approximation. The leading (gaussian)
contribution of HLL was combined with more refined treatment of the LLL
modes recently resulting in reasonably good agreement with \ experimental
data \cite{Li03}.

\emph{\textbf{Fluctuations of magnetic field and the dual theory approach}}

Although it was understood that fluctuations of the magnetic field in
strongly type II superconductors are strongly suppressed \cite%
{Halperin74,Lobb87}, they still play an important role when $\kappa $ is not
large and magnetic field away from $H_{c2}\left( T\right) $ (the situation
mostly not covered in the present review). The main methods are the
numerical simulations \cite{Dasgupta81,Sudbo99,Olsson03} and the dual theory
approach \cite{Tesanovic99,Kovner92,Kovner93}, which was very efficient in
describing the Kosterlitz - Thouless transition in superconducting thin film
and layered materials \cite{Oganesyan06}.
Vortex lines and loops are interpreted as a signal of "inverted $U\left(
1\right) $" or the "magnetic flux" symmetry. The symmetry is spontaneously
broken in the normal phase (with photon as a Goldstone boson), while
restored in the superconductor. Vortices are the worldlines of the flux
symmetry charges. 

\subsection{The effects of quenched disorder}

\emph{\textbf{Vortex glass in the frustrated XY model}}

The original idea of the vortex glass and the continuous glass transition
exhibiting the glass scaling of conductivity diverging in the glass phase
appeared early in the framework of the so called frustrated $XY$ model (the
gauge glass) \cite{Fisher89,Fisher91,Nattermann}. In this approach one fixes
the amplitude of the order parameter retaining the magnetic field with
random component added to the vector potential. It \ was studied by the RG
and variational methods and has been extensively simulated numerically \cite%
{Olsson01,Olsson07,Hu01,Chen03,Kawamura03,Chen08}.%
In analogy to the theory of spin glass the replica symmetry is broken when
crossing the GT line. The model ran into several problems (see Giamarchi and
Bhattacharya in ref.\cite{Giamarchi-Bhattacharya} for a review): for finite
penetration depth $\lambda $ it has no transition \cite{Young95} and there
was a difficulty to explain sharp Bragg peaks observed in experiment at low
fields.

\emph{\textbf{Disordered elastic manifolds. Bragg glass and replica symmetry
breaking}}

To address the last problem another simplified model had been proven to be
more convenient: the elastic medium approach to a collection of interacting
line-like objects subject to both the pinning potential and the thermal bath
Langevin force \cite%
{Reichhardt96,Reichhardt00,Otterlo98,Olson01,Cha94,Cha94b,Faleski96,Fangohr01,Fangohr03,Dodgson00}%
. The resulting theory was treated again using the gaussian approximation
\cite%
{Korshunov93,Giamarchi94,Giamarchi95a,Giamarchi95b,Giamarchi96,Giamarchi97}
and RG \cite{Nattermann90,Bogner01,Nattermann}. The result was that in $%
2<D<4 $ there is a transition to a glassy phase in which the replica
symmetry is broken following the \textquotedblleft hierarchical
pattern\textquotedblright\ (in $D=2$ the breaking is \textquotedblleft one
step\textquotedblright ). The problem of the very fast destruction of the
vortex lattice by disorder was solved with the vortex matter being in the
replica symmetry broken (RSB) phase and it was termed \textquotedblleft
Bragg glass\textquotedblright \cite%
{Giamarchi94,Giamarchi95a,Giamarchi95b,Giamarchi96,Giamarchi97}. A closely
related approach was developed very recently for both 3D and layered
superconductors in which effects of dislocations were incorporated \cite%
{Dietel06,Dietel07,Dietel09}. 

\emph{\textbf{Density functional for a disordered systems, supersymmetry}}

Generalized replicated density functional theory \cite{Menon02} was also
applied resulting in one step RSB solution.%
Although the above approximations to the disordered GL theory are very
useful in more \textquotedblleft fluctuating\textquotedblright\
superconductors like $BSCCO$, a problem arises with their application to $%
YBCO$ at temperature close $T_{c}$ (where most of the experiments mentioned
above are done): vortices are far from being line-like and even their cores
significantly overlap. As a consequence the behavior of the dense vortex
matter is expected to be different from that of a system of line - like
vortices and of the $XY$ model although the elastic medium approximation
might still be meaningful \cite{Brandt}.

One should note the work by Te\v{s}anovi\'{c} and Herbut\ \cite{Tesanovic94b}
on columnar defects in layered materials, which utilizes supersymmetry, as
an alternative to replica or dynamics, to incorporate disorder non -
perturbatively.

\emph{\textbf{Dynamical approach to disorder in the Ginzburg - Landau model}}

The statics and the linear response within disordered GL model has been
discussed in numerous theoretical, numerical and experimental works. The
glass line was first determined, to our knowledge, using the Martin - Siggia
- Rose dynamical approach in gaussian approximation \cite{Dorsey92} and was
claimed to be obtained using resummation of diagram in Kubo formula in \cite%
{Ikeda90}. The glass transition line for the $\Delta T_{c}$ disorder was
obtained using the replica formalism (within similar gaussian approximation)
by Lopatin \cite{Lopatin00} and the result is identical to presented in the
present review. He also extended the discussion beyond the gaussian
approximation employing the Cornwall - Jackiw - Tomboulis variational method
described. This was generalized to other types of disorder (the mean free
path disorder) in ref. \cite{Vinokur06}. The common wisdom is that the
\textquotedblright replica\textquotedblright\ symmetry is generally broken
in the glass (either via \textquotedblright steps\textquotedblright\ or via
\textquotedblright hierarchical\textquotedblright\ continuous process) as in
most of the spin glasses theories \cite{Fischer,Dotsenko}. The divergence of
conductivity on the glass line was obtained in \cite{Zhuravlev07} (it was
assumed in ref. \cite{Dorsey92} and linked phenomenologically to the general
scaling theory of the vortex glass proposed in ref. \cite{Fisher89,Fisher91}%
). Results are consistent with the replica ones presented in the present
review). In this work I-V curves and critical current were derived in
(improved) gaussian approximation and several physical questions related to
the peak effect addressed.

\emph{\textbf{Numerical simulation of the disordered Ginzburg - Landau model}%
}

The theory can be generalized to the 2D case appropriate for description of
thin films or strongly layered superconductors and compared to experiments.
The comparison for organic superconductor $\kappa $ type $BEDT-TTF$ \cite%
{Bel07} and $BSCCO$ \cite{Zeldov05} of the static glass line is quite
satisfactory. There exist, to our knowledge, just two Monte Carlo
simulations of the disordered GL model \cite{Li-Nattermann03,Kienappel97},
both in 2D, in which no clear irreversibility line was found. However the
frustrated XY model was recently extensively simulated \cite%
{Olsson01,Hu01,Chen03,Olsson03,Kawamura03,Olsson07,Chen08} including the
glass transition line and I-V curves
It shares many common features with the GL model although disorder is
introduced in a different way, so that it is difficult to compare the
dependence of pinning. The I-V curves and the glass line resemble the
Ginzburg - Landau results.

\emph{\textbf{Finite electric fields}}

Finite electric fields (namely transport beyond linear response) were also
considered analytically in \cite{Blum97} and our result in the clean limit
agrees with their's. The elastic medium and the vortex dynamics within the
London approximation were discussed beyond linear response in numerous
analytic and numerical works. Although qualitatively the glass lines
obtained here resemble the ones in phenomenological approaches based on
comparison of disorder strength with thermal fluctuations and interaction
\cite{Mikitik01,Mikitik03,Ertas96,Kierfeld00,Radzyner02}, detailed form is
different.

\subsection{Other fields of physics}

There are several physical systems in which the methods described here, in a
slightly modified form, can be applied and indeed appeared under different
names. One area is the superfluidity and the BEC condensate physics (ref.%
\cite{Pethick08} and references therein).
Magnetic field is analogous to the rotation of the superfluid. One can
observe lattice of vortices with properties somewhat reminiscent of those of
the Abrikosov vortices \cite%
{Madison00,Abo-Shaeer01,Engels02,Cooper01,Sinova01,Baym03,Sonin05,Wu07}.
Another closely relate field is the physics of the 2D electron gas in strong
magnetic field \cite{Monarkha04}. 
In some cases the problem can be formulated in a way similar to the present
case with Wigner crystal analogous to the Abrikosov liquid (time playing the
role of the $z$ direction of the fluxon), while quenched disorder appears in
a way similar to the columnar defects in the vortex physics. Some aspects of
the physics of the liquid crystal also can be formulated in a form similar
to the GL equations in magnetic field.

\subsection{Acknowledgments}

We are grateful to many people, who either actively collaborated
with us or discussed various issues related to vortex physics.
Collaborators, colleagues and students include B. Ya. Shapiro, V.
Zhuravlev, V. Vinokur, P.J. Lin, T.J. Yang, I. Shapiro, G. Bel, Z.G.
Wu, B. Tinh, B. Feng, Z.S. Ma, E. Zeldov, E.H. Brandt, R. Lortz, Y.
Yeshurun, P. Lipavsky, and A. Shaulov. We are indebted to T. Maniv,
G. Menon, A.T. Dorsey, C. Reichhardt, A.E. Koshelev, S. Teitel. P.
Olsson, X. Hu, T. Nattermann, T. Giamarchi, G. Blatter, Z.
Te\v{s}anovi\'{c}, R. Mintz, and B. Horowitz for discussions, M.K.
Wu, E. Andrei, C.C. Chi, P.H. Kes, E.M. Forgan, J. Juang, J.Y. Lib,
J.J. Lib, M.R. Eskildsen, H.H. Wen, T. Nishizaki, A. Grover, N.
Kokubo, S. Salem-Sugui, K. Hirata, C. Villard, H. Beidenkopf, J.
Kolacek, C. J. van der
Beek, and M. Konczykowski. 
for discussions and sending experimental results, sometimes prior to
publications. Work supported by NSC of \ R.O.C. grant,
NSC\#952112M009048 and MOE ATU program, and National Science
foundation of China (\#10774005). D.L is grateful to National Chiao
Tung University, while B. R. is grateful to NCTS and University
Center of Samaria for hospitality during sabbatical leave.


\section{Appendices}

\subsection{Integrals of products of the quasimomentum eigenfunctions}

In this appendix a method to calculate space averages of products of the
quasi - momentum eigenfunctions in both static and the dynamic cases.

\subsubsection{Rhombic lattice quasimomentum functions}

Let us specialize to a rhombic lattice with following bases of the direct
and the reciprocal lattices (see Fig. \ref{figII2} for definition of the
angle $\theta $ and the lattice spacing $a_{\theta }$, subject to the flux
quantization relation, eq.(\ref{rhombic_a_II})):%
\begin{eqnarray}
\mathbf{d}_{1} &=&d_{\theta }(1,0),\text{ \ \ \ }\mathbf{d}_{2}=d_{\theta }(%
\frac{1}{2},\frac{1}{2}\tan \theta );  \label{rhombic_A} \\
\widetilde{\mathbf{d}}_{1} &=&d_{\theta }(\frac{1}{2}\tan \theta ,-\frac{1}{2%
}),\text{ \ \ }\widetilde{\mathbf{d}}_{2}=(0,d_{\theta }).  \notag
\end{eqnarray}%
We use here the "LLL" unit of magnetic length.

We start with static LLL functions for an arbitrary rhombic lattice:

\begin{eqnarray}
\varphi _{\mathbf{k}} &=&\sqrt{\frac{2\sqrt{\pi }}{d_{\theta }}}\sum_{l}e^{i[%
\frac{2\pi }{a_{\theta }}\left( x+k_{y}\right) l+\frac{\pi l^{2}}{2}\tan
\theta ]}  \notag \\
&&\text{ \ \ \ \ \ }\times e^{-\frac{1}{2}(y-k_{x}-\frac{2\pi }{d_{\theta }}%
l)^{2}}  \label{LLL_static_A}
\end{eqnarray}

To include higher LL corrections, it is convenient to use rising and
lowering operators introduced in eq.(\ref{creation_anihillation_II})\ to
work with the HLL functions, $\widehat{a}^{\dagger }=\left( 2\right)
^{-1/2}\left( -i\partial _{x}+\partial _{y}-y\right) ,$ \ \ $\widehat{a}%
=-\left( 2\right) ^{-1/2}\left( -i\partial _{x}+\partial _{y}+y\right) $.
The following formula will be frequently used. If $\varphi $ is an LLL
function, then

\begin{equation}
\varphi ^{\ast }a^{+N}f\left( x,y\right) =2^{-\frac{N}{2}}(-i\partial
_{x}+\partial _{y})^{n}\varphi ^{\ast }f\left( x,y\right) .  \label{HLL_A}
\end{equation}

\subsubsection{The basic Fourier transform formulas}

\emph{\textbf{Product of two functions }}

It can be verified by direct calculation of gaussian integrals that

\begin{gather}
\int_{\mathbf{r}}\varphi (\mathbf{r})\varphi _{\mathbf{k}}^{\ast }(\mathbf{r}%
)e^{i\mathbf{r}\cdot \mathcal{K}}=4\pi ^{2}\sum_{K_{1},K_{2}}\delta (%
\mathcal{K}-\mathbf{k-K})F\left( \mathbf{k},\mathbf{K}\right) ;  \notag \\
F\left( \mathbf{k},\mathbf{K}\right) =e^{-\frac{\mathcal{K}^{2}}{4}+i[\frac{%
\pi }{2}K_{1}^{2}-\frac{\mathcal{K}_{x}\mathcal{K}_{y}}{2}+k_{x}\mathcal{K}%
_{y}]},  \label{basic_formula_A}
\end{gather}%
with decomposition of arbitrary momentum $\mathcal{K}$ into its "rational
part" $\mathbf{k}$,which belongs to the Brillouin zone and an "integer part"
$\mathbf{K}$, belonging to the reciprocal lattice%
\begin{equation}
\mathcal{K}=\mathbf{k+K,}\text{ \ \ \ \ \ }\mathbf{K=}K_{1}\mathbf{d}%
_{1}+K_{2}\mathbf{d}_{2}\text{.}  \label{decomposition_A}
\end{equation}%
Its inverse Fourier transform,

\begin{equation}
\varphi (\mathbf{r})\varphi ^{\ast }(\mathbf{r}+\widetilde{\mathbf{k}}%
)=e^{ixk_{x}}\sum\limits_{\mathbf{K}}e^{-i\mathcal{K}\cdot \mathbf{r}%
}F\left( \mathbf{k},\mathbf{K}\right) ,  \label{inverse_Fourier_A}
\end{equation}%
where $\widetilde{k_{i}}=\varepsilon _{ij}k_{j}$, can be generalized into

\begin{gather}
\varphi (\mathbf{r+}\widetilde{\mathbf{l}})\varphi ^{\ast }(\mathbf{r}+%
\widetilde{\mathbf{k}})=e^{i\left( x+l_{y}\right) \left( k-l\right)
_{x}}\times  \label{two_functions_A} \\
\sum\limits_{\mathbf{K}}e^{-i\mathcal{K}\cdot \left( \mathbf{r+}\widetilde{%
\mathbf{l}}\right) \mathbf{+}\frac{\pi i}{2}(K_{1}^{2}+K_{1})-\frac{i%
\mathcal{K}_{x}\mathcal{K}_{y}}{2}+i\left( k-l\right) _{x}\mathcal{K}_{y}-%
\frac{\mathcal{K}^{2}}{4}},  \notag
\end{gather}%
with $\mathcal{K=}\mathbf{K+k}-\mathbf{l}$. This in turn provides a very
useful product representation:%
\begin{gather}
\varphi _{\mathbf{k}}^{\ast }(\mathbf{r})\varphi _{\mathbf{l}}(\mathbf{r}%
)=\sum\limits_{\mathbf{K}}e^{-i\mathcal{K}\cdot \mathbf{r}-\frac{\mathcal{K}%
^{2}}{4}}  \label{product_A} \\
\times e^{i\frac{\pi }{2}K_{1}^{2}-i\frac{\mathcal{K}_{x}\mathcal{K}_{y}}{2}%
+ik_{x}\mathcal{K}_{y}-i\mathcal{K}_{x}l_{y}+il_{y}\left( k-l\right) _{x}}.
\notag
\end{gather}

\emph{\textbf{The four - point vertex function}}

The relation above used twice gives the following expression for the four
point vertex:in the quasi - momentum representation:

\begin{gather}
\int_{\mathbf{r}}\varphi _{\mathbf{k}}^{\ast }(\mathbf{r})\varphi _{\mathbf{l%
}}(\mathbf{r})\varphi _{\mathbf{l}^{\prime }}^{\ast }(\mathbf{r})\varphi _{%
\mathbf{k}^{\prime }}(\mathbf{r})=\left( 2\pi \right) ^{2}\sum\limits_{%
\mathcal{K}}\delta \left[ \mathcal{K-K}^{\prime }\right]  \label{four1_A} \\
\times e^{-\frac{\mathcal{K}^{2}}{2}+i[\frac{\pi }{2}K_{1}^{2}-\frac{\pi }{2}%
K_{1}^{\prime 2}+\left( k_{x}-k_{x}^{\prime }\right) \mathcal{K}_{y}-%
\mathcal{K}_{x}\left( l_{y}-l_{y}^{\prime }\right) +l_{y}\left( k-l\right)
_{x}-l_{y}^{\prime }\left( k^{\prime }-l^{\prime }\right) _{x}]}  \notag
\end{gather}%
The delta function $\delta \left[ \mathcal{K-K}^{\prime }\right] $ implies
that

\begin{equation}
K_{1}+k_{1}-l_{1}=K_{1}^{\prime }+k_{1}^{\prime }-l_{1}^{\prime }\text{.}
\label{delta1_A}
\end{equation}%
As $0\leq k_{1},l_{1},k_{1}^{\prime },l_{1}^{\prime }<1,$ we have only three
possible integer values for each coordinate:

\begin{eqnarray}
k_{1}-l_{1}-k_{1}^{\prime }+l_{1}^{\prime } &=&\mathbf{\Delta }_{1}=0,1,-1,
\label{cases_A} \\
k_{2}-l_{2}-k_{2}^{\prime }+l_{2}^{\prime } &=&\mathbf{\Delta }_{2}=0,1,-1,
\notag
\end{eqnarray}%
which require $K_{1,2}-K_{1,2}^{\prime }=0,-1,1$. Thus
\begin{eqnarray}
\sum_{\mathcal{K},\mathcal{K}^{\prime }}\delta \left[ \mathcal{K-K}^{\prime }%
\right] &=&\sum_{\mathbf{K,\Delta }}\delta \left[ \mathbf{K-K}^{\prime }+%
\mathbf{\Delta }\right]  \label{delta2_A} \\
&&\times \delta \left[ \mathbf{k}-\mathbf{l\mathbf{-}\left( \mathbf{k}%
^{\prime }-\mathbf{l}^{\prime }\right) -\Delta }\right] ;  \notag
\end{eqnarray}%
and the product takes a form:%
\begin{gather}
\int_{\mathbf{r}}\varphi _{\mathbf{k}}^{\ast }(\mathbf{r})\varphi _{\mathbf{l%
}}(\mathbf{r})\varphi _{\mathbf{l}^{\prime }}^{\ast }(\mathbf{r})\varphi _{%
\mathbf{k}^{\prime }}(\mathbf{r})=\left( 2\pi \right) ^{2}\times  \notag \\
\sum_{\mathbf{\Delta }}\delta \left[ \mathbf{k}-\mathbf{l\mathbf{-}\left(
\mathbf{k}^{\prime }-\mathbf{l}^{\prime }\right) -\Delta }\right] f\left[
\mathbf{k},\mathbf{l,\mathbf{k}^{\prime },\mathbf{l}^{\prime },\Delta }%
\right]  \label{four2_A} \\
f\left[ \mathbf{k},\mathbf{l,\mathbf{k}^{\prime },\mathbf{l}^{\prime
},\Delta }\right] =\sum_{\mathbf{K,K}^{\prime }=\mathbf{K}+\mathbf{\Delta }%
}e^{-\frac{\mathcal{K}^{2}}{2}}\times  \notag \\
e^{i[\frac{\pi }{2}K_{1}^{2}-\frac{\pi }{2}K_{1}^{\prime 2}+\left(
k_{x}-k_{x}^{\prime }\right) \mathcal{K}_{y}-\mathcal{K}_{x}\left(
l_{y}-l_{y}^{\prime }\right) +l_{y}\left( k-l\right) _{x}-l_{y}^{\prime
}\left( k^{\prime }-l^{\prime }\right) _{x}]}  \notag
\end{gather}%
where $f\left[ \mathbf{k},\mathbf{l,\mathbf{k}^{\prime },\mathbf{l}^{\prime
},\Delta }\right] =0$ if $\mathbf{k}-\mathbf{l\mathbf{-}\left( \mathbf{k}%
^{\prime }-\mathbf{l}^{\prime }\right) -\Delta \neq 0.}$ The last exponent
in function $f\left[ \mathbf{k},\mathbf{l,\mathbf{k}^{\prime },\mathbf{l}%
^{\prime },\Delta }\right] $ can be also rearranged as%
\begin{eqnarray}
&&-\frac{\mathcal{K}^{2}}{2}-\frac{\pi i}{2}\mathbf{\Delta }_{1}\left(
2K_{1}+\mathbf{\Delta }_{1}\right) +i\left( \mathbf{k-k}^{\prime }\right)
\label{four_exponent_A} \\
&&\wedge \mathcal{K+}i\mathcal{K}_{x}\mathbf{\Delta }_{y}+i\left(
l_{y}-l_{y}^{\prime }\right) \left( k-l\right) _{x}+il_{y}^{\prime }\mathbf{%
\Delta }_{x}.  \notag
\end{eqnarray}%
Using

\begin{equation*}
\sum_{\mathbf{X}=n_{1}\mathbf{d}_{1}+n_{2}\mathbf{d}_{2}}e^{i\mathbf{X}\cdot
\mathbf{q}}=cell\sum_{\mathbf{K}}\delta \left( \mathbf{q}-\mathbf{K}\right) ,
\end{equation*}%
where $cell$ is the volume of the unit cell and in our units is equal to $%
2\pi $, one obtains the Poisson resummation relation,

\begin{equation*}
\sum_{\mathbf{K}}f\left( \mathbf{K}\right) =\frac{1}{cell}\int d\mathbf{q}%
\sum_{\mathbf{X}}\exp \left( i\mathbf{X}\cdot \mathbf{q}\right) f\left(
\mathbf{q}\right) ,
\end{equation*}%
Using Poisson resummation, one rewrites the sum as
\begin{eqnarray}
f\left[ \mathbf{k},\mathbf{l,\mathbf{k}^{\prime },\mathbf{l}^{\prime
},\Delta }\right] &=&\sum_{\mathbf{X}}e^{-\frac{1}{2}\left( \mathbf{X+}%
\widehat{z}\times \left( \mathbf{k-k}^{\prime }\right) \right) ^{2}-i\mathbf{%
X\cdot }\left( \mathbf{k}-\mathbf{l}\right) }\times  \notag \\
&&e^{i[\left( k_{y}\mathbf{-}k_{y}^{\prime }\right) \left(
k_{x}-l_{x}\right) +l_{y}^{\prime }\mathbf{\Delta }_{x}-\frac{\pi }{2}%
\mathbf{\Delta }_{1}^{2}]}.
\end{eqnarray}

\subsubsection{Calculation of the $\protect\beta _{\mathbf{k}},\protect%
\gamma _{\mathbf{k}}$ functions and their$\ $small momentum expansion}

One often encounters the following space averages:

\begin{equation}
\beta _{\mathbf{k}}^{N}=\left\langle |\varphi |^{2}\varphi _{\mathbf{k}%
}\varphi _{\mathbf{k}}^{\ast N}\right\rangle _{r},\text{ }\gamma _{\mathbf{k}%
}^{N}=\left\langle \left( \varphi ^{\ast }\right) ^{2}\varphi _{-\mathbf{k}%
}^{N}\varphi _{\mathbf{k}}\right\rangle _{r}.
\end{equation}%
$\beta _{\mathbf{k}}=\beta _{\mathbf{k}}^{0}$ and $\gamma _{\mathbf{k}%
}=\gamma _{\mathbf{k}}^{N}.$ Using formulas of the previous subsection, one
can write
\begin{gather}
\varphi ^{\ast }\varphi _{\mathbf{k}}^{N}=\frac{1}{2^{N/2}\sqrt{N!}}\left(
-i\partial _{x}+\partial _{y}\right) ^{N}\varphi ^{\ast }\varphi _{\mathbf{k}%
}=  \notag \\
\frac{1}{2^{N/2}\sqrt{N!}}\sum_{\mathbf{Q}}\left( z_{\mathbf{k}}+z_{\mathbf{Q%
}}\right) ^{N}e^{i\left( \mathbf{k+Q}\right) \cdot \mathbf{x}}F^{\ast }(%
\mathbf{k},\mathbf{Q}); \\
\varphi \varphi _{\mathbf{k}}^{\ast N}=\left( \varphi ^{\ast }\varphi _{%
\mathbf{k}}^{N}\right) ^{\ast }=\frac{1}{2^{N/2}\sqrt{N!}}  \notag \\
\sum_{\mathbf{Q}}\left( z_{\mathbf{k}}^{\ast }+z_{\mathbf{Q}}^{\ast }\right)
^{N}e^{-i\left( \mathbf{k+Q}\right) \cdot \mathbf{x}}F(\mathbf{k},\mathbf{Q}%
);  \notag
\end{gather}

\bigskip Therefore\
\begin{equation}
\beta _{k}^{N}=\frac{1}{2^{N/2}\sqrt{N!}}\sum_{\mathbf{X}}\left( iz_{\mathbf{%
X}}^{\ast }\right) ^{N}e^{-\frac{X^{2}}{2}-i\mathbf{k}\cdot \mathbf{X}}.
\label{betaN_A}
\end{equation}%
and%
\begin{equation}
\beta _{k}=\sum_{\mathbf{X}}e^{-\frac{X^{2}}{2}-i\mathbf{k}\cdot \mathbf{X}}.
\end{equation}%
Similarly $\gamma _{\mathbf{k}}^{N}$ can be obtained. for $\gamma _{\mathbf{k%
}}$, we have

\begin{equation}
\gamma _{\mathbf{k}}=e^{-ik_{x}k_{y}-\frac{k^{2}}{2}}\sum_{\mathbf{X}}e^{-%
\frac{X^{2}}{2}-iz_{\mathbf{k}}^{\ast }z_{\mathbf{X}}}.
\end{equation}

\bigskip The above formula is valid for any lattice structure.

\emph{\textbf{Small momentum expansion of the $\beta _{k},\gamma _{k}$
function for the general rhombic lattice}}

Consider the sum%
\begin{equation}
S\left( N,M\right) =\sum_{\mathbf{X}}e^{-\frac{X^{2}}{2}}z_{\mathbf{X}%
}^{N}\left\vert \mathbf{X}\right\vert ^{M}  \label{S}
\end{equation}%
for any integers $N,M$. Due to reflection symmetry
\begin{equation}
\sum_{\mathbf{X}}e^{-\frac{X^{2}}{2}}X_{x}^{l_{1}}X_{y}^{l_{2}}=0
\end{equation}%
for $l_{1},l_{2}$ integers, and $l_{1}+l_{2}$ odd integer. For small $%
\mathbf{k}$:
\begin{eqnarray}
\beta _{\mathbf{k}} &=&\sum_{\mathbf{X}}e^{-\frac{X^{2}}{2}%
}(1+\sum_{l=1}^{\infty }\frac{\left( i\mathbf{k}\cdot \mathbf{X}\right) ^{l}%
}{l!})  \label{beta_expansion_A} \\
&=&\beta _{\Delta }-\sum_{\mathbf{X}}\frac{\left(
k_{x}X_{x}+k_{y}X_{y}\right) ^{2}}{2}e^{-\frac{X^{2}}{2}}+  \notag \\
&&\sum_{\mathbf{X}}e^{-\frac{X^{2}}{2}}\frac{\left(
k_{x}X_{x}+k_{y}X_{y}\right) ^{4}}{24}+O\left( k^{6}\right)  \notag
\end{eqnarray}%
Similarly for the function $\gamma _{k}$ \ can be expanded for small $k^{2}$

\begin{equation}
\gamma _{k}=e^{-ik_{x}k_{y}-\frac{k^{2}}{2}}\sum_{\mathbf{Q}}e^{-\frac{Q^{2}%
}{2}}[1+\sum_{l=1}^{\infty }\frac{\left( \overline{k}\right) ^{2l}Q^{2l}}{%
\left( 2l\right) !}],
\end{equation}%
so that
\begin{eqnarray}
\left\vert \gamma _{k}\right\vert &=&e^{-\frac{k^{2}}{2}}[\sum_{\mathbf{Q}%
}e^{-\frac{Q^{2}}{2}}(1+\sum_{l=1}^{\infty }\frac{z_{k}^{\ast 2l}z_{\mathbf{Q%
}}^{2l}}{\left( 2l\right) !})]^{1/2}  \label{gammaAbs} \\
&&\times \lbrack \sum_{\mathbf{Q}^{\prime }}e^{-\frac{Q^{\prime 2}}{2}%
}(1+\sum_{l^{\prime }=1}^{\infty }\frac{z_{\mathbf{k}}^{2l^{\prime }}z_{%
\mathbf{Q}^{\prime }}^{\ast 2l^{\prime }}}{\left( 2l^{\prime }\right) !}%
)]^{1/2}  \notag
\end{eqnarray}

\bigskip \emph{\textbf{Small momentum expansion of the $\beta _{k},\gamma
_{k}$ function for hexagonal lattice}}

When the symmetry is higher, the expressions simplify. Using the six - fold
symmetry of the hexagonal lattice,%
\begin{equation}
X_{x}^{\prime }+iX_{y}^{\prime }=e^{i\theta }\left( X_{x}+iX_{y}\right) ,\
\theta =\frac{\pi }{3}l;
\end{equation}%
the sum eq.(\ref{S}) transforms into%
\begin{equation}
S\left( N,M\right) =S\left( N,M\right) e^{in\theta }.
\end{equation}%
Thus $S\left( N,M\right) =0$ if $N\neq 6j$. Using $S\left( 2,0\right)
=S\left( 4,0\right) =S\left( 2,2\right) =0,$ one obtains several relations
of different sums to \ simplify expansion of $\beta _{\mathbf{k}}$
\begin{equation}
\beta _{k}=\sum_{\mathbf{X}}e^{-\frac{X^{2}}{2}}(1-\frac{k^{2}X^{2}}{4}+%
\frac{k^{4}X^{4}}{64}).  \label{beta_expantion_A}
\end{equation}%
Similarly

\begin{equation}
\left\vert \gamma _{k}\right\vert =\beta _{\Delta }[1-\frac{k^{2}}{2}+\frac{%
k^{4}}{8}]+O\left( k^{6}\right) ,
\end{equation}%
and its phase\ $\theta _{k}$ has an expansion
\begin{equation}
\frac{\gamma _{\mathbf{k}}}{\left\vert \gamma _{\mathbf{k}}\right\vert }%
=1-ik_{x}k_{y}+O\left( k^{4}\right) ;\theta _{k}=-k_{x}k_{y}+O\left(
k^{4}\right) .
\end{equation}%
In terms of $\ z^{\ast }$ it is an analytic function:%
\begin{equation}
\gamma _{k}=e^{-ik_{x}k_{y}-\frac{k^{2}}{2}}\gamma \left[ z_{\mathbf{k}%
}^{\ast }\right] ;\gamma \left[ z_{\mathbf{k}}^{\ast }\right] =\sum_{\mathbf{%
X}}e^{-\frac{X^{2}}{2}-iz_{\mathbf{k}}^{\ast }X}.
\end{equation}

\bigskip \emph{\textbf{Self duality relation}}

If the lattice is self-dual, one can prove
\begin{equation}
\sum_{\mathbf{X}}X^{2}e^{-\frac{X^{2}}{2}}=\beta _{\Delta }.
\end{equation}%
Thus
\begin{equation}
\beta _{\mathbf{k}}=\beta _{\Delta }-\frac{\beta _{A}}{4}k^{2}+\frac{k^{4}}{%
64}\sum_{\mathbf{X}}X^{4}e^{-\frac{X^{2}}{2}}.
\end{equation}

Using the expansion for $\beta _{k},\gamma _{k},$ one can obtain the
supersoft acoustic phone spectrum:

\begin{equation}
-\beta _{\Delta }+2\beta _{\mathbf{k}}-\left\vert \gamma _{k}\right\vert =(%
\frac{1}{32}\sum_{X}X^{4}e^{-\frac{X^{2}}{2}}-\frac{\beta _{\Delta }}{8}%
)k^{4}+O\left( k^{6}\right) .
\end{equation}

\emph{\textbf{The small momentum expansion of the vertex function}}

For momentum, $\mathbf{k,l,k}^{\prime }$ are not too big to have
Umklapp process
\begin{gather}
\int_{\mathbf{r}}\varphi _{\mathbf{k}}^{\ast }(\mathbf{r})\varphi _{\mathbf{l%
}}(\mathbf{r})\varphi _{\mathbf{0}}^{\ast }(\mathbf{r})\varphi _{\mathbf{k}%
^{\prime }}(\mathbf{r})=\left( 2\pi \right) ^{2}\delta \left[ \mathbf{k}-%
\mathbf{l\mathbf{-k}^{\prime }}\right] \times  \notag \\
\sum_{\mathbf{K}}e^{-\frac{\mathcal{K}^{2}}{2}+i\left( k_{x}-k_{x}^{\prime
}\right) \mathcal{K}_{y}-i\mathcal{K}_{x}\left( l_{y}-l_{y}^{\prime }\right)
+i\left( l_{y}\right) \left( k-l\right) _{x}},
\end{gather}%
where $\mathcal{K}\mathcal{=}\mathbf{K+k}-\mathbf{l.}$ In this case,
we define $f\left[ \mathbf{k},\mathbf{l,\mathbf{k}^{\prime
},\mathbf{l}^{\prime },\Delta }\right]
=[\mathbf{k,\mathbf{l}^{\prime }|l,k}^{\prime }]$ and which has
small momentum expansion:

\begin{equation}
\left[ \mathbf{k,0|l,k}^{\prime }\right] =\beta _{\Delta }(1-\frac{l^{2}+k%
\mathbf{^{\prime }}^{2}}{4}+\frac{i}{2}\left(
k_{x}k_{y}-l_{x}l_{y}-k_{x}^{\prime }k_{y}^{\prime }\right) )
\end{equation}

\emph{\textbf{Another useful identity}}

Any sixfold ($D_{6}$) symmetric function $F(\mathbf{k})$ (namely a function
satisfying $F(\mathbf{k})$ $=F(\mathbf{k}^{^{\prime }}),$ where $\mathbf{k,k}%
^{^{\prime }}$ is related by a $\frac{2\pi }{6}$ rotation) obeys:
\begin{equation}
\int_{\mathbf{k}}F(\mathbf{k})\gamma _{\mathbf{k}}\gamma _{\mathbf{k,l}}=%
\frac{\gamma _{\mathbf{l}}}{\beta _{\Delta }}\int_{\mathbf{k}}F(\mathbf{k}%
)\left\vert \gamma _{\mathbf{k}}\right\vert ^{2},\gamma _{\mathbf{k},\mathbf{%
l}}=<\varphi _{\mathbf{k}}^{\ast }\varphi _{-\mathbf{k}}^{\ast }\varphi _{-%
\mathbf{l}}\varphi _{\mathbf{l}}>
\end{equation}%
This identity can be proved by using the fact $\left( \int_{\mathbf{k}}F(%
\mathbf{k})\gamma _{\mathbf{k}}\gamma _{\mathbf{k,l}}\right) {\Large /}%
\gamma _{\mathbf{l}}$ is a analytic function of $z_{\mathbf{l}}^{\ast }$ and
is a periodic function of reciprocal lattice vectors, \textit{i.g.}

\begin{equation}
\mathbf{l}\rightarrow \mathbf{l+}m_{1}\widetilde{\mathbf{d}}_{1}+m_{2}%
\widetilde{\mathbf{d}}_{2}
\end{equation}%
the function is unchanged. The only solution for a analytic function with
this property is a constant.
\begin{equation}
\frac{\int_{\mathbf{k}}F(\mathbf{k})\gamma _{\mathbf{k}}\gamma _{\mathbf{k,l}%
}}{\gamma _{\mathbf{l}}}=const.
\end{equation}%
The constant can be determined by setting $\mathbf{l}=0$.

\subsection{Parisi algebra for hierarchial matrices}

In this appendix we collect without derivation the formulas used in
calculation of disorder properties. Derivation can be found in ref. \cite%
{Mezard91}. Inverse matrix has the following Parisi parameters:%
\begin{eqnarray}
\widetilde{m^{-1}} &=&\frac{1}{\widetilde{m}-<m>}(1-\int_{0}^{1}\frac{du}{u}%
\frac{[m]_{u}}{\widetilde{m}-<m>-[m]_{u}}  \notag \\
&&-\frac{m_{0}}{\widetilde{m}-<m>}); \\
m_{x}^{-1} &=&-\frac{1}{\widetilde{m}-<m>}(\frac{[m]_{x}}{x\left[ \widetilde{%
m}-<m>-[m]_{x}\right] }  \notag \\
&&+\int_{0}^{x}\frac{dv}{v^{2}}\frac{[m]_{v}}{\widetilde{m}-<m>-[m]_{v}}+%
\frac{m_{0}}{\widetilde{m}-<m>})  \notag
\end{eqnarray}%
Square of matrix can be treated similarly:

\begin{eqnarray}
\widetilde{m_{a,b}^{2}} &=&\left( \widetilde{m}\right) ^{2}-<\left(
m_{x}\right) ^{2}>; \\
\left( m^{2}\right) _{x} &=&2\left( \widetilde{m}-<m>\right)
m_{x}-\int_{0}^{x}dv(m_{x}-m_{v})^{2}  \notag
\end{eqnarray}

\bigskip


\begin{thebibliography}{157}
\expandafter\ifx\csname natexlab\endcsname\relax\def\natexlab#1{#1}\fi
\expandafter\ifx\csname bibnamefont\endcsname\relax
  \def\bibnamefont#1{#1}\fi
\expandafter\ifx\csname bibfnamefont\endcsname\relax
  \def\bibfnamefont#1{#1}\fi
\expandafter\ifx\csname citenamefont\endcsname\relax
  \def\citenamefont#1{#1}\fi
\expandafter\ifx\csname url\endcsname\relax
  \def\url#1{\texttt{#1}}\fi
\expandafter\ifx\csname urlprefix\endcsname\relax\def\urlprefix{URL }\fi
\providecommand{\bibinfo}[2]{#2}
\providecommand{\eprint}[2][]{\url{#2}}


\bibitem[{\citenamefont{ Abo-Shaeer \emph{et~al.}}(2001)}]
  {Abo-Shaeer01}
  \bibinfo{author}{\bibfnamefont{Abo-Shaeer}, \bibnamefont{J.~R.}},
  \bibinfo{author}{\bibfnamefont{C.}~\bibnamefont{Raman}},
  \bibinfo{author}{\bibfnamefont{J.~M.}~\bibnamefont{Vogels}},  and
  \bibinfo{author}{\bibfnamefont{W.}~\bibnamefont{Ketterele}},
  \bibinfo{year}{2001},
  \bibinfo{journal}{Science} \textbf{\bibinfo{volume}{292}},
  \bibinfo{pages}{476}.


\bibitem[{\citenamefont{Abrikosov}(1957)}]
{Abrikosov57}
\bibinfo{author}{\bibnamefont{Abrikosov}, \bibfnamefont{A.~A.}},
  \bibinfo{year}{1957},
  \bibinfo{journal}{Zh. Eksp. Teor. Fiz.} \textbf{\bibinfo{volume}{32}},
  \bibinfo{pages}{1442}.


\bibitem[{\citenamefont{ Adesso \emph{et~al.}}(2006)}]
  {Adesso}
  \bibinfo{author}{\bibfnamefont{Adesso}, \bibnamefont{M.~G.}},
  \bibinfo{author}{\bibfnamefont{D.}~\bibnamefont{Uglietti}},
  \bibinfo{author}{\bibfnamefont{R.}~\bibnamefont{Fl\"{u}kiger}},
  \bibinfo{author}{\bibfnamefont{M.}~\bibnamefont{Polichetti}}, and
  \bibinfo{author}{\bibfnamefont{S.}~\bibnamefont{Pace}},
  \bibinfo{year}{2006},
  \bibinfo{journal}{Phys. Rev. B} \textbf{\bibinfo{volume}{73}},
  \bibinfo{pages}{092513}.

  \bibitem[{\citenamefont{Affleck and Br\'{e}zin}(1985)}]
  {Affleck85}
  \bibinfo{author}{\bibnamefont{Affleck}, \bibfnamefont{I.}},and
  \bibinfo{author}{\bibfnamefont{E.}~\bibnamefont{Br\'{e}zin}},
  \bibinfo{year}{1985},
  \bibinfo{journal}{Nucl. Phys. B} \textbf{\bibinfo{volume}{257}},
  \bibinfo{pages}{451}.


\bibitem[{\citenamefont{Avraham \emph{et~al.}}(2001)}]
  {Avraham01}
  \bibinfo{author}{\bibfnamefont{Avraham}, \bibnamefont{N.}},
  \bibinfo{author}{\bibfnamefont{B.}~\bibnamefont{Khaykovich}},
  \bibinfo{author}{\bibfnamefont{Y.}~\bibnamefont{Myasoedov}},
   \bibinfo{author}{\bibfnamefont{M.}~\bibnamefont{Rappaport}},
   \bibinfo{author}{\bibfnamefont{H.}~\bibnamefont{Shtrikman}},
  \bibinfo{author}{\bibfnamefont{D.~E.}~\bibnamefont{Feldman}},
  \bibinfo{author}{\bibfnamefont{T.}~\bibnamefont{Tamegai}},
  \bibinfo{author}{\bibfnamefont{P.~H.}~\bibnamefont{Kes}},
  \bibinfo{author}{\bibfnamefont{M.}~\bibnamefont{Li}},
   \bibinfo{author}{\bibfnamefont{M.}~\bibnamefont{Konczykowski}},
  \bibinfo{author}{\bibfnamefont{K.}~\bibnamefont{van der Beek}},  and
  \bibinfo{author}{\bibfnamefont{E.}~\bibnamefont{Zeldov}},
  \bibinfo{year}{2001},
  \bibinfo{journal}{Nature (London)} \textbf{\bibinfo{volume}{411}},
  \bibinfo{pages}{451}.

\bibitem[{\citenamefont{Baker}(1990)}]{Baker}
\bibinfo{author}{\bibnamefont{Baker}, \bibfnamefont{G.~A.}},
  \bibinfo{year}{1990},
\emph{\bibinfo{title}{Quantitative theory of critical phenomena}}
  (\bibinfo{publisher}{Academic Press, Boston }).



\bibitem[{\citenamefont{Baraduc \emph{et~al.}}(1994)}]
  {Baraduc94}
  \bibinfo{author}{\bibfnamefont{Baraduc}, \bibnamefont{C.}},
  \bibinfo{author}{\bibfnamefont{E.}~\bibnamefont{Janod}},
  \bibinfo{author}{\bibfnamefont{C.}~\bibnamefont{Ayache}}, and
  \bibinfo{author}{\bibfnamefont{J.~Y.}~\bibnamefont{Henry}},
  \bibinfo{year}{1994},
  \bibinfo{journal}{Physica C} \textbf{\bibinfo{volume}{235}},
  \bibinfo{pages}{1555}.

\bibitem[{\citenamefont{Baym}(2003)}]
  {Baym03}
  \bibinfo{author}{\bibfnamefont{Baym}, \bibnamefont{G.}},
  \bibinfo{year}{2003},
  \bibinfo{journal}{Phys. Rev. Lett.} \textbf{\bibinfo{volume}{91}},
  \bibinfo{pages}{110402}.

\bibitem[{\citenamefont{Bel \emph{et~al.}}(2007)}]
  {Bel07}
  \bibinfo{author}{\bibfnamefont{Bel}, \bibnamefont{G.}},
  \bibinfo{author}{\bibfnamefont{D.}~\bibnamefont{Li}},
  \bibinfo{author}{\bibfnamefont{B.}~\bibnamefont{Rosenstein}},
  \bibinfo{author}{\bibfnamefont{V.}~\bibnamefont{Vinokur}},
  and
  \bibinfo{author}{\bibfnamefont{V.}~\bibnamefont{Zhuravlev}},
  \bibinfo{year}{2007},
  \bibinfo{journal}{Physica C} \textbf{\bibinfo{volume}{460-462}},
  \bibinfo{pages}{1213}.

\bibitem[{\citenamefont{Bellet \emph{et~al.}}(1996a)}]
  {Bellet96a}
  \bibinfo{author}{\bibfnamefont{Bellet}, \bibnamefont{B.}},
  \bibinfo{author}{\bibfnamefont{P.}~\bibnamefont{Garcia}}, and
  \bibinfo{author}{\bibfnamefont{A.}~\bibnamefont{Neveu}},
  \bibinfo{year}{1996a},
  \bibinfo{journal}{Int. J. Mod. Phys. A} \textbf{\bibinfo{volume}{11}},
  \bibinfo{pages}{5587}.



\bibitem[{\citenamefont{Bellet \emph{et~al.}}(1996b)}]
  {Bellet96b}
  \bibinfo{author}{\bibfnamefont{Bellet}, \bibnamefont{B.}},
  \bibinfo{author}{\bibfnamefont{P.}~\bibnamefont{Garcia}}, and
  \bibinfo{author}{\bibfnamefont{A.}~\bibnamefont{Neveu}},
  \bibinfo{year}{1996b},
  \bibinfo{journal}{Int. J. Mod. Phys. A} \textbf{\bibinfo{volume}{11}},
  \bibinfo{pages}{5607}.


\bibitem[{\citenamefont{Bender \emph{et~al.}}(1994)}]
  {Bender94}
  \bibinfo{author}{\bibfnamefont{Bender}, \bibnamefont{C.~M.}},
  \bibinfo{author}{\bibfnamefont{A.}~\bibnamefont{Duncan}}, and
  \bibinfo{author}{\bibfnamefont{H.F.}~\bibnamefont{Jones}},
  \bibinfo{year}{1994},
  \bibinfo{journal}{Phys. Rev. D } \textbf{\bibinfo{volume}{49}},
  \bibinfo{pages}{4219}.


\bibitem[{\citenamefont{Beidenkopf \emph{et~al.}}(2005)}]
  {Zeldov05}
  \bibinfo{author}{\bibfnamefont{Beidenkopf}, \bibnamefont{H.}},
  \bibinfo{author}{\bibfnamefont{N.}~\bibnamefont{Avraham}},
  \bibinfo{author}{\bibfnamefont{Y.}~\bibnamefont{Myasoedov}},
  \bibinfo{author}{\bibfnamefont{H.}~\bibnamefont{Shtrikman}},
  \bibinfo{author}{\bibfnamefont{E.}~\bibnamefont{Zeldov}},
  \bibinfo{author}{\bibfnamefont{B.}~\bibnamefont{Rosenstein}},
  \bibinfo{author}{\bibfnamefont{E.~H.}~\bibnamefont{Brandt}}, and
  \bibinfo{author}{\bibfnamefont{T.}~\bibnamefont{Tamegai}},
  \bibinfo{year}{2005},
  \bibinfo{journal}{Phys. Rev. Lett.} \textbf{\bibinfo{volume}{95}},
  \bibinfo{pages}{257004}.
  
  \bibitem[{\citenamefont{Beidenkopf \emph{et~al.}}(2007)}]
  {Zeldov07}
  \bibinfo{author}{\bibfnamefont{Beidenkopf}, \bibnamefont{H.}},
  \bibinfo{author}{\bibfnamefont{T.}~\bibnamefont{Verdene}},
  \bibinfo{author}{\bibfnamefont{Y.}~\bibnamefont{Myasoedov}},
  \bibinfo{author}{\bibfnamefont{H.}~\bibnamefont{Shtrikman}},
  \bibinfo{author}{\bibfnamefont{E.}~\bibnamefont{Zeldov}},
  \bibinfo{author}{\bibfnamefont{B.}~\bibnamefont{Rosenstein}},
  \bibinfo{author}{\bibfnamefont{D.}~\bibnamefont{Li}}, and
  \bibinfo{author}{\bibfnamefont{T.}~\bibnamefont{Tamegai}},
  \bibinfo{year}{2007},
  \bibinfo{journal}{Phys. Rev. Lett.} \textbf{\bibinfo{volume}{98}},
  \bibinfo{pages}{167004}.
  




\bibitem[{\citenamefont{Blatter} \emph{et~al.}(1994)}]
  {Blatter}
  \bibinfo{author}{\bibnamefont{Blatter}, \bibfnamefont{G.}},
  \bibinfo{author}{\bibfnamefont{M.~V.} \bibnamefont{Feigel'man}},
  \bibinfo{author}{\bibfnamefont{V.~B.} \bibnamefont{Geshkenbein}},
  \bibinfo{author}{\bibfnamefont{A.~I.} \bibnamefont{Larkin}}, and
  \bibinfo{author}{\bibfnamefont{V.~M.} \bibnamefont{Vinokur}},
  \bibinfo{year}{1994},
  \bibinfo{journal}{Rev. Mod. Phys.} \textbf{\bibinfo{volume}{66}},
  \bibinfo{pages}{1125}.


\bibitem[{\citenamefont{Blum and Moore}(1997)}]
{Blum97}
\bibinfo{author}{\bibnamefont{Blum}, \bibfnamefont{T.}}, and
\bibinfo{author}{\bibfnamefont{M.~A.} \bibnamefont{Moore}},
  \bibinfo{year}{1997},
  \bibinfo{journal}{Phys. Rev. B} \textbf{\bibinfo{volume}{56}},
  \bibinfo{pages}{372}.

  \bibitem[{\citenamefont{Bogner \emph{et~al.}}(2001)}]
  {Bogner01}
  \bibinfo{author}{\bibfnamefont{Bogner}, \bibnamefont{S.}},
  \bibinfo{author}{\bibfnamefont{T.}~\bibnamefont{Emig}}, and
  \bibinfo{author}{\bibfnamefont{T.}~\bibnamefont{Nattermann}},
  \bibinfo{year}{2001},
  \bibinfo{journal}{Phys. Rev. B} \textbf{\bibinfo{volume}{63}},
  \bibinfo{pages}{174501}.


  \bibitem[{\citenamefont{Bokil and Young}(1995)}]
  {Young95}
  \bibinfo{author}{\bibfnamefont{Bokil}, \bibnamefont{H.~S.}},  and
  \bibinfo{author}{\bibfnamefont{A.~P.}~\bibnamefont{Young}},
  \bibinfo{year}{1995},
  \bibinfo{journal}{Phys. Rev. Lett.} \textbf{\bibinfo{volume}{74}},
  \bibinfo{pages}{3021}.




\bibitem[{\citenamefont{Bouquet \emph{et~al.}}(2001)}]
  {Bouquet01}
  \bibinfo{author}{\bibfnamefont{Bouquet}, \bibnamefont{F.}},
  \bibinfo{author}{\bibfnamefont{C.}~\bibnamefont{Marcenat}},
  \bibinfo{author}{\bibfnamefont{E.}~\bibnamefont{Steep}},
   \bibinfo{author}{\bibfnamefont{R.}~\bibnamefont{Calemczuk}},
   \bibinfo{author}{\bibfnamefont{W.~K.}~\bibnamefont{Kwok}},
  \bibinfo{author}{\bibfnamefont{U.}~\bibnamefont{Welp}},
  \bibinfo{author}{\bibfnamefont{G.~W.}~\bibnamefont{Crabtree}},
  \bibinfo{author}{\bibfnamefont{R.~A.}~\bibnamefont{Fisher}},
  \bibinfo{author}{\bibfnamefont{N.~E.}~\bibnamefont{Phillips}},
   and
  \bibinfo{author}{\bibfnamefont{A.}~\bibnamefont{Schilling}},
  \bibinfo{year}{2001},
  \bibinfo{journal}{Nature (London)} \textbf{\bibinfo{volume}{411}},
  \bibinfo{pages}{448}.

  \bibitem[{\citenamefont{Brandt}(1977a)}]
{Brandt77}
\bibinfo{author}{\bibnamefont{Brandt}, \bibfnamefont{E.~H.}},
  \bibinfo{year}{1977a},
  \bibinfo{journal}{J. Low Temp.Phys.} \textbf{\bibinfo{volume}{26}},
  \bibinfo{pages}{709}.


  \bibitem[{\citenamefont{Brandt}(1977b)}]
{Brandt77b}
\bibinfo{author}{\bibnamefont{Brandt}, \bibfnamefont{E.~H.}},
  \bibinfo{year}{1977b},
  \bibinfo{journal}{J. Low Temp.Phys.} \textbf{\bibinfo{volume}{26}},
  \bibinfo{pages}{735}.


\bibitem[{\citenamefont{Brandt}(1986)}]
{Brandt86}
\bibinfo{author}{\bibnamefont{Brandt}, \bibfnamefont{E.~H.}},
  \bibinfo{year}{1986},
  \bibinfo{journal}{Phys. Rev. B} \textbf{\bibinfo{volume}{34}},
  \bibinfo{pages}{6514}.

\bibitem[{\citenamefont{Brandt}(1995)}]
{Brandt}
  \bibinfo{author}{\bibnamefont{Brandt}, \bibfnamefont{E.~H.}},
  \bibinfo{year}{1995},
  \bibinfo{journal}{Rep. Prog. Phys.} \textbf{\bibinfo{volume}{58}},
  \bibinfo{pages}{1465}.




  \bibitem[{\citenamefont{Br\'{e}zin \emph{et~al.}}(1985)}]
  {Brezin85}
  \bibinfo{author}{\bibnamefont{Br\'{e}zin}, \bibfnamefont{E.}},
  \bibinfo{author}{\bibfnamefont{D.R.}~\bibnamefont{Nelson}}, and
  \bibinfo{author}{\bibfnamefont{A.}~\bibnamefont{Thiaville}},
  \bibinfo{year}{1985},
  \bibinfo{journal}{Phys. Rev. B} \textbf{\bibinfo{volume}{31}},
  \bibinfo{pages}{7124}.


  \bibitem[{\citenamefont{Br\'{e}zin \emph{et~al.}}(1990)}]
  {Hikami90}
  \bibinfo{author}{\bibfnamefont{Br\'{e}zin}, \bibnamefont{E.}},
  \bibinfo{author}{\bibfnamefont{A.}~\bibnamefont{Fujita}},and
  \bibinfo{author}{\bibfnamefont{S.}~\bibnamefont{Hikami}},
  \bibinfo{year}{1990},
  \bibinfo{journal}{Phys. Rev. Lett.} \textbf{\bibinfo{volume}{65}},
  \bibinfo{pages}{1949}.



  \bibitem[{\citenamefont{Cha and Fertig}(1994a)}]
  {Cha94}
  \bibinfo{author}{\bibfnamefont{Cha}, \bibnamefont{M.-C.}}, and
  \bibinfo{author}{\bibfnamefont{H.~A.}~\bibnamefont{Fertig}},
  \bibinfo{year}{1994a},
  \bibinfo{journal}{Phys. Rev. Lett.} \textbf{\bibinfo{volume}{73}},
  \bibinfo{pages}{870}.

\bibitem[{\citenamefont{Cha and Fertig}(1994b)}]
  {Cha94b}
  \bibinfo{author}{\bibfnamefont{Cha}, \bibnamefont{M.-C.}}, and
  \bibinfo{author}{\bibfnamefont{H.~A.}~\bibnamefont{Fertig}},
  \bibinfo{year}{1994b},
  \bibinfo{journal}{Phys. Rev. B} \textbf{\bibinfo{volume}{50}},
  \bibinfo{pages}{14368}.



  \bibitem[{\citenamefont{Chang \emph{et~al.}}(1998a)}]
  {Chang98}
  \bibinfo{author}{\bibfnamefont{Chang}, \bibnamefont{D.}},
  \bibinfo{author}{\bibfnamefont{C.-Y.}~\bibnamefont{Mou}},
  \bibinfo{author}{\bibfnamefont{B.}~\bibnamefont{Rosenstein}}, and
  \bibinfo{author}{\bibfnamefont{C.~L.}~\bibnamefont{Wu}},
  \bibinfo{year}{1998a},
  \bibinfo{journal}{Phys. Rev. Lett.} \textbf{\bibinfo{volume}{80}},
  \bibinfo{pages}{145}.

\bibitem[{\citenamefont{Chang \emph{et~al.}}(1998b)}]
  {Chang98b}
  \bibinfo{author}{\bibfnamefont{Chang}, \bibnamefont{D.}},
  \bibinfo{author}{\bibfnamefont{C.-Y.}~\bibnamefont{Mou}},
  \bibinfo{author}{\bibfnamefont{B.}~\bibnamefont{Rosenstein}}, and
  \bibinfo{author}{\bibfnamefont{C.~L.}~\bibnamefont{Wu}},
  \bibinfo{year}{1998b},
  \bibinfo{journal}{Phys. Rev. B} \textbf{\bibinfo{volume}{57}},
  \bibinfo{pages}{7955}.



  \bibitem[{\citenamefont{Chauve \emph{et~al.}}(2000)}]
  {Chauve00}
  \bibinfo{author}{\bibfnamefont{Chauve}, \bibnamefont{P.}},
  \bibinfo{author}{\bibfnamefont{T.}~\bibnamefont{Giamarchi}}, and
  \bibinfo{author}{\bibfnamefont{P.}~\bibnamefont{Le Doussal}},
  \bibinfo{year}{2000},
  \bibinfo{journal}{Phys. Rev. B} \textbf{\bibinfo{volume}{62}},
  \bibinfo{pages}{6241}.



  \bibitem[{\citenamefont{Chen and  Hu}(2003)}]
  {Chen03}
  \bibinfo{author}{\bibfnamefont{Chen}, \bibnamefont{Q.~H.}}, and
  \bibinfo{author}{\bibfnamefont{X.}~\bibnamefont{Hu}},
  \bibinfo{year}{2003},
  \bibinfo{journal}{Phys. Rev. Lett.} \textbf{\bibinfo{volume}{90}},
  \bibinfo{pages}{117005}.


\bibitem[{\citenamefont{Chen}(2008)}]
  {Chen08}
  \bibinfo{author}{\bibfnamefont{Chen}, \bibnamefont{Q.~H.}},
  \bibinfo{year}{2008},
  \bibinfo{journal}{Phys. Rev. B} \textbf{\bibinfo{volume}{78}},
  \bibinfo{pages}{104501}.




\bibitem[{\citenamefont{Compagner}(1974)}]
  {Compagner74}
  \bibinfo{author}{\bibfnamefont{Compagner}, \bibnamefont{A.}},
  \bibinfo{year}{1974},
  \bibinfo{journal}{Physica} \textbf{\bibinfo{volume}{72}},
  \bibinfo{pages}{115}.

\bibitem[{\citenamefont{Cooper \emph{et~al.}}(2001)}]
  {Cooper01}
  \bibinfo{author}{\bibfnamefont{Cooper}, \bibnamefont{N.~R.}},
  \bibinfo{author}{\bibfnamefont{N.~K.}~\bibnamefont{Wilkin}}, and
  \bibinfo{author}{\bibfnamefont{J.~M.~F.}~\bibnamefont{Gunn}},
  \bibinfo{year}{2001},
  \bibinfo{journal}{Phys. Rev. Lett. } \textbf{\bibinfo{volume}{87}},
  \bibinfo{pages}{120405}.

\bibitem[{\citenamefont{Cornwall \emph{et~al.}}(1974)}]
  {Cornwall74}
  \bibinfo{author}{\bibfnamefont{Cornwall}, \bibnamefont{J.~M.}},
  \bibinfo{author}{\bibfnamefont{R.}~\bibnamefont{Jackiw}}, and
  \bibinfo{author}{\bibfnamefont{E.}~\bibnamefont{Tomboulis}},
  \bibinfo{year}{1974},
  \bibinfo{journal}{Phys. Rev. D } \textbf{\bibinfo{volume}{10}},
  \bibinfo{pages}{2428}.



 
 
 \bibitem[{\citenamefont{Dasgupta and Halperin}(1981)}]
 {Dasgupta81}
  \bibinfo{author}{\bibfnamefont{Dasgupta}, \bibnamefont{C.}},  and
  \bibinfo{author}{\bibfnamefont{B.~I.}~\bibnamefont{Halperin}},
  \bibinfo{year}{1981},
  \bibinfo{journal}{Phys. Rev. Lett.} \textbf{\bibinfo{volume}{47}},
 \bibinfo{pages}{1556}.



\bibitem[{\citenamefont{David}(1981)}]
  {David81}
  \bibinfo{author}{\bibfnamefont{David}, \bibnamefont{F.}},
  \bibinfo{year}{1981},
  \bibinfo{journal}{Com. Math. Phys.} \textbf{\bibinfo{volume}{81}},
  \bibinfo{pages}{149}.



\bibitem[{\citenamefont{de Alameida and Thouless}(1978)}]
  {Alameida78}
  \bibinfo{author}{\bibfnamefont{de Alameida}, \bibnamefont{J.~R.~L.}},  and
  \bibinfo{author}{\bibfnamefont{D.~J.}~\bibnamefont{Thouless}},
  \bibinfo{year}{1978},
  \bibinfo{journal}{J. Phys. A} \textbf{\bibinfo{volume}{11}},
  \bibinfo{pages}{983}.



\bibitem[{\citenamefont{Deligiannis \emph{et~al.}}(2000)}]
  {Deligiannis00}
  \bibinfo{author}{\bibfnamefont{Deligiannis}, \bibnamefont{K.}},
  \bibinfo{author}{\bibfnamefont{M.}~\bibnamefont{Charalambous}},
  \bibinfo{author}{\bibfnamefont{J.}~\bibnamefont{Chaussy}},
  \bibinfo{author}{\bibfnamefont{R.}~\bibnamefont{Liang}},
  \bibinfo{author}{\bibfnamefont{D.}~\bibnamefont{Bonn}}, and
  \bibinfo{author}{\bibfnamefont{W.~N.}~\bibnamefont{Hardy}},
  \bibinfo{year}{2000},
  \bibinfo{journal}{Physica  C} \textbf{\bibinfo{volume}{341}},
  \bibinfo{pages}{1329}.


\bibitem[{\citenamefont{Dietel and  Kleinert}(2006)}]
  {Dietel06}
  \bibinfo{author}{\bibfnamefont{Dietel}, \bibnamefont{J.}},and
  \bibinfo{author}{\bibfnamefont{H.}~\bibnamefont{Kleinert}},
  \bibinfo{year}{2006},
  \bibinfo{journal}{Phys. Rev. B} \textbf{\bibinfo{volume}{74}},
  \bibinfo{pages}{024515}.
  
  \bibitem[{\citenamefont{Dietel and  Kleinert}(2007)}]
  {Dietel07}
  \bibinfo{author}{\bibfnamefont{Dietel}, \bibnamefont{J.}},and
  \bibinfo{author}{\bibfnamefont{H.}~\bibnamefont{Kleinert}},
  \bibinfo{year}{2007},
  \bibinfo{journal}{Phys. Rev. B} \textbf{\bibinfo{volume}{75}},
  \bibinfo{pages}{144513}.
  
  \bibitem[{\citenamefont{Dietel and  Kleinert}(2009)}]
  {Dietel09}
  \bibinfo{author}{\bibfnamefont{Dietel}, \bibnamefont{J.}},and
  \bibinfo{author}{\bibfnamefont{H.}~\bibnamefont{Kleinert}},
  \bibinfo{year}{2009},
  \bibinfo{journal}{Phys. Rev. B} \textbf{\bibinfo{volume}{79}},
  \bibinfo{pages}{014512}.
  


  \bibitem[{\citenamefont{Divakar \emph{et~al.}}(2004)}]
  {Divakar04}
  \bibinfo{author}{\bibfnamefont{Divakar}, \bibnamefont{U.}},
  \bibinfo{author}{\bibfnamefont{A.~J.}~\bibnamefont{Drew}},
  \bibinfo{author}{\bibfnamefont{S.~L.}~\bibnamefont{Lee}},
  \bibinfo{author}{\bibfnamefont{R.}~\bibnamefont{Gilardi}},
  \bibinfo{author}{\bibfnamefont{J.}~\bibnamefont{Mesot}},
  \bibinfo{author}{\bibfnamefont{F.~Y.}~\bibnamefont{Ogrin}},
  \bibinfo{author}{\bibfnamefont{D.}~\bibnamefont{Charalambous}},
  \bibinfo{author}{\bibfnamefont{E.~M.}~\bibnamefont{Forgan}},
  \bibinfo{author}{\bibfnamefont{G.~I.}~\bibnamefont{Menon}},
  \bibinfo{author}{\bibfnamefont{N.}~\bibnamefont{Momono}},
  \bibinfo{author}{\bibfnamefont{M.}~\bibnamefont{Oda}},
  \bibinfo{author}{\bibfnamefont{C.~D.}~\bibnamefont{Dewhurst}},  and
  \bibinfo{author}{\bibfnamefont{C.}~\bibnamefont{Baines}},
  \bibinfo{year}{2004},
  \bibinfo{journal}{Phys. Rev. Lett.} \textbf{\bibinfo{volume}{92}},
  \bibinfo{pages}{237004}.

\bibitem[{\citenamefont{Dodgson \emph{et~al.}}(2000)}]
  {Dodgson00}
  \bibinfo{author}{\bibfnamefont{Dodgson}, \bibnamefont{M.~J.~W.}},
  \bibinfo{author}{\bibfnamefont{A.~E.}~\bibnamefont{Koshelev}},
  \bibinfo{author}{\bibfnamefont{V.~B.}~\bibnamefont{Geshkenbein}}, and
  \bibinfo{author}{\bibfnamefont{G.}~\bibnamefont{Blatter}},
  \bibinfo{year}{2000},
  \bibinfo{journal}{Phys. Rev. Lett.} \textbf{\bibinfo{volume}{84}},
  \bibinfo{pages}{2698}.




  \bibitem[{\citenamefont{Dorsey \emph{et~al.}}(1992)}]
  {Dorsey92}
  \bibinfo{author}{\bibfnamefont{Dorsey}, \bibnamefont{A.~T.}},
  \bibinfo{author}{\bibfnamefont{M.}~\bibnamefont{Huang}}, and
  \bibinfo{author}{\bibfnamefont{M.~P.~A.}~\bibnamefont{Fisher}},
  \bibinfo{year}{1992},
  \bibinfo{journal}{ Phys. Rev. B} \textbf{\bibinfo{volume}{45}},
  \bibinfo{pages}{523}.

  \bibitem[{\citenamefont{Dotsenko}(2001)}]{Dotsenko}
\bibinfo{author}{\bibnamefont{Dotsenko}, \bibfnamefont{V.}},
  \bibinfo{year}{2001},
\emph{\bibinfo{title}{An Introduction to the Theory of Spin Glasses and Neural Networks}}
  (\bibinfo{publisher}{Cambridge University Press,New York }).

\bibitem[{\citenamefont{Duncan and Jones}(1993)}]
  {Duncan93}
  \bibinfo{author}{\bibfnamefont{Duncan}, \bibnamefont{A.}}, and
  \bibinfo{author}{\bibfnamefont{H.F.}~\bibnamefont{Jones}},
  \bibinfo{year}{1993},
  \bibinfo{journal}{Phys. Rev. D } \textbf{\bibinfo{volume}{47}},
  \bibinfo{pages}{2560}.

  \bibitem[{\citenamefont{Eilenberger}(1967)}]
  {Eilenberger67}
  \bibinfo{author}{\bibfnamefont{Eilenberger},\bibnamefont{G.}},
  \bibinfo{year}{1967},
  \bibinfo{journal}{Phys. Rev. } \textbf{\bibinfo{volume}{164}},
  \bibinfo{pages}{628}.


\bibitem[{\citenamefont{Engels \emph{et~al.}}(2002)}]
  {Engels02}
  \bibinfo{author}{\bibfnamefont{Engels}, \bibnamefont{P.}},
  \bibinfo{author}{\bibfnamefont{I.}~\bibnamefont{Coddington}},
  \bibinfo{author}{\bibfnamefont{P.~C.}~\bibnamefont{Haljian}}, and
  \bibinfo{author}{\bibfnamefont{E.~A.}~\bibnamefont{Cornell}},
  \bibinfo{year}{2002},
  \bibinfo{journal}{Phys. Rev. Lett.} \textbf{\bibinfo{volume}{89}},
  \bibinfo{pages}{100403}.

  \bibitem[{\citenamefont{Ertas and Nelson}(1996)}]
  {Ertas96}
  \bibinfo{author}{\bibfnamefont{Ertas}, \bibnamefont{D.}}, and
  \bibinfo{author}{\bibfnamefont{D.~R.}~\bibnamefont{Nelson}},
  \bibinfo{year}{1996},
  \bibinfo{journal}{Physica C} \textbf{\bibinfo{volume}{272}},
  \bibinfo{pages}{79}.

  \bibitem[{\citenamefont{Eskildsen \emph{et~al.}}(2001)}]
  {Eskildsen01}
  \bibinfo{author}{\bibfnamefont{Eskildsen}, \bibnamefont{M.~R.}},
  \bibinfo{author}{\bibfnamefont{A.~B.}~\bibnamefont{Abrahamsen}},
  \bibinfo{author}{\bibfnamefont{V.~G.}~\bibnamefont{Kogan}},
  \bibinfo{author}{\bibfnamefont{P.~L.}~\bibnamefont{Gammel}},
  \bibinfo{author}{\bibfnamefont{K.}~\bibnamefont{Mortensen}},
  \bibinfo{author}{\bibfnamefont{N.~H.}~\bibnamefont{Andersen}}, and
  \bibinfo{author}{\bibfnamefont{P.~C.}~\bibnamefont{Canfield}},
  \bibinfo{year}{2001},
  \bibinfo{journal}{Phys. Rev. Lett.} \textbf{\bibinfo{volume}{86}},
  \bibinfo{pages}{5148}.

  \bibitem[{\citenamefont{Faleski \emph{et~al.}}(1996)}]
  {Faleski96}
  \bibinfo{author}{\bibfnamefont{Faleski}, \bibnamefont{M.~C.}},
  \bibinfo{author}{\bibfnamefont{M.~C.}~\bibnamefont{Marchetti}},  and
  \bibinfo{author}{\bibfnamefont{A.~A.}~\bibnamefont{Middleton}},
  \bibinfo{year}{1996},
  \bibinfo{journal}{Phys. Rev. B} \textbf{\bibinfo{volume}{54}},
  \bibinfo{pages}{12427}.

 \bibitem[{\citenamefont{Fangohr \emph{et~al.}}(2001)}]
  {Fangohr01}
  \bibinfo{author}{\bibfnamefont{Fangohr}, \bibnamefont{H.}},  and
  \bibinfo{author}{\bibfnamefont{S.~J.}~\bibnamefont{Cox}},
  \bibinfo{year}{2001},
  \bibinfo{journal}{Phys. Rev. B} \textbf{\bibinfo{volume}{64}},
  \bibinfo{pages}{064505}.

  \bibitem[{\citenamefont{Fangohr \emph{et~al.}}(2003)}]
  {Fangohr03}
  \bibinfo{author}{\bibfnamefont{Fangohr}, \bibnamefont{H.}},
  \bibinfo{author}{\bibfnamefont{A.~E.}~\bibnamefont{Koshelev}},  and
  \bibinfo{author}{\bibfnamefont{M.~J.~W.}~\bibnamefont{Dodgson}},
  \bibinfo{year}{2003},
  \bibinfo{journal}{Phys. Rev. B} \textbf{\bibinfo{volume}{67}},
  \bibinfo{pages}{174508}.



  \bibitem[{\citenamefont{Feng \emph{et~al.}}(2009)}]
  {Feng09}
  \bibinfo{author}{\bibfnamefont{Feng}, \bibnamefont{B.}},
  \bibinfo{author}{\bibfnamefont{Z.}~\bibnamefont{Wu}},
  and
  \bibinfo{author}{\bibfnamefont{D.}~\bibnamefont{Li}},
  \bibinfo{year}{2009},
  \bibinfo{journal}{Int. Mod. Phys. B} \textbf{\bibinfo{volume}{23}},
  \bibinfo{pages}{661}.


  \bibitem[{\citenamefont{Fischer and Hertz}(1991)}]{Fischer}
\bibinfo{author}{\bibnamefont{Fischer}, \bibfnamefont{K.~H.}}, and
\bibinfo{author}{\bibfnamefont{J.~A.} \bibnamefont{Hertz}},
  \bibinfo{year}{1991},
\emph{\bibinfo{title}{Spin glasses}}
  (\bibinfo{publisher}{Cambridge University Press,New York }).



  \bibitem[{\citenamefont{Fisher}(1989)}]{Fisher89}
  \bibinfo{author}{\bibfnamefont{Fisher}, \bibnamefont{M.~P.~A.}},
  \bibinfo{year}{1989},
  \bibinfo{journal}{Phys. Rev. Lett.} \textbf{\bibinfo{volume}{62}},
  \bibinfo{pages}{1415}.



\bibitem[{\citenamefont{Fisher \emph{et~al.}}(1991)}]{Fisher91}
  \bibinfo{author}{\bibfnamefont{Fisher}, \bibnamefont{D.~S.}},
  \bibinfo{author}{\bibfnamefont{M.~P.~A.}~\bibnamefont{Fisher}},  and
  \bibinfo{author}{\bibfnamefont{D.~A.}~\bibnamefont{Huse}},
  \bibinfo{year}{1991},
  \bibinfo{journal}{Phys. Rev. B} \textbf{\bibinfo{volume}{43}},
  \bibinfo{pages}{130}.

 \bibitem[{\citenamefont{Fruchter \emph{et~al.}}(1997)}]
  {Fruchter}
  \bibinfo{author}{\bibfnamefont{Fuchs}, \bibnamefont{L.}},
  \bibinfo{author}{\bibfnamefont{A.}~\bibnamefont{Aburto}},  and
  \bibinfo{author}{\bibfnamefont{C.}~\bibnamefont{Pham-Phu}},
  \bibinfo{year}{1997},
  \bibinfo{journal}{Phys. Rev. B} \textbf{\bibinfo{volume}{56}},
  \bibinfo{pages}{R2936}.

\bibitem[{\citenamefont{Fuchs \emph{et~al.}}(1998)}]
  {Fuchs98}
  \bibinfo{author}{\bibfnamefont{Fuchs}, \bibnamefont{D.~T.}},
  \bibinfo{author}{\bibfnamefont{E.}~\bibnamefont{Zeldov}},
  \bibinfo{author}{\bibfnamefont{T.}~\bibnamefont{Tamegai}},
  \bibinfo{author}{\bibfnamefont{S.}~\bibnamefont{Ooi}},
  \bibinfo{author}{\bibfnamefont{M.}~\bibnamefont{Rappaport}}, and
  \bibinfo{author}{\bibfnamefont{H.}~\bibnamefont{Shtrikman}},
  \bibinfo{year}{1998},
  \bibinfo{journal}{Phys. Rev. Lett.} \textbf{\bibinfo{volume}{80}},
  \bibinfo{pages}{4971}.





  \bibitem[{\citenamefont{Giamarchi and Le Doussal}(1994)}]
  {Giamarchi94}
  \bibinfo{author}{\bibfnamefont{Giamarchi}, \bibnamefont{T.}},and
  \bibinfo{author}{\bibfnamefont{P.}~\bibnamefont{Le Doussal}},
  \bibinfo{year}{1994},
  \bibinfo{journal}{Phys. Rev. Lett.} \textbf{\bibinfo{volume}{72}},
  \bibinfo{pages}{1530}.

\bibitem[{\citenamefont{Giamarchi and Le Doussal}(1995a)}]
  {Giamarchi95a}
  \bibinfo{author}{\bibfnamefont{Giamarchi}, \bibnamefont{T.}},and
  \bibinfo{author}{\bibfnamefont{P.}~\bibnamefont{Le Doussal}},
  \bibinfo{year}{1995a},
  \bibinfo{journal}{Phys. Rev. Lett.} \textbf{\bibinfo{volume}{75}},
  \bibinfo{pages}{3372}.



\bibitem[{\citenamefont{Giamarchi and Le Doussal}(1995b)}]
  {Giamarchi95b}
  \bibinfo{author}{\bibfnamefont{Giamarchi}, \bibnamefont{T.}},and
  \bibinfo{author}{\bibfnamefont{P.}~\bibnamefont{Le Doussal}},
  \bibinfo{year}{1995b},
  \bibinfo{journal}{Phys. Rev. B} \textbf{\bibinfo{volume}{52}},
  \bibinfo{pages}{1242}.

\bibitem[{\citenamefont{Giamarchi and Le Doussal}(1996)}]
  {Giamarchi96}
  \bibinfo{author}{\bibfnamefont{Giamarchi}, \bibnamefont{T.}},and
  \bibinfo{author}{\bibfnamefont{P.}~\bibnamefont{Le Doussal}},
  \bibinfo{year}{1996},
  \bibinfo{journal}{Phys. Rev. Lett.} \textbf{\bibinfo{volume}{76}},
  \bibinfo{pages}{3408}.

\bibitem[{\citenamefont{Giamarchi and Le Doussal}(1997)}]
  {Giamarchi97}
  \bibinfo{author}{\bibfnamefont{Giamarchi}, \bibnamefont{T.}},and
  \bibinfo{author}{\bibfnamefont{P.}~\bibnamefont{Le Doussal}},
  \bibinfo{year}{1997},
  \bibinfo{journal}{Phys. Rev. B} \textbf{\bibinfo{volume}{55}},
  \bibinfo{pages}{6577}.

\bibitem[{\citenamefont{Giamarchi and Le Doussal}(1998)}]
  {Giamarchi98}
  \bibinfo{author}{\bibfnamefont{Giamarchi}, \bibnamefont{T.}},and
  \bibinfo{author}{\bibfnamefont{P.}~\bibnamefont{Le Doussal}},
  \bibinfo{year}{1998},
  \bibinfo{journal}{Phys. Rev. B} \textbf{\bibinfo{volume}{57}},
  \bibinfo{pages}{11356}.

\bibitem[{\citenamefont{Giamarchi and Bhattacharya} (2002)}]{Giamarchi-Bhattacharya}
  \bibinfo{author}{\bibnamefont{Giamarchi}, \bibfnamefont{T.}}, and
  \bibinfo{author}{\bibnamefont{S.}, \bibfnamefont{Bhattacharya}},
  \bibinfo{year}{2002}, in
  \emph{\bibinfo{booktitle}{High Magnetic Fields: Applications in Condensed Matter Physics and Spectroscopy}},
  edited by \bibinfo{editor}{\bibfnamefont{C.}~\bibnamefont{Berthier} \emph{et~al.}}
  (\bibinfo{publisher}{Springer-Verlag, Berlin}), p. \bibinfo{pages}{314}.



  \bibitem[{\citenamefont{Gilardi \emph{et~al.}}(2002)}]
  {Gilardi02}
  \bibinfo{author}{\bibfnamefont{Gilardi}, \bibnamefont{R.}},
  \bibinfo{author}{\bibfnamefont{J.}~\bibnamefont{Mesot}},
  \bibinfo{author}{\bibfnamefont{A.}~\bibnamefont{Drew}},
  \bibinfo{author}{\bibfnamefont{U.}~\bibnamefont{Divakar}},
  \bibinfo{author}{\bibfnamefont{S.~L.}~\bibnamefont{Lee}},
  \bibinfo{author}{\bibfnamefont{E.~M.}~\bibnamefont{Forgan}},
  \bibinfo{author}{\bibfnamefont{O.}~\bibnamefont{Zaharko}},
  \bibinfo{author}{\bibfnamefont{K.}~\bibnamefont{Conder}},
  \bibinfo{author}{\bibfnamefont{V.~K.}~\bibnamefont{Aswal}},
  \bibinfo{author}{\bibfnamefont{C.~D.}~\bibnamefont{Dewhurst}},
  \bibinfo{author}{\bibfnamefont{R.}~\bibnamefont{Cubitt}},
  \bibinfo{author}{\bibfnamefont{N.}~\bibnamefont{Momono}}, and
  \bibinfo{author}{\bibfnamefont{M.}~\bibnamefont{Oda}},
  \bibinfo{year}{2002},
  \bibinfo{journal}{Phys. Rev. Lett.} \textbf{\bibinfo{volume}{88}},
  \bibinfo{pages}{217003}.







  \bibitem[{\citenamefont{Ginzburg and Landau}(1950)}]
  {Ginzburg50}
  \bibinfo{author}{\bibnamefont{Ginzburg}, \bibfnamefont{V.~L.}}, and
  \bibinfo{author}{\bibfnamefont{L.~D.} \bibnamefont{Landau}},
  \bibinfo{year}{1950},
  \bibinfo{journal}{Zh. Eksp. Teor. Fiz.} \textbf{\bibinfo{volume}{20}},
  \bibinfo{pages}{1064}.


\bibitem[{\citenamefont{Ginzburg}(1960)}]
{Ginzburg60}
\bibinfo{author}{\bibnamefont{Ginzburg}, \bibfnamefont{V.~L.}},
  \bibinfo{year}{1960},
  \bibinfo{journal}{Sov. Sol. St.} \textbf{\bibinfo{volume}{2}},
  \bibinfo{pages}{1824}.




  
\bibitem[{\citenamefont{Guida \emph{et~al.}}(1995)}]
  {Guida95}
  \bibinfo{author}{\bibfnamefont{Guida}, \bibnamefont{R.}},
  \bibinfo{author}{\bibfnamefont{K.}~\bibnamefont{Konishi}}, and
  \bibinfo{author}{\bibfnamefont{H.}~\bibnamefont{Suzuki}},
  \bibinfo{year}{1995},
  \bibinfo{journal}{Ann. Phys. (NY) } \textbf{\bibinfo{volume}{241}},
  \bibinfo{pages}{152}.

 \bibitem[{\citenamefont{Guida \emph{et~al.}}(1996)}]
  {Guida96}
  \bibinfo{author}{\bibfnamefont{Guida}, \bibnamefont{R.}},
  \bibinfo{author}{\bibfnamefont{K.}~\bibnamefont{Konishi}}, and
  \bibinfo{author}{\bibfnamefont{H.}~\bibnamefont{Suzuki}},
  \bibinfo{year}{1996},
  \bibinfo{journal}{Ann. Phys. (NY) } \textbf{\bibinfo{volume}{249}},
  \bibinfo{pages}{109}.

  \bibitem[{\citenamefont{Halperin \emph{et~al.}}(1974)}]
  {Halperin74}
  \bibinfo{author}{\bibnamefont{Halperin}, \bibfnamefont{B.~I.}},
  \bibinfo{author}{\bibfnamefont{T.~C.}~\bibnamefont{Lubensky}}, and
  \bibinfo{author}{\bibfnamefont{S.-K.}~\bibnamefont{Ma}},
  \bibinfo{year}{1974},
  \bibinfo{journal}{Phys. Rev. Lett.} \textbf{\bibinfo{volume}{32}},
  \bibinfo{pages}{292}.



  \bibitem[{\citenamefont{Herbut and Te\v{s}anovi\'{c}}(1994)}]
  {Herbut94}
  \bibinfo{author}{\bibfnamefont{Herbut}, \bibnamefont{I.~F.}}, and
  \bibinfo{author}{\bibfnamefont{Z.}~\bibnamefont{Te\v{s}anovi\'{c}}},
  \bibinfo{year}{1994},
  \bibinfo{journal}{Phys. Rev. Lett.} \textbf{\bibinfo{volume}{73}},
  \bibinfo{pages}{484}.


\bibitem[{\citenamefont{Herbut and Te\v{s}anovi\'{c}}(1996)}]
  {Herbut96}
  \bibinfo{author}{\bibfnamefont{Herbut}, \bibnamefont{I.~F.}}, and
  \bibinfo{author}{\bibfnamefont{Z.}~\bibnamefont{Te\v{s}anovi\'{c}}},
  \bibinfo{year}{1996},
  \bibinfo{journal}{Phys. Rev. Lett.} \textbf{\bibinfo{volume}{76}},
  \bibinfo{pages}{4588}.

\bibitem[{\citenamefont{Herbut}(2007)}]
{Herbut07}
\bibinfo{author}{\bibnamefont{Herbut}, \bibfnamefont{I.~F.}},
  \bibinfo{year}{2007},
\emph{\bibinfo{title}{A Modern Approach to Critical Phenomena}}
  (\bibinfo{publisher}{Cambridge University Press }).
  


  \bibitem[{\citenamefont{Hikami \emph{et~al.}}(1991)}]
  {Hikami91}
  \bibinfo{author}{\bibfnamefont{Hikami}, \bibnamefont{S.}},
  \bibinfo{author}{\bibfnamefont{A.}~\bibnamefont{Fujita}},and
  \bibinfo{author}{\bibfnamefont{A.~I.}~\bibnamefont{Larkin}},
  \bibinfo{year}{1991},
  \bibinfo{journal}{Phys. Rev. B} \textbf{\bibinfo{volume}{44}},
  \bibinfo{pages}{10400}.


\bibitem[{\citenamefont{Houghton \emph{et~al.}}(1989)}]
  {Houghton89}
  \bibinfo{author}{\bibfnamefont{Houghton}, \bibnamefont{A.}},
  \bibinfo{author}{\bibfnamefont{R.~A.}~\bibnamefont{Pelcovits}}, and
  \bibinfo{author}{\bibfnamefont{A.}~\bibnamefont{Sudb\o}},
  \bibinfo{year}{1989},
  \bibinfo{journal}{Phys. Rev. B} \textbf{\bibinfo{volume}{40}},
  \bibinfo{pages}{6763}.


  \bibitem[{\citenamefont{Hu and MacDonald}(1993)}]
  {Hu-MacDonald93}
  \bibinfo{author}{\bibfnamefont{Hu}, \bibnamefont{J.}}, and
  \bibinfo{author}{\bibfnamefont{A.~H.}~\bibnamefont{MacDonald}},
  \bibinfo{year}{1993},
  \bibinfo{journal}{Phys. Rev. Lett.} \textbf{\bibinfo{volume}{71}},
  \bibinfo{pages}{432}.


 \bibitem[{\citenamefont{Hu \emph{et~al.}}(1994)}]
  {Hu94}
  \bibinfo{author}{\bibfnamefont{Hu}, \bibnamefont{J.}},
  \bibinfo{author}{\bibfnamefont{A.~H.}~\bibnamefont{MacDonald}},and
  \bibinfo{author}{\bibfnamefont{B.~D.}~\bibnamefont{Mckay}},
  \bibinfo{year}{1994},
  \bibinfo{journal}{Phys. Rev. B} \textbf{\bibinfo{volume}{49}},
  \bibinfo{pages}{15263}.

  \bibitem[{\citenamefont{Hu and MacDonald}(1997)}]
  {Hu-MacDonald97}
  \bibinfo{author}{\bibfnamefont{Hu}, \bibnamefont{J.}}, and
  \bibinfo{author}{\bibfnamefont{A.~H.}~\bibnamefont{MacDonald}},
  \bibinfo{year}{1997},
  \bibinfo{journal}{Phys. Rev. B} \textbf{\bibinfo{volume}{56}},
  \bibinfo{pages}{2788}.

 \bibitem[{\citenamefont{Hu \emph{et~al.}}(1997)}]
  {Hu97}
  \bibinfo{author}{\bibfnamefont{Hu}, \bibnamefont{X.}},
  \bibinfo{author}{\bibfnamefont{S.}~\bibnamefont{Miyashita}}, and
  \bibinfo{author}{\bibfnamefont{M.}~\bibnamefont{Tachiki}},
  \bibinfo{year}{1997},
  \bibinfo{journal}{Phys. Rev. Lett.} \textbf{\bibinfo{volume}{79}},
  \bibinfo{pages}{3498}.


\bibitem[{\citenamefont{Hu \emph{et~al.}}(2005)}]
  {Hu05}
 \bibinfo{author}{\bibfnamefont{Hu}, \bibnamefont{X.}},
  \bibinfo{author}{\bibfnamefont{M.~B.}~\bibnamefont{Luo}}, and
  \bibinfo{author}{\bibfnamefont{Y.~Q.}~\bibnamefont{Ma}},
  \bibinfo{year}{2005},
  \bibinfo{journal}{Phys. Rev. B} \textbf{\bibinfo{volume}{72}},
  \bibinfo{pages}{174503}.
 



\bibitem[{\citenamefont{Huh and Finnemore}(2002)}]
  {Huh02}
  \bibinfo{author}{\bibfnamefont{Huh}, \bibnamefont{Y.~M.}}, and
  \bibinfo{author}{\bibfnamefont{D.~K.}~\bibnamefont{Finnemore}},
  \bibinfo{year}{2002},
  \bibinfo{journal}{Phys. Rev. B} \textbf{\bibinfo{volume}{65}},
  \bibinfo{pages}{092506}.

  \bibitem[{\citenamefont{Ikeda \emph{et~al.}}(1990)}]
  {Ikeda90}
  \bibinfo{author}{\bibfnamefont{Ikeda}, \bibnamefont{R.}},
  \bibinfo{author}{\bibfnamefont{T.}~\bibnamefont{Ohmi}}, and
  \bibinfo{author}{\bibfnamefont{T.}~\bibnamefont{Tsuneto}},
  \bibinfo{year}{1990},
  \bibinfo{journal}{J. Phys. Soc. Jpn.} \textbf{\bibinfo{volume}{59}},
  \bibinfo{pages}{1740}.


 \bibitem[{\citenamefont{Ikeda }(1995)}]
  {Ikeda95}
  \bibinfo{author}{\bibfnamefont{Ikeda}, \bibnamefont{R.}},
  \bibinfo{year}{1995},
  \bibinfo{journal}{J. Phys. Soc. Jpn.} \textbf{\bibinfo{volume}{64}},
  \bibinfo{pages}{1683}.






  \bibitem[{\citenamefont{Itzykson and Drouffe}(1991)}]{Itzykson}
\bibinfo{author}{\bibnamefont{Itzykson}, \bibfnamefont{C.}}, and
\bibinfo{author}{\bibfnamefont{J.} \bibnamefont{Drouffe}},
  \bibinfo{year}{1991},
\emph{\bibinfo{title}{ Statistical Field Theory}}
  (\bibinfo{publisher}{Cambridge University Press, Cambridge, New York }).
  
  
\bibitem[{\citenamefont{Jaiswal-Nagar \emph{et~al.}}(2006)}]
  {Jaiswal-Nagar06}
  \bibinfo{author}{\bibfnamefont{Jaiswal-Nagar}, \bibnamefont{D.}},
  \bibinfo{author}{\bibfnamefont{A.~D.}~\bibnamefont{Thakur}}, 
  \bibinfo{author}{\bibfnamefont{S.}~\bibnamefont{Ramakrishnan}},
  \bibinfo{author}{\bibfnamefont{A.~K.}~\bibnamefont{Grover}},
  \bibinfo{author}{\bibfnamefont{D.}~\bibnamefont{Pal}},
  and
  \bibinfo{author}{\bibfnamefont{H.}~\bibnamefont{Takeya}},
  \bibinfo{year}{2006},
  \bibinfo{journal}{Phys. Rev. B} \textbf{\bibinfo{volume}{74}},
  \bibinfo{pages}{184514}.


  \bibitem[{\citenamefont{Jevicki}(1977)}]
  {Jevicki77}
  \bibinfo{author}{\bibfnamefont{Jevicki}, \bibnamefont{A.}},
  \bibinfo{year}{1977},
  \bibinfo{journal}{Phys. Let. B} \textbf{\bibinfo{volume}{71}},
  \bibinfo{pages}{327}.





  \bibitem[{\citenamefont{Johnson \emph{et~al.}}(1999)}]
  {Johnson99}
  \bibinfo{author}{\bibfnamefont{Johnson}, \bibnamefont{S.~T.}},
  \bibinfo{author}{\bibfnamefont{E.~M.}~\bibnamefont{Forgan}},
  \bibinfo{author}{\bibfnamefont{S.~H.}~\bibnamefont{Lloyd}},
  \bibinfo{author}{\bibfnamefont{C.~M.}~\bibnamefont{Aegerter}},
  \bibinfo{author}{\bibfnamefont{S.~L.}~\bibnamefont{Lee}},
  \bibinfo{author}{\bibfnamefont{R.}~\bibnamefont{Cubitt}},
  \bibinfo{author}{\bibfnamefont{P.~G.}~\bibnamefont{Kealey}},
  \bibinfo{author}{\bibfnamefont{C.}~\bibnamefont{Ager}},
  \bibinfo{author}{\bibfnamefont{S.}~\bibnamefont{Tajima}},
   \bibinfo{author}{\bibfnamefont{A.}~\bibnamefont{Rykov}}, and
  \bibinfo{author}{\bibfnamefont{D.}~\bibnamefont{McK. Paul}},
  \bibinfo{year}{1999},
  \bibinfo{journal}{Phys. Rev. Lett.} \textbf{\bibinfo{volume}{82}},
  \bibinfo{pages}{2792}.

   \bibitem[{\citenamefont{Junod \emph{et~al.}}(1998)}]
  {Junod98}
  \bibinfo{author}{\bibfnamefont{Junod}, \bibnamefont{A.}},
  \bibinfo{author}{\bibfnamefont{J.~Y.}~\bibnamefont{Genuod}},
  \bibinfo{author}{\bibfnamefont{G.}~\bibnamefont{Triscone}}, and
  \bibinfo{author}{\bibfnamefont{T.}~\bibnamefont{Schneider}},
  \bibinfo{year}{1998},
  \bibinfo{journal}{Physica C} \textbf{\bibinfo{volume}{294}},
  \bibinfo{pages}{115}.



  \bibitem[{\citenamefont{Kato and Nagaosa}(1993)}]
  {Kato93}
  \bibinfo{author}{\bibfnamefont{Kato}, \bibnamefont{Y.}}, and
  \bibinfo{author}{\bibfnamefont{N.}~\bibnamefont{Nagaosa}},
  \bibinfo{year}{1993},
  \bibinfo{journal}{Phys. Rev. B} \textbf{\bibinfo{volume}{48}},
  \bibinfo{pages}{7383}.


\bibitem[{\citenamefont{Kawamura}(2003)}]
  {Kawamura03}
  \bibinfo{author}{\bibfnamefont{Kawamura}, \bibnamefont{H.}},
  \bibinfo{year}{2003},
  \bibinfo{journal}{Phys. Rev. B} \textbf{\bibinfo{volume}{68}},
  \bibinfo{pages}{220502}.

  \bibitem[{\citenamefont{Keimer \emph{et~al.}}(1994)}]
  {Keimer94}
  \bibinfo{author}{\bibfnamefont{Keimer}, \bibnamefont{B.}},
  \bibinfo{author}{\bibfnamefont{W.~Y.}~\bibnamefont{Shih}},
  \bibinfo{author}{\bibfnamefont{R.~W.}~\bibnamefont{Erwin}},
  \bibinfo{author}{\bibfnamefont{J.-W.}~\bibnamefont{Lynn}},
  \bibinfo{author}{\bibfnamefont{F.}~\bibnamefont{Dogan}}, and
  \bibinfo{author}{\bibfnamefont{I.~A.}~\bibnamefont{Aksay}},
  \bibinfo{year}{1994},
  \bibinfo{journal}{Phys. Rev. Lett.} \textbf{\bibinfo{volume}{73}},
  \bibinfo{pages}{3459}.




  \bibitem[{\citenamefont{Ketterson and Song}(1999)}]{Ketterson}
\bibinfo{author}{\bibnamefont{Ketterson}, \bibfnamefont{J.~B.}}, and
\bibinfo{author}{\bibfnamefont{S.~N.} \bibnamefont{Song}},
  \bibinfo{year}{1999},
\emph{\bibinfo{title}{ Superconductivity}}
  (\bibinfo{publisher}{Cambridge University Press, New York }).




  \bibitem[{\citenamefont{Kienappel and Moore }(1997)}]
  {Kienappel97}
  \bibinfo{author}{\bibfnamefont{Kienappel}, \bibnamefont{A.~K.}}, and
  \bibinfo{author}{\bibfnamefont{M.~A.}~\bibnamefont{Moore}},
  \bibinfo{year}{1997},
  \bibinfo{journal}{Phys. Rev. B} \textbf{\bibinfo{volume}{56}},
  \bibinfo{pages}{8313}.

  \bibitem[{\citenamefont{Kierfeld and Vinokur}(2000)}]
  {Kierfeld00}
  \bibinfo{author}{\bibfnamefont{Kierfeld}, \bibnamefont{J.}}, and
  \bibinfo{author}{\bibfnamefont{V.}~\bibnamefont{Vinokur}},
  \bibinfo{year}{2000},
  \bibinfo{journal}{Phys. Rev. B} \textbf{\bibinfo{volume}{61}},
  \bibinfo{pages}{R14928}.


\bibitem[{\citenamefont{Kim \emph{et~al.}}(1999)}]
  {Lieber99}
  \bibinfo{author}{\bibfnamefont{Kim}, \bibnamefont{P.}},
  \bibinfo{author}{\bibfnamefont{Z.}~\bibnamefont{Yao}},
  \bibinfo{author}{\bibfnamefont{C.~A.}~\bibnamefont{Bolle}}, and
  \bibinfo{author}{\bibfnamefont{C.~M.}~\bibnamefont{Lieber}},
  \bibinfo{year}{1999},
  \bibinfo{journal}{Phys. Rev. B} \textbf{\bibinfo{volume}{60}},
  \bibinfo{pages}{R12589}.


\bibitem[{\citenamefont{Kleinert}(1995)}]{Kleinert}
\bibinfo{author}{\bibnamefont{Kleinert}, \bibfnamefont{H.}},
  \bibinfo{year}{1990},
\emph{\bibinfo{title}{Path Integrals in Quantum Mechanics,
Statistics, and Polymer Physics}}
  (\bibinfo{publisher}{World Scientific, Singapore }).



\bibitem[{\citenamefont{Klironomos and Dorsey}(2003)}]
  {Klironomos03}
  \bibinfo{author}{\bibfnamefont{Klironomos}, \bibnamefont{A.~D.}}, and
  \bibinfo{author}{\bibfnamefont{A.~T.}~\bibnamefont{Dorsey}},
  \bibinfo{year}{2003},
  \bibinfo{journal}{Phys. Rev. Lett.} \textbf{\bibinfo{volume}{91}},
  \bibinfo{pages}{097002}.



 \bibitem[{\citenamefont{ Kobayashi \emph{et~al.}}(2001)}]
  {Kobayashi01}
  \bibinfo{author}{\bibfnamefont{Kobayashi}, \bibnamefont{N.}},
  \bibinfo{author}{\bibfnamefont{T.}~\bibnamefont{Nishizaki}},
  \bibinfo{author}{\bibfnamefont{K.}~\bibnamefont{Shibata}},
  \bibinfo{author}{\bibfnamefont{T.}~\bibnamefont{Sato}},
  \bibinfo{author}{\bibfnamefont{M.}~\bibnamefont{Maki}}, and
  \bibinfo{author}{\bibfnamefont{T.}~\bibnamefont{Sasaki}},
  \bibinfo{year}{2001},
  \bibinfo{journal}{Physica C} \textbf{\bibinfo{volume}{362}},
  \bibinfo{pages}{121}.




\bibitem[{\citenamefont{Kokkaliaris \emph{et~al.}}(2000)}]
  {Kokkaliaris00}
  \bibinfo{author}{\bibfnamefont{Kokkaliaris}, \bibnamefont{S.}},
  \bibinfo{author}{\bibfnamefont{A.~A.}~\bibnamefont{Zhukov}},
  \bibinfo{author}{\bibfnamefont{P.~A.~J.}~\bibnamefont{de Groot}},
  \bibinfo{author}{\bibfnamefont{R.}~\bibnamefont{Gagnon}},
  \bibinfo{author}{\bibfnamefont{L.}~\bibnamefont{Taillefer}}, and
  \bibinfo{author}{\bibfnamefont{T.}~\bibnamefont{Wolf}},
  \bibinfo{year}{2000},
  \bibinfo{journal}{Phys. Rev. B} \textbf{\bibinfo{volume}{61}},
  \bibinfo{pages}{3655}.


\bibitem[{\citenamefont{Kokubo \emph{et~al.}}(2004)}]
  {Kokubo04}
  \bibinfo{author}{\bibfnamefont{Kokubo}, \bibnamefont{N.}},
  \bibinfo{author}{\bibfnamefont{R.}~\bibnamefont{Besseling}}, and
  \bibinfo{author}{\bibfnamefont{P.~H.}~\bibnamefont{Kes}},
  \bibinfo{year}{2004},
  \bibinfo{journal}{Phys. Rev. B} \textbf{\bibinfo{volume}{69}},
  \bibinfo{pages}{064504}.

\bibitem[{\citenamefont{Kokubo \emph{et~al.}}(2005)}]
  {Kokubo05}
  \bibinfo{author}{\bibfnamefont{Kokubo}, \bibnamefont{N.}},
  \bibinfo{author}{\bibfnamefont{K.}~\bibnamefont{Kadowaki}}, and
  \bibinfo{author}{\bibfnamefont{K.}~\bibnamefont{Takita}},
  \bibinfo{year}{2005},
  \bibinfo{journal}{Phys. Rev. Lett.} \textbf{\bibinfo{volume}{95}},
  \bibinfo{pages}{177005}.

  \bibitem[{\citenamefont{ Kokubo \emph{et~al.}}(2007)}]
  {Kokubo07}
  \bibinfo{author}{\bibfnamefont{Kokubo}, \bibnamefont{N.}},
  \bibinfo{author}{\bibfnamefont{T.}~\bibnamefont{Asada}},
  \bibinfo{author}{\bibfnamefont{K.}~\bibnamefont{Kadowaki}},
  \bibinfo{author}{\bibfnamefont{K.}~\bibnamefont{Takita}},
  \bibinfo{author}{\bibfnamefont{T.~G.}~\bibnamefont{Sorop}},   and
  \bibinfo{author}{\bibfnamefont{P.~H.}~\bibnamefont{Kes}},
  \bibinfo{year}{2007},
  \bibinfo{journal}{Phys. Rev. B} \textbf{\bibinfo{volume}{75}},
  \bibinfo{pages}{184512}.


  \bibitem[{\citenamefont{Kopnin}(2001)}]{Kopnin}
\bibinfo{author}{\bibnamefont{Kopnin}, \bibfnamefont{N.~B.}},
  \bibinfo{year}{2001},
\emph{\bibinfo{title}{Theory of Nonequilibrium
Superconductivity}}
  (\bibinfo{publisher}{Oxford University Press}).

\bibitem[{\citenamefont{Korshunov}(1990)}]
  {Korshunov90}
  \bibinfo{author}{\bibfnamefont{Korshunov}, \bibnamefont{S.~E.}},
  \bibinfo{year}{1990},
  \bibinfo{journal}{Europhys. Lett.} \textbf{\bibinfo{volume}{11}},
  \bibinfo{pages}{757}.

\bibitem[{\citenamefont{Korshunov}(1993)}]
  {Korshunov93}
  \bibinfo{author}{\bibfnamefont{Korshunov}, \bibnamefont{S.~E.}},
  \bibinfo{year}{1993},
  \bibinfo{journal}{Phys. Rev. B} \textbf{\bibinfo{volume}{48}},
  \bibinfo{pages}{3969}.



  \bibitem[{\citenamefont{Koshelev}(1994)}]
  {Koshelev94}
  \bibinfo{author}{\bibfnamefont{Koshelev}, \bibnamefont{A.~E.}},
  \bibinfo{year}{1994},
  \bibinfo{journal}{Phys. Rev. B} \textbf{\bibinfo{volume}{50}},
  \bibinfo{pages}{506}.

\bibitem[{\citenamefont{Kovner and Rosenstein}(1992)}]
  {Kovner92}
  \bibinfo{author}{\bibfnamefont{Kovner}, \bibnamefont{A.}}, and
  \bibinfo{author}{\bibfnamefont{B.}~\bibnamefont{Rosenstein}},
  \bibinfo{year}{1992},
  \bibinfo{journal}{J. Phys. Cond. Mat.} \textbf{\bibinfo{volume}{4}},
  \bibinfo{pages}{2903}.


\bibitem[{\citenamefont{Kovner \emph{et~al.}}(1993)}]
  {Kovner93}
 \bibinfo{author}{\bibfnamefont{Kovner}, \bibnamefont{A.}},
  \bibinfo{author}{\bibfnamefont{P.}~\bibnamefont{Kurzepa}},  and
  \bibinfo{author}{\bibfnamefont{B.}~\bibnamefont{Rosenstein}},
  \bibinfo{year}{1993},
  \bibinfo{journal}{Mod. Phys. Lett.} \textbf{\bibinfo{volume}{8}},
  \bibinfo{pages}{1343}.








  \bibitem[{\citenamefont{Larkin}(1970)}]
{Larkin70}
\bibinfo{author}{\bibnamefont{Larkin}, \bibfnamefont{A.~I.}},
  \bibinfo{year}{1970},
  \bibinfo{journal}{Zh. Eksp. Teor. Fiz.} \textbf{\bibinfo{volume}{58}},
  \bibinfo{pages}{1466}.





  \bibitem[{\citenamefont{Larkin and Varlamov}(2005)}]{Larkin-Varlamov}
\bibinfo{author}{\bibnamefont{Larkin}, \bibfnamefont{A.}}, and
  \bibinfo{author}{\bibfnamefont{A.} \bibnamefont{Varlamov}},
  \bibinfo{year}{2005},
\emph{\bibinfo{title}{Theory of fluctuations in superconductors}}
  (\bibinfo{publisher}{Oxford University Press}).



  \bibitem[{\citenamefont{Lascher}(1965)}]
{Lascher65}
\bibinfo{author}{\bibnamefont{Lascher}, \bibfnamefont{G.}},
  \bibinfo{year}{1965},
  \bibinfo{journal}{Phys. Rev.} \textbf{\bibinfo{volume}{140}},
  \bibinfo{pages}{A523}.


  \bibitem[{\citenamefont{Lawrence and Doniach} (1971)}]{Lawrence-Doniach71}
  \bibinfo{author}{\bibnamefont{Lawrence}, \bibfnamefont{W.~E.}},
  \bibinfo{author}{\bibnamefont{S.}, \bibfnamefont{Doniach}},
  \bibinfo{year}{1971}, in
  \emph{\bibinfo{booktitle}{Proc. Twelfth Int. Conf. on Low Temperature Physics, Kyoto, 1970}},
  edited by \bibinfo{editor}{\bibfnamefont{E.}~\bibnamefont{Kanda}}
  (\bibinfo{publisher}{Academic Press, Kyoto}), p. \bibinfo{pages}{361}.


   \bibitem[{\citenamefont{Lawrie }(1994)}]
  {Lawrie94}
  \bibinfo{author}{\bibfnamefont{Lawrie}, \bibnamefont{I.~D.}},
  \bibinfo{year}{1994},
  \bibinfo{journal}{Phys. Rev. B} \textbf{\bibinfo{volume}{50}},
  \bibinfo{pages}{9456}.

  \bibitem[{\citenamefont{Lee and Shenoy}(1972)}]
  {Shenoy72}
  \bibinfo{author}{\bibnamefont{Lee}, \bibfnamefont{P.~A.}}, and
  \bibinfo{author}{\bibfnamefont{S.~R.}~\bibnamefont{Shenoy}},
  \bibinfo{year}{1972},
  \bibinfo{journal}{Phys. Rev. Lett.} \textbf{\bibinfo{volume}{28}},
  \bibinfo{pages}{1025}.

  \bibitem[{\citenamefont{Leote de Carvalho \emph{et~al.}}(1999)}]
  {Carvalho99}
  \bibinfo{author}{\bibfnamefont{Leote de Carvalho}, \bibnamefont{R.~J.~F.}},
  \bibinfo{author}{\bibfnamefont{R.}~\bibnamefont{Evans}}, and
  \bibinfo{author}{\bibfnamefont{Y.}~\bibnamefont{Rosenfeld}},
  \bibinfo{year}{1999},
  \bibinfo{journal}{Phys. Rev. E } \textbf{\bibinfo{volume}{59}},
  \bibinfo{pages}{1435}.

  \bibitem[{\citenamefont{Levanyuk}(1959)}]
{Levanyuk}
\bibinfo{author}{\bibnamefont{Levanyuk}, \bibfnamefont{A.~P.}},
  \bibinfo{year}{1959},
  \bibinfo{journal}{Zh. Eksp. Teor. Fiz.} \textbf{\bibinfo{volume}{36}},
  \bibinfo{pages}{810}.


\bibitem[{\citenamefont{Li and Nattermann }(2003)}]
  {Li-Nattermann03}
  \bibinfo{author}{\bibfnamefont{Li}, \bibnamefont{M.~S.}}, and
  \bibinfo{author}{\bibfnamefont{T.}~\bibnamefont{Nattermann}},
  \bibinfo{year}{2003},
  \bibinfo{journal}{Phys. Rev. B} \textbf{\bibinfo{volume}{67}},
  \bibinfo{pages}{184520}.

\bibitem[{\citenamefont{Li and Rosenstein}(1999a)}]
{Li99}
\bibinfo{author}{\bibnamefont{Li}, \bibfnamefont{D.}}, and
\bibinfo{author}{\bibfnamefont{B.} \bibnamefont{Rosenstein}},
  \bibinfo{year}{1999a},
  \bibinfo{journal}{Phys. Rev. B} \textbf{\bibinfo{volume}{60}},
  \bibinfo{pages}{9704}.

\bibitem[{\citenamefont{Li and Rosenstein}(1999b)}]
  {Li99b}
  \bibinfo{author}{\bibfnamefont{Li}, \bibnamefont{D.}}, and
  \bibinfo{author}{\bibfnamefont{B.}~\bibnamefont{Rosenstein}},
  \bibinfo{year}{1999b},
  \bibinfo{journal}{Phys. Rev. B} \textbf{\bibinfo{volume}{60}},
  \bibinfo{pages}{10460}.

\bibitem[{\citenamefont{Li and Rosenstein}(2001)}]
  {Li01}
  \bibinfo{author}{\bibfnamefont{Li}, \bibnamefont{D.}}, and
  \bibinfo{author}{\bibfnamefont{B.}~\bibnamefont{Rosenstein}},
  \bibinfo{year}{2001},
  \bibinfo{journal}{Phys. Rev. Lett.} \textbf{\bibinfo{volume}{86}},
  \bibinfo{pages}{3618}.


\bibitem[{\citenamefont{Li and Rosenstein}(2002a)}]
  {Li02}
  \bibinfo{author}{\bibfnamefont{Li}, \bibnamefont{D.}}, and
  \bibinfo{author}{\bibfnamefont{B.}~\bibnamefont{Rosenstein}},
  \bibinfo{year}{2002a},
  \bibinfo{journal}{Phys. Rev. B} \textbf{\bibinfo{volume}{65}},
  \bibinfo{pages}{220504}.


\bibitem[{\citenamefont{Li and Rosenstein}(2002b)}]
  {Li02a}
  \bibinfo{author}{\bibfnamefont{Li}, \bibnamefont{D.}}, and
  \bibinfo{author}{\bibfnamefont{B.}~\bibnamefont{Rosenstein}},
  \bibinfo{year}{2002b},
  \bibinfo{journal}{Phys. Rev. B} \textbf{\bibinfo{volume}{65}},
  \bibinfo{pages}{024513}.

\bibitem[{\citenamefont{Li and Rosenstein}(2002c)}]
  {Li02b}
  \bibinfo{author}{\bibfnamefont{Li}, \bibnamefont{D.}}, and
  \bibinfo{author}{\bibfnamefont{B.}~\bibnamefont{Rosenstein}},
  \bibinfo{year}{2002c},
  \bibinfo{journal}{Phys. Rev. B} \textbf{\bibinfo{volume}{65}},
  \bibinfo{pages}{024514}.


\bibitem[{\citenamefont{Li and Rosenstein}(2003)}]
  {Li03}
  \bibinfo{author}{\bibfnamefont{Li}, \bibnamefont{D.}}, and
  \bibinfo{author}{\bibfnamefont{B.}~\bibnamefont{Rosenstein}},
  \bibinfo{year}{2003},
  \bibinfo{journal}{Phys. Rev. Lett.} \textbf{\bibinfo{volume}{90}},
  \bibinfo{pages}{167004}.

\bibitem[{\citenamefont{Li} \emph{et~al.}(2004a)}]
{Li04a}
\bibinfo{author}{\bibnamefont{Li}, \bibfnamefont{D.}}, and
\bibinfo{author}{\bibfnamefont{B.} \bibnamefont{Rosenstein}},
  \bibinfo{year}{2004a},
  \bibinfo{journal}{Phys. Rev. B} \textbf{\bibinfo{volume}{70}},
  \bibinfo{pages}{144521}.


\bibitem[{\citenamefont{Li} \emph{et~al.}(2004b)}]
{Li04}
\bibinfo{author}{\bibnamefont{Li}, \bibfnamefont{D.}},
\bibinfo{author}{\bibfnamefont{A.~M.} \bibnamefont{Malkin}}, and
\bibinfo{author}{\bibfnamefont{B.} \bibnamefont{Rosenstein}},
  \bibinfo{year}{2004b},
  \bibinfo{journal}{Phys. Rev. B} \textbf{\bibinfo{volume}{70}},
  \bibinfo{pages}{214529}.
  
  \bibitem[{\citenamefont{Li} \emph{et~al.}(2006a)}]
{Li06a}
\bibinfo{author}{\bibnamefont{Li}, \bibfnamefont{D.}},
\bibinfo{author}{\bibfnamefont{P.~-J.} \bibnamefont{Lin}},
\bibinfo{author}{\bibfnamefont{B.} \bibnamefont{Rosenstein}},
bibinfo{author}{\bibfnamefont{B.~Ya.} \bibnamefont{Shapiro}},
and
\bibinfo{author}{\bibfnamefont{I.} \bibnamefont{Shapiro}},
  \bibinfo{year}{2006a},
  \bibinfo{journal}{Phys. Rev. B} \textbf{\bibinfo{volume}{74}},
  \bibinfo{pages}{174518}.
  

  \bibitem[{\citenamefont{Li \emph{et~al.}}(2006b)}]
  {Vinokur06}
  \bibinfo{author}{\bibfnamefont{Li}, \bibnamefont{D.}},
  \bibinfo{author}{\bibfnamefont{B.}~\bibnamefont{Rosenstein}}, and
  \bibinfo{author}{\bibfnamefont{V.}~\bibnamefont{Vinokur}},
  \bibinfo{year}{2006b},
  \bibinfo{journal}{J. Supercond. Novel Mag.} \textbf{\bibinfo{volume}{19}},
  \bibinfo{pages}{369}.


\bibitem[{\citenamefont{Liang \emph{et~al.}}(1996)}]
  {Liang96}
  \bibinfo{author}{\bibfnamefont{Liang}, \bibnamefont{R.}},
  \bibinfo{author}{\bibfnamefont{D.~A.}~\bibnamefont{Bonn}}, and
  \bibinfo{author}{\bibfnamefont{W.~N.}~\bibnamefont{Hardy}},
  \bibinfo{year}{1996},
  \bibinfo{journal}{Phys. Rev. Lett.} \textbf{\bibinfo{volume}{76}},
  \bibinfo{pages}{835}.
  

  \bibitem[{\citenamefont{Lin and Rosenstein}(2005)}]
  {Lin05}
  \bibinfo{author}{\bibnamefont{Lin}, \bibfnamefont{P.~-J.}},and
  \bibinfo{author}{\bibfnamefont{B.}~\bibnamefont{Rosenstein}},
  \bibinfo{year}{2005},
  \bibinfo{journal}{Phys. Rev. B} \textbf{\bibinfo{volume}{71}},
  \bibinfo{pages}{172504}.




  \bibitem[{\citenamefont{Lobb}(1987)}]
  {Lobb87}
  \bibinfo{author}{\bibnamefont{Lobb}, \bibfnamefont{C.~J.}},
  \bibinfo{year}{1987},
  \bibinfo{journal}{Phys. Rev. B} \textbf{\bibinfo{volume}{36}},
  \bibinfo{pages}{3930}.



  \bibitem[{\citenamefont{Lopatin and Kotliar}(1999)}]
  {Lopatin99}
  \bibinfo{author}{\bibnamefont{Lopatin}, \bibfnamefont{A.~V.}},and
  \bibinfo{author}{\bibfnamefont{G.}~\bibnamefont{Kotliar}},
  \bibinfo{year}{1999},
  \bibinfo{journal}{Phys. Rev. B} \textbf{\bibinfo{volume}{59}},
  \bibinfo{pages}{3879}.


  \bibitem[{\citenamefont{Lopatin}(2000)}]
  {Lopatin00}
  \bibinfo{author}{\bibfnamefont{Lopatin}, \bibnamefont{A.~V.}},
  \bibinfo{year}{2000},
  \bibinfo{journal}{Europhys. Lett.} \textbf{\bibinfo{volume}{51}},
  \bibinfo{pages}{635}.


\bibitem[{\citenamefont{Lortz \emph{et~al.}}(2006)}]
  {Lortz06}
  \bibinfo{author}{\bibfnamefont{Lortz}, \bibnamefont{R.}},
  \bibinfo{author}{\bibfnamefont{F.}~\bibnamefont{Lin}},
  \bibinfo{author}{\bibfnamefont{N.}~\bibnamefont{Musolino}},
  \bibinfo{author}{\bibfnamefont{Y.}~\bibnamefont{Wang}},
  \bibinfo{author}{\bibfnamefont{A.}~\bibnamefont{Junod}},
  \bibinfo{author}{\bibfnamefont{B.}~\bibnamefont{Rosenstein}},and
  \bibinfo{author}{\bibfnamefont{N.}~\bibnamefont{Toyota}},
  \bibinfo{year}{2006},
  \bibinfo{journal}{Phys. Rev. B} \textbf{\bibinfo{volume}{74}},
  \bibinfo{pages}{104502}.
  
  \bibitem[{\citenamefont{Lortz \emph{et~al.}}(2007)}]
  {Lortz07}
  \bibinfo{author}{\bibfnamefont{Lortz}, \bibnamefont{R.}},
  \bibinfo{author}{\bibfnamefont{N.}~\bibnamefont{Musolino}},
  \bibinfo{author}{\bibfnamefont{Y.}~\bibnamefont{Wang}},
  \bibinfo{author}{\bibfnamefont{A.}~\bibnamefont{Junod}}, and
  \bibinfo{author}{\bibfnamefont{N.}~\bibnamefont{Toyota}},
  \bibinfo{year}{2007},
  \bibinfo{journal}{Phys. Rev. B} \textbf{\bibinfo{volume}{75}},
  \bibinfo{pages}{094503}.

  \bibitem[{\citenamefont{Lovett}(1977)}]
  {Lovett77}
  \bibinfo{author}{\bibfnamefont{Lovett}, \bibnamefont{R.}},
  \bibinfo{year}{1977},
  \bibinfo{journal}{J. Chem. Phys.} \textbf{\bibinfo{volume}{66}},
  \bibinfo{pages}{1225}.


 \bibitem[{\citenamefont{Madison \emph{et~al.}}(2000)}]
  {Madison00}
  \bibinfo{author}{\bibfnamefont{Madison}, \bibnamefont{K.~W.}},
  \bibinfo{author}{\bibfnamefont{F.}~\bibnamefont{Chevy}}, 
  \bibinfo{author}{\bibfnamefont{W.}~\bibnamefont{Wohlleben}},
  and
  \bibinfo{author}{\bibfnamefont{J.}~\bibnamefont{Dalibard}},
  \bibinfo{year}{2000},
  \bibinfo{journal}{Phys. Rev. Lett. } \textbf{\bibinfo{volume}{84}},
  \bibinfo{pages}{806}.

  \bibitem[{\citenamefont{Maki and Takayama}(1971)}]
  {Maki71}
  \bibinfo{author}{\bibfnamefont{Maki},\bibnamefont{K.}}, and
  \bibinfo{author}{\bibfnamefont{H.}~\bibnamefont{Takayama}},
  \bibinfo{year}{1971},
  \bibinfo{journal}{Prog. Theor. Phys.} \textbf{\bibinfo{volume}{46}},
  \bibinfo{pages}{1651}.


  \bibitem[{\citenamefont{Matl \emph{et~al.}}(2002)}]
  {Matl02}
  \bibinfo{author}{\bibfnamefont{Matl}, \bibnamefont{P.}},
  \bibinfo{author}{\bibfnamefont{N.~P.}~\bibnamefont{Ong}}, 
  \bibinfo{author}{\bibfnamefont{R.}~\bibnamefont{Gagnon}},and
  \bibinfo{author}{\bibfnamefont{L.}~\bibnamefont{Taillefer}},
  \bibinfo{year}{2002},
  \bibinfo{journal}{Phys. Rev. B } \textbf{\bibinfo{volume}{65}},
  \bibinfo{pages}{214514}.
  

  \bibitem[{\citenamefont{Mazenko}(2006)}]{Mazenko}
\bibinfo{author}{\bibnamefont{Mazenko}, \bibfnamefont{G.~F.}},
  \bibinfo{year}{2006},
\emph{\bibinfo{title}{Nonequillibrium Statistical
Mechanics}}
  (\bibinfo{publisher}{Wiley-VCH, Weinheim}).


\bibitem[{\citenamefont{McK. Paul \emph{et~al.}}(1998)}]
  {McKPaul98}
  \bibinfo{author}{\bibfnamefont{McK. Paul}, \bibnamefont{D.}},
  \bibinfo{author}{\bibfnamefont{C.~V.}~\bibnamefont{Tomy}},
  \bibinfo{author}{\bibfnamefont{C.~M.}~\bibnamefont{Aegerter}},
  \bibinfo{author}{\bibfnamefont{R.}~\bibnamefont{Cubitt}},
  \bibinfo{author}{\bibfnamefont{S.~H.}~\bibnamefont{Lloyd}},
  \bibinfo{author}{\bibfnamefont{E.~M.}~\bibnamefont{Forgan}},
  \bibinfo{author}{\bibfnamefont{S.~L.}~\bibnamefont{Lee}}, and
  \bibinfo{author}{\bibfnamefont{M.}~\bibnamefont{Yethiraj}},
  \bibinfo{year}{1998},
  \bibinfo{journal}{Phys. Rev. Lett.} \textbf{\bibinfo{volume}{80}},
  \bibinfo{pages}{1517}.


  \bibitem[{\citenamefont{Menon and Dasgupta}(1994)}]
  {Menon94}
  \bibinfo{author}{\bibfnamefont{Menon}, \bibnamefont{G.~I}},and
  \bibinfo{author}{\bibfnamefont{C.}~\bibnamefont{Dasgupta}},
  \bibinfo{year}{1994},
  \bibinfo{journal}{Phys. Rev. Lett.} \textbf{\bibinfo{volume}{73}},
  \bibinfo{pages}{1023}.

  

\bibitem[{\citenamefont{Menon \emph{et~al.}}(1999)}]
  {Menon99}
  \bibinfo{author}{\bibfnamefont{Menon}, \bibnamefont{G.~I.}},
  \bibinfo{author}{\bibfnamefont{C.}~\bibnamefont{Dasgupta}}, and
  \bibinfo{author}{\bibfnamefont{T.~V.}~\bibnamefont{Ramakrishnan}},
  \bibinfo{year}{1999},
  \bibinfo{journal}{Phys. Rev. B} \textbf{\bibinfo{volume}{60}},
  \bibinfo{pages}{7607}.


\bibitem[{\citenamefont{Menon}(2002)}]
  {Menon02}
  \bibinfo{author}{\bibfnamefont{Menon}, \bibnamefont{G.~I.}},
  \bibinfo{year}{2002},
  \bibinfo{journal}{Phys. Rev. B} \textbf{\bibinfo{volume}{65}},
  \bibinfo{pages}{104527}.

  \bibitem[{\citenamefont{Mermin and  Wagner}(1966)}]
  {Mermin66}
  \bibinfo{author}{\bibfnamefont{Mermin}, \bibnamefont{N.~D.}},  and
  \bibinfo{author}{\bibfnamefont{H.}~\bibnamefont{Wagner}},
  \bibinfo{year}{1966},
  \bibinfo{journal}{Phys. Rev. Lett.} \textbf{\bibinfo{volume}{17}},
  \bibinfo{pages}{1133}.

\bibitem[{\citenamefont{Mezard \emph{et~al.}}(1987)}]{Mezard87}
\bibinfo{author}{\bibfnamefont{Mezard}, \bibnamefont{M.}},
  \bibinfo{author}{\bibfnamefont{G.}~\bibnamefont{Parisi}}, and
  \bibinfo{author}{\bibfnamefont{M.A.}~\bibnamefont{Virasoro}},
  \bibinfo{year}{1987},
\emph{\bibinfo{title}{Spin Glass Theory and Beyond}}
  (\bibinfo{publisher}{World Scientific, Singapore, New Jersey, HongKong }).

  \bibitem[{\citenamefont{Mezard}(1991)}]
  {Mezard91}
  \bibinfo{author}{\bibfnamefont{Mezard}, \bibnamefont{M.}},  and
  \bibinfo{author}{\bibfnamefont{G.}~\bibnamefont{Parisi}},
  \bibinfo{year}{1991},
  \bibinfo{journal}{J. De Physique I} \textbf{\bibinfo{volume}{1}},
  \bibinfo{pages}{809}.

\bibitem[{\citenamefont{Mikitik and Brandt}(2001)}]
  {Mikitik01}
  \bibinfo{author}{\bibfnamefont{Mikitik}, \bibnamefont{G.~P.}}, and
  \bibinfo{author}{\bibfnamefont{E.~H.}~\bibnamefont{Brandt}},
  \bibinfo{year}{2001},
  \bibinfo{journal}{Phys. Rev. B} \textbf{\bibinfo{volume}{64}},
  \bibinfo{pages}{184514}.

\bibitem[{\citenamefont{Mikitik and Brandt}(2003)}]
  {Mikitik03}
  \bibinfo{author}{\bibfnamefont{Mikitik}, \bibnamefont{G.~P.}}, and
  \bibinfo{author}{\bibfnamefont{E.~H.}~\bibnamefont{Brandt}},
  \bibinfo{year}{2003},
  \bibinfo{journal}{Phys. Rev. B} \textbf{\bibinfo{volume}{68}},
  \bibinfo{pages}{05450}.


\bibitem[{\citenamefont{Monarkha and Kono}(2004)}]{Monarkha04}
\bibinfo{author}{\bibnamefont{Monarkha}, \bibfnamefont{Y.}}, and
  \bibinfo{author}{\bibfnamefont{K.} \bibnamefont{Kono}},
  \bibinfo{year}{2004},
\emph{\bibinfo{title}{Two-dimensional Coulomb liquids and solids}}
  (\bibinfo{publisher}{Springer , New York}).


  \bibitem[{\citenamefont{Moore}(1989)}]
  {Moore89}
  \bibinfo{author}{\bibfnamefont{Moore}, \bibnamefont{M.~A.}},
  \bibinfo{year}{1989},
  \bibinfo{journal}{Phys. Rev. B} \textbf{\bibinfo{volume}{39}},
  \bibinfo{pages}{136}.

\bibitem[{\citenamefont{Moore}(1992)}]
  {Moore92}
  \bibinfo{author}{\bibfnamefont{Moore}, \bibnamefont{M.~A.}},
  \bibinfo{year}{1992},
  \bibinfo{journal}{Phys. Rev. B} \textbf{\bibinfo{volume}{45}},
  \bibinfo{pages}{7336}.



  \bibitem[{\citenamefont{Moore}(1997)}]
  {Moore97}
  \bibinfo{author}{\bibfnamefont{M.~A.}, \bibnamefont{Moore}},
  \bibinfo{year}{1997},
  \bibinfo{journal}{Phys. Rev. B} \textbf{\bibinfo{volume}{55}},
  \bibinfo{pages}{14136}.

  \bibitem[{\citenamefont{Moore \emph{et~al.}}(1998)}]
  {Moore98}
  \bibinfo{author}{\bibnamefont{Moore}, \bibfnamefont{M.~A.}},
  \bibinfo{author}{\bibfnamefont{T.~J.}~\bibnamefont{Newman}},
  \bibinfo{author}{\bibfnamefont{A.~J.}~\bibnamefont{Bray}},and
  \bibinfo{author}{\bibfnamefont{S.-K.}~\bibnamefont{Chin}},
  \bibinfo{year}{1998},
  \bibinfo{journal}{Phys. Rev. B} \textbf{\bibinfo{volume}{58}},
  \bibinfo{pages}{936}.


\bibitem[{\citenamefont{Mukerjee and  Huse}(2004)}]
  {Mukerjee04}
  \bibinfo{author}{\bibfnamefont{Mukerjee}, \bibnamefont{S.}}, and
  \bibinfo{author}{\bibfnamefont{D.~A.}~\bibnamefont{Huse}},
  \bibinfo{year}{1996},
  \bibinfo{journal}{Phys. Rev. B} \textbf{\bibinfo{volume}{70}},
  \bibinfo{pages}{014506}.




\bibitem[{\citenamefont{Newman and  Moore}(1996)}]
  {Newman96}
  \bibinfo{author}{\bibfnamefont{Newman}, \bibnamefont{T.~J.}}, and
  \bibinfo{author}{\bibfnamefont{M.~A.}~\bibnamefont{Moore}},
  \bibinfo{year}{1996},
  \bibinfo{journal}{Phys. Rev. B} \textbf{\bibinfo{volume}{54}},
  \bibinfo{pages}{6661}.

  \bibitem[{\citenamefont{Nattermann}(1990)}]
  {Nattermann90}
  \bibinfo{author}{\bibfnamefont{Nattermann}, \bibnamefont{T.}},
  \bibinfo{year}{1990},
  \bibinfo{journal}{Phys. Rev. Lett.} \textbf{\bibinfo{volume}{64}},
  \bibinfo{pages}{2454}.

\bibitem[{\citenamefont{Nattermann and Scheidl}(2000)}]{Nattermann}
  \bibinfo{author}{\bibnamefont{Nattermann}, \bibfnamefont{T.}}, and
  \bibinfo{author}{\bibnamefont{Scheidl}, \bibfnamefont{S.}},
  \bibinfo{year}{2000},
  \bibinfo{journal}{Adv. Phys.} \textbf{\bibinfo{volume}{49}},
  \bibinfo{pages}{607}.




  \bibitem[{\citenamefont{Nguyen and Sudb{\o}}(1998)}]
  {Sudbo98}
  \bibinfo{author}{\bibfnamefont{Nguyen}, \bibnamefont{A.~K.}}, and
  \bibinfo{author}{\bibfnamefont{A.}~\bibnamefont{Sudb{\o}}},
  \bibinfo{year}{1998},
  \bibinfo{journal}{Phys. Rev. B} \textbf{\bibinfo{volume}{60}},
  \bibinfo{pages}{15307}.


\bibitem[{\citenamefont{Nikulov \emph{et~al.}}(1995a)}]
  {Nikulov95}
  \bibinfo{author}{\bibfnamefont{Nikulov}, \bibnamefont{A.~V.}},
  \bibinfo{author}{\bibfnamefont{D.~Yu.}~\bibnamefont{Remisov}}, and
  \bibinfo{author}{\bibfnamefont{V.~A.}~\bibnamefont{Oboznov}},
  \bibinfo{year}{1995a},
  \bibinfo{journal}{Phys. Rev. Lett.} \textbf{\bibinfo{volume}{75}},
  \bibinfo{pages}{2586}.

\bibitem[{\citenamefont{Nikulov}(1995b)}]
  {Nikulov95b}
  \bibinfo{author}{\bibfnamefont{Nikulov}, \bibnamefont{A.~V.}},
  \bibinfo{year}{1995b},
  \bibinfo{journal}{Phys. Rev. B} \textbf{\bibinfo{volume}{52}},
  \bibinfo{pages}{10429}.


\bibitem[{\citenamefont{ Nishizaki \emph{et~al.}}(2000)}]
  {Nishizaki00}
  \bibinfo{author}{\bibfnamefont{Nishizaki}, \bibnamefont{T.}},
  \bibinfo{author}{\bibfnamefont{K.}~\bibnamefont{Shibata}},
  \bibinfo{author}{\bibfnamefont{T.}~\bibnamefont{Sasaki}}, and
  \bibinfo{author}{\bibfnamefont{N.}~\bibnamefont{Kobayashi}},
  \bibinfo{year}{2000},
  \bibinfo{journal}{Physica C} \textbf{\bibinfo{volume}{341-348}},
  \bibinfo{pages}{957}.

  \bibitem[{\citenamefont{Nonomura and  Hu}(2001)}]
  {Hu01}
  \bibinfo{author}{\bibfnamefont{Nonomura}, \bibnamefont{Y.}}, and
  \bibinfo{author}{\bibfnamefont{X.}~\bibnamefont{Hu}},
  \bibinfo{year}{2001},
  \bibinfo{journal}{Phys. Rev. Lett.} \textbf{\bibinfo{volume}{86}},
  \bibinfo{pages}{5140}.

Vadim Oganesyan, David A. Huse, and S. L. Sondhi, Phys. Rev. B 73, 094503 (2006)
\bibitem[{\citenamefont{Oganesyan \emph{et~al.}}(2006)}]
  {Oganesyan06}
  \bibinfo{author}{\bibfnamefont{Oganesyan}, \bibnamefont{V.}},
  \bibinfo{author}{\bibfnamefont{D.~A.}~\bibnamefont{Huse}}, and
  \bibinfo{author}{\bibfnamefont{S.~L.}~\bibnamefont{Sondhi}},
  \bibinfo{year}{2006},
  \bibinfo{journal}{Phys. Rev. B} \textbf{\bibinfo{volume}{73}},
  \bibinfo{pages}{094503}.

  \bibitem[{\citenamefont{Okopinska}(1987)}]
  {Okopinska87}
  \bibinfo{author}{\bibfnamefont{Okopinska}, \bibnamefont{A.}},
  \bibinfo{year}{1987},
  \bibinfo{journal}{Phys. Rev. D} \textbf{\bibinfo{volume}{35}},
  \bibinfo{pages}{1835}.


  \bibitem[{\citenamefont{Olson \emph{et~al.}}(2001)}]
  {Olson01}
  \bibinfo{author}{\bibfnamefont{Olson}, \bibnamefont{C.~J.}},
  \bibinfo{author}{\bibfnamefont{C.}~\bibnamefont{Reichhardt}}, and
  \bibinfo{author}{\bibfnamefont{S.}~\bibnamefont{Bhattacharya}},
  \bibinfo{year}{2001},
  \bibinfo{journal}{Phys. Rev. B} \textbf{\bibinfo{volume}{64}},
  \bibinfo{pages}{024518}.

  \bibitem[{\citenamefont{Olsson and Teitel}(2001)}]
  {Olsson01}
  \bibinfo{author}{\bibfnamefont{Olsson}, \bibnamefont{P.}}, and
  \bibinfo{author}{\bibfnamefont{S.}~\bibnamefont{Teitel}},
  \bibinfo{year}{2001},
  \bibinfo{journal}{Phys. Rev. Lett.} \textbf{\bibinfo{volume}{87}},
  \bibinfo{pages}{137001}.

\bibitem[{\citenamefont{Olsson and Teitel}(2003)}]
  {Olsson03}
  \bibinfo{author}{\bibfnamefont{Olsson}, \bibnamefont{P.}}, and
  \bibinfo{author}{\bibfnamefont{S.}~\bibnamefont{Teitel}},
  \bibinfo{year}{2003},
  \bibinfo{journal}{Phys. Rev. B} \textbf{\bibinfo{volume}{67}},
  \bibinfo{pages}{144514}.


\bibitem[{\citenamefont{Olsson}(2007)}]
  {Olsson07}
  \bibinfo{author}{\bibfnamefont{Olsson}, \bibnamefont{P.}},
  \bibinfo{year}{2007},
  \bibinfo{journal}{Phys. Rev. Lett.} \textbf{\bibinfo{volume}{98}},
  \bibinfo{pages}{097001}.

   \bibitem[{\citenamefont{O'Neill and Moore}(1993)}]
  {ONeill93}
  \bibinfo{author}{\bibfnamefont{O'Neill}, \bibnamefont{J.~A.}}, and
  \bibinfo{author}{\bibfnamefont{M.~A.}~\bibnamefont{Moore}},
  \bibinfo{year}{1993},
  \bibinfo{journal}{Phys. Rev. B} \textbf{\bibinfo{volume}{48}},
  \bibinfo{pages}{374}.


\bibitem[{\citenamefont{Pal \emph{et~al.}}(2001)}]
  {Pal01}
  \bibinfo{author}{\bibfnamefont{Pal}, \bibnamefont{D.}},
  \bibinfo{author}{\bibfnamefont{S.}~\bibnamefont{Ramakrishnan}},
  \bibinfo{author}{\bibfnamefont{A.~K.}~\bibnamefont{Grover}},
   \bibinfo{author}{\bibfnamefont{D.}~\bibnamefont{Dasgupta}}, and
   \bibinfo{author}{\bibfnamefont{B.~K.}~\bibnamefont{Sarma}},
  \bibinfo{year}{2001},
  \bibinfo{journal}{Phys. Rev. B} \textbf{\bibinfo{volume}{63}},
  \bibinfo{pages}{132505}.


\bibitem[{\citenamefont{Pal \emph{et~al.}}(2002)}]
  {Pal02}
  \bibinfo{author}{\bibfnamefont{Pal}, \bibnamefont{D.}},
  \bibinfo{author}{\bibfnamefont{S.}~\bibnamefont{Ramakrishnan}}, and
  \bibinfo{author}{\bibfnamefont{A.~K.}~\bibnamefont{Grover}},
  \bibinfo{year}{2002},
  \bibinfo{journal}{Phys. Rev. B} \textbf{\bibinfo{volume}{65}},
  \bibinfo{pages}{096502}.

  \bibitem[{\citenamefont{Paltiel \emph{et~al.}}(2000a)}]
  {Paltiel00}
  \bibinfo{author}{\bibfnamefont{Paltiel}, \bibnamefont{Y.}},
  \bibinfo{author}{\bibfnamefont{E.}~\bibnamefont{Zeldov}},
  \bibinfo{author}{\bibfnamefont{Y.}~\bibnamefont{Myasoedov}},
  \bibinfo{author}{\bibfnamefont{H.}~\bibnamefont{Shtrikman}},
  \bibinfo{author}{\bibfnamefont{S.}~\bibnamefont{Bhattacharya}},
  \bibinfo{author}{\bibfnamefont{M.~J.}~\bibnamefont{Higgins}},
  \bibinfo{author}{\bibfnamefont{Z.~L.}~\bibnamefont{Xiao}},
  \bibinfo{author}{\bibfnamefont{E.~Y.}~\bibnamefont{Andrei}},
  \bibinfo{author}{\bibfnamefont{P.~L.}~\bibnamefont{Gammel}}, and
  \bibinfo{author}{\bibfnamefont{D.~J.}~\bibnamefont{Bishop}},
  \bibinfo{year}{2000a},
  \bibinfo{journal}{Nature} \textbf{\bibinfo{volume}{403}},
  \bibinfo{pages}{398}.

\bibitem[{\citenamefont{Paltiel \emph{et~al.}}(2000b)}]
  {Paltiel00b}
  \bibinfo{author}{\bibfnamefont{Paltiel}, \bibnamefont{Y.}},
  \bibinfo{author}{\bibfnamefont{E.}~\bibnamefont{Zeldov}},
  \bibinfo{author}{\bibfnamefont{Y.}~\bibnamefont{Myasoedov}},
  \bibinfo{author}{\bibfnamefont{M.~L.}~\bibnamefont{Rappaport}},
  \bibinfo{author}{\bibfnamefont{G.}~\bibnamefont{Jung}},
  \bibinfo{author}{\bibfnamefont{S.}~\bibnamefont{Bhattacharya}},
  \bibinfo{author}{\bibfnamefont{M.~J.}~\bibnamefont{Higgins}},
  \bibinfo{author}{\bibfnamefont{Z.~L.}~\bibnamefont{Xiao}},
  \bibinfo{author}{\bibfnamefont{E.~Y.}~\bibnamefont{Andrei}},
  \bibinfo{author}{\bibfnamefont{P.~L.}~\bibnamefont{Gammel}}, and
  \bibinfo{author}{\bibfnamefont{D.~J.}~\bibnamefont{Bishop}},
  \bibinfo{year}{2000b},
  \bibinfo{journal}{Phys. Rev. Lett.} \textbf{\bibinfo{volume}{85}},
  \bibinfo{pages}{3712}.





  \bibitem[{\citenamefont{Parisi}(1980)}]
  {Parisi80}
  \bibinfo{author}{\bibfnamefont{Parisi}, \bibnamefont{G.}},
  \bibinfo{year}{1980},
  \bibinfo{journal}{J. Phy. A} \textbf{\bibinfo{volume}{13}},
  \bibinfo{pages}{1101}.

  \bibitem[{\citenamefont{Park and  Huse}(1998)}]
  {Park98}
  \bibinfo{author}{\bibfnamefont{Park}, \bibnamefont{K.}}, and
  \bibinfo{author}{\bibfnamefont{D.~A.}~\bibnamefont{Huse}},
  \bibinfo{year}{1998},
  \bibinfo{journal}{Phys. Rev. B} \textbf{\bibinfo{volume}{58}},
  \bibinfo{pages}{9427}.

 \bibitem[{\citenamefont{Park and Yeo}(2008)}]
  {Park08}
   \bibinfo{author}{\bibfnamefont{Park}, \bibnamefont{H.}},  and 
   \bibinfo{author}{\bibfnamefont{J.}~\bibnamefont{Yeo}},
  \bibinfo{year}{2008},
  \bibinfo{journal}{J. Korean Phys. Soc.} \textbf{\bibinfo{volume}{52}},
  \bibinfo{pages}{1093}. 

\bibitem[{\citenamefont{Pastoriza \emph{et~al.}}(1994)}]
  {Pastoriza}
  \bibinfo{author}{\bibfnamefont{Pastoriza}, \bibnamefont{H.}},
  \bibinfo{author}{\bibfnamefont{M.~F.}~\bibnamefont{Goffman}},
  \bibinfo{author}{\bibfnamefont{A.}~\bibnamefont{Arrib\^{e}re}}, and
  \bibinfo{author}{\bibfnamefont{F.}~\bibnamefont{de la Cruz}},
  \bibinfo{year}{1994},
  \bibinfo{journal}{Phys. Rev. Lett.} \textbf{\bibinfo{volume}{72}},
  \bibinfo{pages}{2951}.


\bibitem[{\citenamefont{Pethick and Smith}(2008)}]{Pethick08}
\bibinfo{author}{\bibnamefont{Pethick}, \bibfnamefont{C.~J.}}, and
  \bibinfo{author}{\bibfnamefont{H.} \bibnamefont{Smith}},
  \bibinfo{year}{2008},
\emph{\bibinfo{title}{Bose-Einstein Condensation in Dilute Gases}}
  (\bibinfo{publisher}{Cambridge University Press}).

  \bibitem[{\citenamefont{Pierson \emph{et~al.}}(1995)}]
  {Pierson95}
  \bibinfo{author}{\bibfnamefont{Pierson},\bibnamefont{S.~W.}},
  \bibinfo{author}{\bibfnamefont{J.}~\bibnamefont{Buan}},
  \bibinfo{author}{\bibnamefont{B.}, \bibfnamefont{Zhou}},
  \bibinfo{author}{\bibfnamefont{C.~C.}~\bibnamefont{Huang}}, and
  \bibinfo{author}{\bibfnamefont{O.~T.}~\bibnamefont{Valls}},
  \bibinfo{year}{1995},
  \bibinfo{journal}{Phys. Rev. Lett.} \textbf{\bibinfo{volume}{74}},
  \bibinfo{pages}{1887}.

  \bibitem[{\citenamefont{Pierson \emph{et~al.}}(1996)}]
  {Pierson96}
  \bibinfo{author}{\bibfnamefont{Pierson},\bibnamefont{S.~W.}},
  \bibinfo{author}{\bibfnamefont{T.~M}~\bibnamefont{Katona}},
  \bibinfo{author}{\bibfnamefont{Z.}~\bibnamefont{Te\v{s}anovi\'{c}}}, and
  \bibinfo{author}{\bibfnamefont{O.~T.}~\bibnamefont{Valls}},
  \bibinfo{year}{1996},
  \bibinfo{journal}{Phys. Rev. B} \textbf{\bibinfo{volume}{53}},
  \bibinfo{pages}{8638}.

  \bibitem[{\citenamefont{Pierson and Walls}(1998a)}]
  {Pierson98}
  \bibinfo{author}{\bibnamefont{Pierson}, \bibfnamefont{S.~W.}}, and
  \bibinfo{author}{\bibfnamefont{O.~T.}~\bibnamefont{Walls}},
  \bibinfo{year}{1998a},
  \bibinfo{journal}{Phys. Rev. B} \textbf{\bibinfo{volume}{57}},
  \bibinfo{pages}{R8143}.

\bibitem[{\citenamefont{Pierson \emph{et~al.}}(1998b)}]
  {Pierson98b}
  \bibinfo{author}{\bibfnamefont{Pierson},\bibnamefont{S.~W.}},
  \bibinfo{author}{\bibfnamefont{O.~T.}~\bibnamefont{Valls}},
  \bibinfo{author}{\bibfnamefont{Z.}~\bibnamefont{Te\v{s}anovi\'{c}}},and
  \bibinfo{author}{\bibfnamefont{M.~A.}~\bibnamefont{Lindemann}},
  \bibinfo{year}{1998b},
  \bibinfo{journal}{Phys. Rev. B} \textbf{\bibinfo{volume}{57}},
  \bibinfo{pages}{8622}.


\bibitem[{\citenamefont{Pourret \emph{et~al.}}(2006)}]
  {Pourret06}
  \bibinfo{author}{\bibfnamefont{Pourret},\bibnamefont{A.}},
  \bibinfo{author}{\bibfnamefont{H.}~\bibnamefont{Aubin}},
  \bibinfo{author}{\bibfnamefont{J.}~\bibnamefont{Lesueur}},
  \bibinfo{author}{\bibfnamefont{C.~A.}~\bibnamefont{Marrache-Kikuchi}},
  \bibinfo{author}{\bibfnamefont{L.}~\bibnamefont{Berge}},
  \bibinfo{author}{\bibfnamefont{L.}~\bibnamefont{Dumoulin}},
  and
  \bibinfo{author}{\bibfnamefont{K.}~\bibnamefont{Behnia}},
  \bibinfo{year}{2006},
  \bibinfo{journal}{Nat. Phys.} \textbf{\bibinfo{volume}{2}},
  \bibinfo{pages}{683}.


  \bibitem[{\citenamefont{Prange}(1969)}]
  {Prange69}
  \bibinfo{author}{\bibnamefont{Prange}, \bibfnamefont{R.~E.}},
  \bibinfo{year}{1969},
  \bibinfo{journal}{Phys. Rev. B} \textbf{\bibinfo{volume}{1}},
  \bibinfo{pages}{2349}.

\bibitem[{\citenamefont{Radzyner \emph{et~al.}}(2002)}]
  {Radzyner02}
  \bibinfo{author}{\bibfnamefont{Radzyner}, \bibnamefont{Y.}},
  \bibinfo{author}{\bibfnamefont{A.}~\bibnamefont{Shaulov}}, and
  \bibinfo{author}{\bibfnamefont{Y.}~\bibnamefont{Yeshurun}},
  \bibinfo{year}{2002},
  \bibinfo{journal}{Phys. Rev. B} \textbf{\bibinfo{volume}{65}},
  \bibinfo{pages}{100513}.


\bibitem[{\citenamefont{Rajaraman}(1982)}]{Rajaraman}
\bibinfo{author}{\bibnamefont{Rajaraman}, \bibfnamefont{R.}},
  \bibinfo{year}{1982},
\emph{\bibinfo{title}{Solitons and Instantons}}
  (\bibinfo{publisher}{North-Holland Pubishing Company, Amsterdam, New York }).



\bibitem[{\citenamefont{Reichhardt \emph{et~al.}}(1996)}]
  {Reichhardt96}
  \bibinfo{author}{\bibfnamefont{Reichhardt}, \bibnamefont{C.}},
  \bibinfo{author}{\bibfnamefont{C.~J.}~\bibnamefont{Olson}},
  \bibinfo{author}{\bibfnamefont{J.}~\bibnamefont{Groth}},
  \bibinfo{author}{\bibfnamefont{S.}~\bibnamefont{Field}}, and
  \bibinfo{author}{\bibfnamefont{F.}~\bibnamefont{Nori}},
  \bibinfo{year}{1996},
  \bibinfo{journal}{Phys. Rev. B} \textbf{\bibinfo{volume}{53}},
  \bibinfo{pages}{8898}.

  \bibitem[{\citenamefont{Reichhardt \emph{et~al.}}(2000)}]
  {Reichhardt00}
  \bibinfo{author}{\bibfnamefont{Reichhardt}, \bibnamefont{C.}},
  \bibinfo{author}{\bibfnamefont{A.}~\bibnamefont{van Otterlo}}, and
  \bibinfo{author}{\bibfnamefont{G.~T.}~\bibnamefont{Zimanyi}},
  \bibinfo{year}{2000},
  \bibinfo{journal}{Phys. Rev. Lett. } \textbf{\bibinfo{volume}{84}},
  \bibinfo{pages}{1994}.

  \bibitem[{\citenamefont{Revaz \emph{et~al.}}(1998)}]
  {Revaz98}
  \bibinfo{author}{\bibfnamefont{Revaz}, \bibnamefont{B.}},
  \bibinfo{author}{\bibfnamefont{A.}~\bibnamefont{Junod}}, and
  \bibinfo{author}{\bibfnamefont{A.}~\bibnamefont{Erb}},
  \bibinfo{year}{1998},
  \bibinfo{journal}{Phys. Rev. B} \textbf{\bibinfo{volume}{58}},
  \bibinfo{pages}{11153}.


  \bibitem[{\citenamefont{Rosenstein}(1999)}]
  {Rosenstein99}
  \bibinfo{author}{\bibfnamefont{Rosenstein}, \bibnamefont{B.}},
  \bibinfo{year}{1999},
  \bibinfo{journal}{Phys. Rev. B} \textbf{\bibinfo{volume}{60}},
  \bibinfo{pages}{4268}.

  \bibitem[{\citenamefont{Rosenstein and Knigavko}(1999)}]
  {Rosenstein99b}
  \bibinfo{author}{\bibfnamefont{Rosenstein}, \bibnamefont{B.}}, and
  \bibinfo{author}{\bibfnamefont{A.}~\bibnamefont{Knigavko}},
  \bibinfo{year}{1999},
  \bibinfo{journal}{Phys. Rev. Lett.} \textbf{\bibinfo{volume}{83}},
  \bibinfo{pages}{844}.

\bibitem[{\citenamefont{Rosenstein and Zhuravlev}(2007)}]
  {Zhuravlev07}
  \bibinfo{author}{\bibfnamefont{Rosenstein}, \bibnamefont{B.}},  and
  \bibinfo{author}{\bibfnamefont{V.}~\bibnamefont{Zhuravlev}},
  \bibinfo{year}{2007},
  \bibinfo{journal}{Phys. Rev. B} \textbf{\bibinfo{volume}{76}},
  \bibinfo{pages}{014507}.

   \bibitem[{\citenamefont{Rosenstein \emph{et~al.}}(2001)}]
  {Rosenstein01}
  \bibinfo{author}{\bibfnamefont{Rosenstein}, \bibnamefont{B.}},
  \bibinfo{author}{\bibfnamefont{B.~Ya.}~\bibnamefont{Shapiro}},
  \bibinfo{author}{\bibfnamefont{R.}~\bibnamefont{Prozorov}},
  \bibinfo{author}{\bibfnamefont{A.}~\bibnamefont{Shaulov}}, and
  \bibinfo{author}{\bibfnamefont{Y.}~\bibnamefont{Yeshurun}},
  \bibinfo{year}{2001},
  \bibinfo{journal}{Phys. Rev. B} \textbf{\bibinfo{volume}{63}},
  \bibinfo{pages}{134501}.

  \bibitem[{\citenamefont{Roulin \emph{et~al.}}(1996a)}]
  {Roulin96}
  \bibinfo{author}{\bibfnamefont{Roulin}, \bibnamefont{M.}},
  \bibinfo{author}{\bibfnamefont{A.}~\bibnamefont{Junod}}, and
  \bibinfo{author}{\bibfnamefont{E.}~\bibnamefont{Walker}},
  \bibinfo{year}{1996a},
  \bibinfo{journal}{Science} \textbf{\bibinfo{volume}{273}},
  \bibinfo{pages}{1210}.

\bibitem[{\citenamefont{Roulin \emph{et~al.}}(1996b)}]
  {Roulin96b}
  \bibinfo{author}{\bibfnamefont{Roulin}, \bibnamefont{M.}},
  \bibinfo{author}{\bibfnamefont{A.}~\bibnamefont{Junod}},
  \bibinfo{author}{\bibfnamefont{A.}~\bibnamefont{Erb}}, and
  \bibinfo{author}{\bibfnamefont{E.}~\bibnamefont{Walker}},
  \bibinfo{year}{1996b},
  \bibinfo{journal}{J. Low Temp. Phys.} \textbf{\bibinfo{volume}{105}},
  \bibinfo{pages}{1099}.


  \bibitem[{\citenamefont{Ruggeri and Thouless}(1976)}]
{Ruggeri76}
\bibinfo{author}{\bibnamefont{Ruggeri}, \bibfnamefont{G.~J.}}, and
\bibinfo{author}{\bibfnamefont{D.~J.} \bibnamefont{Thouless}},
  \bibinfo{year}{1976},
  \bibinfo{journal}{J. Phys. F: Met. Phys.} \textbf{\bibinfo{volume}{6}},
  \bibinfo{pages}{2063}.

\bibitem[{\citenamefont{Ruggeri}(1978)}]
{Ruggeri78}
\bibinfo{author}{\bibnamefont{Ruggeri}, \bibfnamefont{G.~J.}},
  \bibinfo{year}{1978},
  \bibinfo{journal}{Phys. Rev. B} \textbf{\bibinfo{volume}{20}},
  \bibinfo{pages}{3626}.

  \bibitem[{\citenamefont{Ryu \emph{et~al.}}(1996)}]
  {Ryu96}
  \bibinfo{author}{\bibfnamefont{Ryu}, \bibnamefont{S.}},
  \bibinfo{author}{\bibfnamefont{A.}~\bibnamefont{Kapitulnik}}, and
  \bibinfo{author}{\bibfnamefont{S.}~\bibnamefont{Doniach}},
  \bibinfo{year}{1996},
  \bibinfo{journal}{Phys. Rev. Lett.} \textbf{\bibinfo{volume}{77}},
  \bibinfo{pages}{2300}.



\bibitem[{\citenamefont{Saint-James} \emph{et~al.}(1969)}]{Saint}
\bibinfo{author}{\bibnamefont{Saint-James}, \bibfnamefont{D.}},
\bibinfo{author}{\bibfnamefont{G.} \bibnamefont{Sarma}},and
\bibinfo{author}{\bibfnamefont{E.} \bibnamefont{Thomas}},
\bibinfo{year}{1969},
\emph{\bibinfo{title}{Type II Superconductivity}}
  (\bibinfo{publisher}{Pergamon Press, Oxford }).

   \bibitem[{\citenamefont{Salem-Sugui and Dasilva}(1994)}]
  {SalemSugui94}
  \bibinfo{author}{\bibfnamefont{Salem-Sugui}, \bibnamefont{S.}}, and
  \bibinfo{author}{\bibfnamefont{E.~Z.}~\bibnamefont{Dasilva}},
  \bibinfo{year}{1994},
  \bibinfo{journal}{Physica C} \textbf{\bibinfo{volume}{235}},
  \bibinfo{pages}{1919}.


\bibitem[{\citenamefont{Salem-Sugui \emph{et~al.}}(2002)}]
  {Salem02}
  \bibinfo{author}{\bibfnamefont{Salem-Sugui}, \bibnamefont{S.}},
  \bibinfo{author}{\bibfnamefont{M.}~\bibnamefont{Friesen}},
  \bibinfo{author}{\bibfnamefont{A.~D.}~\bibnamefont{Alvarenga}},
  \bibinfo{author}{\bibfnamefont{F.~G.}~\bibnamefont{Gandra}},
  \bibinfo{author}{\bibfnamefont{M.~M.}~\bibnamefont{Doria}}, and
  \bibinfo{author}{\bibfnamefont{O.~F.}~\bibnamefont{Schilling}},
  \bibinfo{year}{2002},
  \bibinfo{journal}{Phys. Rev. B} \textbf{\bibinfo{volume}{66}},
  \bibinfo{pages}{134521}.


 \bibitem[{\citenamefont{Sasagawa \emph{et~al.}}(2000)}]
  {Sasagawa00}
  \bibinfo{author}{\bibfnamefont{Sasagawa}, \bibnamefont{T.}},
  \bibinfo{author}{\bibfnamefont{Y.}~\bibnamefont{Togawa}},
  \bibinfo{author}{\bibfnamefont{J.}~\bibnamefont{Shimoyama}},
  \bibinfo{author}{\bibfnamefont{A.}~\bibnamefont{Kapitulnik}},
  \bibinfo{author}{\bibfnamefont{K.}~\bibnamefont{Kitazawa}}, and
  \bibinfo{author}{\bibfnamefont{K.}~\bibnamefont{Kishio}},
  \bibinfo{year}{2000},
  \bibinfo{journal}{Phys. Rev. B} \textbf{\bibinfo{volume}{61}},
  \bibinfo{pages}{1610}.

  \bibitem[{\citenamefont{Sasik and Stroud}(1995)}]
  {Sasik95}
  \bibinfo{author}{\bibfnamefont{Sasik}, \bibnamefont{R.}}, and
  \bibinfo{author}{\bibfnamefont{D.}~\bibnamefont{Stroud}},
  \bibinfo{year}{1995},
  \bibinfo{journal}{Phys. Rev. Lett.} \textbf{\bibinfo{volume}{75}},
  \bibinfo{pages}{2582}.


\bibitem[{\citenamefont{Schilling \emph{et~al.}}(1996)}]
  {Schilling96}
  \bibinfo{author}{\bibfnamefont{Schilling}, \bibnamefont{A.}},
  \bibinfo{author}{\bibfnamefont{R.~A.}~\bibnamefont{Fisher}},
  \bibinfo{author}{\bibfnamefont{N.~E.}~\bibnamefont{Phillips}},
  \bibinfo{author}{\bibfnamefont{U.}~\bibnamefont{Welp}},
  \bibinfo{author}{\bibfnamefont{D.}~\bibnamefont{Dasgupta}},
  \bibinfo{author}{\bibfnamefont{W.~K.}~\bibnamefont{Kwok}}, and
  \bibinfo{author}{\bibfnamefont{G.~W.}~\bibnamefont{Crabtree}},
  \bibinfo{year}{1996},
  \bibinfo{journal}{Nature (London)} \textbf{\bibinfo{volume}{382}},
  \bibinfo{pages}{791}.

  \bibitem[{\citenamefont{Schilling \emph{et~al.}}(1997)}]
  {Schilling97}
  \bibinfo{author}{\bibfnamefont{Schilling}, \bibnamefont{A.}},
  \bibinfo{author}{\bibfnamefont{R.~A.}~\bibnamefont{Fisher}},
  \bibinfo{author}{\bibfnamefont{N.~E.}~\bibnamefont{Phillips}},
  \bibinfo{author}{\bibfnamefont{U.}~\bibnamefont{Welp}},
  \bibinfo{author}{\bibfnamefont{W.~K.}~\bibnamefont{Kwok}}, and
  \bibinfo{author}{\bibfnamefont{G.~W.}~\bibnamefont{Crabtree}},
  \bibinfo{year}{1997},
  \bibinfo{journal}{Phys. Rev. Lett.} \textbf{\bibinfo{volume}{78}},
  \bibinfo{pages}{4833}.

  \bibitem[{\citenamefont{Schilling \emph{et~al.}}(2002)}]
  {Schilling02}
  \bibinfo{author}{\bibfnamefont{Schilling}, \bibnamefont{A.}},
  \bibinfo{author}{\bibfnamefont{U.}~\bibnamefont{Welp}},
  \bibinfo{author}{\bibfnamefont{W.~K.}~\bibnamefont{Kwok}},  and
  \bibinfo{author}{\bibfnamefont{G.~W.}~\bibnamefont{Crabtree}},
  \bibinfo{year}{2002},
  \bibinfo{journal}{Phys. Rev. B} \textbf{\bibinfo{volume}{65}},
  \bibinfo{pages}{054505}.

\bibitem[{\citenamefont{ Senatore \emph{et~al.}}(2008)}]
  {Senatore08}
  \bibinfo{author}{\bibfnamefont{Senatore}, \bibnamefont{C.}},
  \bibinfo{author}{\bibfnamefont{R.}~\bibnamefont{Fl¨¹kiger}},
  \bibinfo{author}{\bibfnamefont{M.}~\bibnamefont{Cantoni}}, 
  \bibinfo{author}{\bibfnamefont{G.}~\bibnamefont{Wu}},
  \bibinfo{author}{\bibfnamefont{R.~H.}~\bibnamefont{Liu}},
  and
  \bibinfo{author}{\bibfnamefont{X.~H.}~\bibnamefont{Chen}},
  \bibinfo{year}{2008},
  \bibinfo{journal}{Phys. Rev. B} \textbf{\bibinfo{volume}{78}},
  \bibinfo{pages}{054514}.


\bibitem[{\citenamefont{ Shibata \emph{et~al.}}(2002)}]
  {Nishizaki02}
  \bibinfo{author}{\bibfnamefont{Shibata}, \bibnamefont{K.}},
  \bibinfo{author}{\bibfnamefont{T.}~\bibnamefont{Nishizaki}},
  \bibinfo{author}{\bibfnamefont{T.}~\bibnamefont{Sasaki}}, and
  \bibinfo{author}{\bibfnamefont{N.}~\bibnamefont{Kobayashi}},
  \bibinfo{year}{2002},
  \bibinfo{journal}{Phys. Rev. B} \textbf{\bibinfo{volume}{66}},
  \bibinfo{pages}{214518}.


  \bibitem[{\citenamefont{Shibauchi \emph{et~al.}}(1998)}]
  {Shibauchi98}
  \bibinfo{author}{\bibfnamefont{Shibauchi}, \bibnamefont{T.}},
  \bibinfo{author}{\bibfnamefont{M.}~\bibnamefont{Sato}},
  \bibinfo{author}{\bibfnamefont{S.}~\bibnamefont{Ooi}}, and
  \bibinfo{author}{\bibfnamefont{T.}~\bibnamefont{Tamegai}},
  \bibinfo{year}{1998},
  \bibinfo{journal}{Phys. Rev. B} \textbf{\bibinfo{volume}{57}},
  \bibinfo{pages}{5622}.


\bibitem[{\citenamefont{Sinova \emph{et~al.}}(2001)}]
  {Sinova01}
  \bibinfo{author}{\bibfnamefont{Sinova}, \bibnamefont{J.}},
  \bibinfo{author}{\bibfnamefont{C.~B.}~\bibnamefont{Hanna}}, and
  \bibinfo{author}{\bibfnamefont{A.~H.}~\bibnamefont{MacDonald}},
  \bibinfo{year}{2001},
  \bibinfo{journal}{Phys. Rev. Lett.} \textbf{\bibinfo{volume}{89}},
  \bibinfo{pages}{030403}.

\bibitem[{\citenamefont{Sok \emph{et~al.}}(1995)}]
  {Sok95}
  \bibinfo{author}{\bibfnamefont{Sok}, \bibnamefont{J.}},
  \bibinfo{author}{\bibfnamefont{M.}~\bibnamefont{Xu}},
  \bibinfo{author}{\bibfnamefont{W.}~\bibnamefont{Chen}},
  \bibinfo{author}{\bibfnamefont{B.~J.}~\bibnamefont{Suh}},
  \bibinfo{author}{\bibfnamefont{J.}~\bibnamefont{Gohng}},
  \bibinfo{author}{\bibfnamefont{D.~K.}~\bibnamefont{Finnemore}},
  \bibinfo{author}{\bibfnamefont{M.~J.}~\bibnamefont{Kramer}},
  \bibinfo{author}{\bibfnamefont{L.~A.}~\bibnamefont{Schwartzkopf}},and
  \bibinfo{author}{\bibfnamefont{B.}~\bibnamefont{Dabrowski}},
  \bibinfo{year}{1995},
  \bibinfo{journal}{Phys. Rev. B} \textbf{\bibinfo{volume}{51}},
  \bibinfo{pages}{6035}.


\bibitem[{\citenamefont{Sonin}(2005)}]
 {Sonin05}
 \bibinfo{author}{\bibnamefont{Sonin}, \bibfnamefont{E.~B.}},
  \bibinfo{year}{2005},
 \bibinfo{journal}{Phys. Rev. A} \textbf{\bibinfo{volume}{72}},
  \bibinfo{pages}{021606}.

\bibitem[{\citenamefont{Stevens and Robbins}(1993)}]
  {Stevens93}
  \bibinfo{author}{\bibfnamefont{Stevens}, \bibnamefont{M.}}, and
  \bibinfo{author}{\bibfnamefont{M.}~\bibnamefont{Robbins}},
  \bibinfo{year}{1993},
  \bibinfo{journal}{J. Chem. Phys.} \textbf{\bibinfo{volume}{98}},
  \bibinfo{pages}{2319}.

  \bibitem[{\citenamefont{Stevenson}(1981)}]
  {Stevenson81}
  \bibinfo{author}{\bibfnamefont{Stevenson}, \bibnamefont{P.~W.}},
  \bibinfo{year}{1981},
  \bibinfo{journal}{Phys. Rev. D} \textbf{\bibinfo{volume}{23}},
  \bibinfo{pages}{2916}.

  \bibitem[{\citenamefont{Sudb{\o} and Nguyen  }(1999)}]
  {Sudbo99}
  \bibinfo{author}{\bibfnamefont{Sudb{\o}}, \bibnamefont{A.}}, and
  \bibinfo{author}{\bibfnamefont{A.~K.}~\bibnamefont{Nguyen}},
  \bibinfo{year}{1999},
  \bibinfo{journal}{Phys. Rev. B} \textbf{\bibinfo{volume}{60}},
  \bibinfo{pages}{15307}.






\bibitem[{\citenamefont{Taylor \emph{et~al.}}(2003)}]
  {Taylor03}
  \bibinfo{author}{\bibfnamefont{Taylor}, \bibnamefont{B.~J.}},
  \bibinfo{author}{\bibfnamefont{S.}~\bibnamefont{Li}},
  \bibinfo{author}{\bibfnamefont{M.~B.}~\bibnamefont{Maple}},  and
  \bibinfo{author}{\bibfnamefont{M.~P.}~\bibnamefont{Maley}},
  \bibinfo{year}{2003},
  \bibinfo{journal}{Phys. Rev. B} \textbf{\bibinfo{volume}{68}},
  \bibinfo{pages}{054523}.
  
  \bibitem[{\citenamefont{Taylor and Maple}(2007)}]
  {Taylor07}
  \bibinfo{author}{\bibfnamefont{Taylor}, \bibnamefont{B.~J.}},
    and
  \bibinfo{author}{\bibfnamefont{M.~B.}~\bibnamefont{Maple}},
  \bibinfo{year}{2007},
  \bibinfo{journal}{Phys. Rev. B} \textbf{\bibinfo{volume}{76}},
  \bibinfo{pages}{014517}.
  
  

   \bibitem[{\citenamefont{Te\v{s}anovi\'{c} and  Xing}(1991)}]
  {Tesanovic-Xing91}
  \bibinfo{author}{\bibfnamefont{Te\v{s}anovi\'{c}}, \bibnamefont{Z.}}, and
  \bibinfo{author}{\bibfnamefont{L.}~\bibnamefont{Xing}},
  \bibinfo{year}{1991},
  \bibinfo{journal}{Phys. Rev. Lett.} \textbf{\bibinfo{volume}{67}},
  \bibinfo{pages}{2729}.

  \bibitem[{\citenamefont{Te\v{s}anovi\'{c} \emph{et~al.}}(1992)}]
  {Tesanovic92}
  \bibinfo{author}{\bibfnamefont{Te\v{s}anovi\'{c}},\bibnamefont{Z.}},
  \bibinfo{author}{\bibfnamefont{L.}~\bibnamefont{Xing}},
  \bibinfo{author}{\bibfnamefont{L.}~\bibnamefont{Bulaevskii}},
  \bibinfo{author}{\bibfnamefont{Q.}~\bibnamefont{Li}},and
  \bibinfo{author}{\bibfnamefont{M.}~\bibnamefont{Suenaga}},
  \bibinfo{year}{1992},
  \bibinfo{journal}{Phys. Rev. Lett.} \textbf{\bibinfo{volume}{69}},
  \bibinfo{pages}{3563}.



 \bibitem[{\citenamefont{Te\v{s}anovi\'{c} and Andreev}(1994)}]
  {Tesanovic94}
  \bibinfo{author}{\bibfnamefont{Te\v{s}anovi\'{c}},\bibnamefont{Z.}}, and
  \bibinfo{author}{\bibfnamefont{A.~V.}~\bibnamefont{Andreev}},
  \bibinfo{year}{1994},
  \bibinfo{journal}{Phys. Rev. B} \textbf{\bibinfo{volume}{49}},
  \bibinfo{pages}{4064}.

  \bibitem[{\citenamefont{Te\v{s}anovi\'{c} and Herbut}(1994)}]
  {Tesanovic94b}
  \bibinfo{author}{\bibfnamefont{Te\v{s}anovi\'{c}},\bibnamefont{Z.}}, and
  \bibinfo{author}{\bibfnamefont{I.~F.}~\bibnamefont{Herbut}},
  \bibinfo{year}{1994},
  \bibinfo{journal}{Phys. Rev. B} \textbf{\bibinfo{volume}{50}},
  \bibinfo{pages}{10389}.



\bibitem[{\citenamefont{Te\v{s}anovi\'{c}}(1999)}]
 {Tesanovic99}
 \bibinfo{author}{\bibnamefont{Te\v{s}anovi\'{c}}, \bibfnamefont{Z.}},
  \bibinfo{year}{1999},
 \bibinfo{journal}{Phys. Rev. B} \textbf{\bibinfo{volume}{59}},
  \bibinfo{pages}{6449}.



  \bibitem[{\citenamefont{ Thakur \emph{et~al.}}(2005)}]
  {Thakur}
  \bibinfo{author}{\bibfnamefont{Thakur}, \bibnamefont{A.~D.}},
  \bibinfo{author}{\bibfnamefont{S.~S.}~\bibnamefont{Banerjee}},
  \bibinfo{author}{\bibfnamefont{M.~J.}~\bibnamefont{Higgins}},
  \bibinfo{author}{\bibfnamefont{S.}~\bibnamefont{Ramakrishnan}}, and
  \bibinfo{author}{\bibfnamefont{A.~K.}~\bibnamefont{Grover}},
  \bibinfo{year}{2005},
  \bibinfo{journal}{Phys. Rev. B} \textbf{\bibinfo{volume}{72}},
  \bibinfo{pages}{134524}.


\bibitem[{\citenamefont{Thompson and Hu}(1971)}]
{Thompson-Hu71}
\bibinfo{author}{\bibnamefont{Thompson}, \bibfnamefont{R.~S.}}, and
\bibinfo{author}{\bibfnamefont{C.~R.} \bibnamefont{Hu}},
  \bibinfo{year}{1971},
  \bibinfo{journal}{Phys. Rev. Lett.} \textbf{\bibinfo{volume}{27}},
  \bibinfo{pages}{1352}.


\bibitem[{\citenamefont{Thouless}(1975)}]
{Thouless75}
\bibinfo{author}{\bibnamefont{Thouless}, \bibfnamefont{D.~J.}},
  \bibinfo{year}{1975},
  \bibinfo{journal}{Phys. Rev. Lett.} \textbf{\bibinfo{volume}{34}},
  \bibinfo{pages}{946}.

\bibitem[{\citenamefont{Tinh and Rosenstein}(2009)}]
{Tinh09}
\bibinfo{author}{\bibnamefont{Tinh}, \bibfnamefont{B.~-D.}}, and
\bibinfo{author}{\bibfnamefont{B.} \bibnamefont{Rosenstein}},
  \bibinfo{year}{2009},
  \bibinfo{journal}{Phys. Rev. B} \textbf{\bibinfo{volume}{79}},
  \bibinfo{pages}{024518}.

\bibitem[{\citenamefont{Tinkham}(1996)}]{Tinkham}
\bibinfo{author}{\bibnamefont{Tinkham}, \bibfnamefont{M.}},
  \bibinfo{year}{1996},
\emph{\bibinfo{title}{Introduction to Superconductivity}}
  (\bibinfo{publisher}{McGraw - Hill, New York }).


\bibitem[{\citenamefont{Troy and Dorsey}(1993)}]
{Dorsey93}
\bibinfo{author}{\bibnamefont{Troy}, \bibfnamefont{R.~J.}}, and
\bibinfo{author}{\bibfnamefont{A.~T.} \bibnamefont{Dorsey}},
  \bibinfo{year}{1993},
  \bibinfo{journal}{Phys. Rev. B} \textbf{\bibinfo{volume}{47}},
  \bibinfo{pages}{2715}.




  \bibitem[{\citenamefont{Ullah and Dorsey}(1990)}]
{Ullah90}
\bibinfo{author}{\bibnamefont{Ullah}, \bibfnamefont{S.}}, and
\bibinfo{author}{\bibfnamefont{A.~T.} \bibnamefont{Dorsey}},
  \bibinfo{year}{1990},
  \bibinfo{journal}{Phys. Rev. Lett.} \textbf{\bibinfo{volume}{65}},
  \bibinfo{pages}{2066}.
  
  \bibitem[{\citenamefont{Ullah and Dorsey}(1991)}]
{Ullah91}
\bibinfo{author}{\bibnamefont{Ullah}, \bibfnamefont{S.}}, and
\bibinfo{author}{\bibfnamefont{A.~T.} \bibnamefont{Dorsey}},
  \bibinfo{year}{1991},
  \bibinfo{journal}{Phys. Rev. B} \textbf{\bibinfo{volume}{44}},
  \bibinfo{pages}{262}.


\bibitem[{\citenamefont{Ussishkin \emph{et~al.}}(2002)}]
  {Ussishkin02}
  \bibinfo{author}{\bibfnamefont{Ussishkin}, \bibnamefont{I.}},
  \bibinfo{author}{\bibfnamefont{S.~L.}~\bibnamefont{Sondhi}}, and
  \bibinfo{author}{\bibfnamefont{D.~A.}~\bibnamefont{Huse}},
  \bibinfo{year}{2002},
  \bibinfo{journal}{Phys. Rev. Lett. } \textbf{\bibinfo{volume}{89}},
  \bibinfo{pages}{287001}.
  
  \bibitem[{\citenamefont{Ussishkin}(2003)}]
  {Ussishkin03}
  \bibinfo{author}{\bibfnamefont{Ussishkin}, \bibnamefont{I.}},
  \bibinfo{year}{2003},
  \bibinfo{journal}{Phys. Rev. B } \textbf{\bibinfo{volume}{68}},
  \bibinfo{pages}{024517}.


\bibitem[{\citenamefont{van Otterlo \emph{et~al.}}(1998)}]
  {Otterlo98}
  \bibinfo{author}{\bibfnamefont{van Otterlo}, \bibnamefont{A.}},
  \bibinfo{author}{\bibfnamefont{R.~T.}~\bibnamefont{Scalettar}}, and
  \bibinfo{author}{\bibfnamefont{G.~T.}~\bibnamefont{Zimanyi}},
  \bibinfo{year}{1998},
  \bibinfo{journal}{Phys. Rev. Lett. } \textbf{\bibinfo{volume}{81}},
  \bibinfo{pages}{1497}.

  \bibitem[{\citenamefont{Wang \emph{et~al.}}(2002)}]
  {Wang02}
  \bibinfo{author}{\bibfnamefont{Wang}, \bibnamefont{Y.}},
  \bibinfo{author}{\bibfnamefont{N.~P.}~\bibnamefont{Ong}},
  \bibinfo{author}{\bibfnamefont{Z.~A.}~\bibnamefont{Xu}},
   \bibinfo{author}{\bibfnamefont{T.}~\bibnamefont{Kakeshita}},
  \bibinfo{author}{\bibfnamefont{S.}~\bibnamefont{Uchida}}, 
 \bibinfo{author}{\bibfnamefont{D.~A.}~\bibnamefont{Bonn}},
 \bibinfo{author}{\bibfnamefont{R.}~\bibnamefont{Liang}},
 and
  \bibinfo{author}{\bibfnamefont{W.~N.}~\bibnamefont{Hardy}},
  \bibinfo{year}{2002},
  \bibinfo{journal}{Phys. Rev. Lett.} \textbf{\bibinfo{volume}{88}},
  \bibinfo{pages}{257003}.
  
  \bibitem[{\citenamefont{Wang \emph{et~al.}}(2006)}]
  {Wang06}
  \bibinfo{author}{\bibfnamefont{Wang}, \bibnamefont{Y.}},
 \bibinfo{author}{\bibfnamefont{L.}~\bibnamefont{Li}},
 and
  \bibinfo{author}{\bibfnamefont{N.~P.}~\bibnamefont{Ong}},
  \bibinfo{year}{2006},
  \bibinfo{journal}{Phys. Rev. B} \textbf{\bibinfo{volume}{73}},
  \bibinfo{pages}{024510}.





  \bibitem[{\citenamefont{Welp \emph{et~al.}}(1991)}]
  {Welp91}
  \bibinfo{author}{\bibfnamefont{Welp}, \bibnamefont{U.}},
  \bibinfo{author}{\bibfnamefont{S.}~\bibnamefont{Fleshler}},
  \bibinfo{author}{\bibfnamefont{W.~K.}~\bibnamefont{Kwok}},
  \bibinfo{author}{\bibfnamefont{R.~A.}~\bibnamefont{Klemm}},
  \bibinfo{author}{\bibfnamefont{V.~M.}~\bibnamefont{Vinokur}},
  \bibinfo{author}{\bibfnamefont{J.}~\bibnamefont{Downey}},
  \bibinfo{author}{\bibfnamefont{B.}~\bibnamefont{Veal}}, and
  \bibinfo{author}{\bibfnamefont{G.~W.}~\bibnamefont{Crabtree}},
  \bibinfo{year}{1991},
  \bibinfo{journal}{Phys. Rev. Lett.} \textbf{\bibinfo{volume}{67}},
  \bibinfo{pages}{3180}.

\bibitem[{\citenamefont{Welp \emph{et~al.}}(1996)}]
  {Welp96}
  \bibinfo{author}{\bibfnamefont{Welp}, \bibnamefont{U.}},
  \bibinfo{author}{\bibfnamefont{J.~A.}~\bibnamefont{Fendrich}},
  \bibinfo{author}{\bibfnamefont{W.~K.}~\bibnamefont{Kwok}},
  \bibinfo{author}{\bibfnamefont{G.~W.}~\bibnamefont{Crabtree}}, and
  \bibinfo{author}{\bibfnamefont{B.~W.}~\bibnamefont{Veal}},
  \bibinfo{year}{1996},
  \bibinfo{journal}{Phys. Rev. Lett.} \textbf{\bibinfo{volume}{76}},
  \bibinfo{pages}{4809}.

  \bibitem[{\citenamefont{Wilkin and  Moore}(1993)}]
  {Wilkin93}
  \bibinfo{author}{\bibfnamefont{Wilkin}, \bibnamefont{N.~K.}}, and
  \bibinfo{author}{\bibfnamefont{M.~A.}~\bibnamefont{Moore}},
  \bibinfo{year}{1993},
  \bibinfo{journal}{Phys. Rev. B} \textbf{\bibinfo{volume}{47}},
  \bibinfo{pages}{957}.

  \bibitem[{\citenamefont{Wilkin and Jensen}(1997)}]
  {Wilkin97}
  \bibinfo{author}{\bibfnamefont{Wilkin}, \bibnamefont{N.~K.}}, and
  \bibinfo{author}{\bibfnamefont{H.~J.}~\bibnamefont{Jensen}},
  \bibinfo{year}{1997},
  \bibinfo{journal}{Phys. Rev. Lett.} \textbf{\bibinfo{volume}{79}},
  \bibinfo{pages}{4254}.

  \bibitem[{\citenamefont{ Willemin \emph{et~al.}}(1998)}]
  {Willemin98}
  \bibinfo{author}{\bibfnamefont{Willemin}, \bibnamefont{M.}},
  \bibinfo{author}{\bibfnamefont{A.}~\bibnamefont{Schilling}},
  \bibinfo{author}{\bibfnamefont{H.}~\bibnamefont{Keller}},
  \bibinfo{author}{\bibfnamefont{C.}~\bibnamefont{Rossel}},
  \bibinfo{author}{\bibfnamefont{J.}~\bibnamefont{Hofer}},
  \bibinfo{author}{\bibfnamefont{U.}~\bibnamefont{Welp}},
  \bibinfo{author}{\bibfnamefont{W.~K.}~\bibnamefont{Kwok}},
  \bibinfo{author}{\bibfnamefont{R.~J.}~\bibnamefont{Olsson}},  and
  \bibinfo{author}{\bibfnamefont{G.~W.}~\bibnamefont{Crabtree}},
  \bibinfo{year}{1998},
  \bibinfo{journal}{Phys. Rev. Lett.} \textbf{\bibinfo{volume}{81}},
  \bibinfo{pages}{4236}.

 \bibitem[{\citenamefont{Wu \emph{et~al.}}(2007)}]
  {Wu07}
  \bibinfo{author}{\bibfnamefont{Wu}, \bibnamefont{Z.}},
  \bibinfo{author}{\bibfnamefont{B.}~\bibnamefont{Feng}}, and
  \bibinfo{author}{\bibfnamefont{D.}~\bibnamefont{Li}},
  \bibinfo{year}{2007},
  \bibinfo{journal}{Phys. Rev. A} \textbf{\bibinfo{volume}{75}},
  \bibinfo{pages}{033620}.


\bibitem[{\citenamefont{Xiao \emph{et~al.}}(2002)}]
  {Xiao02}
  \bibinfo{author}{\bibfnamefont{Xiao}, \bibnamefont{Z.~L.}},
  \bibinfo{author}{\bibfnamefont{E.~Y.}~\bibnamefont{Andrei}},
  \bibinfo{author}{\bibfnamefont{Y.}~\bibnamefont{Paltiel}},
  \bibinfo{author}{\bibfnamefont{E.}~\bibnamefont{Zeldov}},
  \bibinfo{author}{\bibfnamefont{P.}~\bibnamefont{Shuk}}, and
  \bibinfo{author}{\bibfnamefont{M.}~\bibnamefont{Greenblatt}},
  \bibinfo{year}{2002},
  \bibinfo{journal}{Phys. Rev. B} \textbf{\bibinfo{volume}{65}},
  \bibinfo{pages}{094511}.



  \bibitem[{\citenamefont{Xiao \emph{et~al.}}(2004)}]
  {Xiao04}
  \bibinfo{author}{\bibfnamefont{Xiao}, \bibnamefont{Z.~L.}},
  \bibinfo{author}{\bibfnamefont{O.}~\bibnamefont{Dogru}},
  \bibinfo{author}{\bibfnamefont{E.~Y.}~\bibnamefont{Andrei}},
  \bibinfo{author}{\bibfnamefont{P.}~\bibnamefont{Shuk}}, and
  \bibinfo{author}{\bibfnamefont{M.}~\bibnamefont{Greenblatt}},
  \bibinfo{year}{2004},
  \bibinfo{journal}{Phys. Rev. Lett.} \textbf{\bibinfo{volume}{92}},
  \bibinfo{pages}{227004}.

  \bibitem[{\citenamefont{Yeo and Moore}(1996a)}]
  {Yeo96}
   \bibinfo{author}{\bibfnamefont{Yeo}, \bibnamefont{J.}}, and
  \bibinfo{author}{\bibfnamefont{M.~A.}~\bibnamefont{Moore}},
  \bibinfo{year}{1996a},
  \bibinfo{journal}{Phys. Rev. Lett.} \textbf{\bibinfo{volume}{76}},
  \bibinfo{pages}{1142}.


  \bibitem[{\citenamefont{Yeo and Moore}(1996b)}]
  {Yeo96b}
   \bibinfo{author}{\bibfnamefont{Yeo}, \bibnamefont{J.}}, and
  \bibinfo{author}{\bibfnamefont{M.~A.}~\bibnamefont{Moore}},
  \bibinfo{year}{1996b},
  \bibinfo{journal}{Phys. Rev. B} \textbf{\bibinfo{volume}{54}},
  \bibinfo{pages}{4218}.

\bibitem[{\citenamefont{Yeo and Moore}(2001)}]
  {Yeo01}
   \bibinfo{author}{\bibfnamefont{Yeo}, \bibnamefont{J.}}, and
  \bibinfo{author}{\bibfnamefont{M.~A.}~\bibnamefont{Moore}},
  \bibinfo{year}{2001},
  \bibinfo{journal}{Phys. Rev. B} \textbf{\bibinfo{volume}{64}},
  \bibinfo{pages}{024514}.

\bibitem[{\citenamefont{Yeo \emph{et~al.}}(2006)}]
  {Yeo06}
   \bibinfo{author}{\bibfnamefont{Yeo}, \bibnamefont{J.}}, 
   \bibinfo{author}{\bibfnamefont{H.}~\bibnamefont{Park}},
   and
  \bibinfo{author}{\bibfnamefont{S.}~\bibnamefont{Yi}},
  \bibinfo{year}{2006},
  \bibinfo{journal}{J. Phys. Cond. Mat.} \textbf{\bibinfo{volume}{18}},
  \bibinfo{pages}{3607}.
 

  

\bibitem[{\citenamefont{Zeldov \emph{et~al.}}(1995)}]
  {Zeldov95}
  \bibinfo{author}{\bibfnamefont{Zeldov}, \bibnamefont{E.}},
  \bibinfo{author}{\bibfnamefont{D.}~\bibnamefont{Majer}},
  \bibinfo{author}{\bibfnamefont{M.}~\bibnamefont{Konczykowski}},
  \bibinfo{author}{\bibfnamefont{V.~B.}~\bibnamefont{Geshkenbein}},
  \bibinfo{author}{\bibfnamefont{V.~M.}~\bibnamefont{Vinokur}}, and
  \bibinfo{author}{\bibfnamefont{H.}~\bibnamefont{Shtrikman}},
  \bibinfo{year}{1995},
  \bibinfo{journal}{Nature (London)} \textbf{\bibinfo{volume}{375}},
  \bibinfo{pages}{373}.


  \bibitem[{\citenamefont{Zhou \emph{et~al.}}(1993)}]
  {Zhou93}
  \bibinfo{author}{\bibnamefont{Zhou}, \bibfnamefont{B.}},
  \bibinfo{author}{\bibfnamefont{J.}~\bibnamefont{Buan}},
  \bibinfo{author}{\bibfnamefont{S.~W.}~\bibnamefont{Pierson}},
  \bibinfo{author}{\bibfnamefont{C.~C.}~\bibnamefont{Huang}},
  \bibinfo{author}{\bibfnamefont{O.~T.}~\bibnamefont{Valls}},
  \bibinfo{author}{\bibfnamefont{J.~Z.}~\bibnamefont{Liu}},  and
  \bibinfo{author}{\bibfnamefont{R.-N.}~\bibnamefont{Shelton}},
  \bibinfo{year}{1993},
  \bibinfo{journal}{Phys. Rev. B} \textbf{\bibinfo{volume}{47}},
  \bibinfo{pages}{11631}.

 \bibitem[{\citenamefont{Zhuravlev and  Maniv}(1999)}]
  {Zhuravlev99}
  \bibinfo{author}{\bibfnamefont{Zhuravlev}, \bibnamefont{V.}}, and
  \bibinfo{author}{\bibfnamefont{T}~\bibnamefont{Maniv}},
  \bibinfo{year}{1999},
  \bibinfo{journal}{Phys. Rev. B} \textbf{\bibinfo{volume}{60}},
  \bibinfo{pages}{4277}.

\bibitem[{\citenamefont{Zhuravlev and  Maniv}(2002)}]
  {Zhuravlev02}
  \bibinfo{author}{\bibfnamefont{Zhuravlev}, \bibnamefont{V.}}, and
  \bibinfo{author}{\bibfnamefont{T}~\bibnamefont{Maniv}},
  \bibinfo{year}{2002},
  \bibinfo{journal}{Phys. Rev. B} \textbf{\bibinfo{volume}{66}},
  \bibinfo{pages}{014529}.







\end{thebibliography}
\end{document}